\documentclass[11pt]{article}
\usepackage{amsbsy,amsfonts,amsmath,amssymb,amsthm,bbm,enumerate,euscript,
 graphicx}

\theoremstyle{plain}
\numberwithin{equation}{section}
\newtheorem{thm}{Theorem}[section]

\newtheorem{defn}[thm]{Definition}

\newtheorem{lmm}[thm]{Lemma}
\newtheorem{prp}[thm]{Proposition}

\newcommand{\bb}{\underline{b}}

\newcommand{\bd}{\partial}

\newcommand{\Bigcup}{\mathop{\Dot{\bigcup}}}
\newcommand{\bh}{{\bf h}}
\newcommand{\bk}{{\bf k}}
\newcommand{\bm}{{\bf m}}
\newcommand{\bN}{{\bf N}}
\newcommand{\bn}{{\bf n}}

\newcommand{\cA}{{\cal A}}
\newcommand{\cB}{{\cal B}}
\newcommand{\cC}{{\cal C}}
\newcommand{\cD}{{\cal D}}
\newcommand{\cE}{{\cal E}}

\newcommand{\cL}{{\cal L}}

\newcommand{\cS}{{\cal S}}
\newcommand{\cU}{{\cal U}}
\newcommand{\cV}{{\cal V}}

\newcommand{\compl}{^{\rm c}}

\newcommand{\Dcup}{\:\Dot{\cup}\:}
\newcommand{\dpst}{\displaystyle}

\newcommand{\eb}{\underline{e}}
\newcommand{\eps}{\epsilon}
\newcommand{\Exists}{{}^\exists}
\newcommand{\Exp}[1]{{\langle #1 \rangle}}

\newcommand{\fS}{\mathfrak{S}}
\newcommand{\Forall}{{}^\forall}

\newcommand{\ind}[1]{\mathbbm{1}{\raisebox{-2pt}{$\scriptstyle \{#1\}$}}}
\newcommand{\indic}[1]{\mathbbm{1}{\raisebox{-2pt}{$\scriptstyle #1$}}}

\newcommand{\lbeq}[1]{\label{eq:#1}}

\newcommand{\mB}{{\mathbb B}}
\newcommand{\mE}{{\mathbb E}}
\newcommand{\mG}{{\mathbb G}}

\newcommand{\mP}{{\mathbb P}}
\newcommand{\mR}{{\mathbb R}}
\newcommand{\mS}{{\mathbb S}}
\newcommand{\mZ}{{\mathbb Z}}

\newcommand{\nn}{\nonumber}

\newcommand{\pc}{p_\text{c}}

\newcommand{\refeq}[1]{(\ref{eq:#1})}
\newcommand{\Rd}{{\mathbb R}^d}

\newcommand{\sss}{\scriptscriptstyle}

\newcommand{\tb}{\overline{b}}
\newcommand{\tbp}{\tb^{\raisebox{-2pt}{$\scriptstyle\prime$}}}

\newcommand{\tbps}{\tb^{\raisebox{-2pt}{$\scriptscriptstyle\prime$}}}

\newcommand{\te}{\overline{e}}
\newcommand{\tind}[1]{\mathbbm{1}{\scriptstyle\{#1\}}}
\newcommand{\tri}{\,\triangle\,}

\newcommand{\vb}{|\!|\!|}
\newcommand{\veee}[1]{\vb#1\vb}

\newcommand{\vno}{\varnothing}
\newcommand{\vtri}{\vartriangle}

\newcommand{\Zd}{{\mathbb Z}^d}
\newcommand{\Zp}{{\mathbb Z}_+}

\newcommand{\cn}[2]{\underset{\raisebox{5pt}{$\sss #1$}}
 {\overset{#2}{\longleftrightarrow}}}
\newcommand{\db}[2]{\underset{\raisebox{5pt}{$\sss #1$}}
 {\overset{#2}{\Longleftrightarrow}}}

\title{Lace expansion for the Ising model}
\author{Akira~Sakai\footnote{Department of Mathematical Sciences,
University of Bath, Bath BA2 7AY, UK.  {\tt a.sakai@bath.ac.uk}}}
\date{October 26, 2005\footnote{Updated: November 13, 2006}}

\begin{document}
\maketitle

\begin{abstract}
The lace expansion has been a powerful tool for investigating mean-field behavior for
various stochastic-geometrical models, such as self-avoiding walk and percolation, above
their respective upper-critical dimension. In this paper, we prove the lace expansion
for the Ising model that is valid for any spin-spin coupling.  For the ferromagnetic
case, we also prove that the expansion coefficients obey certain diagrammatic bounds
that are similar to the diagrammatic bounds on the lace-expansion coefficients for
self-avoiding walk.  As a result, we obtain Gaussian asymptotics of the critical
two-point function for the nearest-neighbor model with $d\gg4$ and for the spread-out
model with $d>4$ and $L\gg1$, without assuming reflection positivity.
\end{abstract}

\tableofcontents

\section{Introduction and results}

\subsection{Model and the motivation}\label{ss:model}
The Ising model is a statistical-mechanical model that was first introduced
in \cite{i25} as a model for magnets.  Consider the $d$-dimensional integer
lattice $\Zd$, and let $\Lambda$ be a finite subset of $\Zd$ containing the
origin $o\in\Zd$.  For example, $\Lambda$ is a $d$-dimensional hypercube
centered at the origin.  At each site $x\in\Lambda$, there is a spin variable
$\varphi_x$ that takes values either $+1$ or $-1$.  The Hamiltonian represents
the energy of the system, and is defined by
\begin{align}\lbeq{hamilton}
H^h_\Lambda(\varphi)=-\sum_{\{x,y\}\subset\Lambda}J_{x,y}\varphi_x\varphi_y
 -h\sum_{x\in\Lambda}\varphi_x,
\end{align}
where $\varphi\equiv\{\varphi_x\}_{x\in\Lambda}$ is a spin configuration,
$\{J_{x,y}\}_{x,y\in\Zd}$ is a collection of spin-spin couplings,
and $h\in\mR$ represents the strength of an external magnetic field uniformly
imposed on $\Lambda$.  We say that the model is ferromagnetic if $J_{x,y}\ge0$
for all pairs $\{x,y\}$; in this case, the Hamiltonian becomes lower as more
spins align.  The partition function $Z_{p,h;\Lambda}$ at the inverse
temperature $p\ge0$ is the expectation of the Boltzmann factor
$e^{-pH^h_\Lambda(\varphi)}$ with respect to the
product measure
$\prod_{x\in\Lambda}(\frac12\tind{\varphi_x=+1}+\frac12\tind{\varphi_x=-1})$:
\begin{align}\lbeq{ZL-def}
Z_{p,h;\Lambda}=2^{-|\Lambda|}\sum_{\varphi\in\{\pm1\}^\Lambda}e^{-pH^h_\Lambda
 (\varphi)}.
\end{align}
Then, we denote the thermal average of a function $f=f(\varphi)$ by
\begin{align}\lbeq{faver}
\Exp{f}_{p,h;\Lambda}=\frac{2^{-|\Lambda|}}{Z_{p,h;\Lambda}}\sum_{\varphi\in
 \{\pm1\}^\Lambda}f(\varphi)\,e^{-pH^h_\Lambda(\varphi)}.
\end{align}

Suppose that the spin-spin coupling is translation-invariant, $\Zd$-symmetric
and finite-range (i.e., there exists an $L<\infty$ such that $J_{o,x}=0$ if
$\|x\|_\infty>L$) and that $J_{o,x}\ge0$ for any
$x\in\Zd$ and $h\ge0$.  Then, there exist monotone infinite-volume limits
of $\Exp{\varphi_x}_{p,h;\Lambda}$ and
$\Exp{\varphi_x\varphi_y}_{p,h;\Lambda}$.  Let
\begin{align}
M_{p,h}=\lim_{\Lambda\uparrow\Zd}\Exp{\varphi_o}_{p,h;\Lambda},&&
G_p(x)=\lim_{\Lambda\uparrow\Zd}\Exp{\varphi_o\varphi_x}_{p,h=0;
 \Lambda},&&
\chi_p=\sum_{x\in\Zd}G_p(x).
\end{align}
When $d\ge2$, there exists a unique critical inverse temperature
$\pc\in(0,\infty)$ such that the spontaneous magnetization
$M^+_p\equiv\lim_{h\downarrow0}M_{p,h}$ equals zero, $G_p(x)$ decays
exponentially as $|x|\uparrow\infty$ (we refer, e.g., to \cite{civ03} for
a sharper Ornstein-Zernike result) and thus the magnetic susceptibility
$\chi_p$ is finite if $p<\pc$, while $M^+_p>0$ and $\chi_p=\infty$ if
$p>\pc$ (see \cite{abf87} and references therein).  We should also
refer to \cite{b05b} for recent results on the phase transition for
the Ising model.

We are interested in the behavior of these observables around $p=\pc$.
The susceptibility $\chi_p$ is known to diverge as $p\uparrow\pc$
\cite{a82,ag83}.  It is generally expected that
$\lim_{p\downarrow\pc}M^+_p=\lim_{h\downarrow0}M_{\pc,h}=0$.  We
believe that there are so-called critical
exponents $\gamma=\gamma(d)$, $\beta=\beta(d)$ and
$\delta=\delta(d)$, which are insensitive to the precise definition
of $J_{o,x}\ge0$ (universality), such that (we use below the limit
notation ``$\approx$'' in some appropriate sense)
\begin{align}
M^+_p\stackrel{p\downarrow\pc}{\approx}(p-\pc)^\beta,&&
\chi_p\stackrel{p\uparrow\pc}{\approx}(\pc-p)^{-\gamma},&&
M_{\pc,h}\stackrel{h\downarrow0}{\approx}h^{1/\delta}.
\end{align}
These exponents (if they exist) are known to obey the mean-field
bounds: $\beta\leq1/2$, $\gamma\ge1$ and $\delta\ge3$.  For
example, $\beta=1/8$, $\gamma=7/4$ and $\delta=15$ for the
nearest-neighbor model on $\mZ^2$ \cite{o44}.  Our ultimate goal is
to identify the values of the critical exponents in other dimensions
and to understand the universality for the Ising model.

There is a sufficient condition, the so-called bubble condition, for
the above critical exponents to take on their respective mean-field
values.  Namely, the finiteness of $\sum_{x\in\Zd}G_{\pc}(x)^2$ (or the
finiteness of $\sum_{x\in\Zd}G_p(x)^2$ uniformly in $p<\pc$) implies
that $\beta=1/2$, $\gamma=1$ and $\delta=3$ \cite{a82,abf87,af86,ag83}.
It is therefore crucial to know how fast $G_{\pc}(x)$ (or $G_p(x)$
near $p=\pc$) decays as $|x|\uparrow\infty$.  We note that the bubble
condition holds for $d>4$ if the anomalous dimension $\eta$ takes on
its mean-field value $\eta=0$, where the anomalous dimension is another
critical exponent formally defined as
\begin{align}\lbeq{eta-formal}
G_{\pc}(x)\stackrel{|x|\uparrow\infty}{\approx}|x|^{-(d-2+\eta)}.
\end{align}

Let $\hat J_k=\sum_{x\in\Zd}J_{o,x}\,e^{ik\cdot x}$ and
$\hat G_p(k)=\sum_{x\in\Zd}G_p(x)\,e^{ik\cdot x}$ for $p<\pc$.  For a class
of models that satisfy the so-called reflection positivity \cite{fss76}, the
following infrared bound\footnote{In \refeq{IRbd-so} and \refeq{IRbd-sokal},
we also use the fact that, for $p<\pc$, our $G_p$ (i.e., the infinite-volume
limit of the two-point function under the free-boundary condition) is equal
to the infinite-volume limit of the two-point function under the
periodic-boundary condition.} holds:
\begin{align}\lbeq{IRbd-so}
0\leq\hat G_p(k)\leq\frac{\text{const.}}{\hat J_0-\hat J_k}\qquad
 \text{uniformly in }p<\pc,
\end{align}
where $d$ is supposed to be large enough to ensure integrability of the
upper bound.  For finite-range models, $d$ has to be bigger than
2, since $\hat J_0-\hat J_k\asymp|k|^2$, where ``$f\asymp g$'' means
that $f/g$ is bounded away from zero and infinity.  By Parseval's identity,
the infrared bound \refeq{IRbd-so} implies the bubble condition for
finite-range reflection-positive models above four dimensions, and
therefore
\begin{align}\lbeq{MFbehavior}
M^+_p\stackrel{p\downarrow\pc}{\asymp}(p-\pc)^{1/2},&&
\chi_p\stackrel{p\uparrow\pc}{\asymp}(\pc-p)^{-1},&&
M_{\pc,h}\stackrel{h\downarrow0}{\asymp}h^{1/3}.
\end{align}
The class of reflection-positive models includes the nearest-neighbor
model, a variant of the next-nearest-neighbor model, Yukawa potentials,
power-law decaying interactions, and their combinations \cite{bcc05}.
For the nearest-neighbor model, we further obtain the following
$x$-space Gaussian bound \cite{s82}: for $x\ne o$,
\begin{align}\lbeq{IRbd-sokal}
G_p(x)\leq\frac{\text{const.}}{|x|^{d-2}}\qquad\text{uniformly in }p<\pc.
\end{align}

The problem in this approach to investigate critical behavior is that, since general
finite-range models do not always satisfy reflection positivity, their mean-field
behavior cannot necessarily be established, even in high dimensions.  If we believe in
universality, we expect that finite-range models exhibit the same mean-field behavior as
soon as $d>4$.  Therefore, it has been desirable to have approaches that do not assume
reflection positivity.

The lace expansion has been used successfully to investigate mean-field behavior for
self-avoiding walk, percolation, lattice trees/animals and the contact process, above
the upper-critical dimension: 4, 6 (4 for oriented percolation), 8 and 4, respectively
(see, e.g., \cite{s04}).  One of the advantages in the application of the lace expansion
is that we do not have to require reflection positivity to prove a Gaussian infrared
bound and mean-field behavior.  Another advantage is the possibility to show an
asymptotic result for the decay of correlation.  Our goal in this paper is to prove the
lace-expansion results for the Ising model.

\subsection{Main results}
From now on, we fix $h=0$ and abbreviate, e.g.,
$\Exp{\varphi_o\varphi_x}_{p,h=0;\Lambda}$ to
$\Exp{\varphi_o\varphi_x}_{p;\Lambda}$.  In this paper, we prove the
following lace expansion for the two-point function, in which we use
the notation
\begin{align}
\tau_{x,y}=\tanh(pJ_{x,y}).
\end{align}

\begin{prp}\label{prp:Ising-lace}
For any $p\ge0$ and any $\Lambda\subset\Zd$, there exist
$\pi_{p;\Lambda}^{\sss(j)}(x)$ and $R_{p;\Lambda}^{\sss(j+1)}(x)$
for $x\in\Lambda$ and $j\ge0$ such that
\begin{align}\lbeq{Ising-lace}
\Exp{\varphi_o\varphi_x}_{p;\Lambda}=\Pi_{p;\Lambda}^{\sss(j)}(x)
 +\sum_{u,v}\Pi_{p;\Lambda}^{\sss(j)}(u)\,\tau_{u,v}\Exp{
 \varphi_v\varphi_x}_{p;\Lambda}+(-1)^{j+1}R_{p;\Lambda}^{\sss(j
 +1)}(x),
\end{align}
where
\begin{align}\lbeq{Pij-def}
\Pi_{p;\Lambda}^{\sss(j)}(x)&=\sum_{i=0}^j(-1)^i\,\pi_{p;
 \Lambda}^{\sss(i)}(x).
\end{align}
For the ferromagnetic case, we have the bounds
\begin{align}\lbeq{pij-Rj-naivebd}
\pi_{p;\Lambda}^{\sss(j)}(x)\ge\delta_{j,0}\delta_{o,x},&&
0\leq R_{p;\Lambda}^{\sss(j+1)}(x)\leq\sum_{u,v}\pi_{p;\Lambda}
 ^{\sss(j)}(u)\,\tau_{u,v}\Exp{\varphi_v\varphi_x}_{p;\Lambda}.
\end{align}
\end{prp}

We defer the display of precise expressions of $\pi_{p;\Lambda}^{\sss(i)}(x)$ and
$R_{p;\Lambda}^{\sss(j+1)}(x)$ to Section~\ref{sss:complexp}, since we need a certain
representation to describe these functions.  We introduce this representation in
Section~\ref{ss:RCrepr} and complete the proof of Proposition~\ref{prp:Ising-lace} in
Section~\ref{ss:derivation}.

It is worth emphasizing that the above proposition holds independently
of the properties of the spin-spin coupling: $J_{u,v}$ does not have
to be translation-invariant or $\Zd$-symmetric.  In particular, the
identity \refeq{Ising-lace} holds independently of the sign of the
spin-spin coupling.  A spin glass, whose spin-spin coupling is randomly
negative, is an extreme example for which \refeq{Ising-lace} holds.

Whether or not the lace expansion \refeq{Ising-lace} is useful
depends on the possibility of good control on the expansion
coefficients and the remainder. As explained below, it is indeed
possible to have optimal bounds on the expansion coefficients for
the nearest-neighbor interaction (i.e., $J_{o,x}=\tind{\|x\|_1=1}$)
and for the following spread-out interaction:
\begin{align}\lbeq{J-def}
J_{o,x}=L^{-d}\mu(L^{-1}x)\qquad(1\leq L<\infty),
\end{align}
where $\mu:[-1,1]^d\setminus\{o\}\mapsto[0,\infty)$ is a bounded
probability distribution, which is symmetric under rotations by
$\pi/2$ and reflections in coordinate hyperplanes, and piecewise
continuous so that the Riemann sum
$L^{-d}\sum_{x\in\Zd}\mu(L^{-1}x)$ approximates
$\int_{\Rd}d^dx\;\mu(x)\equiv1$.  One of the simplest examples would
be
\begin{align}\lbeq{Juniform-def}
J_{o,x}=\frac{\ind{0<\|x\|_\infty\leq L}}{\sum_{z\in\Zd}\ind{0<\|z
 \|_\infty\leq L}}=O(L^{-d})\,\ind{0<\|L^{-1}x\|_\infty\leq1}.
\end{align}

\begin{prp}\label{prp:Pij-Rj-bd}
Let $\rho=2(d-4)>0$.  For the nearest-neighbor model with $d\gg1$ and
for the spread-out model with $L\gg1$, there are finite constants
$\theta$ and $\lambda$ such that
\begin{align}\lbeq{prp-bds}
|\Pi_{p;\Lambda}^{\sss(j)}(x)-\delta_{o,x}|&\leq\theta\delta_{o,x}
 +\frac{\lambda(1-\delta_{o,x})}{|x|^{d+2+\rho}}\quad(j\ge0),&&
|R_{p;\Lambda}^{\sss(j)}(x)|\to0\quad(j\uparrow\infty),
\end{align}
for any $p\leq\pc$, any $\Lambda\subset\Zd$ and any $x\in\Lambda$.
\end{prp}

The proof of Proposition~\ref{prp:Pij-Rj-bd} depends on certain bounds on the expansion
coefficients in terms of two-point functions.  These diagrammatic bounds arise from
counting the number of ``disjoint connections'', corresponding to applications of the BK
inequality in percolation (e.g., \cite{bk85}).  We prove these bounds in
Section~\ref{s:bounds}, and in anticipation of this, in Section~\ref{s:reduction} we
explain how we use their implication to prove Proposition~\ref{prp:Pij-Rj-bd}, with
$\theta=O(d^{-1})$ and $\lambda=O(1)$ for the nearest-neighbor model, and
$\theta=O(L^{-2+\eps})$ and $\lambda=O(\theta^2)$ with a small $\eps>0$ for the
spread-out model.

Let
\begin{align}
\tau\equiv\tau(p)=\sum_x\tau_{o,x},&&
D(x)=\frac{\tau_{o,x}}{\tau},&&
\sigma^2=\sum_x|x|^2D(x).
\end{align}
Due to \refeq{prp-bds} uniformly in $\Lambda\subset\Zd$, there is
a limit $\Pi_p(x)\equiv\lim_{\Lambda\uparrow\Zd}
\lim_{j\uparrow\infty}\Pi_{p;\Lambda}^{\sss(j)}(x)$ such that
\begin{align}\lbeq{Ising-lace-Zdlim}
G_p(x)=\Pi_p(x)+(\Pi_p*\tau D*G_p)(x),&&
|\Pi_p(x)-\delta_{o,x}|\leq\theta\delta_{o,x}+\frac{\lambda(1
 -\delta_{o,x})}{|x|^{d+2+\rho}},
\end{align}
for any $p\leq\pc$ and any $x\in\Zd$, where
$(f*g)(x)=\sum_{y\in\Zd}f(y)\,g(x-y)$.  We note that the identity in
\refeq{Ising-lace-Zdlim} is similar to the recursion equation for the
random-walk Green's function:
\begin{align}
S_r(x)\equiv\sum_{i=0}^\infty r^iD^{*i}(x)=\delta_{o,x}+(rD*S_r)(x)
 \qquad(|r|<1),
\end{align}
where $f^{*i}(x)=(f^{*(i-1)}*f)(x)$, with $f^{*0}(x)=\delta_{o,x}$ by
convention.  The leading asymptotics of $S_1(x)$ for $d>2$ is known as
$\frac{a_d}{\sigma^2}|x|^{-(d-2)}$, where
$a_d=\frac{d}2\pi^{-d/2}\Gamma(\frac{d}2-1)$ (e.g., \cite{h05,hhs03}).
Following the model-independent analysis of the lace expansion in
\cite{h05,hhs03}, we obtain the following asymptotics of the critical
two-point function:

\begin{thm}\label{thm:x-asy}
Let $\rho=2(d-4)>0$ and fix any small $\eps>0$.  For the nearest-neighbor
model with $d\gg1$ and for the spread-out model with $L\gg1$, we have
that, for $x\ne o$,
\begin{align}\lbeq{thm-asy}
G_{\pc}(x)=\frac{A}{\tau(\pc)}\,\frac{a_d}{\sigma^2|x|^{d-2}}
 \times\begin{cases}
 \big(1+O(|x|^{-\frac{(\rho-\eps)\wedge2}d})\big)&(\text{NN model}),\\
 \big(1+O(|x|^{-\rho\wedge2+\eps})\big)
  &(\text{SO model}),
 \end{cases}
\end{align}
where constants in the error terms may vary depending on $\eps$, and
\begin{align}\lbeq{constants}
\tau(\pc)=\bigg(\sum_x\Pi_{\pc}(x)\bigg)^{-1},&&
A=\bigg(1+\frac{\tau(\pc)}{\sigma^2}\sum_x|x|^2\Pi_{\pc}(x)\bigg)^{-1}.
\end{align}
Consequently, \refeq{MFbehavior} holds and $\eta=0$.
\end{thm}

In this paper, we restrict ourselves to the nearest-neighbor model for $d\gg4$ and to
the spread-out model for $d>4$ with $L\gg1$.  However, it is strongly expected that our
method can show the same asymptotics of the critical two-point function for \emph{any}
translation-invariant, $\Zd$-symmetric finite-range model above four dimensions, by
taking the coordination number sufficiently large.

\subsection{Organization}
In the rest of this paper, we focus our attention on the
model-dependent ingredients: the lace expansion for the Ising model
(Proposition~\ref{prp:Ising-lace}) and the bounds on (the alternating
sum of) the expansion coefficients for the ferromagnetic models
(Proposition~\ref{prp:Pij-Rj-bd}).  In Section~\ref{s:laceexp}, we
prove Proposition~\ref{prp:Ising-lace}.  In Section~\ref{s:reduction},
we reduce Proposition~\ref{prp:Pij-Rj-bd} to a few other propositions,
which are then results of the aforementioned diagrammatic bounds on
the expansion coefficients.  We prove these diagrammatic bounds in
Section~\ref{s:bounds}.  As soon as the composition of the diagrams
in terms of two-point functions is understood, it is not so hard to
establish key elements of the above reduced propositions.  We will
prove these elements in Section~\ref{ss:proof-so} for the spread-out
model and in Section~\ref{ss:proof-nn} for the nearest-neighbor model.

\section{Lace expansion for the Ising model}\label{s:laceexp}
The lace expansion was initiated by Brydges and Spencer \cite{bs85}
to investigate weakly self-avoiding walk for $d>4$.  Later, it was
developed for various stochastic-geometrical models, such as strictly
self-avoiding walk for $d>4$ (e.g., \cite{hs92}), lattice trees/animals
for $d>8$ (e.g., \cite{hs90}), unoriented percolation for $d>6$ (e.g.,
\cite{hs90'}), oriented percolation for $d>4$ (e.g., \cite{ny93}) and
the contact process for $d>4$ (e.g., \cite{s01}).  See \cite{s04} for
an extensive list of references.  This is the first lace-expansion paper
for the Ising model.

In this section, we prove the lace expansion \refeq{Ising-lace} for
the Ising model.  From now on, we fix $p\ge0$ and abbreviate, e.g.,
$\pi_{p;\Lambda}^{\sss(i)}(x)$ to $\pi_\Lambda^{\sss(i)}(x)$.

There may be several ways to derive the lace expansion for
$\Exp{\varphi_o\varphi_x}_\Lambda$, using, e.g., the
high-temperature expansion, the random-walk representation (e.g.,
\cite{ffs92}) or the FK random-cluster representation (e.g.,
\cite{fk72}).  In this paper, we use the random-current
representation (Section~\ref{ss:RCrepr}), which applies to models in
the Griffiths-Simon class (e.g., \cite{a82,ag83}).  This
representation is similar in philosophy to the high-temperature
expansion, but it turned out to be more efficient in investigating
the critical phenomena \cite{a82,abf87,af86,ag83}.  The main
advantage in this representation is the source-switching lemma
(Lemma~\ref{lmm:switching} below in Section~\ref{sss:2ndexp}) by
which we have an identity for $\Exp{\varphi_o\varphi_x}_\Lambda
-\Exp{\varphi_o\varphi_x}_\cA$ with ``$\cA\subset\Lambda$'' (the
meaning will be explained in Section~\ref{ss:RCrepr}).  We will
repeatedly apply this identity to complete the lace expansion for
$\Exp{\varphi_o\varphi_x}_\Lambda$ in Section~\ref{sss:complexp}.

\subsection{Random-current representation}\label{ss:RCrepr}
In this subsection, we describe the random-current representation and
introduce some notation that will be essential in the derivation of
the lace expansion.

First we introduce some notions and notation.  We call a pair of
sites $b=\{u,v\}$ with $J_b\ne0$ a \emph{bond}.  So far we have used
the notation $\Lambda\subset\Zd$ for a site set.  However, we will
often abuse this notation to describe a \emph{graph} that consists
of sites of $\Lambda$ and are equipped with a certain bond set,
which we denote by $\mB_\Lambda$.  Note that
``$\{u,v\}\in\mB_\Lambda$'' always implies ``$u,v\in\Lambda$'', but
the latter does not necessarily imply the former.  If we regard
$\cA$ and $\Lambda$ as graphs, then ``$\cA\subset\Lambda$'' means
that $\cA$ is a subset of $\Lambda$ as a site set, and that
$\mB_\cA\subset\mB_\Lambda$.

Now we consider the partition function $Z_\cA$ on
$\cA\subset\Lambda$.  By expanding the Boltzmann factor in
\refeq{ZL-def}, we obtain
\begin{align}\lbeq{ZA-rewr}
Z_\cA&=2^{-|\cA|}\sum_{\varphi\in\{\pm1\}^\cA}\,\prod_{\{u,v\}\in\mB_\cA}\,
 \bigg(\sum_{n_{u,v}\in\Zp}\frac{(p J_{u,v})^{n_{u,v}}}{n_{u,v}!}
 \,\varphi_u^{n_{u,v}}\varphi_v^{n_{u,v}}\bigg)\nn\\
&=\sum_{\bn\in\Zp^{\mB_\cA}}\bigg(\prod_{b\in\mB_\cA}\frac{(p J_b)
 ^{n_b}}{n_b!}\bigg)\prod_{v\in\cA}\bigg(\frac12\sum_{\varphi_v=\pm1}
 \varphi_v^{\sum_{b\ni v}n_b}\bigg),
\end{align}
where we call $\bn=\{n_b\}_{b\in\mB_\cA}$ a \emph{current configuration}.
Note that the single-spin average in the last line equals 1 if
$\sum_{b\ni v}n_b$ is an even integer, and 0 otherwise.
Denoting by $\bd\bn$ the set of \emph{sources} $v\in\Lambda$ at which
$\sum_{b\ni v}n_b$ is an \emph{odd} integer, and defining
\begin{align}\lbeq{weight}
w_\cA(\bn)=\prod_{b\in\mB_\cA}\frac{(p J_b)^{n_b}}{n_b!}\qquad(\bn\in
 \Zp^{\mB_\cA}),
\end{align}
we obtain
\begin{align}\lbeq{ZA-RCrepr1}
Z_\cA=\sum_{\bn\in\Zp^{\mB_\cA}}w_\cA(\bn)\,\prod_{v\in\cA}\ind{
 \sum_{b\ni v}n_b\text{ even}}=\sum_{\bd\bn=\vno}w_\cA(\bn).
\end{align}

The partition function $Z_\cA$ equals the partition function on
$\Lambda$ with $J_b=0$ for all $b\in\mB_\Lambda\setminus\mB_\cA$.
We can also think of $Z_\cA$ as
the sum of $w_\Lambda(\bn)$ over $\bn\in\Zp^{\mB_\Lambda}$
satisfying $\bn|_{\mB_\Lambda\setminus\mB_{\cA}}\equiv0$, where
$\bn|_\mB$ is a projection of $\bn$ over the bonds in a bond set
$\mB$, i.e., $\bn|_\mB=\{n_b:b\in\mB\}$.  By this observation, we
can rewrite \refeq{ZA-RCrepr1} as
\begin{align}\lbeq{ZA-RCrepr2}
Z_\cA=\sum_{\substack{\bd\bn=\vno\\ \bn|_{\mB_\Lambda\setminus
 \mB_{\cA}} \equiv0}}w_\Lambda(\bn).
\end{align}

Following the same calculation, we can rewrite
$Z_\cA\Exp{\varphi_x\varphi_y}_\cA$ for $x,y\in\cA$ as
\begin{align}\lbeq{2pt-rewr}
Z_\cA\Exp{\varphi_x\varphi_y}_\cA
&=\sum_{\bn\in\Zp^{\mB_\cA}}\bigg(\prod_{b\in\mB_\cA}\frac{(p J_b)
 ^{n_b}}{n_b!}\bigg)\prod_{v\in\cA}\bigg(\frac12\sum_{\varphi_v=\pm1}
 \varphi_v^{\indic{\sss\{v\in x\vtri y\}}+\sum_{b\ni v}n_b}\bigg)\nn\\
&=\sum_{\bd\bn=x\vtri y}w_\cA(\bn)=\sum_{\substack{\bd\bn=x\vtri y\\
 \bn|_{\mB_\Lambda\setminus\mB_{\cA}}\equiv0}}w_\Lambda(\bn),
\end{align}
where $x\vtri y$ is an abbreviation for the symmetric difference
$\{x\}\tri\{y\}$:
\begin{align}\lbeq{symmdiff}
x\vtri y\equiv\{x\}\tri\{y\}=\begin{cases}
 \vno&\text{if }x=y,\\
 \{x,y\}&\text{otherwise}.
\end{cases}
\end{align}
If $x$ or $y$ is in $\cA\compl\equiv\Lambda\setminus\cA$, then we
define both sides of \refeq{2pt-rewr} to be zero.  This is
consistent with the above representation when $x\ne y$, since, for
example, if $x\in\cA\compl$, then the leftmost expression of
\refeq{2pt-rewr} is a multiple of
$\frac12\sum_{\varphi_x=\pm1}\varphi_x=0$, while the last expression
in \refeq{2pt-rewr} is also zero because there is no way of
connecting $x$ and $y$ on a current configuration $\bn$ with
$\bn|_{\mB_\Lambda\setminus\mB_{\cA}}\equiv0$.

\begin{figure}[t]
\begin{center}
\includegraphics[scale=0.24]{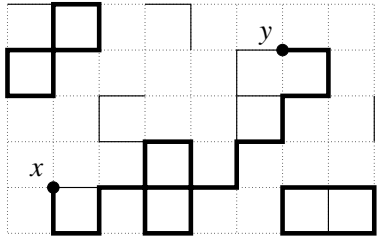}
\caption{\label{fig:RCrepr}A current configuration with sources at $x$ and
$y$.  The thick-solid segments represent bonds with odd currents, while
the thin-solid segments represent bonds with positive even currents, which
cannot be seen in the high-temperature expansion.}
\end{center}
\end{figure}

The key observation in the representation \refeq{2pt-rewr} is that the
right-hand side is nonzero only when $x$ and $y$ are connected by a
chain of bonds with \emph{odd} currents (see Figure~\ref{fig:RCrepr}).
We will exploit this peculiar underlying percolation picture to derive
the lace expansion for the two-point function.

\subsection{Derivation of the lace expansion}\label{ss:derivation}
In this subsection, we derive the lace expansion for
$\Exp{\varphi_o\varphi_x}_\Lambda$ using the random-current
representation.  In Section~\ref{sss:1stexp}, we introduce some
definitions and perform the first stage of the expansion, namely
\refeq{Ising-lace} for $j=0$, simply using inclusion-exclusion.  In
Section~\ref{sss:2ndexp}, we perform the second stage of the expansion,
where the source-switching lemma (Lemma~\ref{lmm:switching}) plays a
significant role to carry on the expansion indefinitely.  Finally, in
Section~\ref{sss:complexp}, we complete the proof of
Proposition~\ref{prp:Ising-lace}.

\subsubsection{The first stage of the expansion}\label{sss:1stexp}
As mentioned in Section~\ref{ss:RCrepr}, the underlying picture in
the random-current representation is quite similar to percolation.
We exploit this similarity to obtain the lace expansion.

First, we introduce some notions and
notation.

\begin{defn}\label{defn:perc}{\rm
\begin{enumerate}[(i)]
\item
Given $\bn\in\Zp^{\mB_\Lambda}$ and $\cA\subset\Lambda$, we say that $x$ is
$\bn$-connected to $y$ in (the graph) $\cA$, and simply write $x\cn{\bn}{}y$ \emph{in}
$\cA$, if either $x=y\in\cA$ or there is a self-avoiding path (or we simply call it a
path) from $x$ to $y$ consisting of bonds $b\in\mB_\cA$ with $n_b>0$.  If
$\bn\in\Zp^{\mB_\cA}$, we omit ``in $\cA$'' and simply write $x\cn{\bn}{}y$.  We also
define
\begin{align}\lbeq{incl/excl}
\{x\cn{\bn}{\cA}y\}=\{x\cn{\bn}{}y\}\setminus\{x\cn{\bn}{}y\text{ in }
 \cA\compl\},
\end{align}
and say that $x$ is $\bn$-connected to $y$ \emph{through} $\cA$.
\item
Given an event $E$ (i.e., a set of current configurations) and a
bond $b$, we define $\{E$ off $b\}$ to be the set of current
configurations $\bn\in E$ such that changing $n_b$ results in a
configuration that is also in $E$.  Let
$\cC_\bn^b(x)=\{y:x\cn{\bn}{}y\text{ off }b\}$.
\item
For a \emph{directed} bond $b=(u,v)$, we write $\bb=u$ and $\tb=v$.
We say that a directed bond $b$ is \emph{pivotal} for $x\cn{\bn}{}y$
from $x$, if $\{x\cn{\bn}{}\bb$ off $b\}\cap\{\tb\cn{\bn}{}y$ in
$\cC_\bn^b(x)\compl\}$ occurs.  If $\{x\cn{\bn}{}y\}$ occurs with no
pivotal bonds, we say that $x$ is \emph{$\bn$-doubly connected to}
$y$, and write $x\db{\bn}{}y$.
\end{enumerate}
}\end{defn}

\begin{figure}[t]
\begin{center}
\[ \raisebox{0.2pc}{\includegraphics[scale=0.17]{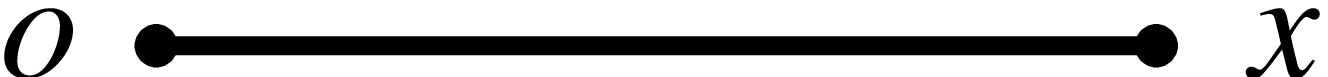}}~~~=~~~
 \raisebox{-0.5pc}{\includegraphics[scale=0.17]{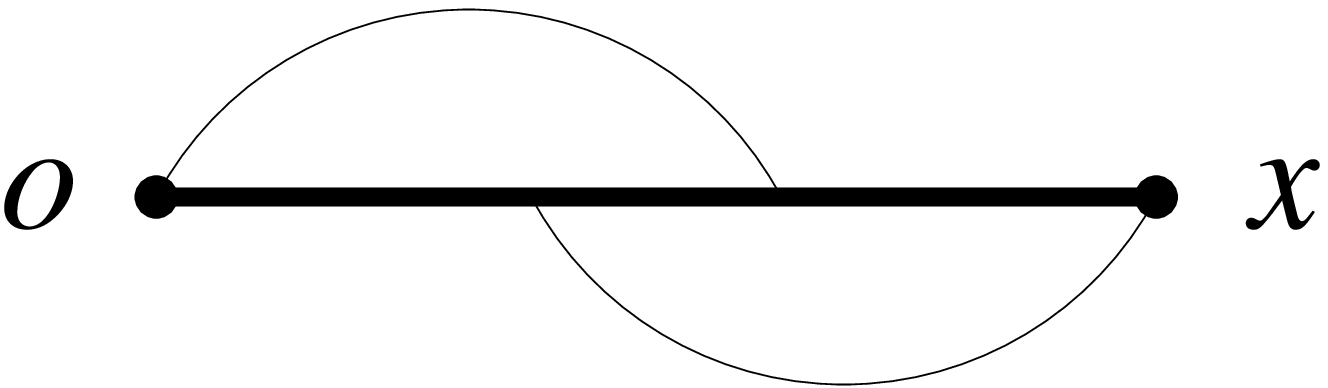}}~~~+~~
 \sum_b~\raisebox{-1.7pc}{\includegraphics[scale=0.17]{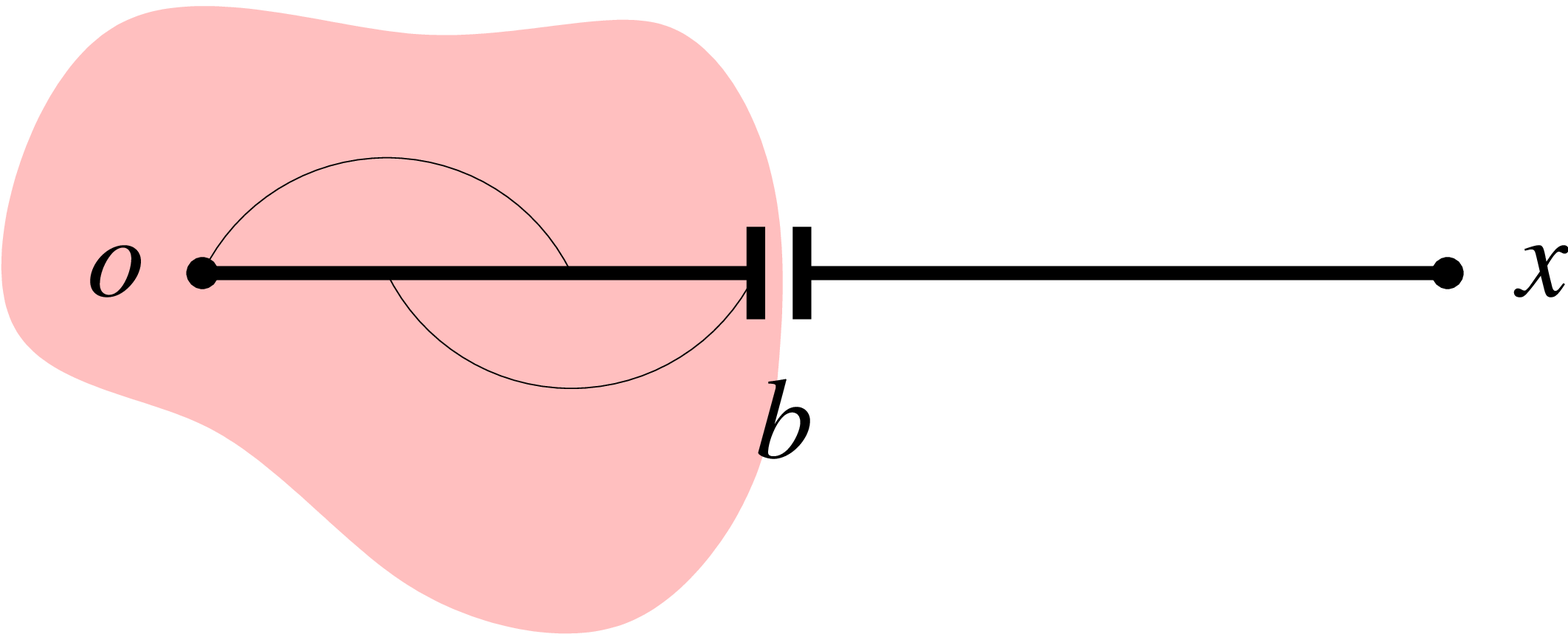}} \]
\caption{\label{fig:1stpiv}A schematic representation of
\refeq{pre-1st-exp}.  The thick lines are connections consisting of bonds
with odd currents, while the thin arcs are connections made of bonds with
positive (not necessarily odd) currents.  The shaded region represents
$\cC_\bn^b(o)$.}
\end{center}
\end{figure}

We begin with the first stage of the lace expansion.  First, by using
the above percolation language, the two-point function can be written as
\begin{align}\lbeq{2pt-perclang}
\Exp{\varphi_o\varphi_x}_\Lambda=\sum_{\bd\bn=o\vtri x}\frac{w_\Lambda
 (\bn)}{Z_\Lambda}\equiv\sum_{\bd\bn=o\vtri x}\frac{w_\Lambda(\bn)}
 {Z_\Lambda}\,{\ind{o\cn{\bn}{}x}}.
\end{align}
We decompose the indicator on the right-hand side into two parts depending
on whether or not there is a pivotal bond for $o\cn{\bn}{}x$ from $o$;
if there is, we take the \emph{first} bond among them.  Then, we have
\begin{align}\lbeq{0th-ind-fact}
\ind{o\cn{\bn}{}x}=\ind{o\db{\bn}{}x}+\sum_{b\in\mB_\Lambda}\ind{o\db{\bn}{}
 \bb\text{ off }b}\,\ind{n_b>0}\,\ind{\tb\cn{\bn}{}x\text{ in }\cC_\bn^b(o)
 \compl}.
\end{align}
Let
\begin{align}\lbeq{pi0-def}
\pi_\Lambda^{\sss(0)}(x)=\sum_{\bd\bn=o\vtri x}\frac{w_\Lambda(\bn)}
 {Z_\Lambda}\,\ind{o\db{\bn}{}x}.
\end{align}
Substituting \refeq{0th-ind-fact} into \refeq{2pt-perclang}, we obtain
(see Figure~\ref{fig:1stpiv})
\begin{align}\lbeq{pre-1st-exp}
\Exp{\varphi_o\varphi_x}_\Lambda=\pi_\Lambda^{\sss(0)}(x)+\sum_{b\in
 \mB_\Lambda}\;\sum_{\bd\bn=o\vtri x}\frac{w_\Lambda(\bn)}{Z_\Lambda}\,
 \ind{o\db{\bn}{}\bb\text{ off }b}\,\ind{n_b>0}\,\ind{\tb\cn{\bn}{}x
 \text{ in }\cC_\bn^b(o)\compl}.
\end{align}

Next, we consider the sum over $\bn$ in \refeq{pre-1st-exp}.  Since
$b$ is pivotal for $o\cn{\bn}{}x$ from $o\,(\ne x$, due to the last
indicator) and $\bd\bn=o\vtri x$, in fact $n_b$ is an \emph{odd}
integer. We alternate the parity of $n_b$ by changing the
source constraint into $o\vtri b\vtri
x\equiv\{o\}\tri\{\bb,\tb\}\tri\{x\}$ and multiplying by
\begin{align}
\frac{\sum_{n\text{ odd}}(p J_b)^n/n!}{\sum_{n\text{ even}}(p
 J_b)^n/n!}=\tanh(p J_b)\equiv\tau_b.
\end{align}
Then, the sum over $\bn$ in \refeq{pre-1st-exp} equals
\begin{align}\lbeq{0th-summand1}
\sum_{\bd\bn=o\vtri b\vtri x}\frac{w_\Lambda(\bn)}{Z_\Lambda}\,\ind{o
 \db{\bn}{}\bb\text{ off }b}\,\tau_b\ind{n_b\text{ even}}\,\ind{\tb
 \cn{\bn}{}x\text{ in }\cC_\bn^b(o)\compl}.
\end{align}
Note that, except for $b$, there are no positive currents on the
boundary bonds of $\cC_\bn^b(o)$.

Now, we condition on $\cC_\bn^b(o)=\cA$ and decouple events
occurring on $\mB_{\cA\compl}$ from events occurring on
$\mB_\Lambda\setminus\mB_{\cA\compl}$, by using the following
notation:
\begin{align}\lbeq{tildew-def}
\tilde w_{\Lambda,\cA}(\bk)=\prod_{b\in\mB_\Lambda\setminus\mB_{\cA
 \compl}}\frac{(pJ_b)^{k_b}}{k_b!}\qquad(\bk\in\Zp^{\mB_\Lambda
 \setminus\mB_{\cA\compl}}).
\end{align}
Conditioning on $\cC_\bn^b(o)=\cA$, multiplying
$Z_{\cA\compl}/Z_{\cA\compl}\equiv1$ (and using the notation
$\bk=\bn|_{\mB_\Lambda\setminus\mB_{\cA\compl}}$ and
$\bm=\bn|_{\mB_{\cA\compl}}$) and then summing over
$\cA\subset\Lambda$, we have
\begin{align}\lbeq{0th-summand2}
\refeq{0th-summand1}&=\sum_{\cA\subset\Lambda}\,\sum_{\substack{\bd\bk
 =o\vtri\bb\\ \bd\bm=\tb\vtri x}}\frac{\tilde w_{\Lambda,\cA}(\bk)\,
 Z_{\cA\compl}}{Z_\Lambda}\,\frac{w_{\cA\compl}(\bm)}{Z_{\cA\compl}}\,
 \ind{o\db{\bk}{}\bb\text{ off }b\}\,\cap\,\{\cC_{\bk}^b(o)=\cA}\,
 \tau_b\ind{k_b\text{ even}}\,\ind{\tb\cn{\bm}{}x\text{ in }\cA\compl}\nn\\
&=\sum_{\cA\subset\Lambda}\;\sum_{\bd\bn=o\vtri\bb}\frac{w_\Lambda(\bn)}
 {Z_\Lambda}\,\ind{o\db{\bn}{}\bb\text{ off }b\}\,\cap\,\{\cC_\bn^b(o)=\cA}
 \,\tau_b\ind{n_b\text{ even}}\underbrace{\sum_{\bd\bm=\tb\vtri x}\frac{
 w_{\cA\compl}(\bm)}{Z_{\cA\compl}}\,\ind{\tb\cn{\bm}{}x\text{ (in }\cA
 \compl)}}_{=\;\Exp{\varphi_{\tb}\varphi_x}_{\cA\compl}}\nn\\
&=\sum_{\bd\bn=o\vtri\bb}\frac{w_\Lambda(\bn)}{Z_\Lambda}\,\ind{o\db{\bn}
 {}\bb\text{ off }b}\,\tau_b\ind{n_b\text{ even}}\,\Exp{\varphi_{\tb}
 \varphi_x}_{\cC_\bn^b(o)\compl}.
\end{align}
Furthermore, ``off $b$'' and $\tind{n_b\text{ even}}$ in the last
line can be omitted, since
$\{o\db{\bn}{}\bb\}\setminus\{o\db{\bn}{}\bb$ off $b\}$ and
$\{\bd\bn=o\vtri\bb\}\cap\{n_b$ odd\} are subsets of
$\{\tb\in\cC_\bn^b(o)\}$, on which
$\Exp{\varphi_{\tb}\varphi_x}_{\cC_\bn^b(o)\compl}=0$.  As a result,
\begin{align}\lbeq{0th-summand3}
\refeq{0th-summand2}~=\sum_{\bd\bn=o\vtri\bb}\frac{w_\Lambda(\bn)}
 {Z_\Lambda}\,\ind{o\db{\bn}{}\bb}\,\tau_b\,\Exp{\varphi_{\tb}
 \varphi_x}_{\cC_\bn^b(o)\compl}.
\end{align}

By \refeq{pre-1st-exp} and \refeq{0th-summand3}, we arrive at
\begin{align}\lbeq{1st-exp}
\Exp{\varphi_o\varphi_x}_\Lambda=\pi_\Lambda^{\sss(0)}(x)+\sum_{b\in
 \mB_\Lambda}\pi_\Lambda^{\sss(0)}(\bb)\,\tau_b\,\Exp{\varphi_{\tb}
 \varphi_x}_\Lambda-R_\Lambda^{\sss(1)}(x),
\end{align}
where
\begin{align}\lbeq{R1-def}
R_\Lambda^{\sss(1)}(x)=\sum_{b\in\mB_\Lambda}\;\sum_{\bd\bn=o\vtri\bb}
 \frac{w_\Lambda(\bn)}{Z_\Lambda}\,\ind{o\db{\bn}{}\bb}\,\tau_b\Big(
 \Exp{\varphi_{\tb}\varphi_x}_\Lambda-\Exp{\varphi_{\tb}\varphi_x}_{
 \cC_\bn^b(o)\compl}\Big).
\end{align}
This completes the proof of \refeq{Ising-lace} for $j=0$, with
$\pi_\Lambda^{\sss(0)}(x)$ and $R_\Lambda^{\sss(1)}(x)$ being defined
in \refeq{pi0-def} and \refeq{R1-def}, respectively.

\subsubsection{The second stage of the expansion}\label{sss:2ndexp}
In the next stage of the lace expansion, we further expand
$R_\Lambda^{\sss(1)}(x)$ in \refeq{1st-exp}.  To do so, we investigate
the difference $\Exp{\varphi_{\tb}\varphi_x}_\Lambda-\Exp{\varphi_{\tb}
\varphi_x}_{\cC_\bn^b(o)\compl}$ in \refeq{R1-def}.  First, we prove
the following key proposition\footnote{The mean-field results in
\cite{a82,abf87,af86,ag83} are based on a couple of differential
inequalities for $M_{p,h}$ and $\chi_p$ (under the periodic-boundary
condition) using a certain random-walk representation.  We can simplify
the proof of the same differential inequalities (under the free-boundary
condition as well) using Proposition~\ref{prp:through}.}:

\begin{prp}\label{prp:through}
For $v,x\in\Lambda$ and $\cA\subset\Lambda$, we have
\begin{align}\lbeq{lmm-through}
\Exp{\varphi_v\varphi_x}_\Lambda-\Exp{\varphi_v\varphi_x}_{\cA\compl}
 =\sum_{\substack{\bd\bm=\vno\\ \bd\bn=v\vtri x}}\frac{w_{\cA\compl}
 (\bm)}{Z_{\cA\compl}}\,\frac{w_\Lambda(\bn)}{Z_\Lambda}\,\ind{v\cn{\bm
 +\bn}{\cA}x}.
\end{align}
Therefore, $\Exp{\varphi_v\varphi_x}_{\cA\compl}\leq
\Exp{\varphi_v\varphi_x}_\Lambda$ for the ferromagnetic case.
\end{prp}

\begin{proof}
Since both sides of \refeq{lmm-through} are equal to $\tind{x\in\cA}$ when
$v=x$ (see below \refeq{symmdiff}), it suffices to prove \refeq{lmm-through}
for $v\ne x$.

First, by using \refeq{ZA-RCrepr1}--\refeq{2pt-rewr}, we obtain
\begin{align}\lbeq{WZ-num}
Z_\Lambda Z_{\cA\compl}\Big(\Exp{\varphi_v\varphi_x}_\Lambda-\Exp{\varphi_v
 \varphi_x}_{\cA\compl}\Big)=\sum_{\bd\bn=\{v,x\}}Z_{\cA\compl}\,w_\Lambda
 (\bn)-\sum_{\bd\bm=\{v,x\}}w_{\cA\compl}(\bm)\,Z_\Lambda\nn\\
=\sum_{\substack{\bd\bm=\vno,\,\bd\bn=\{v,x\}\\ \bm|_{\mB_\Lambda\setminus
 \mB_{\cA\compl}}\equiv0}}w_\Lambda(\bm)\,w_\Lambda(\bn)-\sum_{\substack{
 \bd\bm=\{v,x\},\,\bd\bn=\vno\\ \bm|_{\mB_\Lambda\setminus\mB_{\cA\compl}}
 \equiv0}}w_\Lambda(\bm)\,w_\Lambda(\bn).
\end{align}
Note that the second term is equivalent to the first term if the source
constraints for $\bm$ and $\bn$ are exchanged.

Next, we consider the second term of \refeq{WZ-num}, whose exact expression
is
\begin{gather}\lbeq{2ndterm-expl}
\sum_{\substack{\bd\bm=\{v,x\},\,\bd\bn=\vno\\ \bm|_{\mB_\Lambda\setminus
 \mB_{\cA\compl}}\equiv0}}\bigg(\prod_{b\in\mB_\Lambda\setminus\mB_{\cA
 \compl}}\frac{(pJ_b)^{n_b}}{n_b!}\bigg)\prod_{b\in\mB_{\cA\compl}}\frac{
 (pJ_b)^{m_b+n_b}}{m_b!\,n_b!}=\sum_{\bd\bN=\{v,x\}}w_\Lambda(\bN)\sum_{
 \substack{\bd\bm=\{v,x\}\\ \bm|_{\mB_\Lambda\setminus\mB_{\cA\compl}}
 \equiv0}}\prod_{b\in\mB_{\cA\compl}}\binom{N_b}{m_b}.
\end{gather}
The following is a variant of the source-switching lemma \cite{a82,ghs70}
and allows us to change the source constraints in \refeq{2ndterm-expl}.

\begin{lmm}[\textbf{Source-switching lemma}]
\label{lmm:switching}
\begin{align}\lbeq{switching}
\sum_{\substack{\bd\bm=\{v,x\}\\ \bm|_{\mB_\Lambda\setminus\mB_{\cA
 \compl}}\equiv0}}\,\prod_{b\in\mB_{\cA\compl}}\binom{N_b}{m_b}=\ind{v
 \cn{\bN}{}x\text{ in }\cA\compl}\sum_{\substack{\bd\bm=\vno\\ \bm|_{
 \mB_\Lambda\setminus\mB_{\cA\compl}}\equiv0}}\,\prod_{b\in\mB_{\cA
 \compl}}\binom{N_b}{m_b}.
\end{align}
\end{lmm}

The idea of the proof of \refeq{switching} can easily be extended to
more general cases, in which the source constraint in the left-hand
side of \refeq{switching} is replaced by $\bd\bm=\cV$ for some
$\cV\subset\Lambda$ and that in the right-hand side is replaced by
$\bd\bm=\cV\tri\{v,x\}$ (e.g., \cite{a82}).  We will explain the
proof of \refeq{switching} after completing the proof of
Proposition~\ref{prp:through}.

We continue with the proof of Proposition~\ref{prp:through}.
Substituting \refeq{switching} into \refeq{2ndterm-expl}, we obtain
\begin{align}\lbeq{switching-appl}
\refeq{2ndterm-expl}&=\sum_{\bd\bN=\{v,x\}}w_\Lambda(\bN)\,\ind{v
 \cn{\bN}{}x\text{ in }\cA\compl}\sum_{\substack{\bd\bm=\vno\\
 \bm|_{\mB_\Lambda\setminus\mB_{\cA\compl}}\equiv0}}\,\prod_{b\in
 \mB_{\cA\compl}}\binom{N_b}{m_b}\nn\\
&=\sum_{\substack{\bd\bm=\vno,\,\bd\bn=\{v,x\}\\ \bm|_{\mB_\Lambda
 \setminus\mB_{\cA\compl}}\equiv0}}w_\Lambda(\bm)\,w_\Lambda(\bn)\,
 \ind{v\cn{\bm+\bn}{}x\text{ in }\cA\compl}.
\end{align}
Note that the source constraints for $\bm$ and $\bn$ in the last line
are identical to those in the first term of \refeq{WZ-num}, under
which $\tind{v\cn{\bm+\bn}{}x}$ is always 1.  By \refeq{incl/excl},
we can rewrite \refeq{WZ-num} as
\begin{align}\lbeq{through}
\Exp{\varphi_v\varphi_x}_\Lambda-\Exp{\varphi_v\varphi_x}_{\cA\compl}
&=\sum_{\substack{\bd\bm=\vno,\,\bd\bn=\{v,x\}\\ \bm|_{\mB_\Lambda
 \setminus\mB_{\cA\compl}}\equiv0}}\frac{w_\Lambda(\bm)}{Z_{\cA\compl}}
 \,\frac{w_\Lambda(\bn)}{Z_\Lambda}\,\ind{v\cn{\bm+\bn}{\cA}x}.
\end{align}
Using \refeq{ZA-RCrepr1}--\refeq{ZA-RCrepr2} to omit
``$\bm|_{\mB_\Lambda\setminus\mB_{\cA\compl}}\equiv0$'' and
replace $w_\Lambda(\bm)$ by $w_{\cA\compl}(\bm)$, we arrive
at \refeq{lmm-through}.  This completes the proof of
Proposition~\ref{prp:through}.
\end{proof}

\begin{proof}[Sketch proof of Lemma~\ref{lmm:switching}]
We explain the meaning of the identity \refeq{switching} and the
idea of its proof.  Given $\bN=\{N_b\}_{b\in\mB_\Lambda}$, we denote by
$\mG_\bN$ the graph consisting of $N_b$ \emph{labeled} edges between $\bb$
and $\tb$ for every $b\in\mB_\Lambda$ (see Figure~\ref{fig:switching}).
\begin{figure}[t]
\begin{center}
\begin{align*}
\bN~:&\qquad\includegraphics[scale=0.33]{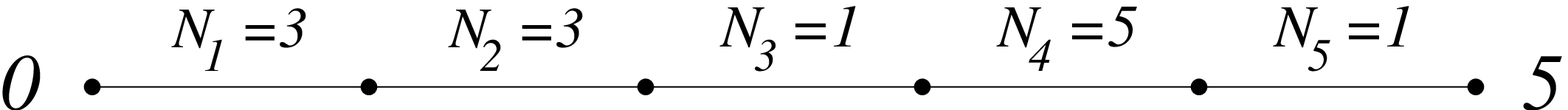}\\[5pt]
\mG_\bN~:&\qquad\raisebox{-1.8pc}{\includegraphics[scale=0.33]
 {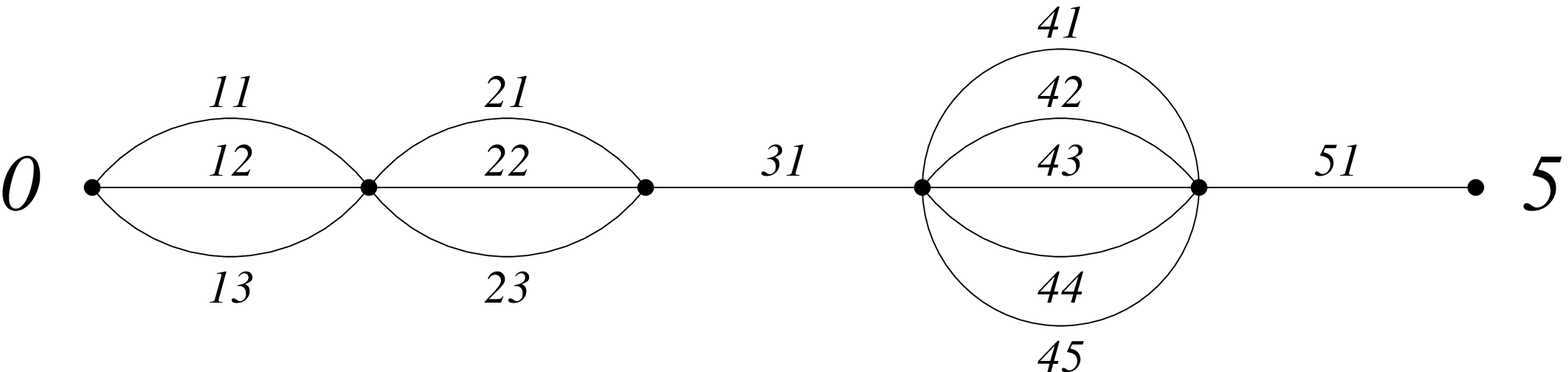}}\\[1pc]
\mS~:&\qquad\raisebox{-1.8pc}{\includegraphics[scale=0.33]
 {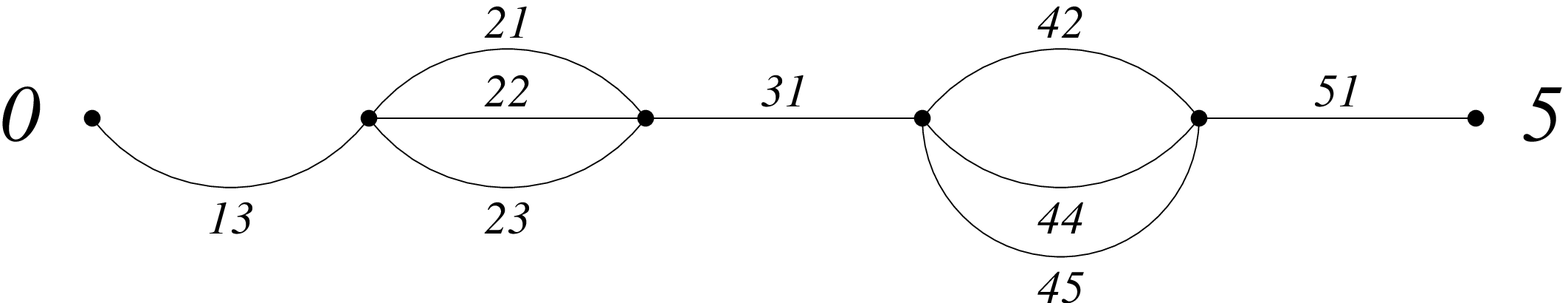}}\\[7pt]
\mS\tri\omega~:&\qquad\raisebox{-1.8pc}{\includegraphics[scale=0.33]
 {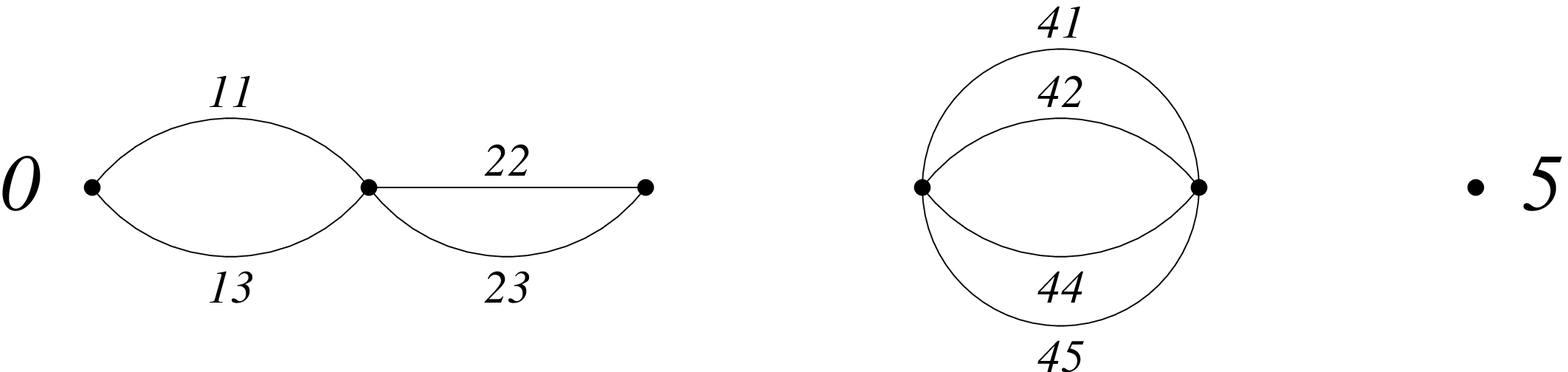}}
\end{align*}
\caption{\label{fig:switching}$\bN=\{N_b\}_{b=1}^5=(3,3,1,5,1)$ is an
example of a current configuration on $[0,5]\cap\Zp$ satisfying
$\bd\bN=\{0,5\} $, and $\mG_\bN$ is the corresponding labeled graph
consisting of edges $e=b\ell_b$, where $\ell_b\in\{1,\dots,N_b\} $.
The third and fourth pictures show the relation between a subgraph $\mS$
with $\bd\mS=\{0,5\} $ and its image $\mS\tri\omega$ of the map defined
in \refeq{bijection}, where $\omega$ is a path of edges
$(11,21,31,41,51)$.}
\end{center}
\end{figure}
For a subgraph $\mS\subset\mG_\bN$, we denote by $\bd\mS$ the set of
vertices at which the number of incident edges in $\mS$ is \emph{odd},
and let $\mS_\cA=\mS\cap\mG_{\bN|_{\mB_\Lambda\setminus\mB_{\cA\compl}}}$.
Then, the left-hand side of \refeq{switching} equals the cardinality
$|\fS|$ of
\begin{align}\lbeq{Sbefore}
\fS=\{\mS\subset\mG_\bN:\bd\mS=\{v,x\},~\mS_\cA=\vno\},
\end{align}
and the sum in the right-hand side of \refeq{switching} equals the
cardinality $|\fS'|$ of
\begin{align}
\fS'=\{\mS\subset\mG_\bN:\bd\mS=\vno,~\mS_\cA=\vno\}.
\end{align}
We note that $|\fS|$ is zero when there are no paths on $\mG_\bN$ between $v$ and $x$
consisting of edges whose endvertices are both in $\cA\compl$, while $|\fS'|$ may not be
zero.  The identity \refeq{switching} reads that $|\fS|$ equals $|\fS'|$ if we
compensate for this discrepancy.

Suppose that there is a path (i.e., a ) $\omega$ from $v$ to $x$ consisting of edges in
$\mG_\bN$ whose endvertices are both in $\cA\compl$.  Then, the map
\begin{align}\lbeq{bijection}
\mS\in\fS~\mapsto~\mS\tri\omega\in\fS',
\end{align}
is a bijection \cite{a82,ghs70}, and therefore $|\fS|=|\fS'|$.  Here and in
the rest of the paper, the symmetric difference between graphs is only in
terms of \emph{edges}.  For example, $\mS\tri\omega$ is the result of adding
or deleting edges (not vertices) contained in $\omega$.  This completes the
proof of \refeq{switching}.
\end{proof}

\begin{figure}[t]
\begin{center}
\[ \raisebox{-1.3pc}{\includegraphics[scale=0.17]{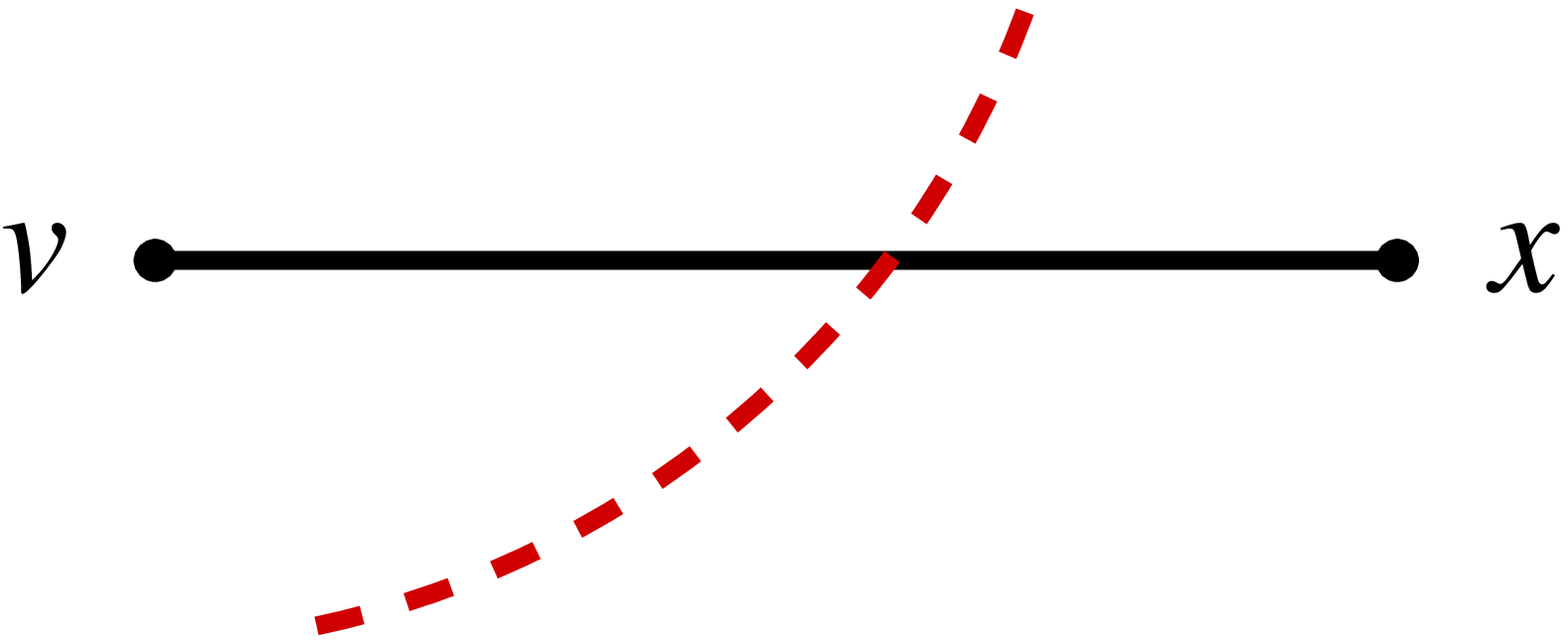}}~~~=~~~
 \raisebox{-1.3pc}{\includegraphics[scale=0.17]{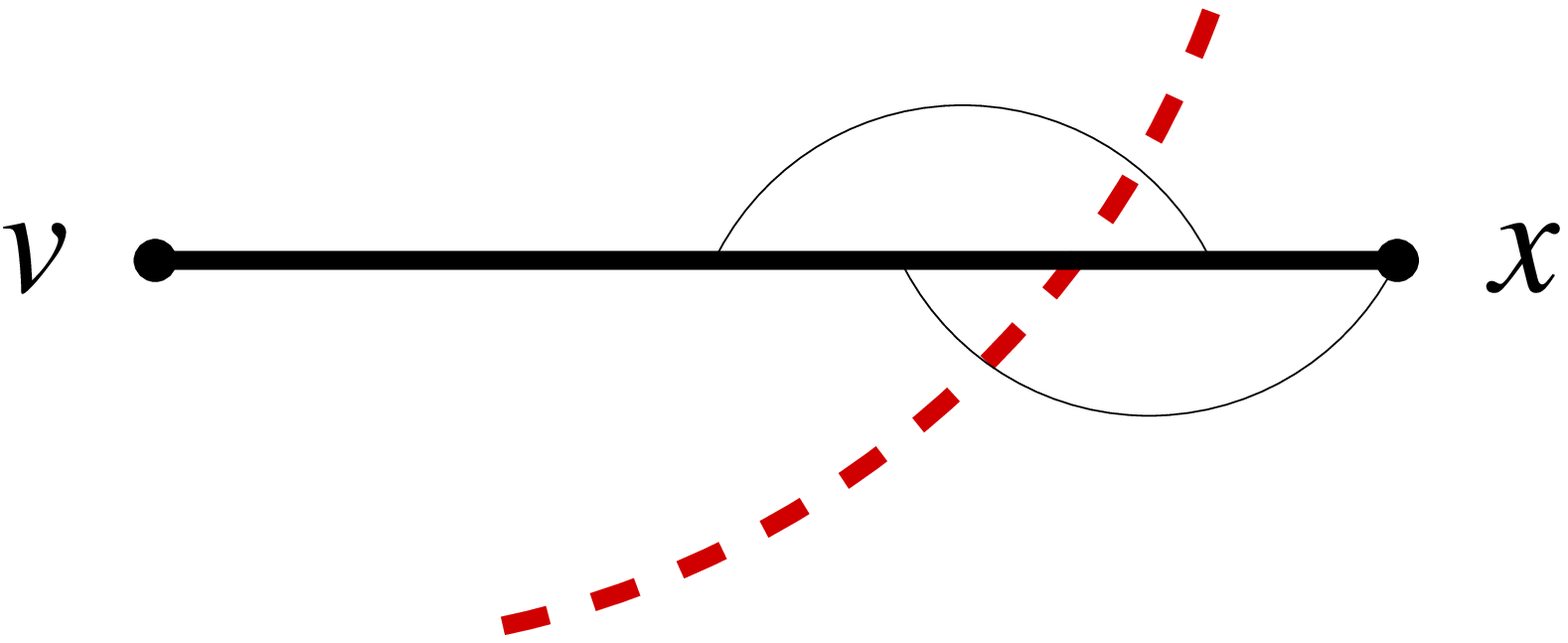}}~~~+~~
 \sum_b~\raisebox{-1.3pc}{\includegraphics[scale=0.17]{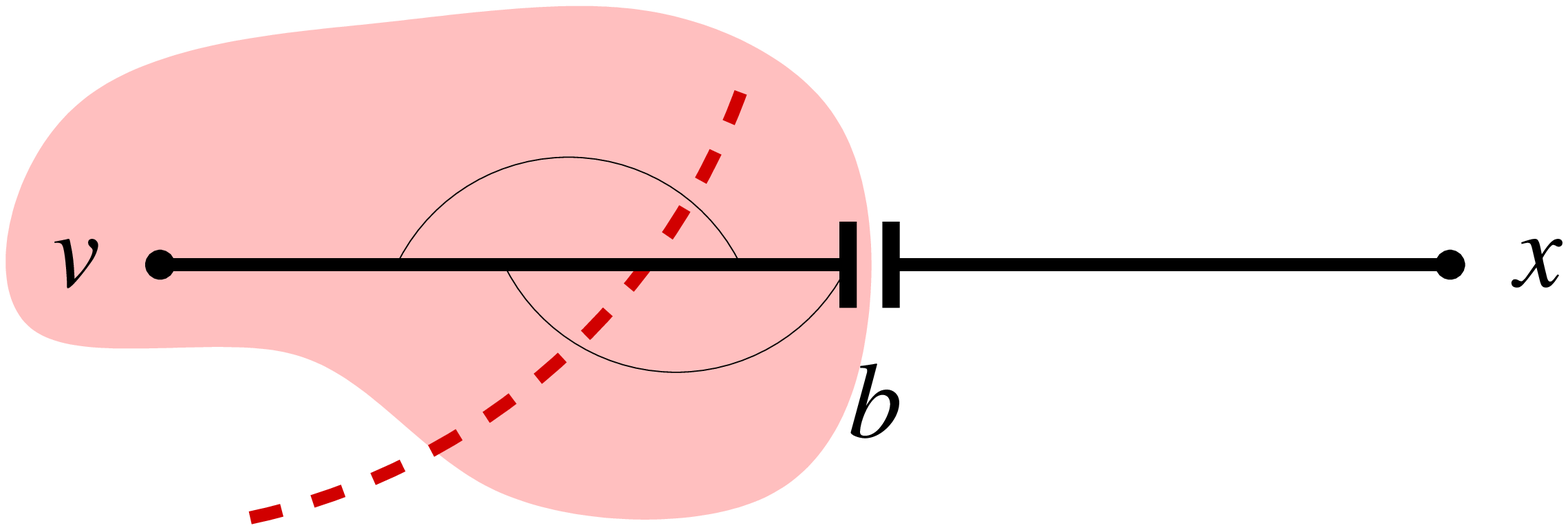}} \]
\caption{\label{fig:through}A schematic representation of
\refeq{2nd-ind-fact}.  The dashed lines represent $\cA$, the
thick-solid lines represent connections consisting of bonds $b_1$
such that $m_{b_1}+n_{b_1}$ is odd, and the thin-solid lines are
connections made of bonds $b_2$ such that $m_{b_2}+n_{b_2}$ is
positive (not necessarily odd). The shaded region represents
$\cC_{\bm+\bn}^b(v)$.}
\end{center}
\end{figure}

We now start with the second stage of the expansion by using
Proposition~\ref{prp:through} and applying inclusion-exclusion as in the
first stage of the expansion in Section~\ref{sss:1stexp}.  First, we
decompose the indicator in \refeq{lmm-through} into two parts depending
on whether or not there is a pivotal bond $b$ for $v\cn{\bm+\bn}{}x$
from $v$ such that $v\cn{\bm+\bn}{\cA}\bb$.  Let
\begin{align}\lbeq{E-def}
E_\bN(v,x;\cA)=\{v\cn{\bN}{\cA}x\}\cap\{\nexists
 \text{ pivotal bond }b\text{ for }v\cn{\bN}{}x
 \text{ from $v$ such that }v\cn{\bN}{\cA}\bb\}.
\end{align}
On the event $\{v\cn{\bm+\bn}{\cA}x\}\setminus
E_{\bm+\bn}(v,x;\cA)$, we take the \emph{first} pivotal bond $b$ for
$v\cn{\bm+\bn}{}x$ from $v$ satisfying $v\cn{\bm+\bn}{\cA}\bb$.
Then, we have (cf., \refeq{0th-ind-fact})
\begin{align}\lbeq{1st-ind-fact}
\ind{v\cn{\bm+\bn}{\cA}x}=\indic{E_{\bm+\bn}(v,x;\cA)}+\sum_{b\in
 \mB_\Lambda}\ind{E_{\bm+\bn}(v,\bb;\cA)\text{ off }b}\,\ind{m_b+n_b>0}
 \,\ind{\tb\cn{\bm+\bn}{}x\text{ in }\cC_{\bm+\bn}^b(v)\compl}.
\end{align}
Let
\begin{align}\lbeq{Theta-def}
\Theta_{v,x;\cA}[X]=\sum_{\substack{\bd\bm=\vno\\ \bd\bn=v\vtri x}}\frac{
 w_{\cA\compl}(\bm)}{Z_{\cA\compl}}\,\frac{w_\Lambda(\bn)}{Z_\Lambda}\,
 \indic{E_{\bm+\bn}(v,x;\cA)}\,X(\bm+\bn),&&
\Theta_{v,x;\cA}=\Theta_{v,x;\cA}[1].
\end{align}
Substituting \refeq{1st-ind-fact} into \refeq{lmm-through}, we
obtain (see Figure~\ref{fig:through})
\begin{align}\lbeq{2nd-ind-fact}
&\Exp{\varphi_v\varphi_x}_\Lambda-\Exp{\varphi_v\varphi_x}_{\cA\compl}\\
&=\Theta_{v,x;\cA}+\sum_{b\in\mB_\Lambda}\sum_{\substack{\bd\bm=\vno\\
 \bd\bn=v\vtri x}}\frac{w_{\cA\compl}(\bm)}{Z_{\cA\compl}}\,\frac{w_\Lambda
 (\bn)}{Z_\Lambda}\,\ind{E_{\bm+\bn}(v,\bb;\cA)\text{ off }b}\,\ind{m_b
 \text{ even, }n_b\text{ odd}}\,\ind{\tb\cn{\bm+\bn}{}x\text{ in }\cC_{
 \bm+\bn}^b(v)\compl},\nn
\end{align}
where we have replaced ``$m_b+n_b>0$'' in \refeq{1st-ind-fact} by
``$m_b$ even, $n_b$ odd'' that is the only possible combination
consistent with the source constraints and the conditions in the
indicators.  As in \refeq{0th-summand1}, we alternate the parity of
$n_b$ by changing the source constraint from $\bd\bn=v\vtri x$ to
$\bd\bn=v\vtri b\vtri x$ and multiplying by $\tau_b$.  Then, the sum
over $\bm$ and $\bn$ in \refeq{2nd-ind-fact} equals
\begin{align}\lbeq{3rd-ind-ppfact}
\sum_{\substack{\bd\bm=\vno\\ \bd\bn=v\vtri b\vtri x}}\frac{w_{\cA
 \compl}(\bm)}{Z_{\cA\compl}}\,\frac{w_\Lambda(\bn)}{Z_\Lambda}\,
 \ind{E_{\bm+\bn}(v,\bb;\cA)\text{ off }b}\,\tau_b\ind{m_b,n_b
 \text{ even}}\,\ind{\tb\cn{\bm+\bn}{}x\text{ in }\cC_{\bm+\bn}^b
 (v)\compl}.
\end{align}
Then, as in \refeq{0th-summand2}, we condition on
$\cC_{\bm+\bn}^b(v)=\cB$ and decouple events occurring on
$\mB_{\cB\compl}$ from events occurring on
$\mB_\Lambda\setminus\mB_{\cB\compl}$. Let
$\bm'=\bm|_{\mB_{\cA\compl}\setminus\mB_{\cA\compl\cap\cB\compl}}$,
$\bm''=\bm|_{\mB_{\cA\compl\cap\cB\compl}}$,
$\bn'=\bn|_{\mB_\Lambda\setminus\mB_{\cB\compl}}$ and
$\bn''=\bn|_{\mB_{\cB\compl}}$.  Note that $\bd\bm'=\bd\bm''=\vno$,
$\bd\bn'=v\vtri\bb$ and $\bd\bn''=\tb\vtri x$.  Multiplying
\refeq{3rd-ind-ppfact} by $(Z_{\cA\compl\cap\cB\compl}/Z_{\cA
\compl\cap\cB\compl})(Z_{\cB \compl}/Z_{\cB\compl})\equiv1$ and
using the notation \refeq{tildew-def}, we obtain
\begin{align}\lbeq{3rd-ind-prefact}
\refeq{3rd-ind-ppfact}&=\sum_{\cB\subset\Lambda}\sum_{\substack{\bd
 \bm'=\vno\\\bd\bn'=v\vtri\bb}}\frac{\tilde w_{\cA\compl,\cB}(\bm')
 \,Z_{\cA\compl\cap\cB\compl}}{Z_{\cA\compl}}\,\frac{\tilde w_{\Lambda,
 \cB}(\bn')\,Z_{\cB\compl}}{Z_\Lambda}\,\ind{E_{\bm'+\bn'}(v,\bb;\cA)
 \text{ off }b\}\,\cap\,\{\cC_{\bm'+\bn'}^b(v)=\cB}\,\nn\\
&\qquad\qquad\times\tau_b\ind{m'_b,n'_b\text{ even}}\sum_{\substack{
 \bd\bm''=\vno\\ \bd\bn''=\tb\vtri x}}\frac{w_{\cA\compl\cap\cB\compl}
 (\bm'')}{Z_{\cA\compl\cap\cB\compl}}\,\frac{w_{\cB\compl}(\bn'')}
 {Z_{\cB\compl}}\,\ind{\tb\cn{\bm''+\bn''}{}x\text{ in }\cB\compl}\nn\\
&=\sum_{\cB\subset\Lambda}\sum_{\substack{\bd\bm=\vno\\ \bd\bn=v\vtri\bb}}
 \frac{w_{\cA\compl}(\bm)}{Z_{\cA\compl}}\,\frac{w_\Lambda(\bn)}{Z_\Lambda}
 \,\ind{E_{\bm+\bn}(v,\bb;\cA)\text{ off }b\}\,\cap\,\{\cC_{\bm+\bn}^b(v)=
 \cB}\,\tau_b\ind{m_b,n_b\text{ even}}\,\Exp{\varphi_{\tb}\varphi_x}_{\cB
 \compl}\nn\\
&=\sum_{\substack{\bd\bm=\vno\\ \bd\bn=v\vtri\bb}}\frac{w_{\cA\compl}(\bm)}
 {Z_{\cA\compl}}\,\frac{w_\Lambda(\bn)}{Z_\Lambda}\,\ind{E_{\bm+\bn}(v,\bb;
 \cA)\text{ off }b}\,\tau_b\ind{m_b,n_b\text{ even}}\,\Exp{\varphi_{\tb}
 \varphi_x}_{\cC_{\bm+\bn}^b(v)\compl},
\end{align}
where we have been able to perform the sum over $\bm''$ and $\bn''$
independently, due to the fact that $\tind{\tb\,\cn{\bm''+\bn''}{}\,x
\text{ in }\cB\compl}\equiv1$ for any $\bn''\in\Zp^{\mB_{\cB\compl}}$
with $\bd\bn''=\tb\vtri x$.  As in the derivation of \refeq{0th-summand3}
from \refeq{0th-summand2}, we can omit ``off $b$'' and $\tind{m_b,n_b
\text{ even}}$ in \refeq{3rd-ind-prefact} using the source constraints and
the fact that $\Exp{\varphi_{\tb}\varphi_x}_{\cC_{\bm+\bn}^b(v)\compl}=0$
whenever $\tb\in\cC_{\bm+\bn}^b(v)$.  Therefore,
\begin{align}\lbeq{3rd-ind-fact}
\refeq{3rd-ind-prefact}~=\sum_{\substack{\bd\bm=\vno\\ \bd\bn=v\vtri\bb}}
 \frac{w_{\cA\compl}(\bm)}{Z_{\cA\compl}}\,\frac{w_\Lambda(\bn)}{Z_\Lambda}
 \,\indic{E_{\bm+\bn}(v,\bb;\cA)}\,\tau_b\,\Exp{\varphi_{\tb}\varphi_x}_{
 \cC_{\bm+\bn}^b(v)\compl}.
\end{align}
By \refeq{Theta-def}--\refeq{3rd-ind-fact}, we arrive at
\begin{align}\lbeq{2nd-exp}
\Exp{\varphi_v\varphi_x}_\Lambda-\Exp{\varphi_v\varphi_x}_{\cA\compl}=
 \Theta_{v,x;\cA}&+\sum_{b\in\mB_\Lambda}\Theta_{v,\bb;\cA}\,\tau_b\,
 \Exp{\varphi_{\tb}\varphi_x}_\Lambda\nn\\
&-\sum_{b\in\mB_\Lambda}\Theta_{v,\bb;\cA}\Big[\tau_b\Big(\Exp{\varphi_{\tb}
 \varphi_x}_\Lambda-\Exp{\varphi_{\tb}\varphi_x}_{\cC^b(v)
 \compl}\Big)\Big],
\end{align}
where $\cC^b(v)\equiv\cC_{\bm+\bn}^b(v)$ is a variable for the
operation $\Theta_{v,\bb;\cA}$.  This completes the second stage of
the expansion.

\subsubsection{Completion of the lace expansion}\label{sss:complexp}
For notational convenience, we define
$w_\vno(\bm)/Z_\vno=\tind{\bm\equiv0}$. Since
$E_\bn(o,x;\Lambda)=\{o\db{\bn}{}x\}$ (cf., \refeq{E-def}), we can
write
\begin{align}\lbeq{pi0-rewr}
\pi_\Lambda^{\sss(0)}(x)=\Theta_{o,x;\Lambda}.
\end{align}
Also, we can write $R_\Lambda^{\sss(1)}(x)$ in \refeq{R1-def} as
\begin{align}\lbeq{R1-rewr}
R_\Lambda^{\sss(1)}(x)=\sum_b\Theta_{o,\bb;\Lambda}\Big[\tau_b\Big(\Exp{
 \varphi_{\tb}\varphi_x}_\Lambda-\Exp{\varphi_{\tb}\varphi_x}_{\cC^b
 (o)\compl}\Big)\Big].
\end{align}
Using \refeq{2nd-exp}, we obtain
\begin{align}\lbeq{R1R2}
R_\Lambda^{\sss(1)}(x)=\sum_b\bigg(&\Theta_{o,\bb;\Lambda}\Big[\tau_b\,
 \Theta_{\tb,x;\cC^b(o)}\Big]+\sum_{b'}\Theta_{o,\bb;\Lambda}\Big[
 \tau_b\,\Theta_{\tb,\bb';\cC^b(o)}\Big]\,\tau_{b'}\Exp{\varphi_{
 \tbps}\varphi_x}_\Lambda\nn\\
&-\sum_{b'}\Theta_{o,\bb;\Lambda}\Big[\tau_b\,\Theta_{\tb,\bb';\cC^b
 (o)}\Big[\tau_{b'}\Big(\Exp{\varphi_{\tbps}\varphi_x}_\Lambda-\Exp{
 \varphi_{\tbps}\varphi_x}_{\cC^{b'}(\tb)\compl}\Big)\Big]
 \Big]\bigg),
\end{align}
where $\cC^b(o)\equiv\cC_\bn^b(o)$ is a variable for the outer operation
$\Theta_{o,\bb;\Lambda}$, and $\cC^{b'}(\tb)\equiv\cC_{\bm'+\bn'}^{b'}(\tb)$
is a variable for the inner operation $\Theta_{\tb,\bb';\cC^b(o)}$.  For
$j\ge1$, we define
\begin{align}
\pi_\Lambda^{\sss(j)}(x)&=\sum_{b_1,\dots,b_j}\Theta^{\sss(0)}_{o,\bb_1;
 \Lambda}\Big[\tau_{b_1}\Theta^{\sss(1)}_{\tb_1,\bb_2;\tilde\cC_0}\Big[
 \cdots\tau_{b_{j-1}}\Theta^{\sss(j-1)}_{\tb_{j-1},\bb_j;\tilde\cC_{j-2}}
 \Big[\tau_{b_j}\Theta^{\sss(j)}_{\tb_j,x;\tilde\cC_{j-1}}\Big]\cdots\Big]
 \Big],\lbeq{pij-def}\\
R_\Lambda^{\sss(j)}(x)&=\sum_{b_1,\dots,b_j}\Theta^{\sss(0)}_{o,\bb_1;
 \Lambda}\Big[\tau_{b_1}\Theta^{\sss(1)}_{\tb_1,\bb_2;\tilde\cC_0}\Big[
 \cdots\tau_{b_{j-1}}\Theta^{\sss(j-1)}_{\tb_{j-1},\bb_j;\tilde\cC_{j-2}}
 \Big[\tau_{b_j}\Big(\Exp{\varphi_{\tb_j}\varphi_x}_\Lambda-\Exp{\varphi_{
 \tb_j}\varphi_x}_{\tilde\cC_{j-1}\compl}\Big)\Big]\cdots\Big]\Big],
 \lbeq{Rj-def}
\end{align}
where the operation $\Theta^{\sss(i)}$ determines the variable
$\tilde\cC_i=\cC_{\bm_i+\bn_i}^{b_{i+1}}(\tb_i)$ (provided that
$\tb_0=o$).  Then, we can rewrite \refeq{R1R2} as
\begin{align}\lbeq{R1R2-rewr}
R_\Lambda^{\sss(1)}(x)=\pi_\Lambda^{\sss(1)}(x)+\sum_{b'}\pi_\Lambda
 ^{\sss(1)}(\bb')\,\tau_{b'}\Exp{\varphi_{\tbps}\varphi_x}_\Lambda
 -R_\Lambda^{\sss(2)}(x).
\end{align}
As a result,
\begin{align}
\Exp{\varphi_o\varphi_x}_\Lambda=\big(\pi_\Lambda^{\sss(0)}(x)-\pi_\Lambda
 ^{\sss(1)}(x)\big)+\sum_b\big(\pi_\Lambda^{\sss(0)}(\bb)-\pi_\Lambda^{\sss
 (1)}(\bb)\big)\,\tau_b\,\Exp{\varphi_{\tb}\varphi_x}_\Lambda+R_\Lambda^{
 \sss(2)}(x).
\end{align}
By repeated applications of \refeq{2nd-exp} to the remainder
$R_\Lambda^{\sss(j)}(x)$, we obtain \refeq{Ising-lace}--\refeq{Pij-def}
in Proposition~\ref{prp:Ising-lace}.

For the ferromagnetic case, $\tau_b$ and $w_{\cA}(\bn)$ for any
$\cA\subset\Lambda$ and $\bn\in\Zp^{\mB_\cA}$ are nonnegative. This
proves the first inequality in \refeq{pij-Rj-naivebd} and, with the
help of Proposition~\ref{prp:through}, the nonnegativity of
$R_\Lambda^{\sss(j+1)}(x)$ .  To prove the upper bound on
$R_\Lambda^{\sss(j+1)}(x)$, we simply ignore
$\Exp{\varphi_{\tb_j}\varphi_x}_{\tilde\cC_{j-1}\compl}$ in
\refeq{Rj-def} and replace $j$ by $j+1$, where $b_{j+1}=\{u,v\}$.
This completes the proof of Proposition~\ref{prp:Ising-lace}.
\qed

\subsection{Comparison to percolation}\label{ss:percolation}
Since we have exploited the underlying percolation picture to derive
the lace expansion \refeq{Ising-lace} for the Ising model, it is not
so surprising that the expansion coefficients \refeq{pi0-rewr} and
\refeq{pij-def} (also recall \refeq{Theta-def}) are quite similar to
the lace-expansion coefficients for unoriented bond-percolation (cf.,
\cite{hs90'}):
\begin{align}\lbeq{pij-perc}
\pi_p^{\sss(j)}(x)=
\begin{cases}
~\dpst\mE_p^{\sss(0)}\big[\ind{o\db{\bn_0}{}x}\big]\equiv\mP_p(o\db{}{}x)
 &(j=0),\\[1pc]
\dpst\sum_{b_1,\dots,b_j}\mE_p^{\sss(0)}\Big[\ind{o\db{\bn_0}{}\bb_1}\,
 p_{b_1}\mE_p^{\sss(1)}\Big[\indic{E_{\bn_1}(\tb_1,\bb_2;\tilde\cC_0)}
 \cdots p_{b_j}\mE_p^{\sss(j)}\Big[\indic{E_{\bn_j}(\tb_j,x;\tilde
 \cC_{j-1})}\Big]\cdots\Big]\Big]&(j\ge1),
\end{cases}
\end{align}
where $p\equiv\sum_xp_{o,x}$ is the bond-occupation parameter, and each
$\mE_p^{\sss(i)}$ denotes the expectation with respect to the product
measure $\prod_b(p_b\tind{\bn_i|_b=1}+(1-p_b)\tind{\bn_i|_b=0})$.  In
particular, the events involved in \refeq{pi0-rewr} and
\refeq{pij-def} are identical to those in \refeq{pij-perc}.

Hoever, there are significant differences between these two models.  The
major differences are the following:
\begin{enumerate}[(a)]
\item
Each current configuration must satisfy not only the conditions
in the indicators, but also its source constraint that is absent in
percolation.
\item
An operation $\Theta$ is not an expectation, since the source
constraints in the numerator and denominator of $\Theta$ in
\refeq{Theta-def} are different.
\item
In each $\Theta^{\sss(i)}$ for $i\ge1$, the sum $\bm_i+\bn_i$ of
two current configurations is coupled with $\bm_{i-1}+\bn_{i-1}$ via
the cluster $\tilde\cC_{i-1}$ determined by $\bm_{i-1}+\bn_{i-1}$.
By contrast, in each $\mE_p^{\sss(i)}$ in \refeq{pij-perc}, a single
percolation configuration $\bn_i$ is coupled with $\bn_{i-1}$ via
$\tilde\cC_{i-1}=\cC_{\bn_{i-1}}^{b_i}(\tb_{i-1})$.  In addition,
$\bm_i$ is nonzero only on bonds in $\mB_{\tilde\cC_{i-1}\compl}$,
while the current configuration $\bn_i$ has no such restriction.
\end{enumerate}

These elements are responsible for the difference in the method of bounding diagrams for
the expansion coefficients.  Take the $0^\text{th}$-expansion coefficient for example.
For percolation, the BK inequality simply tells us that
\begin{align}\lbeq{pi0perc-comp}
\pi_p^{\sss(0)}(x)\leq\mP_p(o\cn{}{}x)^2.
\end{align}
For the ferromagnetic Ising model, on the other hand, we first recall
\refeq{pi0-def}, i.e.,
\begin{align}
\pi_\Lambda^{\sss(0)}(x)=\sum_{\bd\bn=o\vtri x}\frac{w_\Lambda(\bn)}
 {Z_\Lambda}\,\ind{o\db{\bn}{}x},
\end{align}
where $w_\Lambda(\bn)/Z_\Lambda\ge0$. Due to the indicator, every current configuration
$\bn\in\Zp^{\mB_\Lambda}$ that gives nonzero contribution has at least \emph{two
bond-disjoint} paths $\zeta_1,\zeta_2$ from $o$ to $x$ such that $n_b>0$ for all
$b\in\zeta_1\Dcup\zeta_2$.  Also, due to the source constraint, there should be at least
one path $\zeta$ from $o$ to $x$ such that $n_b$ is odd for all $b\in\zeta$. Suppose,
for example, that $\zeta=\zeta_1$ and that $n_b$ for $b\in\zeta_2$ are all
positive-even.  Since a positive-even integer can split into two odd integers, on the
labeled graph $\mG_\bn$ with $\bd\mG_\bn=o\vtri x$ (recall the notation introduced above
\refeq{Sbefore}) there are at least \emph{three edge-disjoint} paths from $o$ to $x$.
This observation naturally leads us to expect that
\begin{align}\lbeq{pi0-comp}
\pi_\Lambda^{\sss(0)}(x)\leq\Exp{\varphi_o\varphi_x}_\Lambda^3
\end{align}
holds for the ferromagnetic Ising model.  This naive argument to justify
\refeq{pi0-comp} will be made rigorous in Section~\ref{s:bounds} by taking
account of partition functions.

The higher-order expansion coefficients are more involved, due to the above
item~(c).  This will also be explained in detail in Section~\ref{s:bounds}.

\section{Bounds on $\Pi_\Lambda^{\sss(j)}(x)$ for the ferromagnetic models}
\label{s:reduction}
From now on, we restrict ourselves to the ferromagnetic models.  In this
section, we explain how to prove Proposition~\ref{prp:Pij-Rj-bd} assuming
a few other propositions
(Propositions~\ref{prp:GimpliesPix}--\ref{prp:exp-bootstrap} below).
These propositions are results of diagrammatic bounds on the expansion
coefficients in terms of two-point functions.  We will show these
diagrammatic bounds in Section~\ref{s:bounds}.

The strategy to prove Proposition~\ref{prp:Pij-Rj-bd} is model-independent,
and we follow the strategy in \cite{h05} for the nearest-neighbor model and
that in \cite{hhs03} for the spread-out model.  Since the latter is simpler,
we first explain the strategy for the spread-out model.  In the rest of
this paper, we will frequently use the notation
\begin{align}
\veee{x}=|x|\vee1.
\end{align}
We also emphasize that constants in the $O$-notation used below (e.g., $O(\theta_0)$ in
\refeq{pi-bd}) are independent of $\Lambda\subset\Zd$.

\subsection{Strategy for the spread-out model}
Using the diagrammatic bounds below in Section~\ref{s:bounds}, we will prove
in detail in Section~\ref{ss:proof-so} that the following proposition holds
for the spread-out model:

\begin{prp}\label{prp:GimpliesPix}
Let $J_{o,x}$ be the spread-out interaction.  Suppose that
\begin{align}\lbeq{IR-xbd}
\tau\leq2,&&
G(x)\leq\delta_{o,x}+\theta_0\veee{x}^{-q}
\end{align}
hold for some $\theta_0\in(0,\infty)$ and $q\in(\frac{d}2,d)$.
Then, for sufficiently small $\theta_0$ (with $\theta_0L^{d-q}$
being bounded away from zero) and any $\Lambda\subset\Zd$, we have
\begin{align}\lbeq{pi-bd}
\pi_\Lambda^{\sss(i)}(x)\leq
\begin{cases}
 O(\theta_0)^i\delta_{o,x}+O(\theta_0^3)\veee{x}^{-3q}&(i=0,1),\\
 O(\theta_0)^i\veee{x}^{-3q}&(i\ge2).
\end{cases}
\end{align}
\end{prp}

The exact value of the assumed upper bound on $\tau$ in \refeq{IR-xbd} is unimportant
and can be any finite number, as long as it is independent of $\theta_0$ and bigger than
the mean-field critical point 1.  We note that the exponent $3q$ in \refeq{pi-bd} is due
to \refeq{pi0-comp} (and diagrammatic bounds on the higher-expansion coefficients), and
is replaced by $2q$ with $q\in(\frac{2d}3,d)$ for percolation, due to, e.g.,
\refeq{pi0perc-comp}.

\begin{proof}[Sketch proof of Proposition~\ref{prp:Pij-Rj-bd} for the spread-out model]
We will show below that, at $p=\pc$,
\begin{align}\lbeq{IR-xbd-so}
\tau\leq2,&& G(x)\leq\delta_{o,x}+O(L^{-2+\epsilon})\veee{x}^{-(d-2)},
\end{align}
for some small $\eps>0$.  Since $\tau$ and $G(x)$ are nondecreasing and continuous in
$p\leq\pc$ for the ferromagnetic models, these bounds imply \refeq{IR-xbd} for all
$p\leq\pc$, with $\theta_0=cL^{-2+\eps}>0$ and $q=d-2$, where $q\in(\frac{d}2,d)$ if
$d>4$ and $\theta_0L^{d-q}=cL^\eps>0$.  Then, by Proposition~\ref{prp:GimpliesPix}, the
bound \refeq{pi-bd} with $\theta_0=O(L^{-2+\eps})$ and $q=d-2$ holds for $d>4$ and
$\theta_0\ll1$ (thus $L\gg1$).  Therefore, by \refeq{pij-Rj-naivebd} with
$\Exp{\varphi_v\varphi_x}_\Lambda\leq1$,
\begin{align}\lbeq{Rj-optSO}
0\leq R_\Lambda^{\sss(j+1)}(x)\leq\tau\sum_u\pi_\Lambda^{\sss(j)}(u)
 =O(\theta_0)^j\to0\qquad(j\uparrow\infty),
\end{align}
and by \refeq{Pij-def} for $j\ge0$,
\begin{align}\lbeq{Pij-optSO}
|\Pi_\Lambda^{\sss(j)}(x)-\delta_{o,x}|\leq O(\theta_0)\delta_{o,x}
 +\frac{O(\theta_0^2)}{\veee{x}^{3(d-2)}}=O(\theta_0)\delta_{o,x}+
 \frac{O(\theta_0^2)(1-\delta_{o,x})}{|x|^{d+2+\rho}},
\end{align}
where $\rho=2(d-4)$.  This completes the proof of Proposition~\ref{prp:Pij-Rj-bd} for
the spread-out model, assuming \refeq{IR-xbd-so} at $p=\pc$.

It thus remains to show the bounds in \refeq{IR-xbd-so} at $p=\pc$.  These bounds are
proved by adapting the model-independent bootstrapping argument in \cite{hhs03} (see the
proof of \cite[Proposition~2.2]{hhs03} for self-avoiding walk and percolation), together
with the fact that $G(x)$ decays exponentially as $|x|\uparrow\infty$ for every $p<\pc$
\cite{l80,s80} so that $\sup_xG(x)$ is continuous in $p<\pc$ \cite{s05}.  We complete
the proof.
\end{proof}

\subsection{Strategy for the nearest-neighbor model}
Since $\sigma^2=O(1)$ for short-range models, we cannot expect that
$\theta_0$ in \refeq{IR-xbd} is small, or that
Proposition~\ref{prp:GimpliesPix} is applicable to bound the
expansion coefficients in this setting.

Under this circumstance, we follow the strategy in \cite{h05}.
The following is the key proposition, whose proof will be
explained in Section~\ref{ss:proof-nn}:

\begin{prp}\label{prp:GimpliesPik}
Let $J_{o,x}$ be the nearest-neighbor or spread-out interaction,
and suppose that
\begin{align}\lbeq{IR-kbd}
\tau-1\leq\theta_0,&& \sup_x(D*G^{*2})(x)\leq\theta_0,&&
\sup_{\substack{x\equiv(x_1,\dots,x_d)\ne o\\ l=1,\dots,d}}
 \bigg(\frac{x_l^2}{\sigma^2}\vee1\bigg)G(x)\leq\theta_0
\end{align}
hold for some $\theta_0\in(0,\infty)$.  Then, for sufficiently small
$\theta_0$ and any $\Lambda\subset\Zd$, we have
\begin{align}\lbeq{pi-sumbd}
\sum_x\pi_\Lambda^{\sss(i)}(x)\leq
 \begin{cases}
 1+O(\theta_0^2)&(i=0),\\ O(\theta_0)^i&(i\ge1),
 \end{cases}&&
\sum_x|x|^2\pi_\Lambda^{\sss(i)}(x)\leq d\sigma^2(i+1)^2O(\theta_0)^{i
 \vee2}.
\end{align}
Furthermore, in addition to \refeq{IR-kbd} with $\theta_0\ll1$, if
\begin{align}\lbeq{IR-xbdNN}
G(x)\leq\lambda_0\veee{x}^{-q}
\end{align}
holds for some $\lambda_0\in[1,\infty)$ and $q\in(0,d)$, then we have
for $i\ge0$
\begin{align}\lbeq{pi-kbd}
\pi_\Lambda^{\sss(i)}(x)\leq O(\theta_0)^i\delta_{o,x}+\frac{\lambda_0^3
 (i+1)^{3q+2}O(\theta_0)^{(i-2)\vee0}}{|x|^{3q}}(1-\delta_{o,x}).
\end{align}
\end{prp}

\begin{proof}[Sketch proof of Proposition~\ref{prp:Pij-Rj-bd} (primarily) for
the nearest-neighbor model]~~ First we claim that the assumed bounds
in \refeq{IR-kbd} indeed hold for any $p\leq\pc$ if $d>4$ and
$\theta_0\ll1$, where $\theta_0=O(d^{-1})$ for the nearest-neighbor
model and $\theta_0=O(L^{-d})$ for the spread-out model.  The proof
is based on the orthodox model-independent bootstrapping argument
in, e.g., \cite{ms93} (see also \cite{hs02} for improved random-walk
estimates; bootstrapping assumptions that are different from, but
philosophically similar to, \refeq{IR-kbd} are used in \cite{hhs?}).
Therefore, \refeq{pi-sumbd} holds for $p\leq\pc$ and hence ensures
the existence of an infinite-volume limit
$\Pi(x)=\lim_{\Lambda\uparrow\Zd}\lim_{j\uparrow\infty}\Pi_\Lambda^{\sss
(j)}(x)$ that satisfies
\begin{align}\lbeq{Pi-bdNN}
\sum_x|\Pi(x)|=1+O(\theta_0),&& \sum_x|x|^2|\Pi(x)|=d\sigma^2O(\theta_0^2).
\end{align}
As a byproduct, we obtain the identity in \refeq{constants} for $\tau(\pc)$ for both
models.  Suppose that
\begin{align}\lbeq{IR-xbd-nn}
G(x)\leq\lambda_0\veee{x}^{-(d-2)}
\end{align}
holds at $p=\pc$.  Then, by Proposition~\ref{prp:GimpliesPik}, we obtain \refeq{pi-kbd}
with $q=d-2$.  Using this in \refeq{Rj-optSO}--\refeq{Pij-optSO}, we can prove
Proposition~\ref{prp:Pij-Rj-bd}.

To complete the proof, it thus remains to show \refeq{IR-xbd-nn} at $p=\pc$.  To show
this, we use the following proposition:

\begin{prp}\label{prp:exp-bootstrap}
Let
\begin{align}\lbeq{GbarWbar}
\bar G^{\sss(s)}=\sup_x|x|^sG(x),&& \bar
W^{\sss(t)}=\sup_x\sum_y|y|^tG(y)\,G(x-y),
\end{align}
and suppose that the bounds in \refeq{IR-kbd} hold with $\theta_0\ll1$.
\begin{enumerate}[(i)]
\item
If $\sum_x\Pi(x)=\tau^{-1}$ and $|\Pi(x)|\leq O(\veee{x}^{-(d+2)})$, then we have
\begin{align}\lbeq{prp-asy}
G(x)\sim\frac{\sum_x\Pi(x)}{\tau\sum_x|x|^2(D*\Pi)(x)}\,\frac{a_d}
 {|x|^{d-2}}\qquad\text{as }|x|\uparrow\infty.
\end{align}
\item
If $\sum_x|x|^r|\Pi(x)|<\infty$ for some $r>0$, then, for $s,t>0$ which are not odd
integers, we have
\begin{align}
\begin{cases}
\bar G^{\sss(s)}<\infty&\text{if}~~~s\leq r~~\text{and}~~s<d-2,\\
\bar W^{\sss(t)}<\infty&\text{if}~~~t\leq\lfloor r\rfloor~~\text{and}
 ~~t<d-4.
\end{cases}
\end{align}
\item
If $\bar W^{\sss(t)}<\infty$ for some $t\ge0$, then
$\sum_x|x|^{t+2}|\Pi(x)|<\infty$.
\end{enumerate}
\end{prp}

The above proposition is a summary of key elements in \cite[Proposition~1.3 and
Lemmas~1.5--1.6]{h05} that are sufficient to prove \refeq{IR-xbd-nn} in the current
setting.  The proofs of Propositions~\ref{prp:exp-bootstrap}(i) and
\ref{prp:exp-bootstrap}(ii) are model-independent and can be found in \cite[Sections~2
and 4]{h05}, respectively.  The proof of Proposition~\ref{prp:exp-bootstrap}(iii) is
similar to that of the first statement of Proposition~\ref{prp:GimpliesPik}:
\refeq{IR-kbd} implies \refeq{pi-sumbd}.  We will explain this in
Section~\ref{ss:proof-nn}.

Now we continue with the proof of \refeq{IR-xbd-nn}.  Fix $p=\pc$.  Since the asymptotic
behavior \refeq{prp-asy} is good enough for the bound \refeq{IR-xbd-nn}, it suffices to
check the assumptions of Proposition~\ref{prp:exp-bootstrap}(i).  The first assumption
on the sum of $\Pi(x)$ is satisfied at $p=\pc$, as mentioned below \refeq{Pi-bdNN}.  The
second assumption is also satisfied if $\bar G^{({\sss\frac{d+2}3})}<\infty$, because of
the second statement of Proposition~\ref{prp:GimpliesPik}: \refeq{IR-xbdNN} implies
\refeq{pi-kbd}.  By Proposition~\ref{prp:exp-bootstrap}(ii), it thus suffices to show
that $\sum_x|x|^{\sss\frac{d+2}3}|\Pi(x)|$ is finite if $d>4$.

To show this, we let
\begin{align}
r_0=2,&& r_{i+1}=\Big((d-2)\wedge\big(\lfloor r_i\rfloor+2\big)
 \Big)-\eps,
\end{align}
where $0<\eps\leq\frac{2}3(d-4)$.  Note that, by this definition, $r_i$ for $i\ge1$
equals $((d-2)\wedge(i+3))-\eps$ and increases until it reaches $d-2-\eps$.  We prove
below by induction that $\sum_x|x|^{r_i}|\Pi(x)|$ is finite for all $i\ge0$. This is
sufficient for the finiteness of $\sum_x|x|^{\sss\frac{d+2}3}|\Pi(x)|$, since
\begin{align}
\lim_{i\uparrow\infty}r_i=d-2-\eps\ge d-2-\tfrac{2}3(d-4)
 =\tfrac{d+2}3.
\end{align}

Note that, by \refeq{Pi-bdNN}, $\sum_x|x|^{r_0}|\Pi(x)|<\infty$.  Suppose
$\sum_x|x|^{r_i}|\Pi(x)|<\infty$ for some $i\ge0$.  Then, by
Proposition~\ref{prp:exp-bootstrap}(ii), $\bar W^{\sss(t)}$ is finite for
$t\in(0,\lfloor r_i\rfloor]\cap(0,d-4)$.  Since $\lfloor r_0\rfloor=2$ and $\lfloor
r_i\rfloor=(d-3)\wedge(i+2)$ for $i\ge1$, $\bar W^{\sss(T)}$ with
$T=(i+2)\wedge(d-4-\eps)$ is finite.  Then, by Proposition~\ref{prp:exp-bootstrap}(iii),
$\sum_x|x|^{T+2}|\Pi(x)|$ is finite.  Since
\begin{align}
T+2=(i+4)\wedge(d-2-\eps)\ge\big((d-2)\wedge(i+4)\big)-\eps
 =r_{i+1},
\end{align}
we obtain that $\sum_x|x|^{r_{i+1}}|\Pi(x)|<\infty$.  This completes the induction and
the proof of \refeq{IR-xbd-nn}.  The proof of Proposition~\ref{prp:Pij-Rj-bd} is now
completed.
\end{proof}

\section{Diagrammatic bounds on $\pi_\Lambda^{\sss(j)}(x)$}\label{s:bounds}
In this section, we prove diagrammatic bounds on the expansion coefficients.
In Section~\ref{ss:diagram}, we construct diagrams in terms of two-point
functions and state the bounds.  In Section~\ref{ss:pi0bd}, we prove a key
lemma for the diagrammatic bounds and show how to apply this lemma to prove
the bound on $\pi_\Lambda^{\sss(0)}(x)$.  In Section~\ref{ss:pijbd}, we
prove the bounds on $\pi_\Lambda^{\sss(j)}(x)$ for $j\ge1$.

\subsection{Construction of diagrams}\label{ss:diagram}
\begin{figure}[t]
\begin{center}
\begin{gather*}
P_\Lambda^{\sss(1)}(v_1,v'_1)=\raisebox{-7pt}{\includegraphics[scale=.1]
 {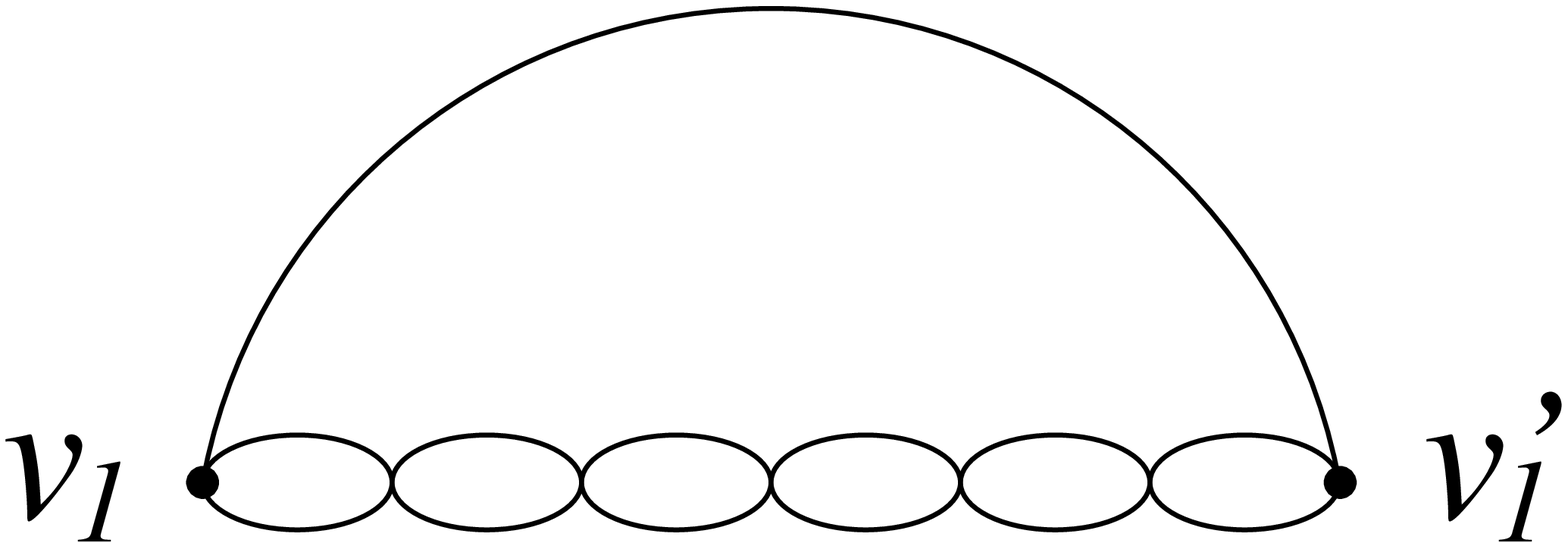}}\qquad
P_\Lambda^{\sss(2)}(v_1,v'_2)=\raisebox{-9pt}{\includegraphics[scale=.1]
 {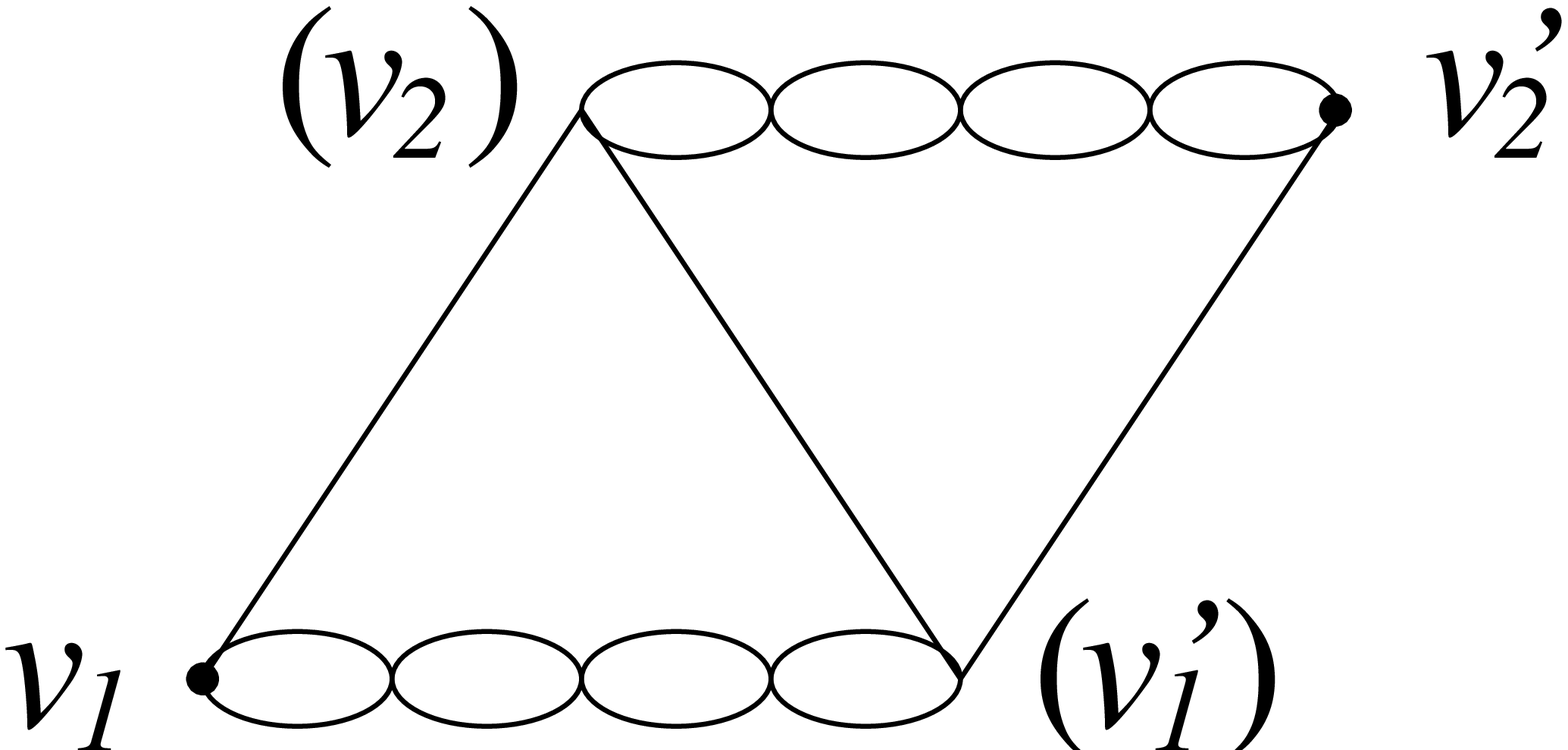}}\qquad
P_\Lambda^{\sss(3)}(v_1,v'_3)=\raisebox{-15pt}{\includegraphics[scale=.1]
 {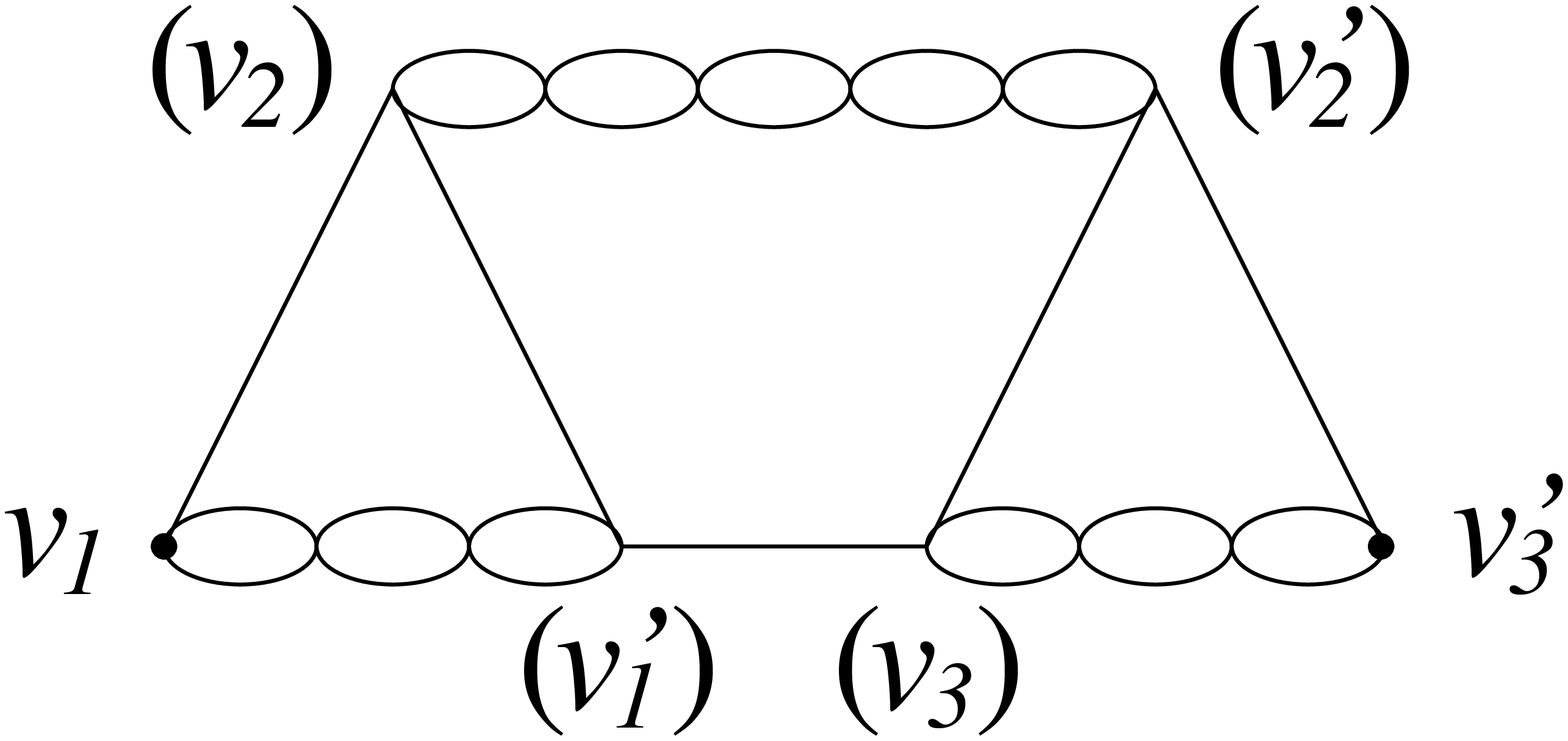}}\\[5pt]
P_{\Lambda;u}^{\prime{\sss(1)}}(v_1,v'_1)=\raisebox{-7pt}{\includegraphics
 [scale=.1]{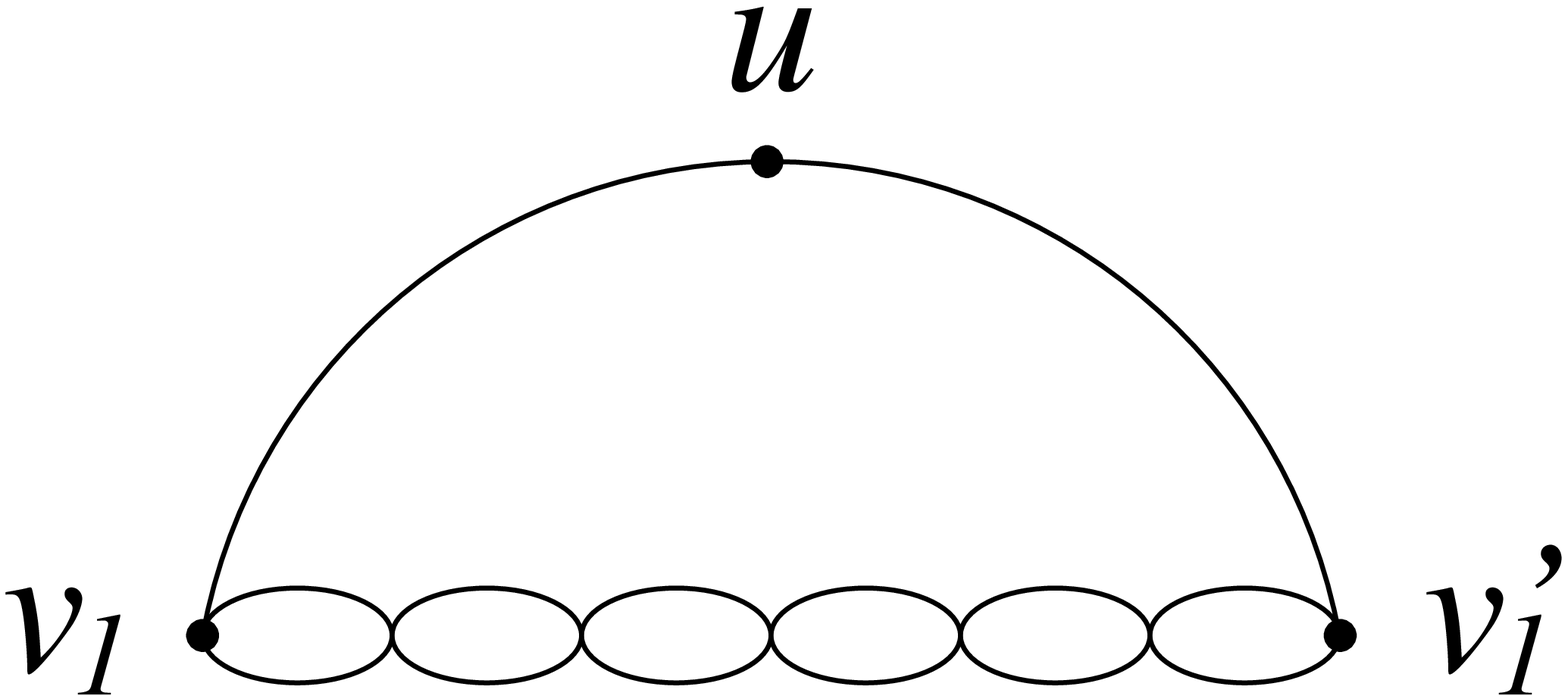}}\hspace{5pc}
P_{\Lambda;u,v}^{\prime\prime{\sss(1)}}(v_1,v'_1)=\raisebox{-18pt}
 {\includegraphics[scale=.1]{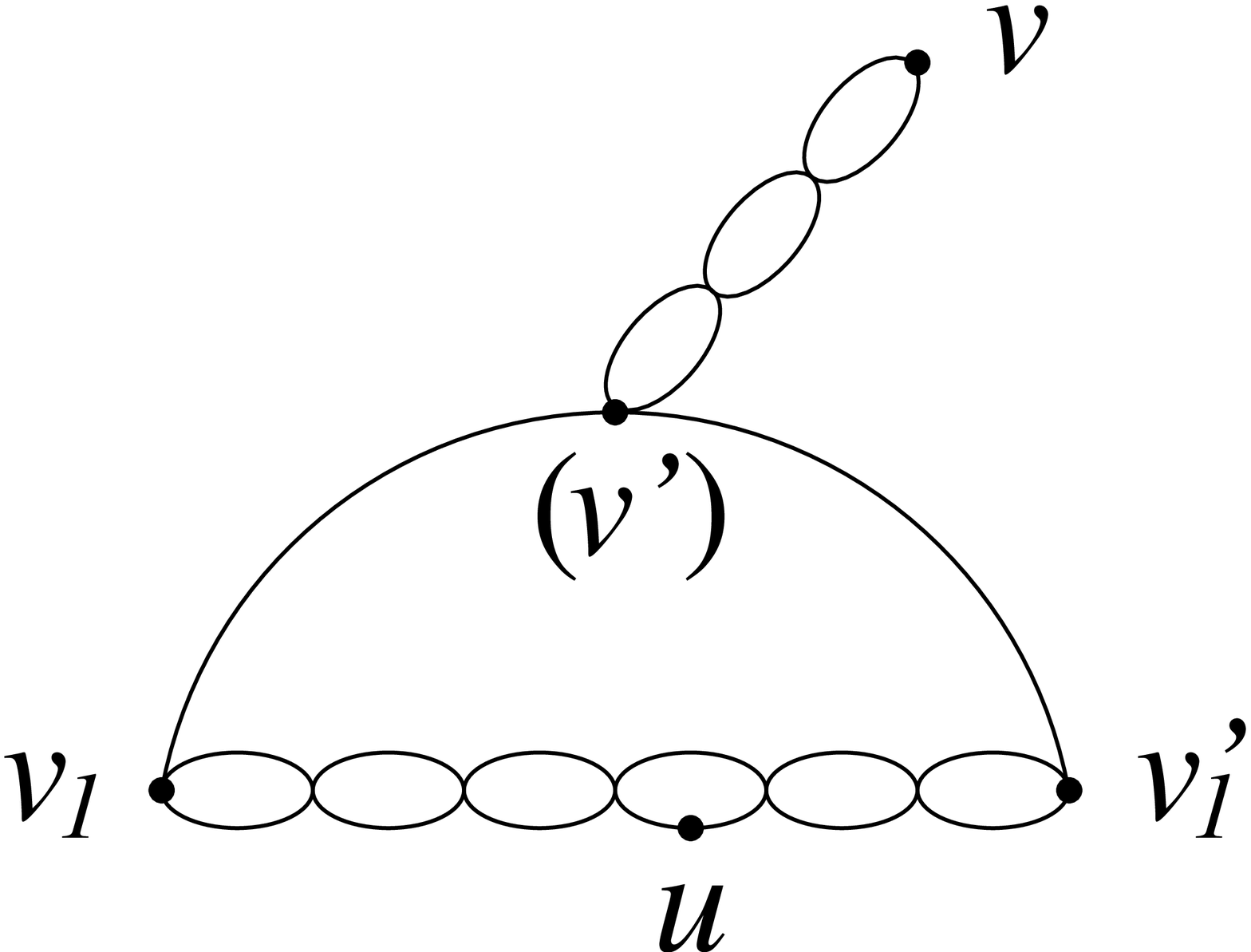}}+~
 \raisebox{-18pt}{\includegraphics[scale=.1]{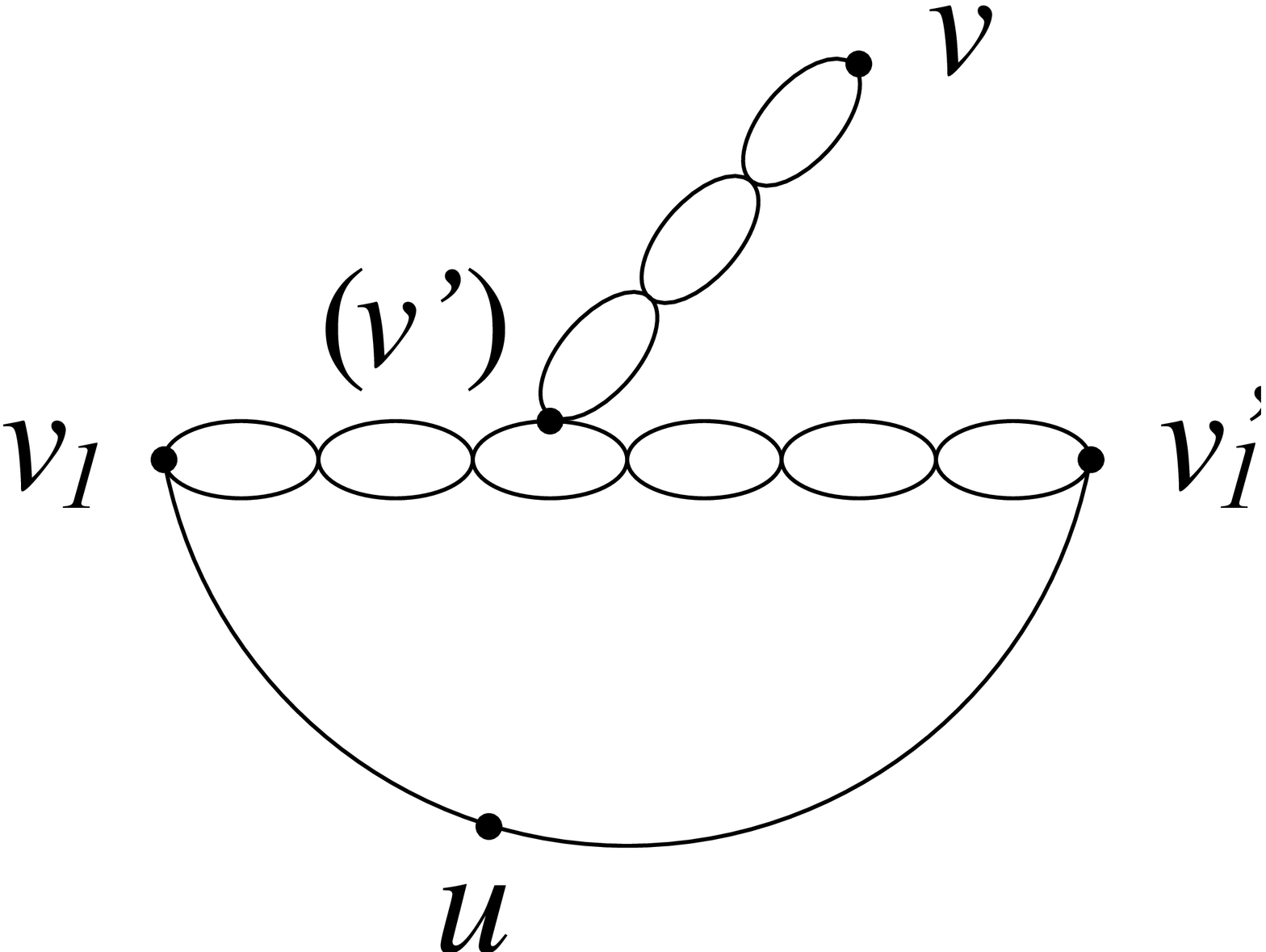}}\\[5pt]
P_{\Lambda;u}^{\prime{\sss(0)}}(y,x)=\raisebox{-12pt}{\includegraphics
 [scale=.1]{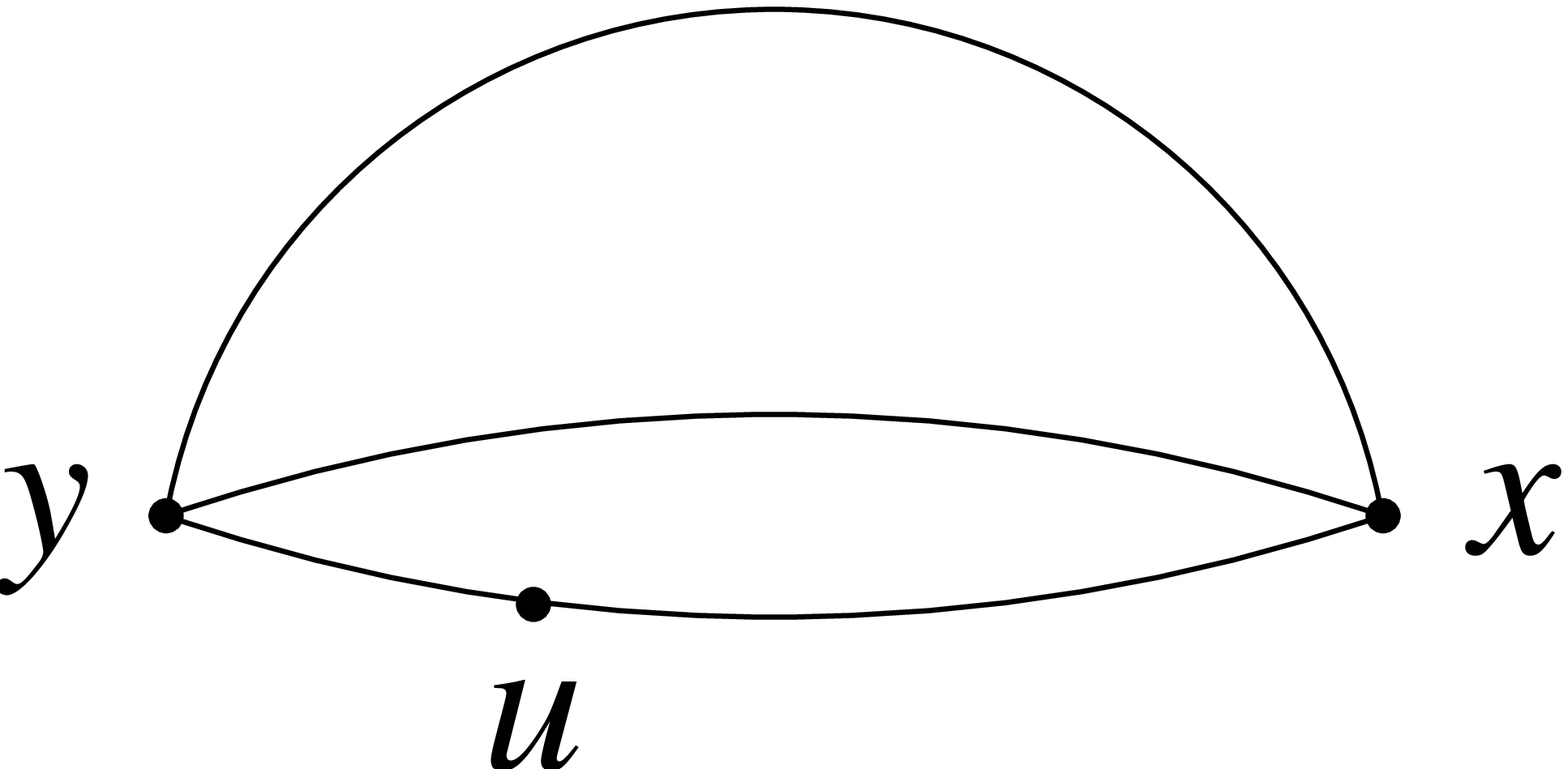}}\hspace{5pc}
P_{\Lambda;u,v}^{\prime\prime{\sss(0)}}(y,x)=\raisebox{-18pt}
 {\includegraphics[scale=.1]{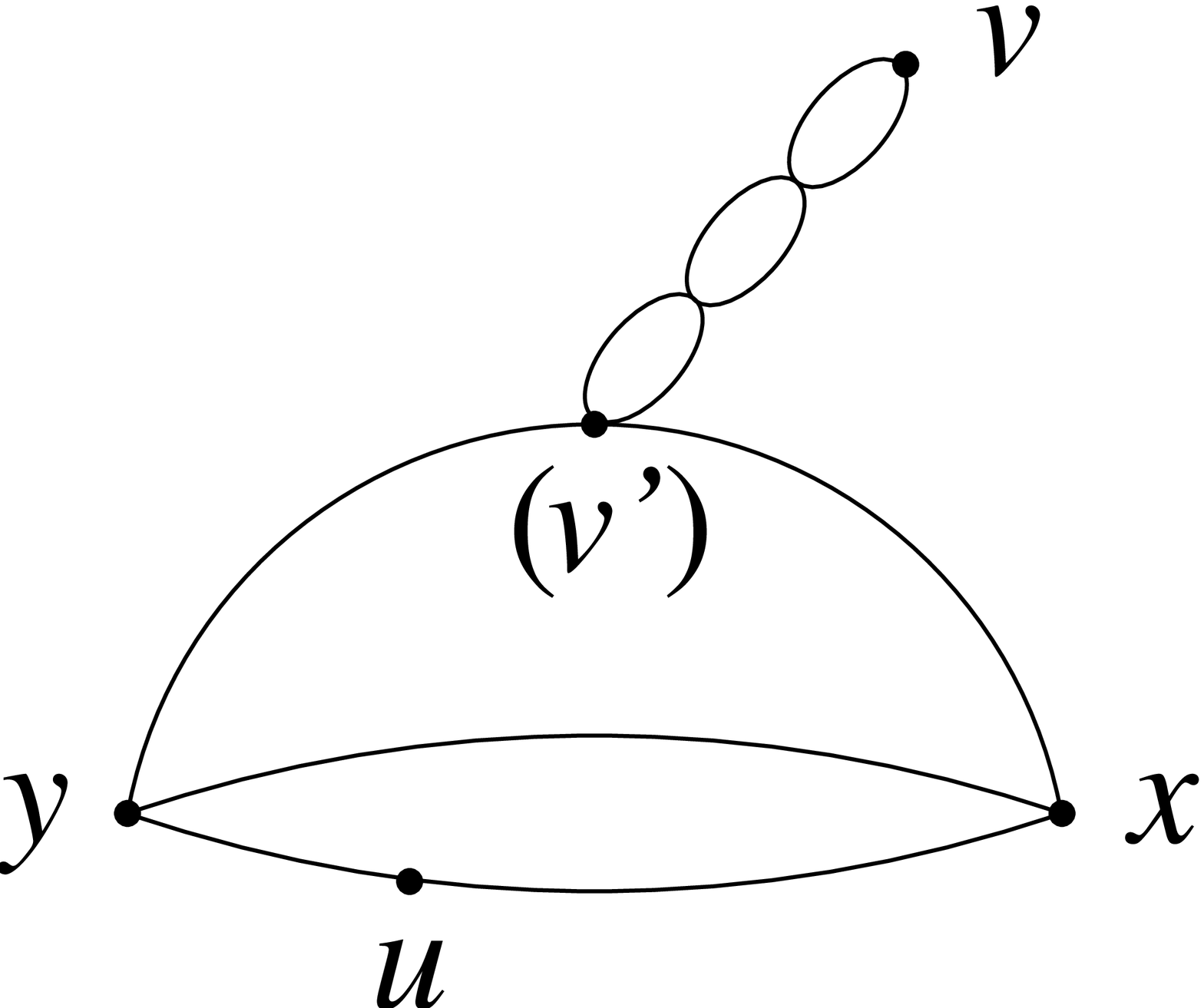}}
\end{gather*}
\caption{\label{fig:P-def}Schematic representations of
$P_\Lambda^{\sss(j)}(v_1,v'_j)$ for $j=1,2,3$,
$P_{\Lambda;u}^{\prime{\sss(1)}}(v_1,v'_1)$,
$P_{\Lambda;u,v}^{\prime\prime{\sss(1)}}(v_1,v'_1)$,
$P_{\Lambda;u}^{\prime{\sss(0)}}(y,x)$ and
$P_{\Lambda;u,v}^{\prime\prime{\sss(0)}}(y,x)$.  The labels in the
parentheses represent vertices that are summed over, each sequence
of bubbles from $v_i$ and $v'_i$ represents
$\psi_\Lambda(v_i,v'_i)-\delta_{v_i,v'_i}$, and the sequence of
bubbles from $v'$ to $v$ represents $\psi_\Lambda(v',v)$.}
\end{center}
\end{figure}

To state bounds on the expansion coefficients (as in
Proposition~\ref{prp:diagram-bd} below), we first define
diagrammatic functions consisting of two-point functions.  Let
\begin{align}\lbeq{tildeG-def}
\tilde G_\Lambda(y,x)
 =\sum_{b:\tb=x}\Exp{\varphi_y\varphi_{\bb}}_\Lambda\tau_b,
\end{align}
which satisfies\footnote{Repeated applications of \refeq{G-delta-bd} to the
translation-invariant models result in the random-walk bound:
$\Exp{\varphi_o\varphi_x}_\Lambda\leq S_\tau(x)$ for $\Lambda\subset\Zd$ and
$\tau\leq1$.}
\begin{align}\lbeq{G-delta-bd}
\Exp{\varphi_y\varphi_x}_\Lambda\leq\delta_{y,x}+\sum_{b:\tb=x}\,
 \sum_{\substack{\bd\bn=y\vtri x\\ n_b\text{ odd}}}\frac{w_\Lambda
 (\bn)}{Z_\Lambda}=\delta_{y,x}+\sum_{b:\tb=x}\tau_b\sum_{\substack{
 \bd\bn=y\vtri\bb\\ n_b\text{ even}}}\frac{w_\Lambda(\bn)}{Z_\Lambda}
 \leq\delta_{y,x}+\tilde G_\Lambda(y,x).
\end{align}
Let
\begin{align}\lbeq{psi-def}
\psi_\Lambda(y,x)=\sum_{j=0}^\infty\big(\tilde G_\Lambda^2\big)^{*j}(y,x)
 &\equiv\delta_{y,x}+\sum_{j=1}^\infty\sum_{\substack{u_0,\dots,u_j\\ u_0=
 y,\;u_j=x}}\prod_{l=1}^j\tilde G_\Lambda(u_{l-1},u_l)^2,
\end{align}
and define (see the first line in Figure~\ref{fig:P-def})
\begin{align}
P_\Lambda^{\sss(1)}(v_1,v'_1)&=2\big(\psi_\Lambda(v_1,v'_1)-
 \delta_{v_1,v'_1}\big)\,\Exp{\varphi_{v_1}\varphi_{v'_1}}_\Lambda,
 \lbeq{P1-def}\\[5pt]
P_\Lambda^{\sss(j)}(v_1,v'_j)&=\sum_{\substack{v_2,\dots,v_j\\
 v'_1,\dots,v'_{j-1}}}\bigg(\prod_{i=1}^j\big(\psi_\Lambda(v_i,
 v'_i)-\delta_{v_i,v'_i}\big)\bigg)\Exp{\varphi_{v_1}\varphi_{
 v_2}}_\Lambda\Exp{\varphi_{v_2}\varphi_{v'_1}}_\Lambda\nn\\
&\qquad\qquad\times\bigg(\prod_{i=2}^{j-1}\Exp{\varphi_{v'_{i
 -1}}\varphi_{v_{i+1}}}_\Lambda\Exp{\varphi_{v_{i+1}}\varphi_{
 v'_i}}_\Lambda\bigg)\Exp{\varphi_{v'_{j-1}}\varphi_{v'_j}}
 _\Lambda\qquad(j\ge2),\lbeq{Pj-def}
\end{align}
where the empty product for $j=2$ is regarded as 1.

Next, we define $P_{\Lambda;u}^{\prime{\sss(j)}}(v_1,v'_j)$ by
replacing one of the $2j-1$ two-point functions on the right-hand
side of \refeq{P1-def}--\refeq{Pj-def} by the product of \emph{two}
two-point functions, such as replacing
$\Exp{\varphi_z\varphi_{z'}}_\Lambda$ by
$\Exp{\varphi_z\varphi_u}_\Lambda\Exp{\varphi_u\varphi_{z'}}_\Lambda$,
and then summing over all $2j-1$ choices of this replacement. For
example, we define (see the second line in Figure~\ref{fig:P-def})
\begin{align}\lbeq{P'1-def}
P_{\Lambda;u}^{\prime{\sss(1)}}(v_1,v'_1)=2\big(\psi_\Lambda(v_1,v'_1)-
 \delta_{v_1,v'_1}\big)\Exp{\varphi_{v_1}\varphi_u}_\Lambda\Exp{\varphi_u
 \varphi_{v'_1}}_\Lambda,
\end{align}
and
\begin{align}
P_{\Lambda;u}^{\prime{\sss(2)}}(v_1,v'_2)=\sum_{v_2,v'_1}\bigg(\prod_{i
 =1}^2\big(\psi_\Lambda(v_i,v'_i)-\delta_{v_i,v'_i}\big)\bigg)\Big(\Exp{
 \varphi_{v_1}\varphi_u}_\Lambda\Exp{\varphi_u\varphi_{v_2}}_\Lambda
 \Exp{\varphi_{v_2}\varphi_{v'_1}}_\Lambda\Exp{\varphi_{v'_1}\varphi_{
 v'_2}}_\Lambda&\nn\\
+\Exp{\varphi_{v_1}\varphi_{v_2}}_\Lambda\Exp{\varphi_{v_2}\varphi_u
 }_\Lambda\Exp{\varphi_u\varphi_{v'_1}}_\Lambda\Exp{\varphi_{v'_1}
 \varphi_{v'_2}}_\Lambda&\nn\\[7pt]
+\Exp{\varphi_{v_1}\varphi_{v_2}}_\Lambda\Exp{\varphi_{v_2}\varphi_{
 v'_1}}_\Lambda\Exp{\varphi_{v'_1}\varphi_u}_\Lambda\Exp{\varphi_u
 \varphi_{v'_2}}_\Lambda&\Big).
\end{align}

We define $P_{\Lambda;u,v}^{\prime\prime{\sss(j)}}(v_1,v'_j)$
similarly as follows. First we take \emph{two} two-point functions
in $P_\Lambda^{\sss(j)}(v_1,v'_j)$, one of which (say,
$\Exp{\varphi_{z_1}\varphi_{z'_1}}_\Lambda$ for some $z_1,z'_1$) is
among the aforementioned $2j-1$ two-point functions, and the other
(say, $\tilde G_\Lambda(z_2,z'_2)$ for some $z_2,z'_2$) is among
those of which $\psi_\Lambda(v_i,v'_i)-\delta_{v_i,v'_i}$ for
$i=1,\dots,j$ are composed.  The product
$\Exp{\varphi_{z_1}\varphi_{z'_1}}_\Lambda\tilde
G_\Lambda(z_2,z'_2)$ is then replaced by
\begin{align}
&\bigg(\sum_{v'}\Exp{\varphi_{z_1}\varphi_{v'}}_\Lambda\Exp{\varphi_{v'}
 \varphi_{z'_1}}_\Lambda\,\psi_\Lambda(v',v)\bigg)\Big(\Exp{\varphi_{z_2}
 \varphi_u}_\Lambda\tilde G_\Lambda(u,z'_2)+\tilde G_\Lambda(z_2,z'_2)
 \,\delta_{u,z'_2}\Big)\nn\\
&+\Exp{\varphi_{z_1}\varphi_u}_\Lambda\Exp{\varphi_u\varphi_{z'_1}
 }_\Lambda\sum_{v'}\Big(\Exp{\varphi_{z_2}\varphi_{v'}}_\Lambda\tilde
 G_\Lambda(v',z'_2)+\tilde G_\Lambda(z_2,z'_2)\,\delta_{v',z'_2}\Big)
 \,\psi_\Lambda(v',v).
\end{align}
Finally, we define $P_{\Lambda;u,v}^{\prime\prime{\sss(j)}}(v_1,v'_j)$ by taking account
of all possible combinations of $\Exp{\varphi_{z_1} \varphi_{z'_1}}_\Lambda$ and $\tilde
G_\Lambda(z_2,z'_2)$.  For example, we define
$P_{\Lambda;u,v}^{\prime\prime{\sss(1)}}(v_1,v'_1)$ as (see Figure~\ref{fig:P-def})
\begin{align}\lbeq{P''1-def}
P_{\Lambda;u,v}^{\prime\prime{\sss(1)}}(v_1,v'_1)\nn\\
=\sum_{u',u'',v'}\bigg(&2\psi_\Lambda(v_1,u')\,\tilde G_\Lambda(u',u'')
 \Big(\Exp{\varphi_{u'}\varphi_u}_\Lambda\tilde G_\Lambda(u,u'')+\tilde
 G_\Lambda(u',u'')\,\delta_{u,u''}\Big)\,\psi_\Lambda(u'',v'_1)\nn\\
&\times\Exp{\varphi_{v_1}\varphi_{v'}}_\Lambda\Exp{\varphi_{v'}\varphi_{
 v'_1}}_\Lambda\psi_\Lambda(v',v)+(\text{permutation of $u$ and }v')
 \bigg),
\end{align}
where the permutation term corresponds to the second term for
$P_{\Lambda;u,v}^{\prime\prime{\sss(1)}}(v_1,v'_1)$ in Figure~\ref{fig:P-def}.

In addition to the above quantities, we define (see the third line in
Figure~\ref{fig:P-def})
\begin{align}
P_{\Lambda;u}^{\prime{\sss(0)}}(y,x)&=\Exp{\varphi_y\varphi_x}_\Lambda^2
 \Exp{\varphi_y\varphi_u}_\Lambda\Exp{\varphi_u\varphi_x}_\Lambda,
 \lbeq{P'0-def}\\[5pt]
P_{\Lambda;u,v}^{\prime\prime{\sss(0)}}(y,x)&=\Exp{\varphi_y\varphi_x}
 _\Lambda\Exp{\varphi_y\varphi_u}_\Lambda\Exp{\varphi_u\varphi_x}
 _\Lambda\sum_{v'}\Exp{\varphi_y\varphi_{v'}}_\Lambda\Exp{\varphi_{v'}
 \varphi_x}_\Lambda\,\psi_\Lambda(v',v),\lbeq{P''0-def}
\end{align}
and let
\begin{align}\lbeq{P'P''-def}
P'_{\Lambda;u}(y,x)=\sum_{j\ge0}P_{\Lambda;u}^{\prime{\sss(j)}}(y,x),&&
P''_{\Lambda;u,v}(y,x)&=\sum_{j\ge0}P_{\Lambda;u,v}^{\prime\prime{\sss
 (j)}}(y,x),
\end{align}
where $P_{\Lambda;u}^{\prime{\sss(0)}}(y,x)$ and
$P_{\Lambda;u,v}^{\prime\prime{\sss(0)}}(y,x)$ are the leading
contributions to $P'_{\Lambda;u}(y,x)$ and $P''_{\Lambda;u,v}(y,x)$,
respectively.

Finally, we define
\begin{align}
Q'_{\Lambda;u}(y,x)&=\sum_z\big(\delta_{y,z}+\tilde G_\Lambda(y,z)\big)
 P'_{\Lambda;u}(z,x),\lbeq{Q'-def}\\
Q''_{\Lambda;u,v}(y,x)&=\sum_z\big(\delta_{y,z}+\tilde G_\Lambda(y,z)
 \big)P''_{\Lambda;u,v}(z,x)\nn\\
&\quad+\sum_{v',z}\big(\delta_{y,v'}+\tilde G_\Lambda(y,v')\big)\,\tilde
 G_\Lambda(v',z)\,P'_{\Lambda;u}(z,x)\,\psi_\Lambda(v',v).\lbeq{Q''-def}
\end{align}

\begin{figure}[t]
\begin{center}
\begin{align*}
\pi^{\sss(1)}_\Lambda(x)\lesssim\raisebox{-11pt}{\includegraphics[scale=
 0.18]{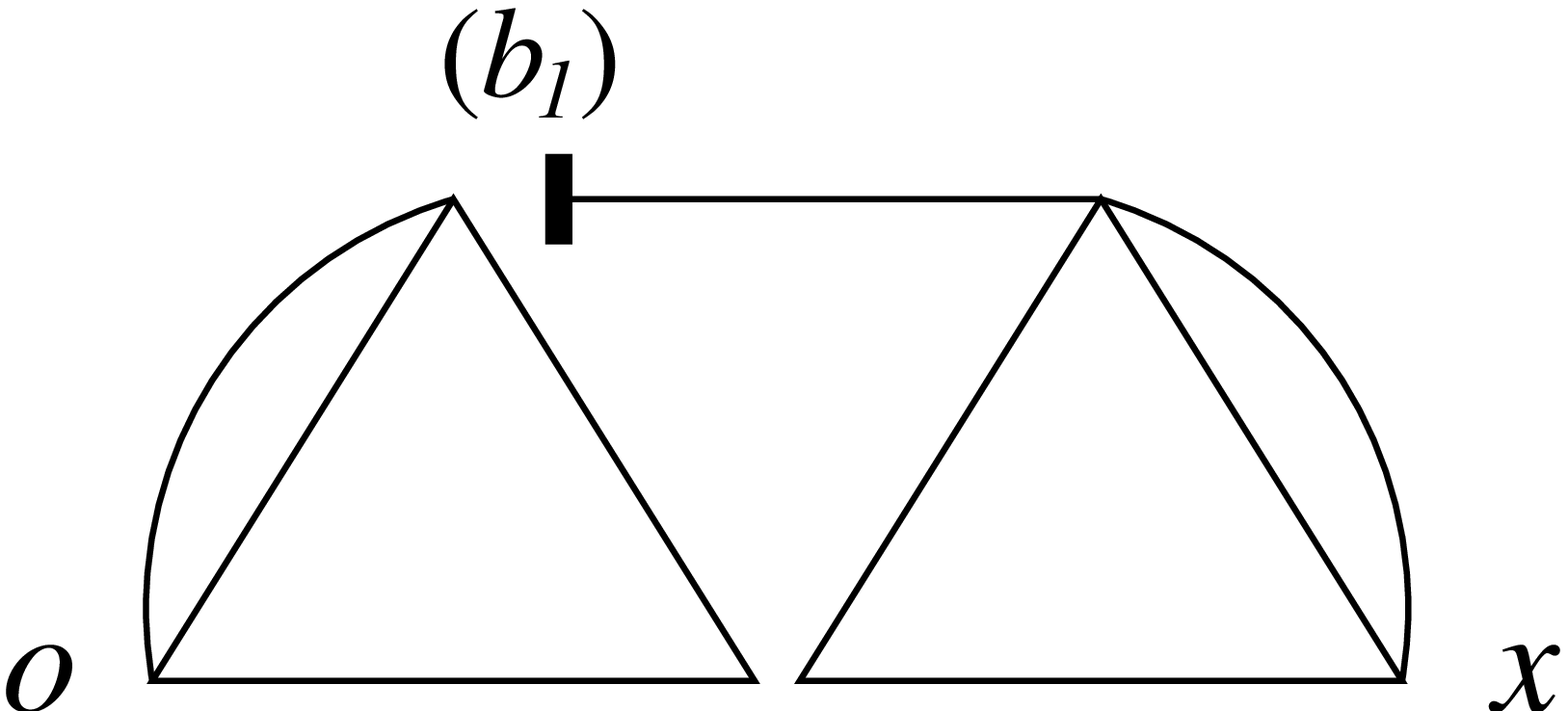}}\qquad
\pi^{\sss(2)}_\Lambda(x)\lesssim\raisebox{-20pt}{\includegraphics[scale=
 0.18]{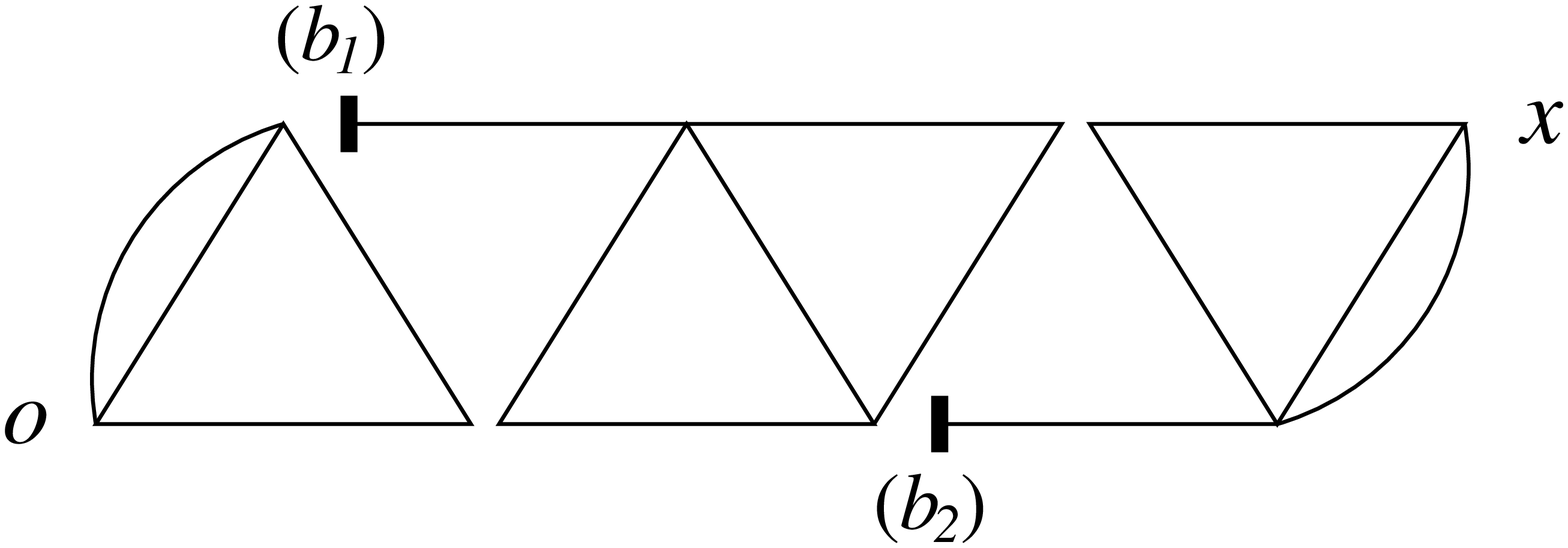}}+\raisebox{-20pt}{\includegraphics[scale=0.18]{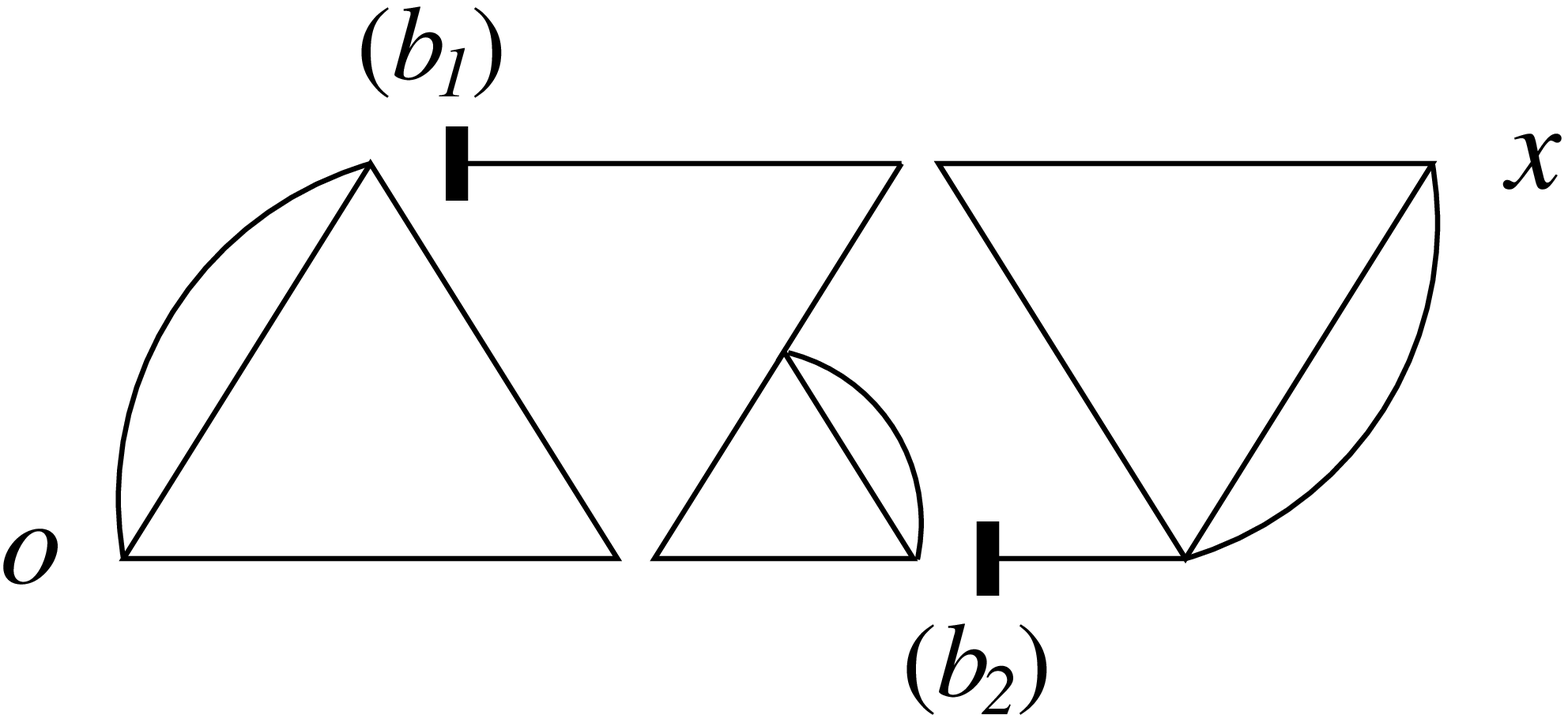}}
\end{align*}
\caption{\label{fig:piN-bd}The leading diagrams for
$\pi_\Lambda^{\sss(1)}(x)$ and $\pi_\Lambda^{\sss(2)}(x)$. The
segments that terminate with $b_i$ for $i=1,2$ represent
$\delta+\tilde G_\Lambda$ (cf., \refeq{Q'-def}--\refeq{Q''-def}).
The labels in the parentheses represent bonds that are summed over.
There are artificial gaps in the figures to distinguish different
building blocks.}
\end{center}
\end{figure}

The following are the diagrammatic bounds on the expansion coefficients
(see Figure~\ref{fig:piN-bd}):

\begin{prp}[\textbf{Diagrammatic bounds}]\label{prp:diagram-bd}
For the ferromagnetic Ising model, we have
\begin{align}
\pi_\Lambda^{\sss(j)}(x)\leq
\begin{cases}\lbeq{piNbd}
P_{\Lambda;o}^{\prime{\sss(0)}}(o,x)\equiv\Exp{\varphi_o\varphi_x}_\Lambda^3
 &(j=0),\\[5pt]
\dpst\sum_{\substack{b_1,\dots,b_j\\ v_1,\dots,v_j}}P_{\Lambda;v_1}^{\prime
 {\sss(0)}}(o,\bb_1)\,\bigg(\prod_{i=1}^{j-1}\tau_{b_i}Q''_{\Lambda;v_i,v_{
 i+1}}(\tb_i,\bb_{i+1})\bigg)\,\tau_{b_j}Q'_{\Lambda;v_j}(\tb_j,x)&(j\ge1),
\end{cases}
\end{align}
where, as well as in the rest of the paper, the empty product is regarded
as 1 by convention.
\end{prp}

\subsection{Bound on $\pi_\Lambda^{\sss(0)}(x)$}
\label{ss:pi0bd} The key ingredient of the proof of
Proposition~\ref{prp:diagram-bd} is Lemma~\ref{lmm:GHS-BK} below,
which is an extension of the GHS idea used in the proof of
Lemma~\ref{lmm:switching}.  In this subsection, we demonstrate how
this extension works to prove the bound on
$\pi_\Lambda^{\sss(0)}(x)$ and the inequality
\begin{align}\lbeq{pi0'-bd}
\sum_{\bd\bn=o\vtri x}\frac{w_\Lambda(\bn)}{Z_\Lambda}\,\ind{o\db{\bn}{}
 x\}\,\cap\,\{o\cn{\bn}{}y}\leq P_{\Lambda;y}^{\prime{\sss(0)}}(o,x),
\end{align}
which will be used in Section~\ref{ss:pijbd} to obtain the bounds on
$\pi_\Lambda^{\sss(j)}(x)$ for $j\ge1$.

\begin{proof}[Proof of \refeq{piNbd} for $j=0$]
Since the inequality is trivial if $x=o$, we restrict our attention
to the case of $x\ne o$.

First we note that, for each current configuration $\bn$ with
$\bd\bn=\{o,x\}$ and $\tind{o\db{\bn}{}x}=1$, there are at least
\emph{three edge-disjoint} paths on $\mG_\bn$ between $o$ and $x$.
See, for example, the first term on the right-hand side in
Figure~\ref{fig:1stpiv}.  Suppose that the thick line in that
picture, referred to as $\zeta_1$ and split into
$\zeta_{11}\Dcup\zeta_{12}\Dcup\zeta_{13}$ from $o$ to $x$,
consists of bonds $b$ with $n_b=1$, and that the thin lines,
referred to as $\zeta_2$ and $\zeta_3$ that terminate at $o$ and
$x$ respectively, consist of bonds $b'$ with $n_{b'}=2$.  Let
$\zeta'_i$, for $i=2,3$, be the duplication of $\zeta_i$. Then,
the three paths $\zeta_2\Dcup\zeta_{13}$,
$\zeta'_2\Dcup\zeta_{12}\Dcup\zeta_3$ and
$\zeta_{11}\Dcup\zeta'_3$ are edge-disjoint.

Then, by multiplying $\pi_\Lambda^{\sss(0)}(x)$ by \emph{two}
dummies $(Z_\Lambda/Z_\Lambda)^2\,(\equiv1)$, we obtain
\begin{align}\lbeq{pi0*Z2}
\pi_\Lambda^{\sss(0)}(x)&=\sum_{\substack{\bd\bn=\{o,x\}\\ \bd\bm'=\bd\bm''
 =\vno}}\frac{w_\Lambda(\bn)}{Z_\Lambda}\,\frac{w_\Lambda(\bm')}{Z_\Lambda}
 \,\frac{w_\Lambda(\bm'')}{Z_\Lambda}\,\ind{o\db{\bn}{}x}\nn\\
&=\sum_{\bd\bN=\{o,x\}}\frac{w_\Lambda(\bN)}{Z_\Lambda^3}\sum_{\substack{
 \bd\bn=\{o,x\}\\ \bd\bm'=\bd\bm''=\vno\\ \bN\equiv\bn+\bm'+\bm''}}\ind{o
 \db{\bn}{}x}\prod_b\frac{N_b!}{n_b!\;m'_b!\;m''_b!},
\end{align}
where the sum over $\bn,\bm',\bm''$ in the second line equals the
cardinality of the following set of partitions:
\begin{align}\lbeq{Ssubset}
\fS_0=\bigg\{(\mS_0,\mS_1,\mS_2):\mG_\bN=\Bigcup_{i=0,1,2}\mS_i,\;
 \bd\mS_0=\{o,x\},\;\bd\mS_1=\bd\mS_2=\vno,\;o\db{}{}x\text{ in }
 \mS_0\bigg\},
\end{align}
where ``$o\db{}{}x$ in $\mS_0$'' means that there are at least two
\emph{bond}-disjoint paths in $\mS_0$.  We will show
$|\fS_0|\leq|\fS'_0|$, where
\begin{align}\lbeq{Ssupset}
\fS'_0=\bigg\{(\mS_0,\mS_1,\mS_2):\mG_\bN=\Bigcup_{i=0,1,2}\mS_i,\;
 \bd\mS_0=\bd\mS_1=\bd\mS_2=\{o,x\}\bigg\}.
\end{align}
This implies \refeq{piNbd} for $j=0$, because
\begin{align}
|\fS'_0|=\sum_{\substack{\bd\bn=\bd\bm'=\bd\bm''=\{o,x\}\\ \bN\equiv
 \bn+\bm'+\bm''}}\prod_b\frac{N_b!}{n_b!\,m'_b!\,
 m''_b!},
\end{align}
and
\begin{align}
\sum_{\bd\bN=\{o,x\}}\frac{w_\Lambda(\bN)}{Z_\Lambda^3}\sum_{\substack{
 \bd\bn=\bd\bm'=\bd\bm''=\{o,x\}\\ \bN\equiv\bn+\bm'+\bm''}}\prod_b
 \frac{N_b!}{n_b!\;m'_b!\;m''_b!}=\bigg(\sum_{\bd\bn=\{
 o,x\}}\frac{w_\Lambda(\bn)}{Z_\Lambda}\bigg)^3.
\end{align}

It remains to show $|\fS_0|\leq|\fS'_0|$.  To do so, we use the following lemma, in
which we denote by $\Omega_{z\to z'}^{\bN}$ the set of paths on $\mG_\bN$ from $z$ to
$z'$ and write $\omega\cap\omega'=\vno$ to mean that $\omega$ and $\omega'$ are
\emph{edge}-disjoint (not necessarily \emph{bond}-disjoint).

\begin{lmm}\label{lmm:GHS-BK}
Given a current configuration $\bN\in\Zp^{\mB_\Lambda}$, $k\ge1$,
$\cV\subset\Lambda$ and $z_i\ne z'_i\in\Lambda$ for $i=1,\dots,k$,
we let
\begin{align}\lbeq{fS-gen}
\fS=\left\{(\mS_0,\mS_1,\dots,\mS_k):
\begin{array}{r}
 \mG_\bN=\Bigcup_{i=0}^{\raisebox{-3pt}{$\scriptstyle k$}}
  \mS_i,\;\bd\mS_0=\cV,\;\bd\mS_i=\vno~(i=1,\dots,k),\;\\
 \Exists\omega_i\in\Omega^\bN_{z_i\to z'_i}~(i=1,\dots,k)~
  \text{\rm such that }\omega_i\subset\mS_0\Dcup\mS_i\\
 \text{\rm and }\omega_i\cap\omega_j=\vno~(i\ne j)
\end{array}\right\},
\end{align}
and define $\fS'$ to be the right-hand side of \refeq{fS-gen} with
``$\bd\mS_0=\cV$, $\bd\mS_i=\vno$'' being replaced by
``$\bd\mS_0=\cV\tri\{z_1,z'_1\}\tri\cdots\tri\{z_k,z'_k\}$,
$\bd\mS_i=\{z_i,z'_i\}$''.  Then, $|\fS|=|\fS'|$.
\end{lmm}

We will prove this lemma at the end of this subsection.

Now we use Lemma~\ref{lmm:GHS-BK} with $k=2$ and
$\cV=\{z_1,z'_1\}=\{z_2,z'_2\}=\{o,x\}$.  Note that $\fS_0$ in
\refeq{Ssubset} is a subset of $\fS$, since $\fS$ includes
partitions $(\mS_0,\mS_1,\mS_2)$ in which there does not exist
two \emph{bond}-disjoint paths on $\mS_0$.  In addition, $\fS'$
is trivially a subset of $\fS'_0$ in \refeq{Ssupset}.  Therefore,
we have $|\fS_0|\leq|\fS'_0|$.  This completes the proof of
\refeq{piNbd} for $j=0$.
\end{proof}

Here, we summarize the basic steps that we have followed to bound
$\pi_\Lambda^{\sss(0)}(x)$ and which we generalize to prove \refeq{pi0'-bd}
below and the bounds on $\pi_\Lambda^{\sss(j)}(x)$ for $j\ge1$ in
Section~\ref{sss:dbconn}.

\begin{enumerate}[(i)]
\item
Count the (minimum) number, say, $k+1$, of \emph{edge-disjoint} paths on $\mG_\bn$ that
satisfy the source constraint (as well as other additional conditions, if there are) of
the considered function $f(x)$.  For example, $k=2$ for
$\pi_\Lambda^{\sss(0)}(x)\equiv\frac1{Z_\Lambda}\sum_{\bd\bn=\{o,x\}}
w_\Lambda(\bn)\,\tind{o\db{\bn}{}x}$.
\item
Multiply $f(x)$ by $(\frac{Z_\Lambda}{Z_\Lambda})^k=\prod_{i=1}^k
(\frac1{Z_\Lambda}\sum_{\bd\bm_i=\vno}w_\Lambda(\bm_i)) \,(\equiv1)$
and then overlap the $k$~dummies $\bm_1,\dots,\bm_k$ on the original
current configuration $\bn$.  Choose $k$ paths
$\omega_1,\dots,\omega_k$ among $k+1$~edge-disjoint paths on
$\mG_{\bn+\sum_{i=1}^k\bm_i}$.
\item Use Lemma~\ref{lmm:GHS-BK} to exchange the occupation status of edges
on $\omega_i$ between $\mG_\bn$ and $\mG_{\bm_i}$ for every
$i=1,\dots,k$.  The current configurations after the mapping,
denoted by $\tilde\bn,\tilde\bm_1,\dots,\tilde\bm_k$, satisfy
$\bd\tilde\bn=\bd\bn\vtri\bd\omega_1\vtri\cdots\vtri\bd\omega_k$ and
$\bd\tilde\bm_i=\bd\omega_i$ for $i=1,\dots,k$.
\end{enumerate}

\begin{proof}[Proof of \refeq{pi0'-bd}]
If $y=o$ or $x$, then \refeq{pi0'-bd} is reduced to the inequality for
$\pi_\Lambda^{\sss(0)}(x)$.  Also, if $y\ne o=x$, then the left-hand side
of \refeq{pi0'-bd} multiplied by
$Z_\Lambda/Z_\Lambda=\sum_{\bd\bm=\vno}w_\Lambda(\bm)/Z_\Lambda\equiv1$
equals
\begin{align}\lbeq{dbbd}
\sum_{\bd\bn=\bd\bm=\vno}\frac{w_\Lambda(\bn)}{Z_\Lambda}\,\frac{w_\Lambda
 (\bm)}{Z_\Lambda}\,\ind{o\cn{\bn}{}y}&\leq\sum_{\bd\bn=\bd\bm=\vno}\frac{
 w_\Lambda(\bn)}{Z_\Lambda}\,\frac{w_\Lambda(\bm)}{Z_\Lambda}\,\ind{o\cn{
 \bn+\bm}{}y}\nn\\
&=\sum_{\bd\bn=\bd\bm=\{o,y\}}\frac{w_\Lambda(\bn)}{Z_\Lambda}\,\frac{
 w_\Lambda(\bm)}{Z_\Lambda}\;=\Exp{\varphi_o\varphi_y}_\Lambda^2,
\end{align}
where the first equality is due to Lemma~\ref{lmm:switching}.  Therefore,
we can assume $o\ne x\ne y\ne o$.

We follow the three steps described above.

(i) Since $y\notin\bd\bn=\{o,x\}$ and $\tind{o\db{\bn}{}x\}\,\cap\,\{o\cn{\bn}{}y}=1$,
it is not hard to see that there is an \emph{edge}-disjoint cycle (closed path) $o\to
y\to x\to o$. Since a cycle does not have a source, there must be another edge-disjoint
connection from $o$ to $x$, due to the source constraint $\bd\bn=\{o,x\}$. Therefore,
there are at least $4\,(=k+1)$ edge-disjoint paths on $\mG_\bn$: one is between $o$ and
$y$, another is between $y$ and $x$, and the other two are between $o$ and $x$.

(ii) Multiplying both sides of \refeq{pi0'-bd} by $(Z_\Lambda/Z_\Lambda)^3$ is
equivalent to
\begin{align}\lbeq{pi0'-equiv}
&\sum_{\bd\bN=\{o,x\}}\frac{w_\Lambda(\bN)}{Z_\Lambda^4}\sum_{\substack{
 \bd\bn=\{o,x\}\\ \bd\bm_i=\vno~\Forall i=1,2,3\\ \bN=\bn+\sum_{i=1}^3
 \bm_i}}\ind{o\db{\bn}{}x\}\,\cap\,\{o\cn{\bn}{}y}\prod_b
 \frac{N_b!}{n_b!\;m^{\sss(1)}_b!\;m^{\sss(2)}_b!\;m^{\sss(3)}_b!}\nn\\
&\quad\leq\sum_{\bd\bN=\{o,x\}}\frac{w_\Lambda(\bN)}{Z_\Lambda^4}\sum_{
 \substack{\bd\bn=\bd\bm_3=\{o,x\}\\ \bd\bm_1=\{o,y\},~\bd\bm_2=\{y,x\}\\
 \bN=\bn+\sum_{i=1}^3\bm_i}}\prod_b\frac{N_b!}{n_b!\;
 m^{\sss(1)}_b!\;m^{\sss(2)}_b!\;m^{\sss(3)}_b!},
\end{align}
where we have used the notation $m_b^{\sss(i)}=\bm_i|_b$.  Note that
the second sum on the left-hand side equals the cardinality of
\begin{align}\lbeq{S03sub}
\bigg\{(\mS_0,\mS_1,\mS_2,\mS_3):
 \begin{array}{c}
 \mG_\bN=\Bigcup_{i=0}^{\raisebox{-3pt}{$\scriptstyle3$}}
  \mS_i,\;\bd\mS_0=\{o,x\},\;\bd\mS_1=\bd\mS_2=\bd\mS_3=\vno\\
 o\db{}{}x\text{ in }\mS_0,~o\cn{}{}y\text{ in }\mS_0
 \end{array}\bigg\},
\end{align}
and the second sum on the right-hand side of \refeq{pi0'-equiv} equals
the cardinality of
\begin{align}\lbeq{S03sup}
\textstyle\Big\{(\mS_0,\mS_1,\mS_2,\mS_3):\mG_\bN=\Bigcup_{i=0}^{
 \raisebox{-3pt}{$\scriptstyle3$}}\mS_i,\;\bd\mS_0=\bd\mS_3=\{o,x\},\;\bd
 \mS_1=\{o,y\},\;\bd\mS_2=\{y,x\}\Big\}.
\end{align}
Therefore, to prove \refeq{pi0'-equiv}, it is sufficient to show that the
cardinality of \refeq{S03sub} is not bigger than that of \refeq{S03sup}.

(iii) Now we use Lemma~\ref{lmm:GHS-BK} with $k=3$ and
$\cV=\{z_3,z'_3\}=\{o,x\}$, $\{z_1,z'_1\}=\{o,y\}$ and
$\{z_2,z'_2\}=\{y,x\}$.  Since \refeq{S03sub} is a subset of $\fS$
in the current setting, while $\fS'$ is a subset of \refeq{S03sup},
we obtain \refeq{pi0'-equiv}.  This completes the proof of
\refeq{pi0'-bd}.
\end{proof}

\begin{proof}[Proof of Lemma~\ref{lmm:GHS-BK}]
We prove Lemma~\ref{lmm:GHS-BK} by decomposing $\fS^{(\prime)}$ into
$\Bigcup_{\vec\omega_k}\fS_{\vec\omega_k}^{(\prime)}$ (described in
detail below) and then constructing a bijection from
$\fS_{\vec\omega_k}$ to $\fS'_{\vec\omega_k}$ for every
$\vec\omega_k$.  To do so, we first introduce some notation.

\begin{enumerate}
\item
For every $i=1,\dots,k$, we introduce an arbitrarily fixed order
among elements in $\Omega_{z_i\to z'_i}^\bN$.  For
$\omega,\omega'\in\Omega_{z_i\to z'_i}^\bN$, we write
$\omega\prec\omega'$ if $\omega$ is earlier than $\omega'$ in this
order.  Let $\tilde\Omega_{z_1\to z'_1}^\bN$ be the set of paths
$\zeta\in\Omega_{z_1\to z'_1}^\bN$ such that there are $k-1$
edge-disjoint paths on $\mG_{\bN}\setminus\zeta$ (= the resulting
graph by removing the edges in $\zeta$) each of which connects $z_i$
and $z'_i$ for every $i=2,\dots,k$.
\item
Then, for $\omega_1\in\tilde\Omega_{z_1\to z'_1}^\bN$, we define
$\Xi_{z_2\to z'_2}^{\bN;\omega_1}$ to be the set of paths
$\zeta\in\Omega_{z_2\to z'_2}^\bN$ on $\mG_\bN\setminus\omega_1$
such that $\zeta\not\supset\xi$ for any $\xi\in\tilde\Omega_{z_1\to
z'_1}^\bN$ earlier than $\omega_1$.  Then, we define
$\tilde\Omega_{z_2\to z'_2}^{\bN;\omega_1}$ to be the set of paths
$\zeta\in\Xi_{z_2\to z'_2}^{\bN;\omega_1}$ such that there are $k-2$
edge-disjoint paths on $\mG_{\bN}\setminus(\omega_1\Dcup\zeta)$ each
of which is from $z_i$ to $z'_i$ for $i=3,\dots,k$.
\item
More generally, for $l<k$ and $\vec\omega_l=(\omega_1,
\dots,\omega_l)$ with $\omega_1\in\tilde\Omega_{z_1\to z'_1}^\bN$,
$\omega_2\in\tilde\Omega_{z_2\to z'_2}^{\bN;\omega_1},\dots$,
$\omega_l\in\tilde\Omega_{z_l\to z'_l}^{\bN;\vec\omega_{l-1}}$, we
define $\Xi_{z_{l+1}\to z'_{l+1}}^{\bN;\vec\omega_l}$ to be the set
of paths $\zeta\in\Omega_{z_{l+1}\to z'_{l+1}}^\bN$ on
$\mG_\bN\setminus\Bigcup_{i=1}^{\raisebox{-3pt}{$\scriptstyle l$}}
\omega_i$ such that $\zeta\not\supset\xi$ for any
$\xi\in\tilde\Omega_{z_i\to z'_i}^{\bN;\vec\omega_{i-1}}$ earlier
than $\omega_i$, for every $i=1,\dots,l$.  Then, we define
$\tilde\Omega_{z_{l+1}\to z'_{l+1}}^{\bN;\vec\omega_l}$ to be the
set of paths $\zeta\in\Xi_{z_{l+1}\to z'_{l+1}}^{\bN;\vec\omega_l}$
such that there are $k-(l+1)$~edge-disjoint paths on $\mG_{
\bN}\setminus(\Bigcup_{i=1}^{\raisebox{-3pt}{$\scriptstyle
l$}}\omega_i \Dcup\zeta)$ each of which is from $z_i$ to $z'_i$ for
$i=l+2,\dots,k$.
\item
If $l=k-1$, then we simply define $\tilde\Omega_{z_k\to z'_k}^{\bN;\vec
\omega_{k-1}}=\Xi_{z_k\to z'_k}^{\bN;\vec\omega_{k-1}}$.  We will
also abuse the notation to denote $\tilde\Omega_{z_1\to z'_1}^\bN$ by
$\tilde\Omega_{z_1\to z'_1}^{\bN;\vec\omega_0}$.
\end{enumerate}

Using the above notation, we can decompose $\fS^{(\prime)}$ disjointly as follows.  For
a collection $\omega_i\in\tilde\Omega_{z_i\to z'_i}^{\bN;\vec\omega_{i-1}}$ for
$i=1,\dots,k$, we denote by $\fS_{\vec\omega_k}^{(\prime)}$ the set of partitions
$\vec\mS_k\equiv(\mS_0,\mS_1,\dots,\mS_k)\in\fS^{(\prime)}$ such that, for every
$i=1,\dots,k$, the earliest element of $\tilde\Omega_{z_i\to
z'_i}^{\bN;\vec\omega_{i-1}}$ contained in $\mS_0\Dcup\mS_i$ is $\omega_i$.  Then,
$\fS^{(\prime)}$ is decomposed as
\begin{align}\lbeq{SS'-dec}
\fS^{(\prime)}=\Bigcup_{\omega_1\in\tilde\Omega_{z_1\to z'_1}^\bN}\,
 \Bigcup_{\omega_2\in\tilde\Omega_{z_2\to z'_2}^{\bN;\omega_1}}\cdots
 \Bigcup_{\omega_k\in\tilde\Omega_{z_k\to z'_k}^{\bN;\vec\omega_{k-1}}}
 \fS_{\vec\omega_k}^{(\prime)}.
\end{align}

To complete the proof of Lemma~\ref{lmm:GHS-BK}, it suffices to construct
a bijection from $\fS_{\vec\omega_k}$ to $\fS'_{\vec\omega_k}$ for every
$\vec\omega_k$.  For $\vec\mS_k\in\fS_{\vec\omega_k}$, we define
\begin{align}\lbeq{Fdef}
\textstyle\vec F_{\vec\omega_k}(\vec\mS_k)\equiv\big(F_{\vec\omega_k}^{
 \sss(0)}(\mS_0),\dots,F_{\vec\omega_k}^{\sss(k)}(\mS_k)\big)=\Big(\mS_0
 \tri\Bigcup_{i=1}^{\raisebox{-3pt}{$\scriptstyle k$}}\omega_i,\,\mS_1
 \tri\omega_1,\dots,\mS_k\tri\omega_k\Big),
\end{align}
where $\bd F_{\vec\omega_k}^{\sss(0)}(\mS_0)=\cV\tri\{z_1,z'_1\}
\tri\cdots\tri\{z_k,z'_k\}$ and $\bd
F_{\vec\omega_k}^{\sss(i)}(\mS_i)= \{z_i,z'_i\}$ for $i=1,\dots,k$.
Note that, by definition using symmetric difference, we have $\vec
F_{ \vec\omega_k}(\vec F_{\vec\omega_k}(\vec\mS_k))=\vec\mS_k$.
Also, by simple combinatorics using
$\omega_i\cap\omega_j=\mS_i\cap\mS_j=\vno$ and
$\omega_j\subset\mS_0\Dcup\mS_j$ for $1\leq j\leq k$ and $i\ne j$,
we have
\begin{align}\lbeq{F0DcupFi}
F_{\vec\omega_k}^{\sss(i)}(\mS_i)\cap F_{\vec\omega_k}^{\sss(j)}(\mS_j)=
 \vno,&&
F_{\vec\omega_k}^{\sss(0)}(\mS_0)\Dcup F_{\vec\omega_k}^{\sss(j)}
 (\mS_j)\textstyle=\Big(\mS_0\tri\Bigcup_{i\ne j}\omega_i\Big)\Dcup
 \mS_j.
\end{align}
Since $\omega_j\subset\mS_0\Dcup\mS_j$ and $\omega_j\cap\Bigcup_{i\ne j}\omega_i=\vno$,
we have $\omega_j\subset F_{\vec\omega_k}^{\sss(0)}(\mS_0)\Dcup
F_{\vec\omega_k}^{\sss(j)} (\mS_j)$.

It remains to show that $\omega_j$ is the earliest element of $\tilde\Omega_{z_j\to
z'_j}^{\bN;\vec\omega_{j-1}}$ in $F_{\vec\omega_k}^{\sss(0)}(\mS_0)\Dcup
F_{\vec\omega_k}^{\sss(j)} (\mS_j)$.  To see this, we first recall that
$\tilde\Omega_{z_j\to z'_j}^{\bN;\vec \omega_{j-1}}$ is a set of paths on
$\mG_\bN\setminus\Bigcup_{i<j} \omega_i$, so that its earliest element contained in
$(\mS_0\tri\Bigcup_{i<j}\omega_i)\Dcup\mS_j$ is still $\omega_j$. Furthermore, since
each $\tilde\Omega_{z_i\to z'_i}^{\bN;\vec \omega_{i-1}}$ for $i>j$ is a set of paths
that do not fully contain $\omega_j$ or any earlier element of $\tilde\Omega_{z_j\to
z'_j}^{\bN;\vec\omega_{j-1}}$ as a subset, $\omega_j$ is still the earliest element of
\begin{align}
\bigg(\textstyle\Big(\mS_0\tri\Bigcup_{i<j}\omega_i\Big)\Dcup\mS_j\bigg)
 \tri\Big(\Bigcup_{i>j}\omega_i\Big)=\Big(\mS_0\tri\Bigcup_{i\ne j}
 \omega_i\Big)\Dcup\mS_j\equiv F_{\vec\omega_k}^{\sss(0)}(\mS_0)\Dcup
 F_{\vec\omega_k}^{\sss(j)}(\mS_j).
\end{align}
Therefore, $\vec F_{\vec\omega_k}$ is a bijection from
$\fS_{\vec\omega_k}$ to $\fS'_{\vec\omega_k}$.  This completes the
proof of Lemma~\ref{lmm:GHS-BK}.
\end{proof}

\subsection{Bounds on $\pi_\Lambda^{\sss(j)}(x)$ for $j\ge1$}
\label{ss:pijbd} First we prove \refeq{piNbd} for $j\ge1$ assuming the following two
lemmas, in which we recall \refeq{Theta-def} and use
\begin{align}
&E'_\bN(z,x;\cA)=\{z\cn{\bN}{\cA}x\}\cap\{z\db{\bN}{}x\},&
&E''_\bN(z,x,v;\cA)=E'_\bN(z,x;\cA)\cap\{z\cn{\bN}{}v\},\lbeq{E'E''-def}\\
&\Theta'_{z,x;\cA}=\!\sum_{\substack{\bd\bm=\vno\\ \bd\bn=z\vtri x}}\!\!
 \frac{w_{\cA\compl}(\bm)}{Z_{\cA\compl}}\,\frac{w_\Lambda(\bn)}{Z_\Lambda}
 \,\indic{E'_{\bm+\bn}(z,x;\cA)},\quad&
&\Theta''_{z,x,v;\cA}=\!\sum_{\substack{\bd\bm=\vno\\ \bd\bn=z\vtri x}}\!\!
 \frac{w_{\cA\compl}(\bm)}{Z_{\cA\compl}}\,\frac{w_\Lambda(\bn)}{Z_\Lambda}
 \,\indic{E''_{\bm+\bn}(z,x,v;\cA)}.\lbeq{Theta'Theta''-def}
\end{align}

\begin{lmm}\label{lmm:Thetabds}
For the ferromagnetic Ising model, we have
\begin{align}
\Theta_{y,x;\cA}&\leq\sum_z\big(\delta_{y,z}+\tilde G_\Lambda(y,z)\big)
 \,\Theta'_{z,x;\cA},\lbeq{Theta[1]-bd}\\[5pt]
\Theta_{y,x;\cA}\big[\ind{y\cn{}{}v}\big]&\leq\sum_z\big(\delta_{y,z}
 +\tilde G_\Lambda(y,z)\big)\,\Theta''_{z,x,v;\cA}\nn\\
&\quad+\sum_{v',z}\big(\delta_{y,v'}+\tilde G_\Lambda(y,v')\big)\,\tilde G_\Lambda(v',z)\,\Theta'_{z,x;
 \cA}\,\psi_\Lambda(v',v).\lbeq{Theta[I]-bd}
\end{align}
\end{lmm}

\begin{lmm}\label{lmm:Theta'Theta''bd}
For the ferromagnetic Ising model, we have
\begin{align}\lbeq{Theta'Theta''bd}
\Theta'_{y,x;\cA}\leq\sum_{u\in\cA}P'_{\Lambda;u}(y,x),&&
\Theta''_{y,x,v;\cA}\leq\sum_{u\in\cA}P''_{\Lambda;u,v}(y,x).
\end{align}
\end{lmm}

We prove Lemma~\ref{lmm:Thetabds} in Section~\ref{sss:chopping-off},
and Lemma~\ref{lmm:Theta'Theta''bd} in Section~\ref{sss:dbconn}.

\begin{proof}[Proof of \refeq{piNbd} for $j\ge1$ assuming
Lemmas~\ref{lmm:Thetabds}--\ref{lmm:Theta'Theta''bd}]
Recall \refeq{pij-def}.  By
\refeq{Theta[1]-bd}, \refeq{Theta'Theta''bd} and \refeq{Q'-def}, we obtain
\begin{align}\lbeq{nest-diagbd}
\Theta^{\sss(j-1)}_{\tb_{j-1},\bb_j;\tilde\cC_{j-2}}\Big[\tau_{b_j}
 \Theta^{\sss(j)}_{\tb_j,x;\tilde\cC_{j-1}}\Big]&\leq\Theta^{
 \sss(j-1)}_{\tb_{j-1},\bb_j;\tilde\cC_{j-2}}\bigg[\sum_z\tau_{b_j}\big(
 \delta_{\tb_j,z}+\tilde G_\Lambda(\tb_j,z)\big)\sum_{v_j\in
 \tilde\cC_{j-1}}P'_{\Lambda;v_j}(z,x)\bigg]\nn\\
&\leq\sum_{v_j}\Theta^{\sss(j-1)}_{\tb_{j-1},\bb_j;\tilde\cC_{j-2}}
 \big[\ind{\tb_{j-1}\cn{}{}v_j}\big]\,\tau_{b_j}Q'_{\Lambda;v_j}
 (\tb_j,x).
\end{align}
For $j=1$, we use \refeq{pi0'-bd} and \refeq{nest-diagbd} to obtain
\begin{align}\lbeq{pi0'-bd-appl}
&\pi_\Lambda^{\sss(1)}(x)\equiv\sum_{b_1}\Theta^{\sss(0)}_{o,\bb_1;\Lambda}
 \Big[\tau_{b_1}\,\Theta^{\sss(1)}_{\tb_1,x;\tilde\cC_0}\Big]\leq\sum_{b_1,
 v_1}\Theta^{\sss(0)}_{o,\bb_1;\Lambda}\big[\ind{o\cn{}{}v_1}\big]\,\tau_{
 b_1}Q'_{\Lambda;v_1}(\tb_1,x)\\
&~=\sum_{b_1,v_1}\bigg(\sum_{\bd\bn=o\vtri\bb_1}\frac{w_\Lambda(\bn)}
 {Z_\Lambda}\,\ind{o\db{\bn}{}\bb_1\}\,\cap\,\{o\cn{\bn}{}v_1}\bigg)\tau_{
 b_1}Q'_{\Lambda;v_1}(\tb_1,x)\leq\sum_{b_1,v_1}P_{\Lambda;v_1}^{\prime{
 \sss(0)}}(o,\bb_1)\,\tau_{b_1}Q'_{\Lambda;v_1}(\tb_1,x).\nn
\end{align}
For $j\ge2$, we use \refeq{Theta[I]-bd}--\refeq{Theta'Theta''bd} and then
\refeq{Q'-def}--\refeq{Q''-def} to obtain
\begin{align}\lbeq{Theta-bd-appl}
&\Theta^{\sss(j-1)}_{\tb_{j-1},\bb_j;\tilde\cC_{j-2}}\Big[\tau_{b_j}\Theta^{
 \sss(j)}_{\tb_j,x;\tilde\cC_{j-1}}\Big]\leq\sum_{v_j}\Theta^{\sss(j-1)}_{
 \tb_{j-1},\bb_j;\tilde\cC_{j-2}}\big[\ind{\tb_{j-1}\cn{}{}v_j}\big]\,\tau_{
 b_j}Q'_{\Lambda;v_j}(\tb_j,x)\nn\\
&\quad\leq\sum_{v_j}\tau_{b_j}Q'_{\Lambda;v_j}(\tb_j,x)\Bigg(\sum_z\big(
 \delta_{\tb_{j-1},z}+\tilde G_\Lambda(\tb_{j-1},z)\big)\sum_{v_{j-1}\in
 \tilde\cC_{j-2}}P''_{\Lambda;v_{j-1},v_j}(z,\bb_j)\nn\\
&\qquad\qquad+\sum_{v',z}\big(\delta_{\tb_{j-1},v'}+\tilde G_\Lambda(\tb_{
 j-1},v')\big)\,\tilde G_\Lambda(v',z)\,\sum_{v_{j-1}\in\tilde\cC_{j-2}}
 P'_{\Lambda;v_{j-1}}(z,\bb_j)\,\psi_\Lambda(v',v_j)\Bigg)\nn\\
&\quad\leq\sum_{v_{j-1},v_j}\ind{v_{j-1}\in\tilde\cC_{j-2}}\,Q''_{\Lambda;
 v_{j-1},v_j}(\tb_{j-1},\bb_j)\,\tau_{b_j}Q'_{\Lambda;v_j}(\tb_j,x).
\end{align}
We repeatedly use \refeq{Theta[I]-bd}--\refeq{Theta'Theta''bd} to
bound $\Theta^{\sss(i)}_{\tb_i,\bb_{i+1};\tilde\cC_{i-1}}
[\tind{\tb_i \cn{}{}v_{i+1}}]$ for $i=j-2,\dots,1$ as in
\refeq{Theta-bd-appl}, and then at the end we apply \refeq{pi0'-bd}
as in \refeq{pi0'-bd-appl} to obtain \refeq{piNbd}. This completes
the proof.
\end{proof}

\subsubsection{Proof of Lemma~\ref{lmm:Thetabds}}\label{sss:chopping-off}
\begin{proof}[Proof of \refeq{Theta[1]-bd}]
Recall \refeq{Theta-def} and \refeq{Theta'Theta''-def}.  Then, to
prove \refeq{Theta[1]-bd}, it suffices to bound the contribution
from $\indic{E_{\bm+\bn}(y,x;\cA)\setminus E'_{\bm+\bn}(y,x;\cA)}$
by $\sum_z\tilde G_\Lambda(y,z)\,\Theta'_{z,x;\cA}$.

First we recall \refeq{E-def} and \refeq{E'E''-def}.  Then, we have
\begin{align}
E_{\bm+\bn}(y,x;\cA)\setminus E'_{\bm+\bn}(y,x;\cA)=E_{\bm+\bn}(y,x;\cA)
 \cap\big\{\{y\cn{\bm+\bn}{}x\}\setminus\{y\db{\bm+\bn}{}x\}\big\}.
\end{align}
On $\{y\cn{\bm+\bn}{}x\}\setminus\{y\db{\bm+\bn}{}x\}$, there is at
least one pivotal bond for $y\cn{\bm+\bn}{}x$ from $y$.  Let $b$ be
the last pivotal bond among them.  Then, we have
$\tb\db{\bm+\bn}{}x\text{ off }b$, $m_b+n_b>0$, and
$y\cn{\bm+\bn}{}\bb$ in $\cC_{\bm+\bn}^b(x)\compl$.  Moreover, on
the event $E_{\bm+\bn}(y,x;\cA)$, we have that $y\cn{\bm+\bn}{}\bb$
in $\cA\compl$ and $\tb\cn{\bm+\bn}{\cA}x$.  Since
$\{\tb\db{\bm+\bn}{}x\text{ off }b\}\cap\{\tb
\cn{\bm+\bn}{\cA}x\}=\{E'_{\bm+\bn}(\tb,x;\cA)\text{ off }b\}$ on
the event that $b$ is pivotal for $y\cn{\bm+\bn}{}x$ from $y$, we
have
\begin{align}\lbeq{EE'-dec}
&E_{\bm+\bn}(y,x;\cA)\setminus E'_{\bm+\bn}(y,x;\cA)\nn\\
&=\Bigcup_b\Big\{\{E'_{\bm+\bn}(\tb,x;\cA)\text{ off }b\}\cap\{m_b+n_b>0
 \}\cap\big\{y\cn{\bm+\bn}{}\bb\text{ in }\cA\compl\cap\cC_{\bm+\bn}^b(x)
 \compl\big\}\Big\}.
\end{align}
Therefore, we obtain
\begin{align}\lbeq{Theta[1]-rewr}
&\Theta_{y,x;\cA}-\Theta'_{y,x;\cA}\nn\\
&=\sum_b\sum_{\substack{\bd\bm=\vno\\ \bd\bn=y\vtri x}}\frac{
 w_{\cA\compl}(\bm)}{Z_{\cA\compl}}\,\frac{w_\Lambda(\bn)}
 {Z_\Lambda}\,\ind{E'_{\bm+\bn}(\tb,x;\cA)\text{ off }b}\,
 \ind{m_b+n_b>0}\,\ind{y\cn{\bm+\bn}{}\bb\text{ in }\cA\compl
 \cap\cC_{\bm+\bn}^b(x)\compl}.
\end{align}

It remains to bound the right-hand side of \refeq{Theta[1]-rewr},
which is nonzero only if $m_b$ is even and $n_b$ is odd, due to the
source constraints and the conditions in the indicators.  First, as
in \refeq{2nd-ind-fact}, we alternate the parity of $n_b$ by
changing the source constraint into $\bd\bn=y\vtri b\vtri x$ and
multiplying by $\tau_b$.  Then, by conditioning on
$\cC_{\bm+\bn}^b(x)$ as in \refeq{3rd-ind-prefact} (i.e.,
conditioning on $\cC_{\bm+\bn}^b(x)=\cB$, letting
$\bm'=\bm|_{\mB_{\cA\compl}\setminus\mB_{\cA\compl\cap\cB\compl}}$,
$\bm''=\bm|_{\mB_{\cA\compl\cap\cB\compl}}$,
$\bn'=\bn|_{\mB_\Lambda\setminus\mB_{\cB\compl}}$ and
$\bn''=\bn|_{\mB_{\cB\compl}}$, and then summing over
$\cB\subset\Lambda$), we obtain
\begin{align}\lbeq{2ndind-contr}
\sum_{\cB\subset\Lambda}\sum_{\substack{\bd\bm'=\vno\\ \bd\bn'=
 \tb\vtri x}}\frac{\tilde w_{\cA\compl,\cB}(\bm')\,Z_{\cA\compl
 \cap\cB\compl}}{Z_{\cA\compl}}\,\frac{\tilde w_{\Lambda,\cB}
 (\bn')\,Z_{\cB\compl}}{Z_\Lambda}\,\ind{E'_{\bm'+\bn'}(\tb,x;
 \cA)\text{ off }b\}\,\cap\,\{\cC^b_{\bm'+\bn'}(x)=\cB}\nn\\
\times\tau_b\ind{m'_b,n'_b\text{ even}}\underbrace{\sum_{
 \substack{\bd\bm''=\vno\\ \bd\bn''=y\vtri\bb}}\frac{w_{\cA
 \compl\cap\cB\compl}(\bm'')}{Z_{\cA\compl\cap\cB\compl}}\,
 \frac{w_{\cB\compl}(\bn'')}{Z_{\cB\compl}}\,\ind{y\cn{\bm''+
 \bn''}{}\bb~\text{ in }\cA\compl\cap\cB\compl}}_{\stackrel{
 \because\refeq{switching-appl}\;}=\Exp{\varphi_y\varphi_{\bb}
 }_{\cA\compl\cap\,\cB\compl}}\nn\\
=\sum_{\substack{\bd\bm=\vno\\ \bd\bn=\tb\vtri x}}\frac{
 w_{\cA\compl}(\bm)}{Z_{\cA\compl}}\,\frac{w_\Lambda(\bn)}
 {Z_\Lambda}\,\ind{E'_{\bm+\bn}(\tb,x;\cA)\text{ off }b}\,
 \tau_b\ind{m_b,n_b\text{ even}}\,\Exp{\varphi_y\varphi_{\bb}
 }_{\cA\compl\cap\,\cC_{\bm+\bn}^b(x)\compl}.
\end{align}
Since $\Exp{\varphi_y\varphi_{\bb}}_{\cA\compl\cap\,\cC_{\bm
+\bn}^b(x)\compl}=0$ on
$E'_{\bm+\bn}(\tb,x;\cA)\setminus\{E'_{\bm+\bn}(\tb,x;\cA)$ off
$b\}\subset\{\bb\in\cC_{\bm+\bn}^b(x)\}$ and on the event that $m_b$
or $n_b$ is odd (see below \refeq{0th-summand2} or above
\refeq{3rd-ind-fact}), we can omit ``off $b$'' and
$\tind{m_b,n_b\text{ even}}$ in \refeq{2ndind-contr}.  Since
$\Exp{\varphi_y\varphi_{\bb}}_{\cA\compl\cap\,\cC_{\bm+ \bn}^b(x)
\compl}\leq\Exp{\varphi_y\varphi_{\bb}}_\Lambda$ due to
Proposition~\ref{prp:through}, we have
\begin{align}\lbeq{2ndind-contrbd}
\refeq{2ndind-contr}\leq\Exp{\varphi_y\varphi_{\bb}}_\Lambda\tau_b
 \sum_{\substack{\bd\bm=\vno\\ \bd\bn=\tb\vtri x}}\frac{w_{\cA
 \compl}(\bm)}{Z_{\cA\compl}}\,\frac{w_\Lambda(\bn)}{Z_\Lambda}\,
 \indic{E'_{\bm+\bn}(\tb,x;\cA)}=\Exp{\varphi_y\varphi_{\bb}
 }_\Lambda\tau_b\,\Theta'_{\tb,x;\cA}.
\end{align}
Therefore, \refeq{Theta[1]-rewr} is bounded by $\sum_b\Exp{\varphi_y
\varphi_{\bb}}_\Lambda\tau_b\,\Theta'_{\tb,x;\cA}\equiv\sum_z\tilde
G_\Lambda(y,z)\,\Theta'_{z,x;\cA}$.  This completes the proof of
\refeq{Theta[1]-bd}.
\end{proof}

\begin{proof}[Proof of \refeq{Theta[I]-bd}]
Recall \refeq{Theta-def} and \refeq{Theta'Theta''-def}.  To prove
\refeq{Theta[I]-bd}, we investigate
\begin{align}
L&\equiv\big\{E_{\bm+\bn}(y,x;\cA)\cap\{y\cn{\bm+\bn}{}v\}\big\}
 \setminus E''_{\bm+\bn}(y,x,v;\cA)\nn\\
&=\{E_{\bm+\bn}(y,x;\cA)\setminus E'_{\bm+\bn}(y,x;\cA)\}\cap
 \{y\cn{\bm+\bn}{}v\},
\end{align}
where $\Theta_{y,x;\cA}[\indic{L}]=\Theta_{y,x;\cA}[\tind{y\cn{}{}
v}]-\Theta''_{y,x,v;\cA}$.

First we recall \refeq{EE'-dec}, in which $b$ is the last pivotal
bond for $y\cn{\bm+\bn}{}x$ from $y$, and define
\begin{align}
R_1(b)&=\{E''_{\bm+\bn}(\tb,x,v;\cA)\text{ off }b\}\cap\{m_b
 +n_b>0\}\cap\big\{y\cn{\bm+\bn}{}\bb\text{ in }\cA\compl\cap
 \cC_{\bm+\bn}^b(x)\compl\big\},\lbeq{R1b-def}\\
R_2(b)&=\{E'_{\bm+\bn}(\tb,x;\cA)\text{ off }b\}
 \cap\{m_b+n_b>0\}\cap\big\{y\cn{\bm+\bn}{}\bb\text{ in }\cA
 \compl\cap\cC_{\bm+\bn}^b(x)\compl,\;y\cn{\bm+\bn}{}v\big\},
 \lbeq{R2b-def}
\end{align}
where $v\in\cC_{\bm+\bn}^b(x)$ on $R_1(b)$, while
$v\in\cC_{\bm+\bn}^b(y)$ on $R_2(b)$.  Since
\begin{align}\lbeq{EE'E''-predec}
L=\Bigcup_b\{R_1(b)\Dcup R_2(b)\},
\end{align}
we have
\begin{align}\lbeq{EE'E''predec2}
\Theta_{y,x;\cA}[\tind{y\cn{}{} v}]-\Theta''_{y,x,v;\cA}=\sum_b
 \Big(\Theta_{y,x;\cA}\big[\indic{R_1(b)}\big]+\Theta_{y,x;\cA}
 \big[\indic{R_2(b)}\big]\Big).
\end{align}
Following the same argument as in
\refeq{2ndind-contr}--\refeq{2ndind-contrbd}, we easily obtain
\begin{align}\lbeq{EE'E''predec3}
\Theta_{y,x;\cA}\big[\indic{R_1(b)}\big]&=\sum_{\substack{\bd
 \bm=\vno\\ \bd\bn=\tb\vtri x}}\frac{w_{\cA\compl}(\bm)}{Z_{
 \cA\compl}}\,\frac{w_\Lambda(\bn)}{Z_\Lambda}\,\ind{E''_{\bm
 +\bn}(\tb,x,v;\cA)\text{ off }b}\,\tau_b\ind{m_b,n_b
 \text{ even}}\,\Exp{\varphi_y\varphi_{\bb}}_{\cA\compl\cap\,
 \cC_{\bm+\bn}^b(x)\compl}\nn\\
&\leq\Exp{\varphi_y\varphi_{\bb}}_\Lambda\tau_b\sum_{\substack{
 \bd\bm=\vno\\ \bd\bn=\tb\vtri x}}\frac{w_{\cA\compl}(\bm)}{Z_{
 \cA\compl}}\,\frac{w_\Lambda(\bn)}{Z_\Lambda}\,\indic{E''_{\bm
 +\bn}(\tb,x,v;\cA)}=\Exp{\varphi_y\varphi_{\bb}}_\Lambda\tau_b
 \,\Theta''_{\tb,x,v;\cA}.
\end{align}
Similarly, we have
\begin{align}\lbeq{EE'E''decpre3}
\Theta_{y,x;\cA}\big[\indic{R_2(b)}\big]&=\sum_{\cB\subset
 \Lambda}\sum_{\substack{\bd\bm=\vno\\ \bd\bn=\tb\vtri x}}
 \frac{w_{\cA\compl}(\bm)}{Z_{\cA\compl}}\,\frac{w_\Lambda
 (\bn)}{Z_\Lambda}\,\ind{E'_{\bm+\bn}(\tb,x;\cA)\text{ off }
 b\}\,\cap\,\{\cC^b_{\bm+\bn}(x)=\cB}\,\tau_b\ind{m_b,n_b
 \text{ even}}\nn\\
&\qquad\qquad\times\sum_{\substack{\bd\bh=\vno\\ \bd\bk=y\vtri\bb}}
 \frac{w_{\cA\compl\cap\cB\compl}(\bh)}{Z_{\cA\compl\cap\cB
 \compl}}\,\frac{w_{\cB\compl}(\bk)}{Z_{\cB\compl}}\,\ind{y
 \cn{\bh+\bk}{}\bb\text{ in }\cA\compl\cap\cB\compl,~y\cn{
 \bh+\bk}{}v\text{ (in }\cB\compl)}\nn\\
&\leq\sum_{\substack{\bd\bm=\vno\\ \bd\bn=\tb\vtri x}}\frac{
 w_{\cA\compl}(\bm)}{Z_{\cA\compl}}\,\frac{w_\Lambda(\bn)}
 {Z_\Lambda}\,\ind{E'_{\bm+\bn}(\tb,x;\cA)\text{ off }b}\,
 \tau_b\ind{m_b,n_b\text{ even}}\,\Psi_{y,\bb,v;\cA,\cC_{\bm
 +\bn}^b(x)},
\end{align}
where
\begin{align}\lbeq{Psi-def}
\Psi_{y,z,v;\cA,\cB}=\sum_{\substack{\bd\bh=\vno\\ \bd\bk=
 y\vtri z}}\frac{w_{\cA\compl\cap\,\cB\compl}(\bh)}{Z_{\cA
 \compl\cap\,\cB\compl}}\,\frac{w_{\cB\compl}(\bk)}{Z_{\cB
 \compl}}\,\ind{y\cn{\bh+\bk}{}v}.
\end{align}
We note that, by ignoring the indicator in \refeq{Psi-def}, we have
$0\leq\Psi_{y,z,v;\cA,\cB}\leq\Exp{\varphi_y\varphi_z}_{\cB\compl}$,
which is zero whenever $z\in\cB$.  Therefore, we can omit ``off
$b$'' and $\tind{m_b,n_b\text{ even}}$ in \refeq{EE'E''decpre3} to
obtain
\begin{align}\lbeq{EE'E''dec3}
\Theta_{y,x;\cA}\big[\indic{R_2(b)}\big]\leq\sum_{\substack{\bd\bm=
 \vno\\ \bd\bn=\tb\vtri x}}\frac{w_{\cA\compl}(\bm)}{Z_{\cA\compl}}
 \,\frac{w_\Lambda(\bn)}{Z_\Lambda}\,\indic{E'_{\bm+\bn}(\tb,x;\cA)}
 \,\tau_b\,\Psi_{y,\bb,v;\cA,\cC_{\bm+\bn}^b(x)}.
\end{align}
Substituting \refeq{EE'E''predec3} and \refeq{EE'E''dec3} to
\refeq{EE'E''predec2}, we arrive at
\begin{align}\lbeq{EE'E''dec2}
\Theta_{y,x;\cA}\big[\ind{y\cn{}{}v}\big]&\leq\sum_z\big(\delta_{y,
 z}+\tilde G_\Lambda(y,z)\big)\,\Theta''_{z,x,v;\cA}\nn\\
&\quad+\sum_b\sum_{\substack{\bd\bm=\vno\\ \bd\bn=\tb\vtri x}}\frac{
 w_{\cA\compl}(\bm)}{Z_{\cA\compl}}\,\frac{w_\Lambda(\bn)}{Z_\Lambda}
 \,\indic{E'_{\bm+\bn}(\tb,x;\cA)}\,\tau_b\,\Psi_{y,\bb,v;\cA,\cC_{
 \bm+\bn}^b(x)}.
\end{align}
The proof of \refeq{Theta[I]-bd} is completed by using
\begin{align}\lbeq{Psi-bd}
\Psi_{y,z,v;\cA,\cB}\leq\sum_{v'}\Exp{\varphi_y\varphi_{v'}}_\Lambda
 \Exp{\varphi_{v'}\varphi_z}_\Lambda\,\psi_\Lambda(v',v),
\end{align}
and replacing $\Exp{\varphi_y\varphi_{v'}}_\Lambda$ in
\refeq{Psi-bd} by $\delta_{y,v'}+\tilde G_\Lambda(y,v')$, due to
\refeq{G-delta-bd}.

To complete the proof of \refeq{Theta[I]-bd}, it thus remains to
show \refeq{Psi-bd}. First we note that, if $\cA\subset\cB$, then by
Lemma~\ref{lmm:switching} we have
\begin{align}
\Psi_{y,z,v;\cA,\cB}=\sum_{\substack{\bd\bh=\vno\\ \bd\bk=y\vtri z}}
 \frac{w_{\cB\compl}(\bh)}{Z_{\cB\compl}}\,\frac{w_{\cB\compl}(\bk)}
 {Z_{\cB\compl}}\,\ind{y\cn{\bh+\bk}{}v}=\Exp{\varphi_y\varphi_v}_{\cB
 \compl}\Exp{\varphi_v\varphi_z}_{\cB\compl}\leq\Exp{\varphi_y
 \varphi_v}_\Lambda\Exp{\varphi_v\varphi_z}_\Lambda.
\end{align}
However, to prove \refeq{Psi-bd} for a general $\cA$ that does not
necessarily satisfy $\cA\subset\cB$, we use
\begin{align}\lbeq{Psi-ind-dec}
\{y\cn{\bh+\bk}{}v\}=\{y\cn{\bk}{}v\}\Dcup\big\{\{y\cn{\bh+\bk}{}v\}
 \setminus\{y\cn{\bk}{}v\}\big\},
\end{align}
and consider the two events on the right-hand side separately.  The
contribution to $\Psi_{y,z,v;\cA,\cB}$ from $\{y\cn{\bk}{}v\}$ is easily
bounded, similarly to \refeq{dbbd}, as
\begin{align}\lbeq{psi-delta}
\sum_{\bd\bk=y\vtri z}\frac{w_{\cB\compl}(\bk)}{Z_{\cB\compl}}\,\ind{y\cn{
 \bk}{}v}\leq\sum_{\substack{\bd\bk=y\vtri z\\ \bd\bk'=\vno}}\frac{w_{\cB
 \compl}(\bk)}{Z_{\cB\compl}}\,\frac{w_{\cB\compl}(\bk')}{Z_{\cB\compl}}\,
 \ind{y\cn{\bk+\bk'}{}v}&=\Exp{\varphi_y\varphi_v}_{\cB\compl}\Exp{
 \varphi_v\varphi_z}_{\cB\compl}\nn\\
&\leq\Exp{\varphi_y\varphi_v}_\Lambda\Exp{\varphi_v\varphi_z}_\Lambda.
\end{align}

Next we consider the contribution to $\Psi_{y,z,v;\cA,\cB}$ from
$\{y\cn{\bh+\bk}{}v\}\setminus\{y\cn{\bk}{}v\}$ in
\refeq{Psi-ind-dec}.  We denote by $\cC_\bk(y)$ the set of sites
$\bk$-connected from $y$. Since
$v\in\cC_{\bh+\bk}(y)\setminus\cC_\bk(y)$, there is a \emph{nonzero}
alternating chain of mutually-disjoint $\bh$-connected clusters and
mutually-disjoint $\bk$-connected clusters, from some
$u_0\in\cC_\bk(y)$ to $v$. Therefore, we have
\begin{align}\lbeq{ind-bd}
\ind{y\cn{\bh+\bk}{}v\}\setminus\{y\cn{\bk}{}v}\leq\sum_{j=1}^\infty
 \sum_{\substack{u_0,\dots,u_j\\ u_l\ne u_{l'}\,\Forall l\ne l'\\ u_j
 =v}}\ind{y\cn{\bk}{}u_0}\bigg(\prod_{l\ge0}\ind{u_{2l}\cn{\bh}{}u_{2l
 +1}}\bigg)\bigg(\prod_{l\ge1}\ind{u_{2l-1}\cn{\bk}{}u_{2l}}\bigg)\nn\\
\times\bigg(\prod_{\substack{l,l'\ge0\\ l\ne l'}}\ind{\cC_\bh(u_{2l})
 \,\cap\,\cC_\bh(u_{2l'})=\vno}\,\ind{\cC_\bk(u_{2l})\,\cap\,\cC_\bk
 (u_{2l'})=\vno}\bigg),
\end{align}
where we regard an empty product as 1.  Using this bound, we can
perform the sums over $\bh$ and $\bk$ in \refeq{Psi-def}
independently.

For $j=1$ and given $u_0\ne u_1=v$, the summand of \refeq{ind-bd}
equals $\tind{y\cn{\bk}{}u_0}\tind{u_0\cn{\bh}{}v}$, which is simply
equal to $\tind{y\cn{\bh}{}v}$ if $u_0=y$.  Then, by
\refeq{psi-delta} and \refeq{G-delta-bd}, the contribution from this
to $\Psi_{y,z,v;\cA,\cB}$ is
\begin{align}\lbeq{psi-delta-G2}
\sum_{\bd\bk=y\vtri z}\frac{w_{\cB\compl}(\bk)}{Z_{\cB\compl}}\,\ind{y
 \cn{\bk}{}u_0}\sum_{\bd\bh=\vno}\frac{w_{\cA\compl\cap\,\cB\compl}(\bh)}
 {Z_{\cA\compl\cap\,\cB\compl}}\,\ind{u_0\cn{\bh}{}v}\leq\Exp{\varphi_y
 \varphi_{u_0}}_\Lambda\Exp{\varphi_{u_0}\varphi_z}_\Lambda\,\tilde
 G_\Lambda(u_0,v)^2.
\end{align}

Fix $j\ge2$ and a sequence of distinct sites $u_0,\dots,u_j\,(=v)$,
and first consider the contribution to the sum over $\bk$ in
\refeq{Psi-def} from the relevant indicators in the right-hand side
of \refeq{ind-bd}, which is
\begin{align}\lbeq{nsum-0thbd}
&\sum_{\bd\bk=y\vtri z}\frac{w_{\cB\compl}(\bk)}{Z_{\cB\compl}}\,\ind{y
 \cn{\bk}{}u_0}\bigg(\prod_{l\ge1}\ind{u_{2l-1}\cn{\bk}{}u_{2l}}\bigg)
 \prod_{\substack{l,l'\ge0\\ l\ne l'}}\ind{\cC_\bk(u_{2l})\,\cap\,
 \cC_\bk(u_{2l'})=\vno}\\
&=\sum_{\bd\bk=y\vtri z}\frac{w_{\cB\compl}(\bk)}{Z_{\cB\compl}}
 \bigg(\prod_{l\ge1}\ind{u_{2l-1}\cn{\bk}{}u_{2l}}\bigg)\bigg(
 \prod_{\substack{l,l'\ge1\\ l\ne l'}}\ind{\cC_\bk(u_{2l})\,\cap\,
 \cC_\bk(u_{2l'})=\vno}\bigg)\ind{y\cn{\bk}{}u_0\}\,\cap\,\{\cC_\bk
 (u_0)\,\cap\,\cU_{\bk;1}=\vno}\nn,
\end{align}
where $\cU_{\bk;1}=\Bigcup_{l\ge1}\cC_\bk(u_{2l})$.  Conditioning
on $\cU_{\bk;1}$, we obtain that
\begin{gather}
\refeq{nsum-0thbd}=\sum_{\bd\bk=\vno}\frac{w_{\cB\compl}(\bk)}
 {Z_{\cB\compl}}\bigg(\prod_{l\ge1}\ind{u_{2l-1}\cn{\bk}{}u_{2l}}
 \bigg)\bigg(\prod_{\substack{l,l'\ge1\\ l\ne l'}}\ind{\cC_\bk
 (u_{2l})\,\cap\,\cC_\bk(u_{2l'})=\vno}\bigg)\nn\\
\times\underbrace{\sum_{\bd\bk'=y\vtri z}\frac{w_{\cB\compl\cap\,
 \cU_{\bk;1}\compl}(\bk')}{Z_{\cB\compl\cap\,\cU_{\bk;1}\compl}}
 \ind{y\cn{\bk'}{}u_0}}_{\stackrel{\because\refeq{psi-delta}\;}\leq
 \Exp{\varphi_y\varphi_{u_0}}_\Lambda\Exp{\varphi_{u_0}\varphi_z
 }_\Lambda}.\lbeq{nsum-1stbd}
\end{gather}
Then, by conditioning on
$\cU_{\bk;2}\equiv\Bigcup_{l\ge2}\cC_\bk(u_{2l})$, following the
same computation as above and using \refeq{G-delta-bd}, we further
obtain that
\begin{gather}
\refeq{nsum-0thbd}\leq\Exp{\varphi_y\varphi_{u_0}}_\Lambda\Exp{
 \varphi_{u_0}\varphi_z}_\Lambda\sum_{\bd\bk=\vno}\frac{w_{\cB\compl}
 (\bk)}{Z_{\cB\compl}}\bigg(\prod_{l\ge2}\ind{u_{2l-1}\cn{\bk}{}u_{2
 l}}\bigg)\bigg(\prod_{\substack{l,l'\ge2\\ l\ne l'}}\ind{\cC_\bk
 (u_{2l})\,\cap\,\cC_\bk(u_{2l'})=\vno}\bigg)\nn\\
\times\underbrace{\sum_{\bd\bk'=\vno}\frac{w_{\cB\compl\cap\,\cU_{\bk
 ;2}\compl}(\bk')}{Z_{\cB\compl\cap\,\cU_{\bk;2}\compl}}\ind{u_1\cn{
 \bk'}{}u_2}}_{\leq\;\tilde G_\Lambda(u_1,u_2)^2}.\lbeq{nsum-2ndbd}
\end{gather}
We repeat this computation until all indicators for $\bk$ are used
up.  We also apply the same argument to the sum over $\bh$ in
\refeq{Psi-def}.  Summarizing these bounds with \refeq{psi-delta}
and \refeq{psi-delta-G2}, and replacing $u_0$ in
\refeq{ind-bd}--\refeq{nsum-1stbd} by $v'$, we obtain
\refeq{Psi-bd}.  This completes the proof of \refeq{Theta[I]-bd}.
\end{proof}

\subsubsection{Proof of Lemma~\ref{lmm:Theta'Theta''bd}}\label{sss:dbconn}
We note that the common factor $\tind{y\db{\bm+\bn}{}x}$ in
$\Theta'_{y,x;\cA}$ and $\Theta''_{y,x,v;\cA}$ can be decomposed as
\begin{align}\lbeq{Theta'-evdec}
\ind{y\db{\bm+\bn}{}x}=\ind{y\db{\bn}{}x}+\ind{y\db{\bm+\bn}{}x\}
 \setminus\{y\db{\bn}{}x}.
\end{align}
We estimate the contributions from $\tind{y\db{\bn}{}x}$ to
$\Theta'_{y,x;\cA}$ and $\Theta''_{y,x,v;\cA}$ in the following
paragraphs~(a) and (b), respectively.  Then, in the paragraphs~(c)
and (d) below, we will estimate the contributions from
$\tind{y\db{\bm+\bn}{}x\}\setminus\{y\db{\bn}{}x}$ in
\refeq{Theta'-evdec} to $\Theta'_{y,x;\cA}$ and
$\Theta''_{y,x,v;\cA}$, respectively.

\bigskip

\textbf{(a)} First we investigate the contribution to
$\Theta'_{y,x;\cA}$ from $\tind{y\db{\bn}{}x}$:
\begin{align}\lbeq{contr-(a)}
\sum_{\substack{\bd\bm=\vno\\ \bd\bn=y\vtri x}}\frac{w_{\cA\compl}
 (\bm)}{Z_{\cA\compl}}\,\frac{w_\Lambda(\bn)}{Z_\Lambda}\,\ind{y
 \cn{\bm+\bn}{\cA}x\}\,\cap\,\{y\db{\bn}{}x}.
\end{align}
For a set of events $E_1,\dots,E_N$, we define
$E_1\circ\cdots\circ E_N$ to be the event that $E_1,\dots,E_N$ occur
\emph{bond}-disjointly.  Then, we have
\begin{align}
\ind{y\cn{\bm+\bn}{\cA}x\}\,\cap\,\{y\db{\bn}{}x}\leq\ind{y\cn{\bn}
 {\cA}x\}\,\cap\,\{y\db{\bn}{}x}\leq\sum_{u\in\cA}\ind{y\cn{\bn}{}u
 \}\,\circ\,\{u\cn{\bn}{}x\}\,\circ\,\{y\cn{\bn}{}x},
\end{align}
where the right-hand side does not depend on $\bm$.  Therefore, the
contribution to $\Theta'_{y,x;\cA}$ is bounded by
\begin{align}\lbeq{Theta'-bd1stbd}
\refeq{contr-(a)}\leq\sum_{u\in\cA}\,\sum_{\bd\bn=y\vtri x}\frac{
 w_\Lambda(\bn)}{Z_\Lambda}\,\ind{y\cn{\bn}{}u\}\,\circ\,\{u\cn
 {\bn}{}x\}\,\circ\,\{y\cn{\bn}{}x}\leq\sum_{u\in\cA}P_{\Lambda;
 u}^{\prime{\sss(0)}}(y,x),
\end{align}
where we have applied the same argument as in the proof of
\refeq{pi0'-bd}, which is around \refeq{dbbd}--\refeq{S03sup}.
\qed

\bigskip

\textbf{(b)} Next we investigate the contribution to
$\Theta''_{y,x,v;\cA}$ from $\tind{y\db{\bn}{}x}$ in
\refeq{Theta'-evdec}:
\begin{align}\lbeq{contr-(b)}
\sum_{\substack{\bd\bm=\vno\\ \bd\bn=y\vtri x}}\frac{w_{\cA\compl}
 (\bm)}{Z_{\cA\compl}}\,\frac{w_\Lambda(\bn)}{Z_\Lambda}\,\ind{y
 \cn{\bm+\bn}{\cA}x\}\,\cap\,\{y\db{\bn}{}x\}\,\cap\,\{y\cn{\bm+
 \bn}{}v}.
\end{align}
Note that, by using \refeq{Psi-ind-dec} and
$\tind{y\cn{\bm+\bn}{\cA}x}\leq\tind{y\cn{\bn}{\cA}x}$, we have
\begin{align}\lbeq{Theta''-1stindbd}
\ind{y\cn{\bm+\bn}{\cA}x\}\,\cap\,\{y\db{\bn}{}x\}\,\cap\,\{y\cn{\bm+\bn}{}
 v}\leq\ind{y\cn{\bn}{\cA}x\}\,\cap\,\{y\db{\bn}{}x}\,\Big(\ind{y\cn{\bn}{}
 v}+\ind{y\cn{\bm+\bn}{}v\}\setminus\{y\cn{\bn}{}v}\Big).
\end{align}
We investigate the contributions from the two indicators in the parentheses
separately.

We begin with the contribution from $\tind{y\cn{\bn}{}v}$, which is
independent of $\bm$.  Since
\begin{align}
\{y\cn{\bn}{\cA}x\}\cap\{y\db{\bn}{}x\}\cap\{y\cn{\bn}{}v\}&\subset\{y
 \cn{\bn}{\cA}x\}\circ\{y\cn{\bn}{}x,~y\cn{\bn}{}v\},\\
\{y\cn{\bn}{\cA}x\}&\subset\bigcup_{u\in\cA}\{y\cn{\bn}{}u\}\circ\{u
 \cn{\bn}{}x\},\lbeq{Theta''-1stind1st-pcontr}
\end{align}
the contribution to \refeq{contr-(b)} from
$\tind{y\cn{\bn}{}v}$
in \refeq{Theta''-1stindbd} is bounded by
\begin{align}\lbeq{Theta''-1stind1stcontr}
\sum_{u\in\cA}\,\sum_{\bd\bn=y\vtri x}\frac{w_\Lambda(\bn)}{Z_\Lambda}\,
 \ind{y\cn{\bn}{}u\}\,\circ\,\{u\cn{\bn}{}x\}\,\circ\,\{y\cn{\bn}{}x,~y
 \cn{\bn}{}v}.
\end{align}
We follow Steps~(i)--(iii) described above \refeq{dbbd} in
Section~\ref{ss:pi0bd}.  Without loss of generality, we can assume
that $y,u,x$ and $v$ are all different; otherwise, the following
argument can be simplified.  (i) Since $y$ and $x$ are sources, but
$u$ and $v$ are not, there is an edge-disjoint cycle $y\to u\to x\to
v\to y$, with an extra edge-disjoint path from $y$ to $x$.
Therefore, we have in total at least $5\,(=4+1)$ edge-disjoint
paths.  (ii) Multiplying by $(Z_\Lambda/Z_\Lambda)^4$, we have
\begin{align}\lbeq{Theta''-bd1stprebd}
\refeq{Theta''-1stind1stcontr}=\sum_{u\in\cA}\,\sum_{\bd\bN=y\vtri
x}\frac{w_\Lambda(\bN)}
 {Z_\Lambda^5}\sum_{\substack{\bd\bn=y\vtri x\\ \bd\bm_i=\vno~
 \Forall i=1,\dots,4\\ \bN=\bn+\sum_{i=1}^4\bm_i}}\ind{y\cn{\bn}
 {}u\}\,\circ\,\{u\cn{\bn}{}x\}\,\circ\,\{y\cn{\bn}{}x,~y\cn{\bn}
 {}v}\prod_b\frac{N_b!}{n_b!\prod_{i=1}^4m^{\sss(i)}_b!},
\end{align}
where we have used the notation $m_b^{\sss(i)}=\bm_i|_b$.  (iii) The
sum over $\bn,\bm_1,\dots,\bm_4$ in \refeq{Theta''-bd1stprebd} is
bounded by the cardinality of $\fS$ in Lemma~\ref{lmm:GHS-BK} with
$k=4$, $\cV=\{y,x\}$, $\{z_1,z'_1\}=\{y,u\}$, $\{z_2,z'_2\}=\{u,x\}$,
$\{z_3,z'_3\}=\{y,v\}$ and $\{z_4,z'_4\}=\{v,x\}$.  Bounding the
cardinality of $\fS'$ in Lemma~\ref{lmm:GHS-BK} for this setting,
we obtain
\begin{align}\lbeq{Theta''-bd1stbd1}
\refeq{Theta''-bd1stprebd}&\leq\sum_{u\in\cA}\,\sum_{\bd\bN=y\vtri
 x}\frac{w_\Lambda(\bN)}{Z_\Lambda^5}\sum_{\substack{\bd\bn=y\vtri
 x\\ \bd\bm_1=y\vtri u,~\bd\bm_2=u\vtri x\\ \bd\bm_3=y\vtri v,~\bd
 \bm_4=v\vtri x\\ \bN=\bn+\sum_{i=1}^4\bm_i}}\prod_b\frac{N_b!}{n_b!
 \prod_{i=1}^4m^{\sss(i)}_b!}\nn\\
&\leq\sum_{u\in\cA}\Exp{\varphi_y\varphi_x}_\Lambda\Exp{\varphi_y
 \varphi_u}_\Lambda\Exp{\varphi_u\varphi_x}_\Lambda\Exp{\varphi_y
 \varphi_v}_\Lambda\Exp{\varphi_v\varphi_x}_\Lambda.
\end{align}

Next we investigate the contribution to \refeq{contr-(b)} from
$\tind{y\cn{\bm+\bn}{}v\}\setminus\{y\cn{\bn}{}v}$ in
\refeq{Theta''-1stindbd}.  On the event $\{y\db{\bn}{}x\}\cap\{\{
y\cn{\bm+\bn}{}v\}\setminus\{y\cn{\bn}{}v\}\}$, there exists a
$v_0\ne v$ such that
$\{y\cn{\bn}{}x\}\circ\{y\cn{\bn}{}x,~y\cn{\bn}{}v_0\}$ occurs and
that $v_0$ and $v$ are connected via a nonzero alternating chain of
mutually-disjoint $\bm$-connected clusters and mutually-disjoint
$\bn$-connected clusters.  Therefore, by \refeq{ind-bd} and
\refeq{Theta''-1stind1st-pcontr} (see also
\refeq{Theta''-1stind1stcontr}), we obtain
\begin{align}\lbeq{Theta''-1stindbd2}
&\ind{y\cn{\bn}{\cA}x\}\,\cap\,\{y\db{\bn}{}x\}\,\cap\,\{\{y\cn{\bm+\bn}{}v
 \}\setminus\{y\cn{\bn}{}v\}}\nn\\
&\leq\sum_{u\in\cA}\,\sum_{j\ge1}\sum_{\substack{v_0,\dots,v_j\\ v_l\ne
 v_{l'}\,\Forall l\ne l'\\ v_j=v}}\ind{y\cn{\bn}{}u\}\,\circ\,\{u\cn{\bn}{}
 x\}\,\circ\,\{y\cn{\bn}{}x,~y\cn{\bn}{}v_0}\,\bigg(\prod_{l\ge0}\ind{v_{2
 l}\cn{\bm}{}v_{2l+1}}\bigg)\nn\\
&\qquad\times\bigg(\prod_{l\ge1}\ind{v_{2l-1}\cn{\bn}{}v_{2l}}\bigg)\bigg(
 \prod_{\substack{l,l'\ge0\\ l\ne l'}}\ind{\cC_\bm(v_{2l})\,\cap\,\cC_\bm
 (v_{2l'})=\vno}\;\ind{\cC_\bn(v_{2l})\,\cap\,\cC_\bn(v_{2l'})=\vno}\bigg).
\end{align}
For the three products of indicators, we repeate the same argument
as in \refeq{psi-delta-G2}--\refeq{nsum-2ndbd} to derive the factor
$\psi_\Lambda(v_0,v)-\delta_{v_0,v}$.  As a result, we have
\begin{align}\lbeq{Theta''-prebd1stbd2}
&\sum_{\substack{\bd\bm=\vno\\ \bd\bn=y\vtri x}}\frac{w_{\cA\compl}(\bm)}
 {Z_{\cA\compl}}\,\frac{w_\Lambda(\bn)}{Z_\Lambda}\,\ind{y\cn{\bn}{\cA}x\}
 \,\cap\,\{y\db{\bn}{}x\}\,\cap\,\{\{y\cn{\bm+\bn}{}v\}\setminus\{y\cn{\bn}
 {}v\}}\nn\\
&~\leq\sum_{v_0}\big(\psi_\Lambda(v_0,v)-\delta_{v_0,v}\big)\sum_{u\in\cA}
 \,\sum_{\bd\bn=y\vtri x}\frac{w_\Lambda(\bn)}{Z_\Lambda}\,\ind{y\cn{\bn}{}
 u\}\,\circ\,\{u\cn{\bn}{}x\}\,\circ\,\{y\cn{\bn}{}x,~y\cn{\bn}{}v_0}.
\end{align}
Following the same argument as in
\refeq{Theta''-1stind1stcontr}--\refeq{Theta''-bd1stbd1}, we obtain
\begin{align}\lbeq{Theta''-bd1stbd2}
\refeq{Theta''-prebd1stbd2}&\leq\sum_{u\in\cA,\;v_0}\big(\psi_\Lambda(v_0,
 v)-\delta_{v_0,v}\big)\,\Exp{\varphi_y\varphi_x}_\Lambda\Exp{\varphi_y
 \varphi_u}_\Lambda\Exp{\varphi_u\varphi_x}_\Lambda\Exp{\varphi_y
 \varphi_{v_0}}_\Lambda\Exp{\varphi_{v_0}\varphi_x}_\Lambda\nn\\
&\leq\sum_{u\in\cA}\Big(P_{\Lambda;u,v}^{\prime\prime{\sss(0)}}(y,x)-
 \Exp{\varphi_y\varphi_x}_\Lambda\Exp{\varphi_y\varphi_u}_\Lambda
 \Exp{\varphi_u\varphi_x}_\Lambda\Exp{\varphi_y\varphi_v}_\Lambda
 \Exp{\varphi_v\varphi_x}_\Lambda\Big).
\end{align}

Summarizing \refeq{Theta''-1stindbd}, \refeq{Theta''-bd1stbd1} and
\refeq{Theta''-bd1stbd2}, we arrive at
\begin{align}\lbeq{Theta''-0bdfin}
\refeq{contr-(b)}\leq\sum_{u\in\cA}P_{\Lambda;u,v}^{\prime\prime
 {\sss(0)}}(y,x).
\end{align}
This completes the bound on the contribution to $\Theta''_{y,x,v;\cA}$
from $\tind{y\db{\bn}{}x}$ in \refeq{Theta'-evdec}.
\qed

\bigskip

\textbf{(c)} The contribution to $\Theta'_{y,x;\cA}$ from
$\tind{y\db{\bm+\bn}{}x\}\setminus\{y\db{\bn}{}x}$ in
\refeq{Theta'-evdec} equals
\begin{align}\lbeq{contr-(c)}
\sum_{\substack{\bd\bm=\vno\\ \bd\bn=y\vtri x}}\frac{w_{\cA\compl}(\bm)}
 {Z_{\cA\compl}}\,\frac{w_\Lambda(\bn)}{Z_\Lambda}\,\ind{y\cn{\bm+\bn}
 {\cA}x\}\,\cap\,\{\{y\db{\bm+\bn}{}x\}\setminus\{y\db{\bn}{}x\}}.
\end{align}
Note that, if $\tind{\bd\bn=y\vtri x\}\setminus\{y\db{\bn}{}x}=1$,
then $y$ is $\bn$-connected, but not $\bn$-doubly connected, to $x$,
and therefore there exists at least one pivotal bond for
$y\cn{\bn}{}x$.  Given an ordered set of bonds $\vec
b_T=(b_1,\dots,b_T)$, we define
\begin{align}\lbeq{H-def}
H_{\bn;\vec b_T}(y,x)=\{y\db{\bn}{}\bb_1\}\cap\bigcap_{i=1}^T\Big\{\{\tb_i
 \db{\bn}{}\bb_{i+1}\}\cap\big\{n_{b_i}>0,~b_i\text{ is pivotal for }y\cn
 {\bn}{}x\big\}\Big\},
\end{align}
where, by convention, $\bb_{T+1}=x$.  Then, by
$\ind{y\cn{\bm+\bn}{\cA}x}\leq\ind{y\cn{\bn}{\cA}x}$, we obtain
\begin{align}\lbeq{Theta'-2ndindbd2}
\refeq{contr-(c)}&=\sum_{T\ge1}\sum_{\vec b_T}\sum_{\substack{\bd\bm=\vno\\
 \bd\bn=y\vtri x}}\frac{w_{\cA\compl}(\bm)}{Z_{\cA\compl}}\,\frac{w_\Lambda
 (\bn)}{Z_\Lambda}\,\ind{y\cn{\bm+\bn}{\cA}x\}\,\cap\,H_{\bn;\vec b_T}(y,x)
 \,\cap\,\{y\db{\bm+\bn}{}x}\nn\\
&\leq\sum_{T\ge1}\sum_{\vec b_T}\sum_{\substack{\bd\bm=\vno\\ \bd\bn=y
 \vtri x}}\frac{w_{\cA\compl}(\bm)}{Z_{\cA\compl}}\,\frac{w_\Lambda(\bn)}
 {Z_\Lambda}\,\ind{y\cn{\bn}{\cA}x\}\,\cap\,H_{\bn;\vec b_T}(y,x)\,\cap\,
 \{y\db{\bm+\bn}{}x}.
\end{align}

\begin{figure}[t]
\begin{center}
\includegraphics[scale=0.28]{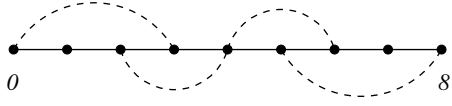}
\caption{\label{fig:lace-edges}An element in $\cL_{[0,8]}^{\sss(4)}$,
which consists of $s_1t_1=\{0,3\}$, $s_2t_2=\{2,4\}$, $s_3t_3=\{4,6\}$
and $s_4t_4=\{5,8\}$.}
\end{center}
\end{figure}

On the event $H_{\bn;\vec b_T}(y,x)$, we denote the $\bn$-double
connections between the pivotal bonds $b_1,\dots,b_T$ by
\begin{align}
\cD_{\bn;i}=\begin{cases}
 \cC_\bn^{b_1}(y)&(i=0),\\
 \cC_\bn^{b_{i+1}}(y)\setminus\cC_\bn^{b_i}(y)&(i=1,\dots,T-1),\\
 \cC_\bn(y)\setminus\cC_\bn^{b_T}(y)&(i=T).
 \end{cases}
\end{align}
As in Figure~\ref{fig:lace-edges}, we can think of $\cC_\bn(y)$ as
the interval $[0,T]$, where each integer $i\in[0,T]$ corresponds to
$\cD_{\bn;i}$ and the unit interval $(i-1,i)\subset[0,T]$
corresponds to the pivotal bond $b_i$.  Since $y\db{\bm+\bn}{}x$,
we see that, for every $b_i$, there must be an $(\bm+\bn)$-bypath
(i.e., an $(\bm+\bn)$-connection that does not go through $b_i$) from
some $z\in\cD_{\bn;s}$ with $s<i$ to some $z'\in\cD_{\bn;t}$ with
$t\ge i$.  We abbreviate $\{s,t\}$ to $st$ if there is no confusion.
Let $\cL_{[0,T]}^{\sss(1)}=\{\{0T\}\}$,
$\cL_{[0,T]}^{\sss(2)}=\{\{0t_1,s_2T\}:0<s_2\leq t_1<T\}$ and
generally for $j\leq T$ (see Figure~\ref{fig:lace-edges}),
\begin{align}
\cL_{[0,T]}^{\sss(j)}=\big\{\{s_it_i\}_{i=1}^j:0=s_1<s_2\leq t_1<s_3
 \leq\cdots\leq t_{j-2}<s_j\leq t_{j-1}<t_j=T\big\}.
\end{align}
For every $j\in\{1,\dots,T\}$, we have $\bigcup_{st\in\Gamma}[s,t]=
[0,T]$ for any $\Gamma\in\cL_{[0,T]}^{\sss(j)}$, which implies double
connection.  Conditioning on $\cC_\bn(y)\equiv\bigcup_{i=0}^{
\raisebox{-3pt}{$\scriptstyle T$}}\cD_{\bn;i}=\cB$ (and denoting
$\bk=\bn|_{\mB_{\cB\compl}}$, $\bh=\bn|_{\mB_\Lambda\setminus\mB_{\cB
\compl}}$ and $\cD_{\bn;i}\equiv\cD_{\bh;i}=\cB_i$) and multiplying
by $Z_{\cB\compl}/Z_{\cB\compl}$, we obtain
\begin{align}\lbeq{Theta'-2ndindbd2.2}
\refeq{Theta'-2ndindbd2}=\sum_{\cB\subset\Lambda}\,\sum_{T\ge1}
 \sum_{\vec b_T}\sum_{\substack{\bd\bm=\bd\bk=\vno\\ \bd\bh=y\vtri
 x}}\frac{w_{\cA\compl}(\bm)}{Z_{\cA\compl}}\,\frac{\tilde w_{\Lambda,
 \cB}(\bh)\,Z_{\cB\compl}}{Z_\Lambda}\,\frac{w_{\cB\compl}(\bk)}
 {Z_{\cB\compl}}\,\ind{y\cn{\bh}{\cA}x\}\,\cap\,H_{\bh;\vec b_T}(y,x)
 \,\cap\,\{\cC_\bh(y)=\cB}\nn\\
\times\sum_{j=1}^T\sum_{\{s_it_i\}_{i=1}^j\in\cL_{[0,T]}^{(j)}}
 \,\sum_{\substack{z_1,\dots,z_j\\ z'_1,\dots,z'_j}}\bigg(\prod_{i=1}^j
 \ind{z_i\in\cB_{s_i},~z'_i\in\cB_{t_i}\}\,\cap\,\{z_i\cn{\bm+\bk}{}
 z'_i}\bigg)\prod_{i\ne l}\ind{\cC_{\bm+\bk}(z_i)\,\cap\,\cC_{\bm+\bk}
 (z_l)=\vno}.
\end{align}
Reorganizing this expression and then summing over $\cB\subset\cA$,
we obtain
\begin{align}\lbeq{Theta'-2ndindbd3}
\refeq{Theta'-2ndindbd2.2}&=\sum_{T\ge1}\sum_{\vec
b_T}\sum_{\bd\bn=y\vtri x}\frac{w_\Lambda
 (\bn)}{Z_\Lambda}\,\indic{\{y\cn{\bn}{\cA}x\}\,\cap\,H_{\bn;\vec b_T}
 (y,x)}\nn\\
&\quad\times\sum_{j=1}^T\sum_{\{s_it_i\}_{i=1}^j\in\cL_{[0,T]}^{(j)}}\,
 \sum_{\substack{z_1,\dots,z_j\\ z'_1,\dots,z'_j}}\bigg(\prod_{i=1}^j
 \ind{z_i\in\cD_{\bn;s_i},~z'_i\in\cD_{\bn;t_i}}\bigg)\nn\\
&\quad\times\sum_{\bd\bm=\bd\bk=\vno}\frac{w_{\cA\compl}(\bm)}
 {Z_{\cA\compl}}\,\frac{w_{\tilde\cD\compl}(\bk)}{Z_{\tilde\cD\compl}}
 \bigg(\prod_{i=1}^j\ind{z_i\cn{\bm+\bk}{}z'_i}\bigg)\prod_{i\ne l}
 \ind{\cC_{\bm+\bk}(z_i)\,\cap\,\cC_{\bm+\bk}(z_l)=\vno},
\end{align}
where we have denoted $\cC_\bn(y)$ by $\tilde\cD$.  In the rightmost
expression, the first line determines $\tilde\cD$ that contains vertices
$z_i,z'_i$ for all $i=1,\dots,j$ in a specific manner, while the second
line determines the bypaths $\cC_{\bm+\bk}(z_i)$ joining $z_i$ and
$z'_i$ for every $i=1,\dots,j$.  We first derive $\bn$-independent
bounds on these bypaths in the following paragraph (c-1).  Then, in
(c-2) below, we will bound the first two lines of the rightmost
expression in \refeq{Theta'-2ndindbd3}.

\smallskip

\textbf{(c-1)} For $j=1$, the last line of the rightmost expression
in \refeq{Theta'-2ndindbd3} simply equals
\begin{align}\lbeq{Theta'-2ndindbd3:j=1}
\sum_{\bd\bm=\bd\bk=\vno}\frac{w_{\cA\compl}(\bm)}{Z_{\cA\compl}}\,
 \frac{w_{\tilde\cD\compl}(\bk)}{Z_{\tilde\cD\compl}}\,\ind{z_1
 \cn{\bm+\bk}{}z'_1}.
\end{align}
Since $z_1,z'_1\in\tilde\cD$ and $z_1\ne z'_1$, these two vertices
are connected via a nonzero alternating chain of mutually-disjoint
$\bm$-connected clusters and mutually-disjoint $\bk$-connected
clusters.  Moreover, since $z_1,z'_1\in\tilde\cD$ and
$\bk\in\Zp^{\mB_{\tilde\cD\compl}}$, this chain of bubbles starts
and ends with $\bm$-connected clusters (possibly with a single
$\bm$-connected cluster), not with $\bk$-connected clusters.
Therefore, by following the argument around
\refeq{ind-bd}--\refeq{nsum-2ndbd}, we can easily show
\begin{align}\lbeq{Theta'-2ndindbd3:j=1bd}
\refeq{Theta'-2ndindbd3:j=1}\leq\sum_{l\ge1}\big(\tilde G_\Lambda^2
 \big)^{*(2l-1)}(z_1,z'_1).
\end{align}

For $j\ge2$, since $\cC_{\bm+\bk}(z_i)$ for $i=1,\dots,j$ are
mutually-disjoint due to the last product of the indicators in
\refeq{Theta'-2ndindbd3}, we can treat each bypath separately by the
conditioning-on-clusters argument.  By conditioning on
$\cV_{\bm+\bk}\equiv\Bigcup_{i\ge2}\cC_{\bm+\bk}(z_i)$, the last
line in the rightmost expression of \refeq{Theta'-2ndindbd3} equals
\begin{gather}
\sum_{\bd\bm=\bd\bk=\vno}\frac{w_{\cA\compl}(\bm)}{Z_{\cA\compl}}\,
 \frac{w_{\tilde\cD\compl}(\bk)}{Z_{\tilde\cD\compl}}\bigg(\prod_{i=
 2}^j\ind{z_i\cn{\bm+\bk}{}z'_i}\bigg)\bigg(\prod_{\substack{i,l\ge
 2\\ i\ne l}}\ind{\cC_{\bm+\bk}(z_i)\,\cap\,\cC_{\bm+\bk}(z_l)=\vno}
 \bigg)\nn\\
\times\sum_{\bd\bm'=\bd\bk'=\vno}\frac{w_{\cA\compl\cap\,\cV_{\bm+
 \bk}\compl}(\bm')}{Z_{\cA\compl\cap\,\cV_{\bm+\bk}\compl}}\,\frac{
 w_{\tilde\cD\compl\cap\,\cV_{\bm+\bk}\compl}(\bk')}{Z_{\tilde\cD
 \compl\cap\,\cV_{\bm+\bk}\compl}}\,\ind{z_1\cn{\bm'+\bk'}{}z'_1}.
 \lbeq{lace-edges}
\end{gather}
By using \refeq{Theta'-2ndindbd3:j=1bd} (and replacing $\cA\compl$
and $\tilde\cD\compl$ in \refeq{Theta'-2ndindbd3:j=1} by
$\cA\compl\cap\cV_{\bm+\bk}\compl$ and
$\tilde\cD\compl\cap\cV_{\bm+\bk}\compl$, respectively),
the second line of \refeq{lace-edges} is bounded by
$\sum_{l\ge1}(\tilde G_\Lambda^2)^{*(2l-1)}(z_1,z'_1)$.  Repeating
the same argument until the remaining products of the indicators are
used up, we obtain
\begin{align}\lbeq{lace-edgesbd}
\refeq{lace-edges}&\leq\prod_{i=1}^j\sum_{l\ge1}\big(\tilde
 G_\Lambda^2\big)^{*(2l-1)}(z_i,z'_i).
\end{align}

We have proved that
\begin{align}\lbeq{Theta'-2ndindbd4}
\refeq{Theta'-2ndindbd3}\leq\sum_{j\ge1}\sum_{\substack{z_1,\dots,
 z_j\\ z'_1,\dots,z'_j}}\bigg(\prod_{i=1}^j\sum_{l\ge1}\big(\tilde
 G_\Lambda^2\big)^{*(2l-1)}(z_i,z'_i)\bigg)\sum_{\bd\bn=y\vtri x}
 \frac{w_\Lambda(\bn)}{Z_\Lambda}\,\ind{y\cn{\bn}{\cA}x}\nn\\
\times\sum_{T\ge j}\sum_{\vec b_T}\sum_{\{s_it_i\}_{i=1}^j\in\cL_{[0,
 T]}^{(j)}}\indic{H_{\bn;\vec b_T}(y,x)}\prod_{i=1}^j\ind{z_i\in\cD_{
 \bn;s_i},\,z'_i\in\cD_{\bn;t_i}}.
\end{align}

\smallskip

\textbf{(c-2)} Since \refeq{Theta'-2ndindbd4} depends only on a single
current configuration, we may use Lemma~\ref{lmm:GHS-BK} to obtain an
upper bound.  To do so, we first simplify the second line of
\refeq{Theta'-2ndindbd4}, which is, by definition, equal to the
indicator of the disjoint union
\begin{align}\lbeq{fin-ind}
&\Bigcup_{T\ge j}\,\Bigcup_{\vec b_T}\Bigcup_{\{s_it_i\}_{i=1}^j\in
 \cL_{[0,T]}^{\sss(j)}}\bigg\{H_{\bn;\vec b_T}(y,x)\cap\bigcap_{i=1}^j
 \big\{z_i\in\cD_{\bn;s_i},\,z'_i\in\cD_{\bn;t_i}\big\}\bigg\}\\
&=\Bigcup_{e_1,\dots,e_j}\Bigg\{\Bigcup_{T\ge j}\,\Bigcup_{\vec b_T}
 \Bigcup_{\substack{\{s_it_i\}_{i=1}^j\in\cL_{[0,T]}^{\sss(j)}\\ b_{t_i
 +1}=e_{i+1}\;\Forall i=0,\dots,j-1}}\bigg\{H_{\bn;\vec b_T}(y,x)\cap
 \bigcap_{i=1}^j\big\{z_i\in\cD_{\bn;s_i},\,z'_i\in\cD_{\bn;t_i}\big\}
 \bigg\}\Bigg\},\nn
\end{align}
where $t_0=0$ by convention.  On the left-hand side of
\refeq{fin-ind}, the first two unions identify the number and
location of the pivotal bonds for $y\cn{\bn}{}x$, and the third
union identifies the indices of double connections associated with
the bypaths between $z_i$ and $z'_i$, for every $i=1,\dots,j$.  The
union over $e_1,\dots,e_j$ on the right-hand side identifies some of
the pivotal bonds $b_1,\dots,b_T$ that are essential to decompose
the chain of double connections $H_{\bn;\vec b_T}(y,x)$ into the
following building blocks (see Figure~\ref{fig:I-def}):
\begin{figure}[t]
\begin{center}
\begin{gather*}
I_1(y,z,x)=~~\raisebox{-12pt}{\includegraphics[scale=0.12]{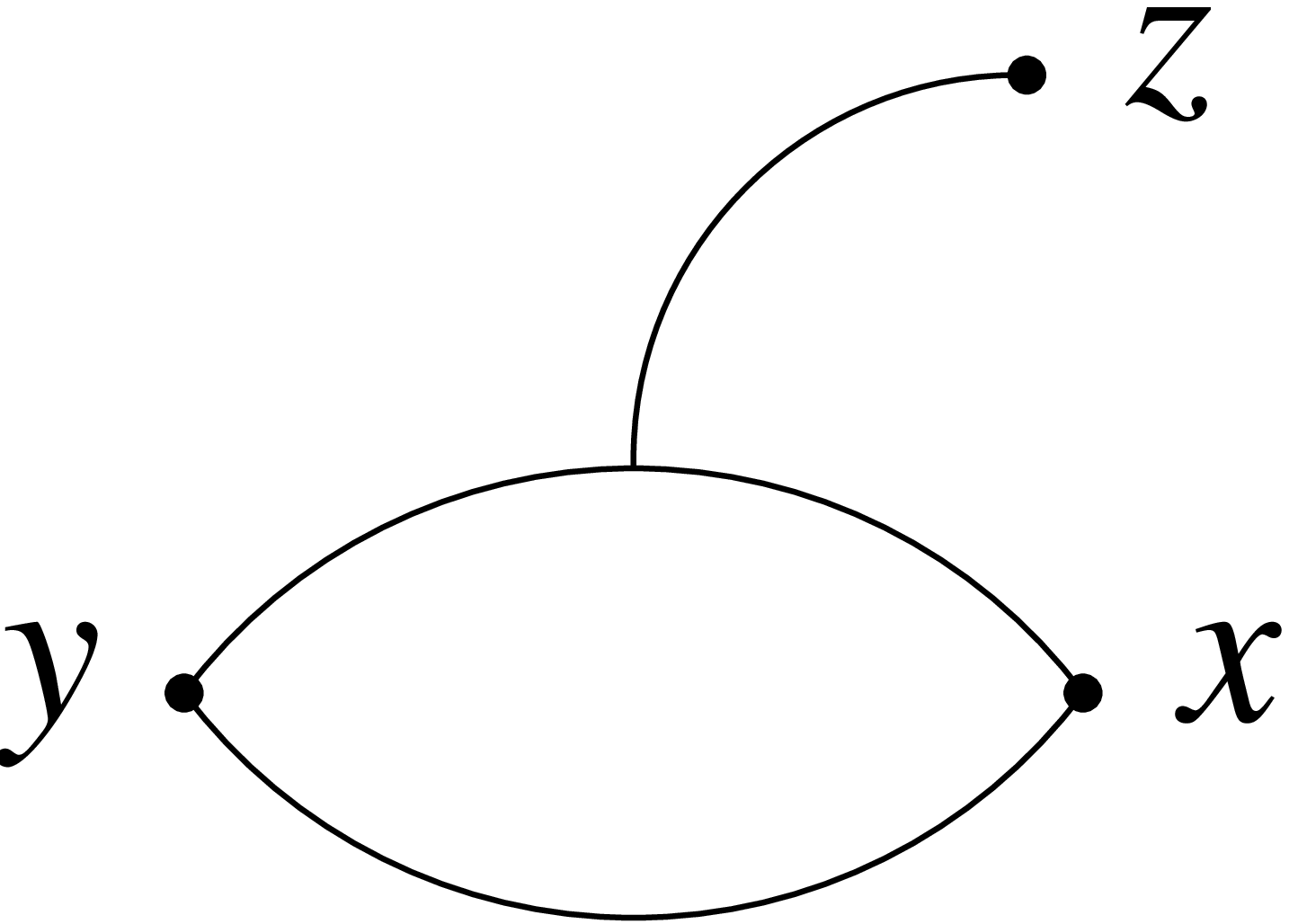}}\hspace{7pc}
I_2(y,z',x)=~~\raisebox{-12pt}{\includegraphics[scale=0.12]{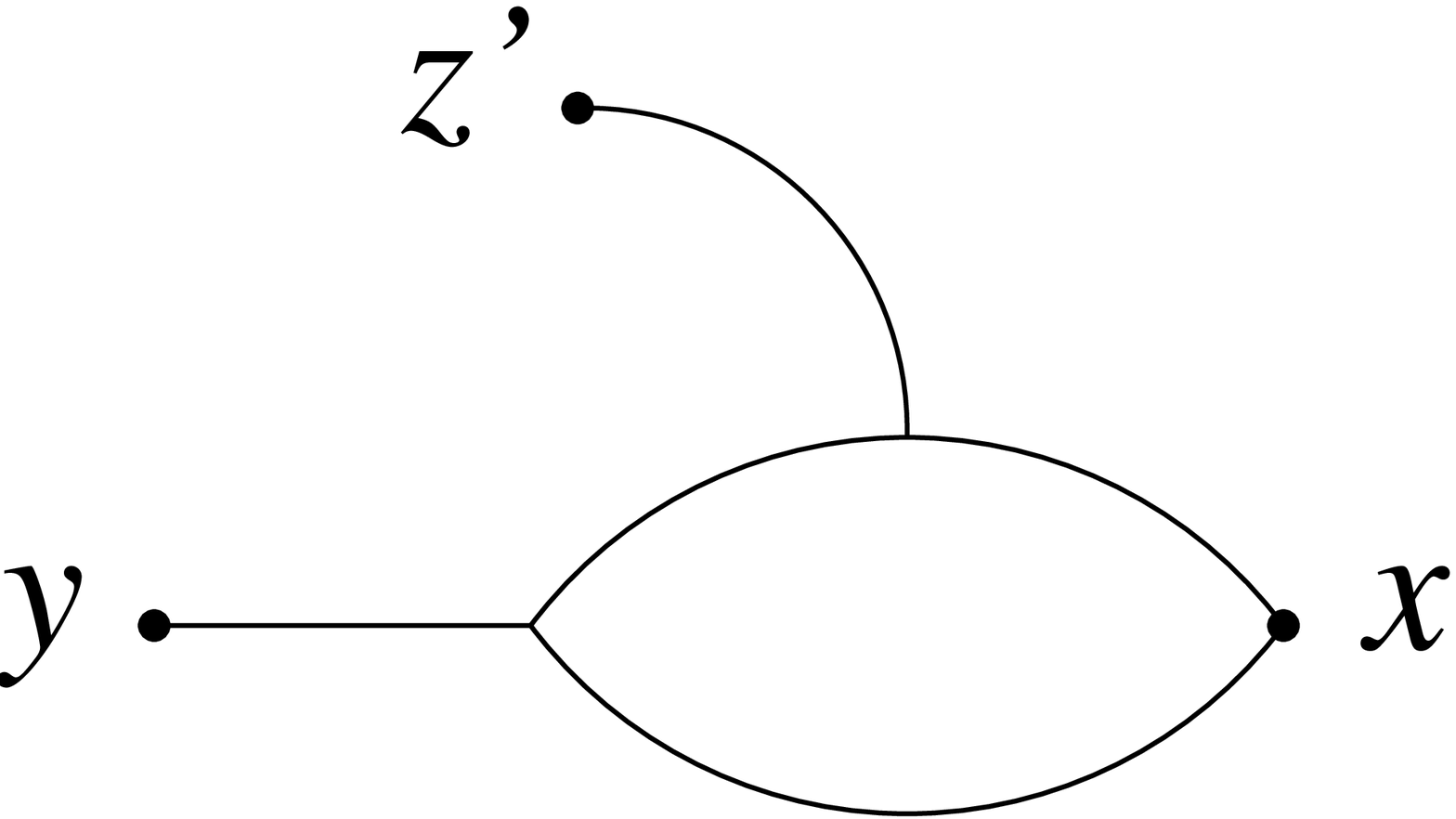}}\\[1pc]
I_3(y,z,z',x)=~~\raisebox{-12pt}{\includegraphics[scale=0.12]{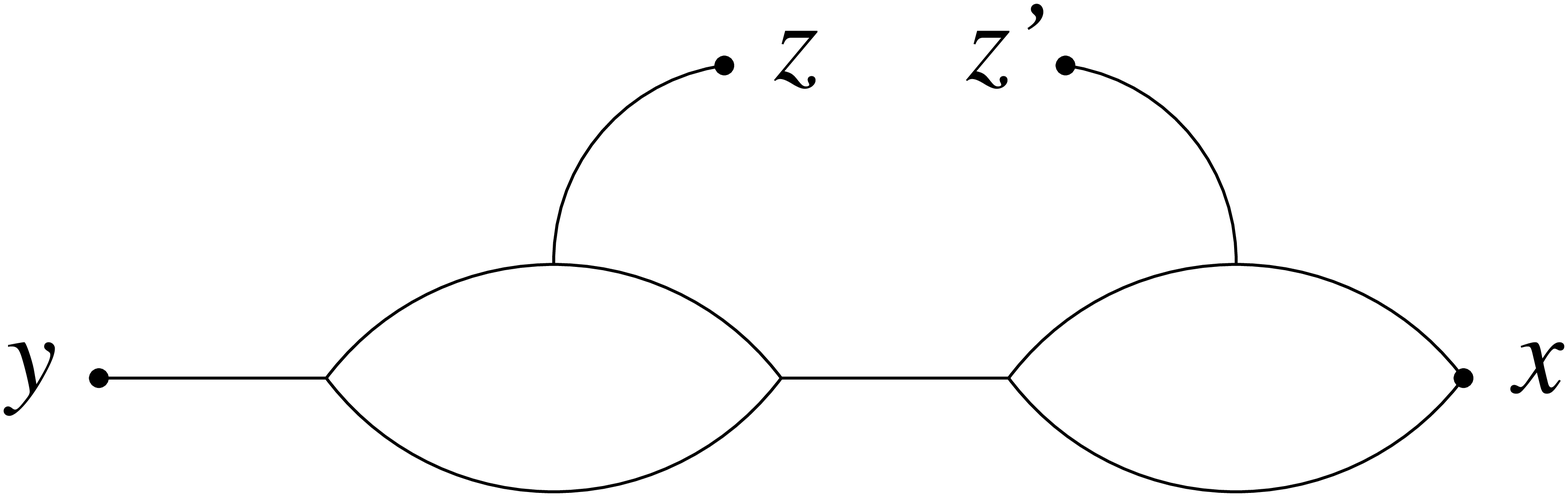}}\quad~
 \cup\quad~\raisebox{-12pt}{\includegraphics[scale=0.12]{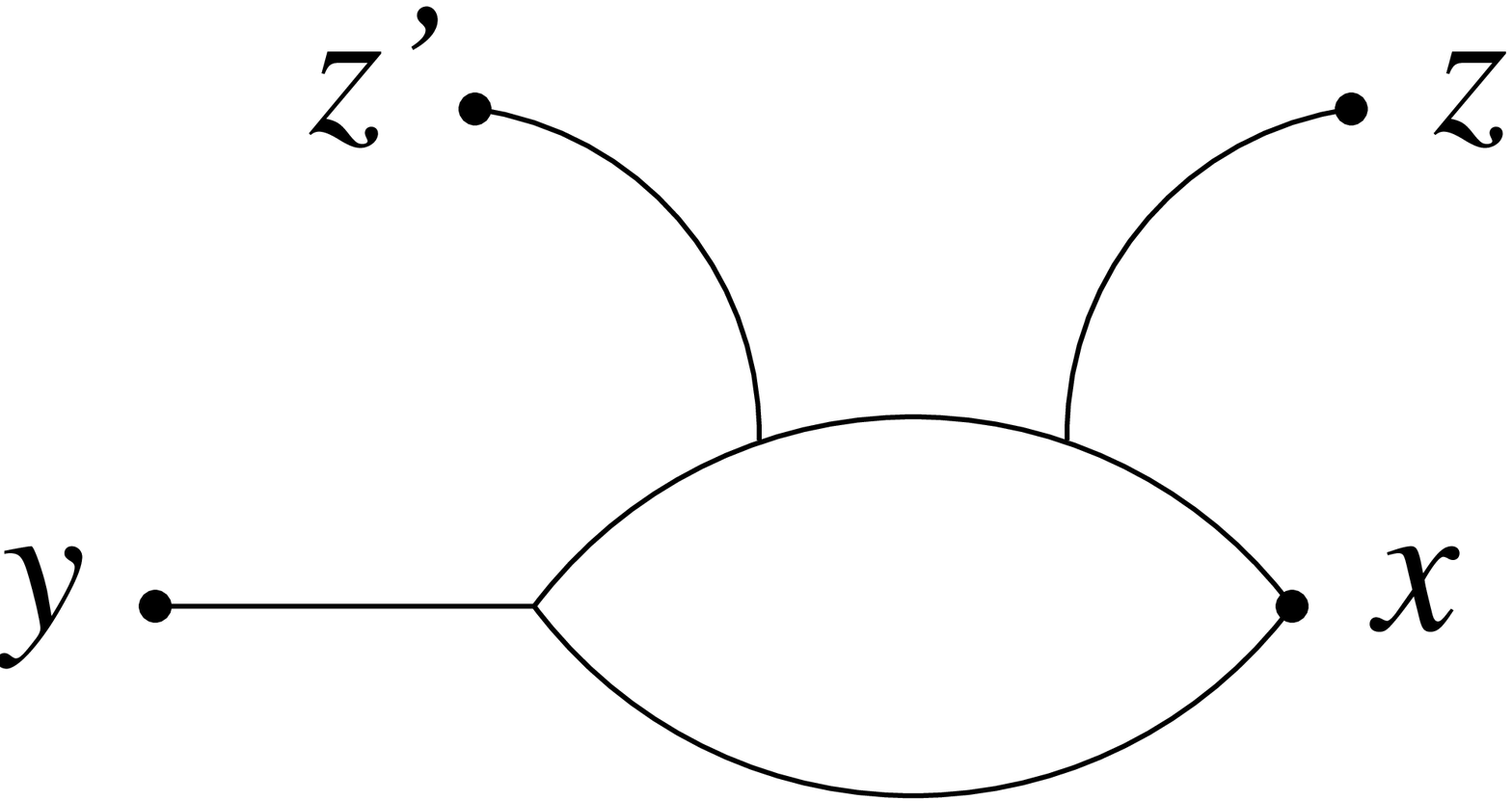}}
\end{gather*}
\caption{\label{fig:I-def}Schematic representations of $I_1(y,z,x)$,
$I_2(y,z',x)$ and $I_3(y,z,z',x)$.}
\end{center}
\end{figure}
\begin{gather}
I_1(y,z,x)=\{y\db{\bn}{}x,~y\cn{\bn}{}z\},\qquad
I_2(y,z',x)=\bigcup_u\big\{\{y\cn{\bn}{}u\}\circ I_1(u,z',x)\big\},
 \lbeq{I12-def}\\
I_3(y,z,z',x)=\bigcup_u\Big\{\{I_2(y,z,u)\circ I_2(u,z',x)\}\cup\big\{
 \{y\cn{\bn}{}u\}\circ\{I_1(u,z,x)\cap I_1(u,z',x)\}\big\}\Big\}.
 \lbeq{I3-def}
\end{gather}
For example, since $\cL_{[0,T]}^{\sss(1)}=\{\{0T\}\}$, we have
\begin{align}\lbeq{fin-ind:=1}
&(\refeq{fin-ind}\text{ for }j=1)=\Bigcup_{e_1}\Bigcup_{T\ge1}\,
 \Bigcup_{\vec b_T:b_1=e_1}\Big\{H_{\bn;\vec b_T}(y,x)\cap\big\{
 z_1\in\cD_{\bn;0},\,z'_1\in\cD_{\bn;T}\big\}\Big\}\nn\\
&\qquad\subset\Bigcup_{e_1}\Big\{\big\{I_1(y,z_1,\eb_1)\circ I_2
 (\te_1,z'_1,x)\big\}\cap\big\{n_{e_1}>0,~e_1\text{ is pivotal for }
 y\cn{\bn}{}x\big\}\Big\}.
\end{align}
It is not hard to see in general that
\begin{align}\lbeq{fin-ind:geq2}
&(\refeq{fin-ind}\text{ for }j\ge2)\nn\\
&\quad\subset\Bigcup_{e_1,\dots,e_j}\bigg\{\Big\{I_1(y,z_1,\eb_1)
 \circ I_3(\te_1,z_2,z'_1,\eb_2)\circ\cdots\circ I_3(\te_{j-1},
 z_j,z'_{j-1},\eb_j)\circ I_2(\te_j,z'_j,x)\Big\}\nn\\
&\hspace{5pc}\cap\bigcap_{i=1}^j\big\{n_{e_i}>0,~e_i
 \text{ is pivotal for }y\cn{\bn}{}x\big\}\bigg\}.
\end{align}

To bound \refeq{Theta'-2ndindbd4} using Lemma~\ref{lmm:GHS-BK}, we further consider an
event that includes \refeq{fin-ind:=1}--\refeq{fin-ind:geq2} as subsets.  Without losing
generality, we can assume that $y\ne\eb_1$, $\te_{i-1}\ne\eb_i$ for $i=2,\dots,j$, and
$\te_j\ne x$; otherwise, the following argument can be simplified.  We consider each
event $I_i$ in \refeq{fin-ind:=1}--\refeq{fin-ind:geq2} individually, and to do so, we
assume that $y$ and $\eb_1$ are the only sources for $I_1(y,z_1,\eb_1)$, that
$\te_{i-1}$ and $\eb_i$ are the only sources for $I_3(\te_{i-1},z_i,z'_{i-1},\eb_i)$ for
every $i=2,\dots,j$, and that $\te_j$ and $x$ are the only sources for
$I_2(\te_j,z'_j,x)$. This is because $y$ and $x$ are the only sources for the entire
event \refeq{fin-ind:geq2}, and every $e_i$ is pivotal for $y\cn{\bn}{}x$.

On $I_1(y,z,x)$ with $y,x$ being the only sources, according to
the observation in Step~(i) described below \refeq{dbbd}, we have two
edge-disjoint connections from $y$ to $z$, one of which may go through
$x$, and another edge-disjoint connection from $y$ to $x$ (cf.,
$I_1(y,z,x)$ in Figure~\ref{fig:I-def}).  Therefore,
\begin{align}\lbeq{I1-supset}
I_1(y,z,x)\subset\big\{\Exists\omega_1,\omega_2\in\Omega_{y\to z}^\bn
 \,\Exists\omega_3\in\Omega_{y\to x}^\bn\text{ such that }~\omega_i
 \cap\omega_l=\vno~(i\ne l)\big\}.
\end{align}
Similarly, for $I_2(y,z',x)$ with $y,x$ being the only sources (cf.,
$I_2(y,z',x)$ in Figure~\ref{fig:I-def}),
\begin{align}\lbeq{I2-supset}
I_2(y,z',x)\subset\big\{\Exists\omega_1,\omega_2\in\Omega_{x\to
 z'}^\bn\,\Exists\omega_3\in\Omega_{y\to x}^\bn\text{ such that }
 ~\omega_i\cap\omega_l=\vno~(i\ne l)\big\}.
\end{align}

On $I_3(y,z,z',x)$ with $y,x$ being the only sources, there are at
least three edge-disjoint paths, one from $y$ to $z$, another one
from $z$ to $z'$, and another one from $z'$ to $x$.  It is not hard
to see this from $\bigcup_u\{I_2(y,z,u)\circ I_2(u,z',x)\}$ in
\refeq{I3-def}, which corresponds to the first event depicted in
Figure~\ref{fig:I-def}.
It is also possible to extract such three edge-disjoint paths from the
remaining event in \refeq{I3-def}.  See the second event depicted in
Figure~\ref{fig:I-def} for one of the worst topological situations.
Since there are at least three edge-disjoint paths between $u$ and $x$,
say, $\zeta_1,\zeta_2$ and $\zeta_3$, we can go from $y$ to $z$ via
$\zeta_1$ and a part of $\zeta_2$, and go from $z$ to $z'$ via the
middle part of $\zeta_2$, and then go from $z'$ to $x$ via the remaining
part of $\zeta_2$ and $\zeta_3$.  The other cases can be dealt with
similarly.  As a result, we have
\begin{align}\lbeq{I3-supset}
I_3(y,z,z',x)\subset\big\{\Exists\omega_1\in\Omega_{y\to z}^\bn\,\Exists
 \omega_2\in\Omega_{z\to z'}^\bn\,\Exists\omega_3\in\Omega_{z'\to x}^\bn
 \text{ such that }\omega_i\cap\omega_l=\vno~(i\ne l)\big\}.
\end{align}

\begin{figure}[t]
\begin{center}
\includegraphics[scale=0.19]{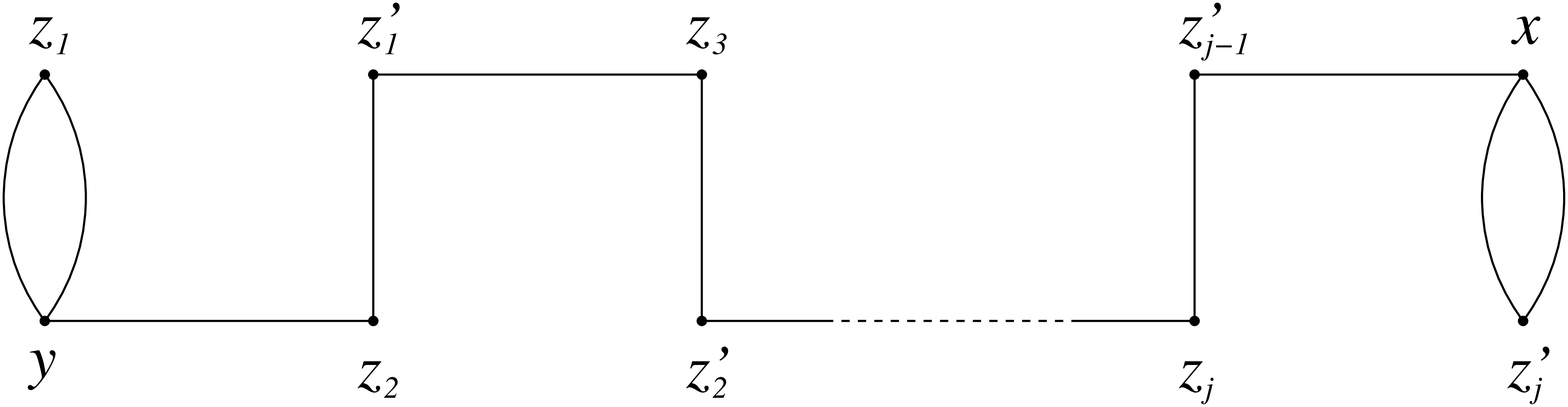}
\caption{\label{fig:eventI}A schematic representation of
$\tilde I_{\vec z_j,\vec z'_j}^{\sss(j)}(y,x)$ for $j\ge2$
consisting of $2j+3$ edge-disjoint paths on $\mG_\bn$.}
\end{center}
\end{figure}

Since
\begin{align}
\bigcup_e\Big\{\big\{\{\Exists\omega\in\Omega_{z\to\eb}^\bn\}\circ\{\Exists
 \omega\in\Omega_{\te\to z'}^\bn\}\big\}\cap\{n_e>0\}\Big\}\subset\{\Exists
 \omega\in\Omega_{z\to z'}^\bn\},
\end{align}
we see that \refeq{fin-ind:=1} is a subset of
\begin{align}\lbeq{tildeI-def:=1}
\tilde I_{z_1,z'_1}^{\sss(1)}(y,x)=\left\{\!
\begin{array}{c}
\Exists\omega_1,\omega_2\in\Omega_{z_1\to y}^\bn\;\Exists\omega_3\in
 \Omega_{y\to x}^\bn\;\Exists\omega_4,\omega_5\in\Omega_{x\to z'_1}^\bn\\
\text{such that }~\omega_i\cap\omega_l=\vno~(i\ne l)
\end{array}\!\right\},
\end{align}
and that \refeq{fin-ind:geq2} is a subset of (see
Figure~\ref{fig:eventI})
\begin{align}\lbeq{tildeI-def:geq2}
\tilde I_{\vec z_j,\vec z'_j}^{\sss(j)}(y,x)=\left\{\!
\begin{array}{c}
\Exists\omega_1,\omega_2\in\Omega_{z_1\to y}^\bn\;\Exists\omega_3\in
 \Omega_{y\to z_2}^\bn\;\Exists\omega_4\in\Omega_{z_2\to z'_1}^\bn\;
 \Exists\omega_5\in\Omega_{z'_1\to z_3}^\bn\cdots\\
\cdots\Exists\omega_{2j}\in\Omega_{z_j\to z'_{j-1}}^\bn\,\Exists\omega_{
 2j+1}\in\Omega_{z'_{j-1}\to x}^\bn\;\Exists\omega_{2j+2},\omega_{2j+3}
 \in\Omega_{x\to z'_j}^\bn\\
\text{such that }~\omega_i\cap\omega_l=\vno~(i\ne l)
\end{array}\!\right\},
\end{align}
where $\vec z_j^{(\prime)}=(z_1^{(\prime)},\dots,z_j^{(\prime)})$.
Therefore,
\begin{align}\lbeq{Theta'-2ndindbd5}
\refeq{Theta'-2ndindbd4}\leq\sum_{j\ge1}\sum_{\substack{z_1,\dots,z_j\\
 z'_1,\dots,z'_j}}\bigg(\prod_{i=1}^j\sum_{l\ge1}\big(\tilde G_\Lambda^2
 \big)^{*(2l-1)}(z_i,z'_i)\bigg)\sum_{\bd\bn=y\vtri x}\frac{w_\Lambda
 (\bn)}{Z_\Lambda}\,\ind{y\cn{\bn}{\cA}x}\,\indic{\tilde I_{\vec z_j,
 \vec z'_j}^{\sss(j)}(y,x)}.
\end{align}

Now we apply Lemma~\ref{lmm:GHS-BK} to bound
\refeq{Theta'-2ndindbd5}.  To clearly understand how it is applied,
for now we ignore $\tind{y\cn{\bn}{\cA}x}$ in
\refeq{Theta'-2ndindbd5} and only consider the contribution from
$\indic{\tilde I_{\vec z_j, \vec z'_j}^{\sss(j)}(y,x)}$.  Without
losing generality, we assume that $y,x,z_i,z'_i$ for $i=1,\dots,j$
are all different. Since there are $2j+3$ edge-disjoint paths on
$\mG_\bn$ as in \refeq{tildeI-def:=1}--\refeq{tildeI-def:geq2} (see
also Figure~\ref{fig:eventI}), we multiply \refeq{Theta'-2ndindbd5}
by $(Z_\Lambda/Z_\Lambda)^{2j+2}$, following Step~(ii) of the
strategy described in Section~\ref{ss:pi0bd}.  Overlapping the
$2j+3$ current configurations and using Lemma~\ref{lmm:GHS-BK} with
$\cV=\{y,x\}$ and $k=2j+2$, we obtain
\begin{align}\lbeq{Theta'-2ndindbd6}
\sum_{\bd\bn=y\vtri x}\frac{w_\Lambda(\bn)}{Z_\Lambda}&\,\indic{\tilde
 I_{\vec z_j,\vec z'_j}^{\sss(j)}(y,x)}\leq\Exp{\varphi_{z_1}\varphi_y
 }_\Lambda^2\Exp{\varphi_x\varphi_{z'_j}}_\Lambda^2\\
&\times\begin{cases}
  \dpst\Exp{\varphi_y\varphi_x}_\Lambda&(j=1),\\
  \dpst\Exp{\varphi_y\varphi_{z_2}}_\Lambda\Exp{\varphi_{z_2}\varphi_{
   z'_1}}_\Lambda\bigg(\prod_{i=2}^{j-1}\Exp{\varphi_{z'_{i-1}}\varphi_{
   z_{i+1}}}_\Lambda\Exp{\varphi_{z_{i+1}}\varphi_{z'_i}}_\Lambda\bigg)
   \Exp{\varphi_{z'_{j-1}}\varphi_x}_\Lambda&(j\ge2).
 \end{cases}\nn
\end{align}
Note that, by \refeq{G-delta-bd}, we have
\begin{align}
\left.\begin{array}{r}
\sum_{l\ge1}(\tilde G_\Lambda^2)^{*(2l-1)}(y,x)\\[5pt]
\sum_z\Exp{\varphi_z\varphi_y}_\Lambda^2\sum_{l\ge1}(\tilde
 G_\Lambda^2)^{*(2l-1)}(z,x)\\[5pt]
\sum_{z'}\Exp{\varphi_x\varphi_{z'}}_\Lambda^2\sum_{l\ge1}
 (\tilde G_\Lambda^2)^{*(2l-1)}(y,z')
\end{array}\right\}&\leq\psi_\Lambda(y,x)-\delta_{y,x},\\[5pt]
\sum_{z,z'}\Exp{\varphi_z\varphi_y}_\Lambda^2\Exp{\varphi_x
 \varphi_{z'}}_\Lambda^2\sum_{l\ge1}\big(\tilde G_\Lambda^2
 \big)^{*(2l-1)}(z,z')&\leq2\big(\psi_\Lambda(y,x)-\delta_{y,x}\big).
\end{align}
Therefore, \refeq{Theta'-2ndindbd5} without
$\tind{y\cn{\bn}{\cA}x}$ is bounded by
\begin{align}\lbeq{Theta'-2ndindbd7}
&\Exp{\varphi_y\varphi_x}_\Lambda\sum_{z_1,z'_1}\Exp{\varphi_{z_1}
 \varphi_y}_\Lambda^2\Exp{\varphi_x\varphi_{z'_1}}_\Lambda^2\sum_{l
 \ge1}\big(\tilde G_\Lambda^2\big)^{*(2l-1)}(z_1,z'_1)\nn\\
&+\sum_{j\ge2}\sum_{\substack{z_2,\dots,z_j\\ z'_1,\dots,z'_{j-1}}}
 \bigg(\prod_{i=2}^{j-1}\big(\psi_\Lambda(z_i,z'_i)-\delta_{z_i,z'_i}
 \big)\bigg)\bigg(\sum_{z_1}\Exp{\varphi_y\varphi_{z_1}}_\Lambda^2
 \sum_{l\ge1}\big(\tilde G_\Lambda^2\big)^{*(2l-1)}(z_1,z'_1)\bigg)
 \nn\\
&\hspace{4pc}\times\bigg(\sum_{z'_j}\Exp{\varphi_x\varphi_{z'_j}
 }_\Lambda^2\sum_{l\ge1}\big(\tilde G_\Lambda^2\big)^{*(2l-1)}(z_j,
 z'_j)\bigg)\Exp{\varphi_y\varphi_{z_2}}_\Lambda\Exp{\varphi_{z_2}
 \varphi_{z'_1}}_\Lambda\nn\\
&\hspace{4pc}\times\bigg(\prod_{i=2}^{j-1}\Exp{\varphi_{z'_{i-1}}
 \varphi_{z_{i+1}}}_\Lambda\Exp{\varphi_{z_{i+1}}\varphi_{z'_i}}_\Lambda
 \bigg)\Exp{\varphi_{z'_{j-1}}\varphi_x}_\Lambda\leq\sum_{j\ge1}
 P_\Lambda^{\sss(j)}(y,x).
\end{align}

If $\tind{y\cn{\bn}{\cA}x}$ is present in the above argument, then
at least one of the paths $\omega_i$ for $i=3,\dots,2j+1$ has to go
through $\cA$.  For example, if $\omega_3~(\in\Omega_{y\to
z_2}^\bn)$ goes through $\cA$, then we can split it into two
edge-disjoint paths at some $u\in\cA$, such as
$\omega'_3\in\Omega_{y\to u}^\bn$ and $\omega''_3\in\Omega_{u\to
z_2}^\bn$.  The contribution from this case is bounded, by following
the same argument as above, by \refeq{Theta'-2ndindbd6} with
$\Exp{\varphi_y\varphi_{z_2}}_\Lambda$ being replaced by
$\sum_{u\in\cA}
\Exp{\varphi_y\varphi_u}_\Lambda\Exp{\varphi_u\varphi_{z_2}}_\Lambda$.
Bounding the other $2j-2$ cases similarly and summing these bounds
over $j\ge1$, we obtain
\begin{align}\lbeq{Theta'-2ndindbd8}
\refeq{Theta'-2ndindbd5}\leq\sum_{u\in\cA}\sum_{j\ge1}P_{\Lambda;
 u}^{\prime{\sss(j)}}(y,x).
\end{align}

This together with \refeq{Theta'-bd1stbd} in the above paragraph~(a)
complete the proof of the bound on $\Theta'_{y,x;\cA}$ in
\refeq{Theta'Theta''bd}.
\qed

\bigskip

\textbf{(d)} Finally, we investigate the contribution to
$\Theta''_{y,x,v;\cA}$ from $\tind{y\db{\bm+\bn}{}x\}\setminus
\{y\db{\bn}{}x}$ in \refeq{Theta'-evdec}:
\begin{align}\lbeq{contr-(d)}
\sum_{\substack{\bd\bm=\vno\\ \bd\bn=y\vtri x}}\frac{w_{\cA\compl}(\bm)}
 {Z_{\cA\compl}}\,\frac{w_\Lambda(\bn)}{Z_\Lambda}\,\ind{y\cn{\bm+\bn}
 {\cA}x\}\,\cap\,\{\{y\db{\bm+\bn}{}x\}\setminus\{y\db{\bn}{}x\}\}\,\cap
 \,\{y\cn{\bm+\bn}{}v}.
\end{align}
Using $H_{\bn;\vec b_T}(y,x)$ defined in \refeq{H-def}, we can write
\refeq{contr-(d)} as (cf., \refeq{Theta'-2ndindbd2})
\begin{align}\lbeq{Theta''-2ndindrewr}
\refeq{contr-(d)}=\sum_{T\ge1}\sum_{\vec b_T}\sum_{\substack{\bd\bm=\vno\\
 \bd\bn=y\vtri x}}\frac{w_{\cA\compl}(\bm)}{Z_{\cA\compl}}\,\frac{w_\Lambda
 (\bn)}{Z_\Lambda}\,\ind{y\cn{\bm+\bn}{\cA}x\}\,\cap\,H_{\bn;\vec b_T}(y,x)
 \,\cap\,\{y\db{\bm+\bn}{}x\}\,\cap\,\{y\cn{\bm+\bn}{}v}.
\end{align}
To bound this, we will also use a similar expression to
\refeq{Theta'-2ndindbd3}, in which
$\bk=\bn|_{\mB_{\tilde\cD\compl}}$ with $\tilde\cD=\cC_\bn^b(y)$. We
investigate \refeq{Theta''-2ndindrewr} separately (in the following
paragraphs~(d-1) and (d-2)) depending on whether or not there is a
bypath $\cC_{\bm+\bk}(z_i)$ for some $i\in\{1,\dots,j\}$ containing
$v$.

\smallskip

\textbf{(d-1)} If there is such a bypath, then we use
$\tind{y\cn{\bm+\bn}{\cA}x}\leq\tind{y\cn{\bn}{\cA}x}$ as in
\refeq{Theta'-2ndindbd2} to bound the contribution from this case to
\refeq{Theta''-2ndindrewr} by
\begin{align}\lbeq{Theta''-2ndindbd1}
\sum_{T\ge1}\sum_{\vec b_T}\sum_{\bd\bn=y\vtri x}\frac{w_\Lambda(\bn)}
 {Z_\Lambda}\,\indic{\{y\cn{\bn}{\cA}x\}\,\cap\,H_{\bn;\vec b_T}(y,x)}
 \sum_{j=1}^T\,\sum_{\{s_it_i\}_{i=1}^j\in\cL_{[0,T]}^{(j)}}\,\sum_{
 \substack{z_1,\dots,z_j\\ z'_1,\dots,z'_j}}\bigg(\prod_{i=1}^j\ind{z_i
 \in\cD_{\bn;s_i},~z'_i\in\cD_{\bn;t_i}}\bigg)\nn\\
\times\sum_{\substack{\bd\bm=\vno\\ \bd\bk=\vno}}\frac{w_{\cA\compl}
 (\bm)}{Z_{\cA\compl}}\,\frac{w_{\tilde\cD\compl}(\bk)}{Z_{\tilde\cD
 \compl}}\bigg(\prod_{i=1}^j\ind{z_i\cn{\bm+\bk}{}z'_i}\bigg)\bigg(
 \prod_{i\ne l}\ind{\cC_{\bm+\bk}(z_i)\,\cap\,\cC_{\bm+\bk}(z_l)=\vno}
 \bigg)\sum_{i=1}^j\ind{v\in\cC_{\bm+\bk}(z_i)}.
\end{align}
Note that the last sum of the indicators is the only difference from
\refeq{Theta'-2ndindbd3}.

When $j=1$, the second line of \refeq{Theta''-2ndindbd1} equals
\begin{align}\lbeq{Theta''-2ndindbd1:j=1}
\sum_{\bd\bm=\bd\bk=\vno}\frac{w_{\cA\compl}(\bm)}{Z_{\cA\compl}}\,
 \frac{w_{\tilde\cD\compl}(\bk)}{Z_{\tilde\cD\compl}}\,\ind{z_1
 \cn{\bm+\bk}{}z'_1}\,\ind{z_1\cn{\bm+\bk}{}v}.
\end{align}
As described in \refeq{Theta'-2ndindbd3:j=1}--\refeq{Theta'-2ndindbd3:j=1bd}, we can
bound \refeq{Theta''-2ndindbd1:j=1} without $\tind{z_1\cn{\bm+\bk}{}v}$ by a chain of
bubbles $\sum_{l\ge1}(\tilde G_\Lambda^2)^{*(2l-1)}(z_1,z'_1)$.  If
$\tind{z_1\cn{\bm+\bk}{}v}=1$, then, by the argument around
\refeq{ind-bd}--\refeq{nsum-2ndbd}, one of the bubbles has an extra vertex $v'$ that is
further connected to $v$ with another chain of bubbles $\psi_\Lambda(v',v)$.  That is,
the effect of $\tind{z_1\cn{\bm+\bk}{}v}$ is to replace one of the $\tilde G_\Lambda$'s
in the chain of bubbles, say, $\tilde G_\Lambda(a,a')$, by $\sum_{v'}(\Exp{\varphi_{a}
\varphi_{v'}}_\Lambda\tilde G_\Lambda(v',a')+\tilde G_\Lambda
(a,a')\delta_{v',a'})\,\psi_\Lambda(v',v)$.  Let
\begin{align}\lbeq{g-def}
g_{\Lambda;y}(z,z')=\sum_{l\ge1}\sum_{i=1}^{2l-1}\sum_{a,a'}\big(
 \tilde G_\Lambda^2\big)^{*(i-1)}(z,a)\,\tilde G_\Lambda(a,a')\,
 \big(\tilde G_\Lambda^2\big)^{*(2l-1-i)}(a',z')\nn\\
\times\Big(\Exp{\varphi_a\varphi_y}_\Lambda\tilde G_\Lambda(y,a')
 +\tilde G_\Lambda(a,a')\,\delta_{y,a'}\Big).
\end{align}
Then, we have
\begin{align}\lbeq{Theta''-2ndindbd1:j=1bd}
\refeq{Theta''-2ndindbd1:j=1}\leq\sum_{v'}g_{\Lambda;v'}(z_1,z'_1)
 \,\psi_\Lambda(v',v).
\end{align}

Let $j\ge2$ and consider the contribution to
\refeq{Theta''-2ndindbd1} from $\tind{v\in\cC_{\bm+\bk}(z_1)}$; the
contribution from $\tind{v\in\cC_{\bm+\bk}(z_i)}$ with $i\ne1$ can
be estimated in the same way.  By conditioning on
$\cV_{\bm+\bk}\equiv\Bigcup_{i\ge2}\cC_{\bm+\bk}(z_i)$ as in
\refeq{lace-edges}, the contribution to the second line of
\refeq{Theta''-2ndindbd1} from
$\tind{v\in\cC_{\bm+\bk}(z_1)}\equiv\tind{z_1\cn{\bm+\bk}{}v}$ equals
\begin{align}\lbeq{Theta''-2ndindbd2}
&\sum_{\bd\bm=\bd\bk=\vno}\frac{w_{\cA\compl}(\bm)}{Z_{\cA\compl}}\,
 \frac{w_{\tilde\cD\compl}(\bk)}{Z_{\tilde\cD\compl}}\bigg(\prod_{i=
 2}^j\ind{z_i\cn{\bm+\bk}{}z'_i}\bigg)\bigg(\prod_{\substack{i,i'\ge
 2\\ i\ne i'}}\ind{\cC_{\bm+\bk}(z_i)\,\cap\,\cC_{\bm+\bk}(z_{i'})=
 \vno}\bigg)\nn\\
&\qquad\times\sum_{\bd\bm'=\bd\bk'=\vno}\frac{w_{\cA\compl\cap\,\cV_{
 \bm+\bk}\compl}(\bm')}{Z_{\cA\compl\cap\,\cV_{\bm+\bk}\compl}}\,
 \frac{w_{\tilde\cD\compl\cap\,\cV_{\bm+\bk}\compl}(\bk')}{Z_{\tilde
 \cD\compl\cap\,\cV_{\bm+\bk}\compl}}\,\ind{z_1\cn{\bm'+\bk'}{}z'_1}
 \,\ind{z_1\cn{\bm'+\bk'}{}v},
\end{align}
where the second line is bounded by \refeq{Theta''-2ndindbd1:j=1bd}
for $j=1$, and then the first line is bounded by
$\prod_{i=2}^j\sum_{l\ge1}(\tilde G_\Lambda^2)^{*(2l-1)}(z_i,z'_i)$,
due to \refeq{lace-edges}--\refeq{lace-edgesbd}.

Summarizing the above bounds, we have (cf., \refeq{Theta'-2ndindbd5})
\begin{align}\lbeq{Theta''-2ndindbd2.2}
\refeq{Theta''-2ndindbd1}\leq\sum_{j\ge1}\sum_{\substack{z_1,\dots,
 z_j\\ z'_1,\dots,z'_j}}\bigg(&\sum_{h=1}^j\sum_{v'}g_{\Lambda;v'}
 (z_h,z'_h)\,\psi_\Lambda(v',v)\prod_{i\ne h}\sum_{l\ge1}\big(
 \tilde G_\Lambda^2\big)^{*(2l-1)}(z_i,z'_i)\bigg)\nn\\
&\times\sum_{\bd\bn=y\vtri x}\frac{w_\Lambda(\bn)}{Z_\Lambda}\,\ind{
 y\cn{\bn}{\cA}x}\,\indic{\tilde I_{\vec z_j,\vec z'_j}^{\sss(j)}(y,
 x)},
\end{align}
to which we can apply the bound discussed between
\refeq{Theta'-2ndindbd2} and \refeq{Theta'-2ndindbd8}.

\smallskip

\textbf{(d-2)} If $v\notin\cC_{\bm+\bk}(z_i)$ for any $i=1,\dots,j$,
then there exists a $v'\in\cD_{\bn;l}$ for some $l\in\{0,\dots,T\}$
such that $v'\cn{\bm+\bk}{}v$ and
$\cC_{\bm+\bk}(v')\cap\cC_{\bm+\bk} (z_i)=\vno$ for any $i$.  In
addition, since all connections from $y$ to $x$ on the graph
$\tilde\cD\cup\Bigcup_{i=1}^{\raisebox{-2pt} {$\scriptstyle
j$}}\cC_{\bm+\bk} (z_i)$ have to go through $\cA$, there is an
$h\in\{1,\dots,j\}$ such that $z_h\cn{\bm+\bk}{\cA}z'_h$. Therefore,
the contribution from this case to \refeq{Theta''-2ndindrewr} is
bounded by
\begin{gather}
\sum_{T\ge1}\sum_{\vec b_T}\sum_{\bd\bn=y\vtri x}\!\frac{w_\Lambda(\bn)}
 {Z_\Lambda}\,\indic{H_{\bn;\vec b_T}(y,x)}\sum_{j=1}^T\sum_{\{s_it_i\}_{i
 =1}^j\in\cL_{[0,T]}^{(j)}}\sum_{\substack{v',z_1,\dots,z_j\\ z'_1,\dots,
 z'_j}}\!\bigg(\prod_{i=1}^j\ind{z_i\in\cD_{\bn;s_i},\;z'_i\in\cD_{\bn;
 t_i}}\bigg)\sum_{l=0}^T\ind{v'\in\cD_{\bn;l}}\nn\\
\times\sum_{\substack{\bd\bm=\vno\\ \bd\bk=\vno}}\frac{w_{\cA\compl}(\bm)}
 {Z_{\cA\compl}}\,\frac{w_{\tilde\cD\compl}(\bk)}{Z_{\tilde\cD\compl}}
 \bigg(\sum_{h=1}^j\ind{z_h\cn{\bm+\bk}{\cA}z'_h}
 \prod_{i=1}^j\ind{z_i\cn{\bm+\bk}{}z'_i}\bigg)\bigg(\prod_{i\ne i'}
 \ind{\cC_{\bm+\bk}(z_i)\,\cap\,\cC_{\bm+\bk}(z_{i'})=\vno}\bigg)\nn\\
\times\ind{v'\cn{\bm+\bk}{}v}\prod_{i=1}^j\ind{\cC_{\bm+\bk}(v')\,\cap\,
 \cC_{\bm+\bk}(z_i)=\vno},\lbeq{Theta''-2ndindbd3}
\end{gather}
where, by conditioning on $\cS_{\bm+\bk}\equiv\Bigcup_{i=1}^{\raisebox
{-2pt}{$\scriptstyle j$}}\cC_{\bm+\bk}(z_i)$, the last two lines are
(see below \refeq{lace-edges})
\begin{gather}
\sum_{\bd\bm=\bd\bk=\vno}\frac{w_{\cA\compl}(\bm)}{Z_{\cA\compl}}\,
 \frac{w_{\tilde\cD\compl}(\bk)}{Z_{\tilde\cD\compl}}\bigg(\sum_{h=1}^j
 \ind{z_h\cn{\bm+\bk}{\cA}z'_h}\prod_{i=1}^j\ind{z_i\cn{\bm+\bk}{}z'_i}
 \bigg)\bigg(\prod_{i\ne i'}\ind{\cC_{\bm+\bk}(z_i)\,\cap\,\cC_{\bm+\bk}
 (z_{i'})=\vno}\bigg)\nn\\
\times\underbrace{\sum_{\bd\bm''=\bd\bk''=\vno}\frac{w_{\cA\compl\cap\,
 \cS_{\bm+\bk}\compl}(\bm'')}{Z_{\cA\compl\cap\,\cS_{\bm+\bk}\compl}}\,
 \frac{w_{\tilde\cD\compl\cap\,\cS_{\bm+\bk}\compl}(\bk'')}{Z_{\tilde\cD
 \compl\cap\,\cS_{\bm+\bk}\compl}}\,\ind{v'\cn{\bm''+\bk''}{}v}}_{\leq\;
 \psi_\Lambda(v',v)}.\lbeq{Theta''-2ndindbd3-l2,3}
\end{gather}

When $j=1$, we have
\begin{align}\lbeq{Theta''-2ndindbd3-l2,3:j=1}
(\refeq{Theta''-2ndindbd3-l2,3}\text{ for }j=1)\leq\psi_\Lambda
 (v',v)\sum_{\bd\bm=\bd\bk=\vno}\frac{w_{\cA\compl}(\bm)}{Z_{\cA
 \compl}}\,\frac{w_{\tilde\cD\compl}(\bk)}{Z_{\tilde\cD\compl}}\,
 \ind{z_1\cn{\bm+\bk}{\cA}z'_1}.
\end{align}
If we ignore the ``through $\cA$''-condition in the last indicator,
then the sum is bounded, as in \refeq{Theta'-2ndindbd3:j=1bd}, by a
chain of bubbles $\sum_{l\ge1}(\tilde
G_\Lambda^2)^{*(2l-1)}(z_1,z'_1)$.  However, because of this
condition, one of the $\tilde G_\Lambda$'s in the bound, say,
$\tilde G_\Lambda(a,a')$, is replaced by
$\sum_{u\in\cA}(\Exp{\varphi_{a}\varphi_u}_\Lambda\tilde
G_\Lambda(u,a') +\tilde G_\Lambda(a,a')\delta_{u,a'})$.  Using
\refeq{g-def}, we have
\begin{align}\lbeq{Theta''-2ndindbd3-l2,3:j=1bd}
\refeq{Theta''-2ndindbd3-l2,3:j=1}\leq\psi_\Lambda(v',v)\sum_{y\in\cA}
 g_{\Lambda;y}(z_1,z'_1).
\end{align}

Let $j\ge2$ and consider the contribution to
\refeq{Theta''-2ndindbd3-l2,3} from $\tind{z_1\cn{\bm+\bk}{\cA}z'_1}$;
the contributions from $\tind{z_h\cn{\bm+\bk}{\cA}z'_h}$ with $h\ne1$
can be estimated similarly.  By conditioning on
$\cV_{\bm+\bk}\equiv\Bigcup_{i\ge2}\cC_{\bm+\bk}(z_i)$, the
contribution to \refeq{Theta''-2ndindbd3-l2,3} from
$\tind{z_1\cn{\bm+\bk}{\cA}z'_1}$ equals
\begin{align}\lbeq{Theta''-2ndindbd4}
\sum_{\bd\bm=\bd\bk=\vno}&\frac{w_{\cA\compl}(\bm)}{Z_{\cA\compl}}\,
 \frac{w_{\tilde\cD\compl}(\bk)}{Z_{\tilde\cD\compl}}\bigg(\prod_{i=
 2}^j\ind{z_i\cn{\bm+\bk}{}z'_i}\bigg)\bigg(\prod_{\substack{i,i'\ge
 2\\ i\ne i'}}\ind{\cC_{\bm+\bk}(z_i)\,\cap\,\cC_{\bm+\bk}(z_{i'})=
 \vno}\bigg)\nn\\
&\times\psi_\Lambda(v',v)\sum_{\bd\bm'=\bd\bk'=\vno}\frac{w_{\cA\compl
 \cap\,\cV_{\bm+\bk}\compl}(\bm')}{Z_{\cA\compl\cap\,\cV_{\bm+\bk}
 \compl}}\,\frac{w_{\tilde\cD\compl\cap\,\cV_{\bm+\bk}\compl}(\bk')}
 {Z_{\tilde\cD\compl\cap\,\cV_{\bm+\bk}\compl}}\,\ind{z_1\cn{\bm'+
 \bk'}{\cA}z'_1},
\end{align}
where the second line is bounded by
\refeq{Theta''-2ndindbd3-l2,3:j=1bd} for $j=1$, and then the first
line is bounded by $\prod_{i=2}^j\sum_{l\ge1}(\tilde G_\Lambda^2)^{
*(2l-1)}(z_i,z'_i)$, as described below \refeq{Theta''-2ndindbd2}.

As a result, \refeq{Theta''-2ndindbd3} is bounded by
\begin{align}\lbeq{Theta''-2ndindbd5}
&\sum_{j\ge1}\sum_{\substack{v'\!,z_1,\dots,z_j\\ z'_1,\dots,z'_j}}
 \psi_\Lambda(v',v)\bigg(\sum_{h=1}^j\sum_{y\in\cA}g_{\Lambda;y}(z_h,
 z'_h)\prod_{i\ne h}\sum_{l\ge1}\big(\tilde G_\Lambda^2\big)^{*(2l-
 1)}(z_i,z'_i)\bigg)\nn\\
&\times\sum_{\bd\bn=y\vtri x}\frac{w_\Lambda(\bn)}{Z_\Lambda}\sum_{T
 \ge j}\sum_{\vec b_T}\sum_{\{s_it_i\}_{i=1}^j\in\cL_{[0,T]}^{(j)}}
 \indic{H_{\bn;\vec b_T}(y,x)}\bigg(\prod_{i=1}^j\ind{z_i\in\cD_{\bn;
 s_i},\,z'_i\in\cD_{\bn;t_i}}\bigg)\sum_{l=0}^T\ind{v'\in\cD_{\bn;l}}.
\end{align}
The second line can be bounded by following the argument between
\refeq{Theta'-2ndindbd4} and \refeq{Theta'-2ndindbd7}; note that the
sum of the indicators in \refeq{Theta''-2ndindbd5}, except for the
last factor $\sum_{l=0}^T\tind{v'\in\cD_{\bn;l}}$, is identical to
that in \refeq{Theta'-2ndindbd4}.  First, we rewrite the sum of the
indicators in \refeq{Theta''-2ndindbd5} as a single indicator of an
event $\cE$ similar to \refeq{fin-ind}.  Then, we construct another
event similar to $\tilde I^{\sss(j)}_{\vec z_j,\vec z'_j}(y,x)$ in
\refeq{tildeI-def:=1}--\refeq{tildeI-def:geq2}, of which $\cE$ is a
subset.  Due to $\sum_{l=0}^T\tind{v'\in\cD_{\bn;l}}$ in
\refeq{Theta''-2ndindbd5}, one of the paths in the definition of
$\tilde I^{\sss(j)}_{\vec z_j,\vec z'_j}(y,x)$, say,
$\omega_i\in\Omega^{\bn}_{a\to a'}$ for some $a,a'$ (depending on
$i$) is split into two edge-disjoint paths
$\omega'_i\in\Omega^{\bn}_{a\to v'}$ and
$\omega''_i\in\Omega^{\bn}_{v'\to a'}$, followed by the summation
over $i=3,\dots,2j+1$ (cf., Figure~\ref{fig:eventI}).  Finally, we
apply Lemma~\ref{lmm:GHS-BK} to obtain the desired bound on the last
line of \refeq{Theta''-2ndindbd5}.

Summarizing the above (d-1) and (d-2), we obtain
\begin{align}
\refeq{Theta''-2ndindrewr}\leq\sum_{j\ge1}\sum_{u\in\cA}P_{
 \Lambda;u,v}^{\prime\prime{\sss
 (j)}}(y,x).
\end{align}

This together with \refeq{Theta''-0bdfin} in the above paragraph~(b)
complete the proof of the bound on $\Theta''_{y,x,v;\cA}$ in
\refeq{Theta'Theta''bd}.
\qed

\section{Bounds on $\pi_\Lambda^{\sss(j)}(x)$ assuming the decay of
$G(x)$}
Using the diagrammatic bounds proved in the previous section, we prove
Proposition~\ref{prp:GimpliesPix} in Section~\ref{ss:proof-so}, and
Propositions~\ref{prp:GimpliesPik} and \ref{prp:exp-bootstrap}(iii) in
Section~\ref{ss:proof-nn}.

\subsection{Bounds for the spread-out model}\label{ss:proof-so}
We prove Proposition~\ref{prp:GimpliesPix} for the spread-out model
using the following convolution bounds:

\begin{prp}\label{prp:conv-star}
\begin{enumerate}[(i)]
\item Let $a\ge b>0$ and $a+b>d$.  There is a $C=C(a,b,d)$ such that
\begin{align}\lbeq{conv}
\sum_y\frac1{\veee{y-v}^a}\,\frac1{\veee{x-y}^b}\leq\frac{C}
 {\veee{x-v}^{(a\wedge d+b)-d}}.
\end{align}
\item Let $q\in(\frac{d}2,d)$.  There is a $C'=C'(d,q)$ such that
\begin{align}
\sum_z\frac1{\veee{x-z}^q}\,\frac1{\veee{x'-z}^q}\,\frac1{\veee{z-y}^q}\,
 \frac1{\veee{z-y'}^q}\leq\frac{C'}{\veee{x-y}^q\veee{x'-y'}^q}.\lbeq{star}
\end{align}
\end{enumerate}
\end{prp}

\begin{figure}[t]
\begin{center}
\begin{align*}
\begin{array}{cc}
\text{(a)}&\dpst\sum_z~~\raisebox{-1.4pc}{\includegraphics[scale=0.2]
 {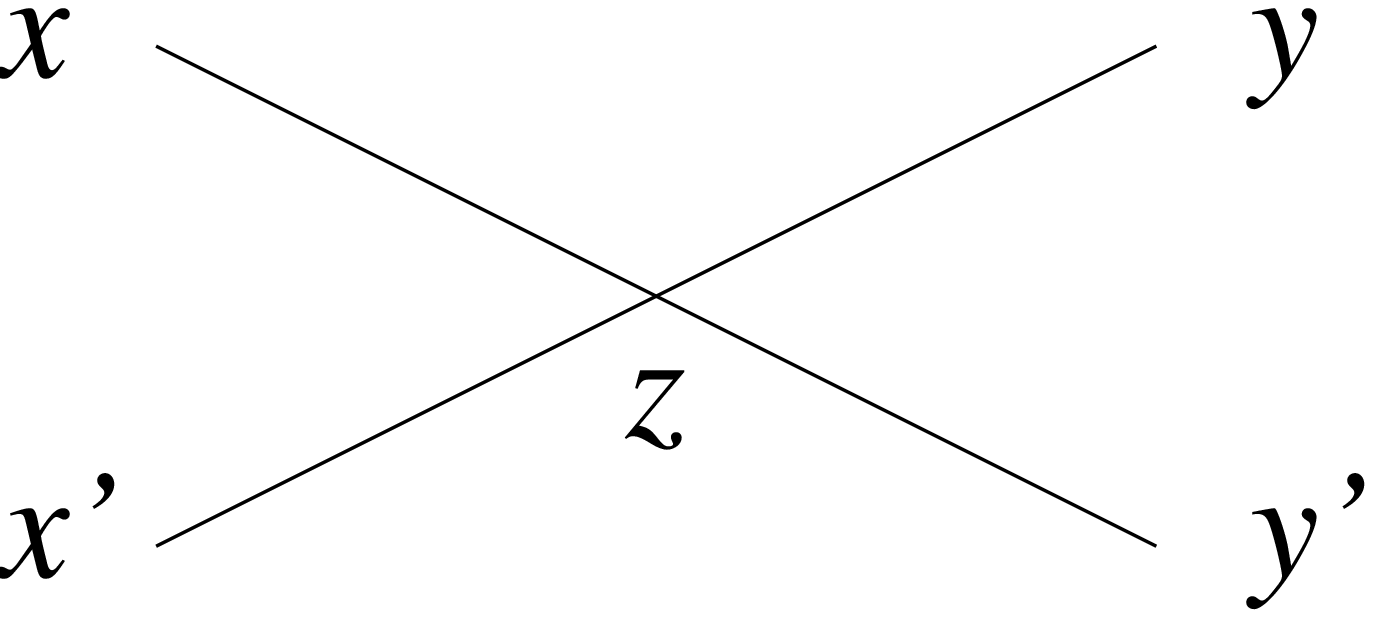}}~~~~\lesssim~~~~\raisebox{-1.4pc}{\includegraphics[scale=0.2]
 {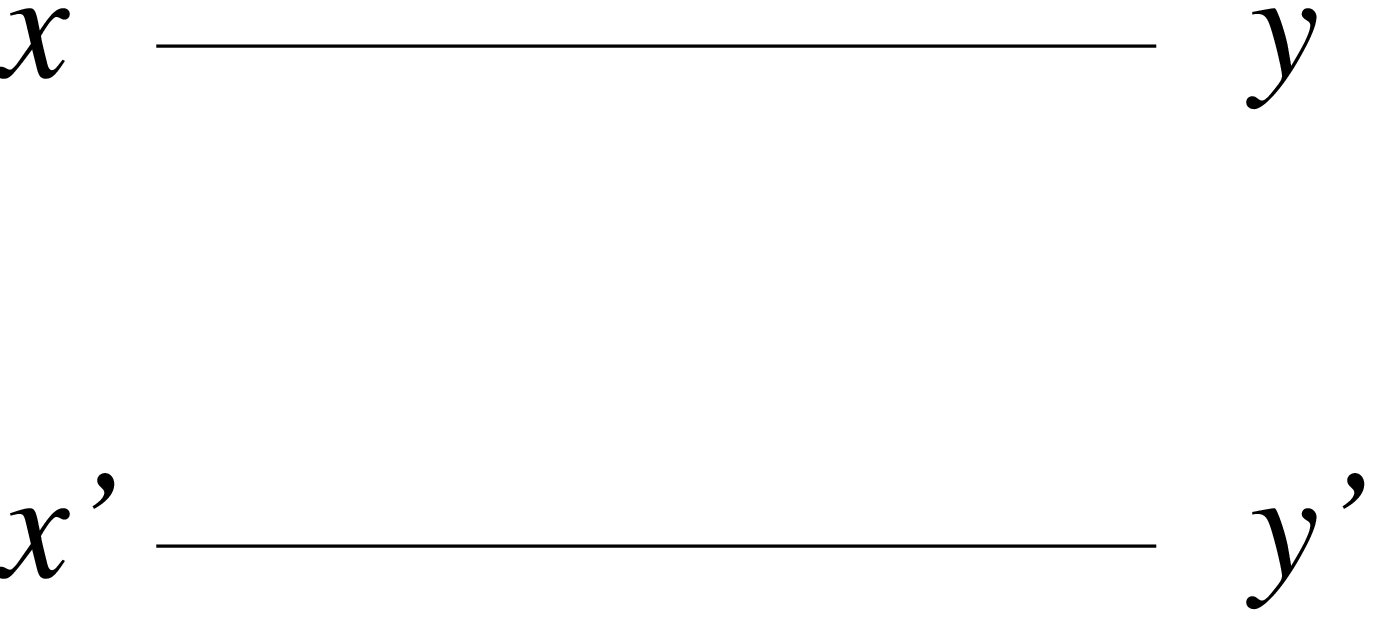}}\\[2pc]
\text{(b)}&\qquad\dpst\sum_{u_j,v_j}~~\raisebox{-21pt}{\includegraphics
 [scale=0.2]{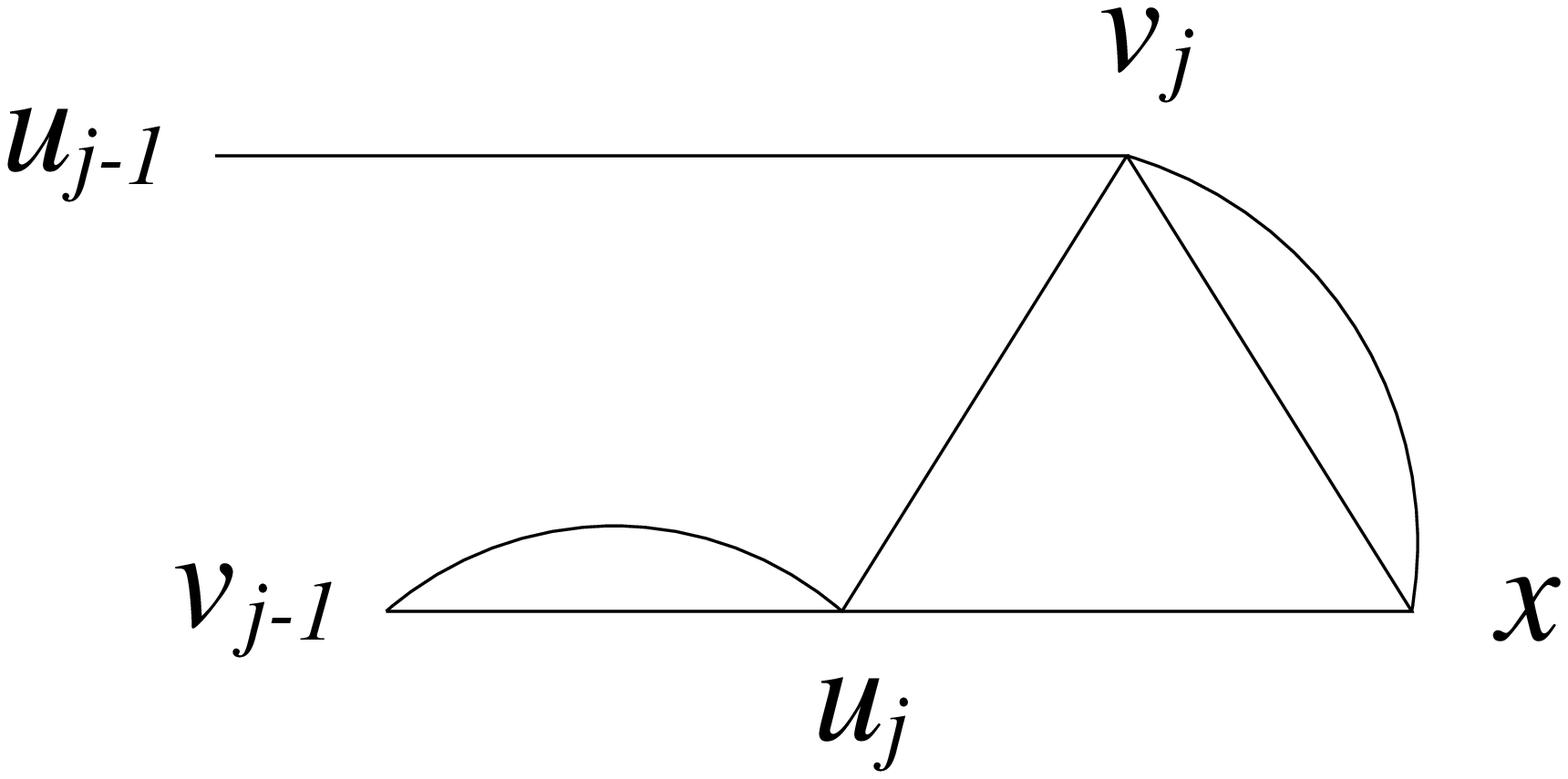}}~~~~\lesssim~~~~\sum_{v_j}~~\raisebox{-14pt}{
 \includegraphics[scale=0.2]{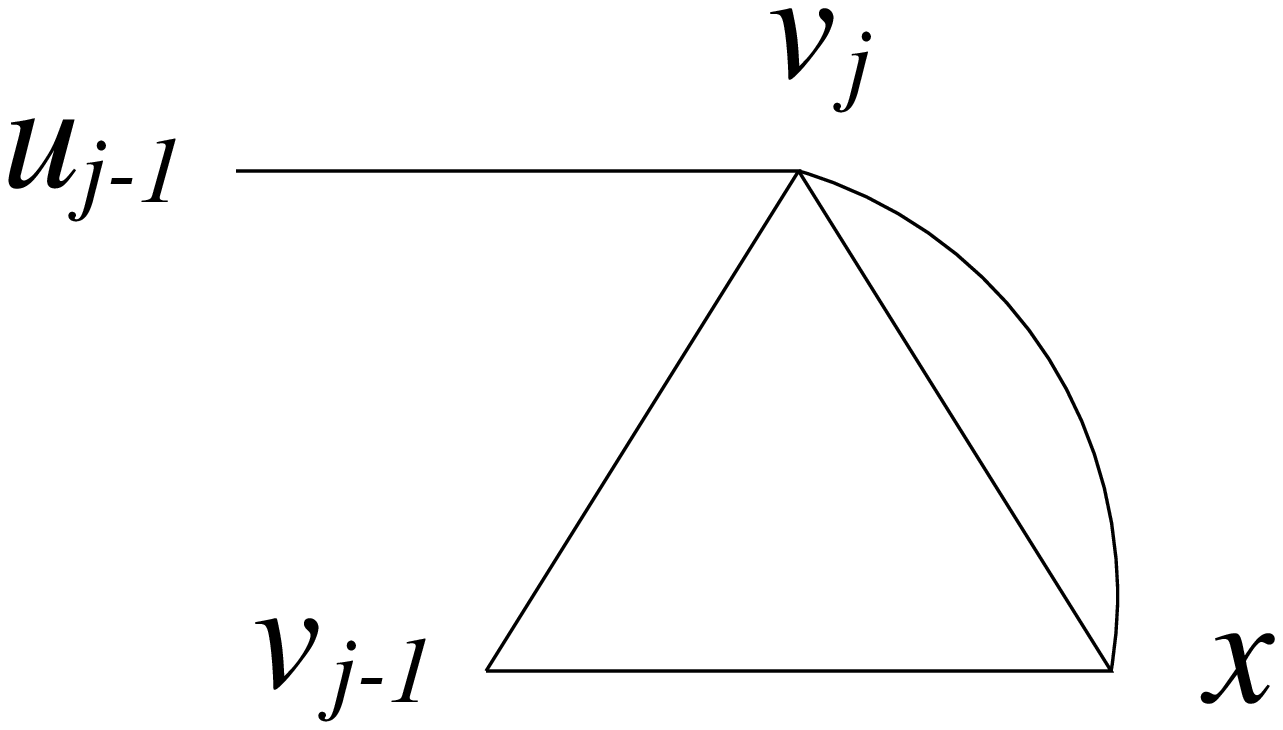}}~~~~\lesssim~~~\raisebox{-14pt}{
 \includegraphics[scale=0.2]{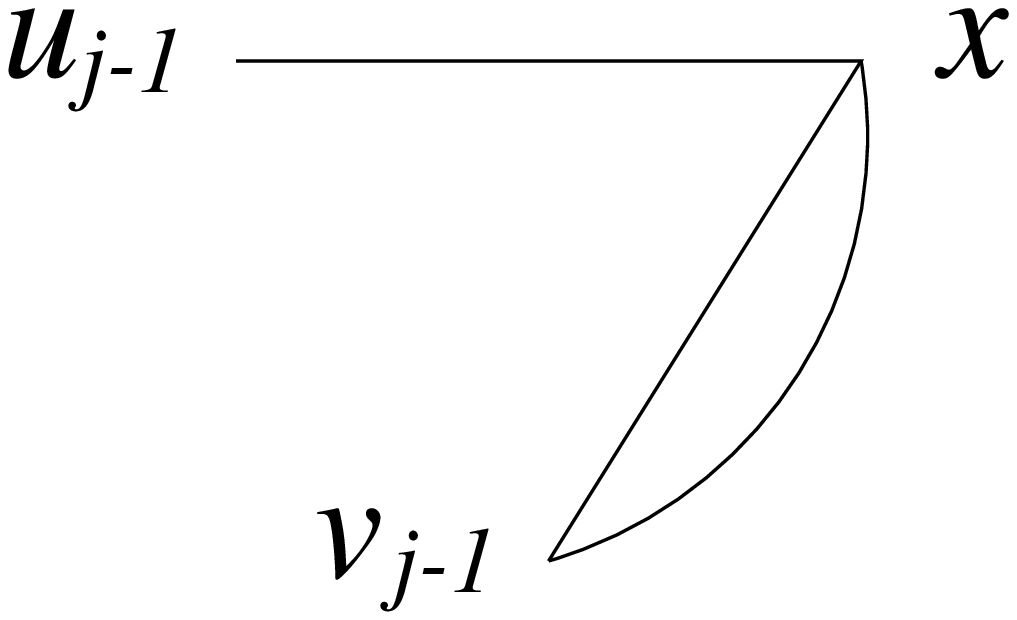}}
\end{array}
\end{align*}
\caption{\label{fig:star}(a) A schematic representation of
Proposition~\ref{prp:conv-star}(i), where each segment, say,
from $x$ to $y$ represent $\veee{x-y}^{-q}$.  (b) A schematic
representation of \refeq{succ-appl}, which is a result of
successive applications of Proposition~\ref{prp:conv-star}(ii)
with $x=x'$ or $y=y'$.}
\end{center}
\end{figure}

\begin{proof}
The inequality \refeq{conv} is identical to
\cite[Proposition~1.7(i)]{hhs03}.  We use this to prove \refeq{star}.
By the triangle inequality, we have
$\frac12\veee{x-y}\leq\veee{x-z}\vee\veee{z-y}$ and
$\frac12\veee{x'-y'}\leq\veee{x'-z}\vee\veee{z-y'}$.  Suppose that
$\veee{x-z}\leq\veee{z-y}$ and $\veee{x'-z}\leq\veee{z-y'}$.  Then, by
\refeq{conv} with $a=b=q$, the contribution from this case is bounded by
\begin{align}
\frac{2^{2q}}{\veee{x-y}^q\veee{x'-y'}^q}\sum_z\frac1{\veee{x-z}^q}\,
 \frac1{\veee{x'-z}^q}\leq\frac{2^{2q}c\veee{x-x'}^{d-2q}}{\veee{x-y}^q
 \veee{x'-y'}^q},
\end{align}
for some $c<\infty$, where we note that $\veee{x-x'}^{d-2q}\leq1$
because of $\frac12d<q$.  The other three possible
cases can be estimated similarly (see Figure~\ref{fig:star}(a)).  This
completes the proof of Proposition~\ref{prp:conv-star}.
\end{proof}

Before going into the proof of Proposition~\ref{prp:GimpliesPix}, we
summarize prerequisites.  Recall that \refeq{Q'-def}--\refeq{Q''-def}
involve $\tilde G_\Lambda$, and note that, by \refeq{G-delta-bd},
\begin{align}\lbeq{pi0-1stbd}
\Exp{\varphi_o\varphi_x}_\Lambda^3\leq\delta_{o,x}+\tilde G_\Lambda(o,x)^3.
\end{align}
We first show that
\begin{align}\lbeq{tildeG-bd}
\tilde G_\Lambda(o,x)\leq\frac{O(\theta_0)}{\veee{x}^q},&&
\sum_{b:\bb=o}\tau_b\big(\delta_{\tb,x}+\tilde G_\Lambda(\tb,x)\big)
 \leq\frac{O(\theta_0)}{\veee{x}^q}
\end{align}
hold assuming the bounds in \refeq{IR-xbd}.

\begin{proof}
By the assumed bound $\tau\leq2$ in \refeq{IR-xbd}, we have
\begin{align}\lbeq{tildeG-1stbd}
\tilde G_\Lambda(o,x)=\tau D(x)+\sum_{y\ne x}\tau D(y)\,\Exp{\varphi_y
 \varphi_x}_\Lambda\leq2D(x)+\sum_{y\ne x}2D(y)\,G(x-y),
\end{align}
where, and from now on without stating explicitly, we use the
translation invariance of $G(x)$ and the fact that $G(x-y)$ is an
increasing limit of $\Exp{\varphi_y\varphi_x}_\Lambda$ as
$\Lambda\uparrow\Zd$.  By \refeq{J-def} and the assumption in
Proposition~\ref{prp:GimpliesPix} that $\theta_0L^{d-q}$, with
$q<d$, is bounded away from zero, we obtain
\begin{align}\lbeq{Dbd}
D(x)\leq O(L^{-d})\ind{0<\|x\|_\infty\leq L}\leq\frac{O(L^{-d+q})}
 {\veee{x}^q}\leq\frac{O(\theta_0)}{\veee{x}^q}.
\end{align}
For the last term in \refeq{tildeG-1stbd}, we consider the cases for
$|x|\leq2\sqrt{d}L$ and $|x|\ge2\sqrt{d}L$ separately.

When $|x|\leq2\sqrt{d}L$, we use \refeq{Dbd}, \refeq{IR-xbd} and
\refeq{conv} with $\frac12d<q<d$ to obtain
\begin{align}
\sum_{y\ne x}D(y)\,G(x-y)\leq\sum_y\frac{O(L^{-d+q})}{\veee{y}^q}\,
 \frac{\theta_0}{\veee{x-y}^q}\leq\frac{O(\theta_0L^{-d+q})}
 {\veee{x}^{2q-d}}\leq\frac{O(\theta_0)}{\veee{x}^q}.
\end{align}

When $|x|\ge2\sqrt{d}L$, we use the triangle inequality $|x-y|\ge|x|-|y|$
and the fact that $D(y)$ is nonzero only when $0<\|y\|_\infty\leq L$ (so
that $|y|\leq\sqrt{d}\|y\|_\infty\leq\sqrt{d}L\leq\frac12|x|$).  Then, we
obtain
\begin{align}
\sum_{y\ne x}D(y)\,G(x-y)\leq\sum_yD(y)\,\frac{2^q\theta_0}{\veee{x}^q}
 =\frac{2^q\theta_0}{\veee{x}^q}.
\end{align}

This completes the proof of the first inequality in \refeq{tildeG-bd}.
The second inequality can be proved similarly.
\end{proof}

By repeated use of \refeq{tildeG-bd} and Proposition~\ref{prp:conv-star}(i)
with $a=b=2q$ (or Proposition~\ref{prp:conv-star}(ii) with $x=x'$ and
$y=y'$), we obtain
\begin{align}\lbeq{psi-bd}
\psi_\Lambda(v',v)\leq\delta_{v',v}+\frac{O(\theta_0^2)}{\veee{v-v'}^{2q}}.
\end{align}
Together with the naive bound $G(x)\leq O(1)\veee{x}^{-q}$ (cf.,
\refeq{IR-xbd}) as well as Proposition~\ref{prp:conv-star}(ii) (with
$x=x'$ or $y=y'$), we also obtain
\begin{align}\lbeq{GGpsi-bd}
\sum_{v'}G(v'-y)\,G(z-v')\,\psi_\Lambda(v',v)&\leq G(v-y)\,G(z-v)+
 \sum_{v'}\frac{O(\theta_0^2)}{\veee{v'-y}^q\veee{z-v'}^q\veee{v-v'}^{2q}}
 \nn\\
&\leq\frac{O(1)}{\veee{v-y}^q\veee{z-v}^q}.
\end{align}
The $O(1)$ term in the right-hand side is replaced by $O(\theta_0)$
or $O(\theta_0^2)$ depending on the number of $G$'s on the left being
replaced by $\tilde G_\Lambda$'s.

\begin{proof}[Proof of Proposition~\ref{prp:GimpliesPix}]
Since \refeq{pi0-1stbd}--\refeq{tildeG-bd} immediately imply the bound on
$\pi_\Lambda^{\sss(0)}(x)$, it suffices to prove the bounds on
$\pi_\Lambda^{\sss(i)}(x)$ for $i\ge1$.  To do so, we first estimate the
building blocks of the diagrammatic bound \refeq{piNbd}:
$\sum_{b:\bb=y}\tau_b\,Q'_{\Lambda;u}(\tb,x)$ and
$\sum_{b:\bb=y}\tau_b\,Q''_{\Lambda;u,v}(\tb,x)$.

Recall \refeq{P'0-def}--\refeq{Q''-def}.  First, by using
$G(x)\leq O(1)\veee{x}^{-q}$ and \refeq{GGpsi-bd}, we obtain
\begin{align}
P_{\Lambda;u}^{\prime{\sss(0)}}(y,x)&\leq\frac{O(1)}{\veee{x-y}^{2q}
 \veee{u-y}^q\veee{x-u}^q},\lbeq{P'0-bd}\\
P_{\Lambda;u,v}^{\prime\prime{\sss(0)}}(y,x)&\leq\frac{O(1)}{\veee{x
 -y}^q\veee{u-y}^q\veee{x-u}^q\veee{v-y}^q\veee{x-v}^q}.\lbeq{P''0-bd}
\end{align}
We will show at the end of this subsection that, for $j\ge1$,
\begin{align}
P_{\Lambda;u}^{\prime{\sss(j)}}(y,x)&\leq\frac{O(j)\,O(\theta_0^2)^j}
 {\veee{x-y}^{2q}\veee{u-y}^q\veee{x-u}^q},\lbeq{P'j-bd}\\
P_{\Lambda;u,v}^{\prime\prime{\sss(j)}}(y,x)&\leq\frac{O(j^2)\,O
 (\theta_0^2)^j}{\veee{x-y}^q\veee{u-y}^q\veee{x-u}^q\veee{v-y}^q
 \veee{x-v}^q}.\lbeq{P''j-bd}
\end{align}
As a result, $P_{\Lambda;u}^{\prime{\sss(0)}}(y,x)$ (resp.,
$P_{\Lambda;u,v}^{\prime\prime{\sss(0)}}(y,x)$) is the leading term of
$P'_{\Lambda;u}(y,x)$ (resp., $P''_{\Lambda;u,v}(y,x)$), which thus obeys
the same bound as in \refeq{P'0-bd} (resp., \refeq{P''0-bd}), with
a different constant in $O(1)$.  Combining these bounds with
\refeq{tildeG-bd} and \refeq{GGpsi-bd} (with both $G$ in the left-hand
side being replace by $\tilde G_\Lambda$) and then using
Proposition~\ref{prp:conv-star}(ii), we obtain
\begin{align}\lbeq{bb1-bd}
\sum_{b:\bb=y}\tau_b\,Q'_{\Lambda;u}(\tb,x)&\leq\sum_z\frac{O(\theta_0)}
 {\veee{z-y}^q}\,\frac1{\veee{x-z}^{2q}\veee{u-z}^q\veee{x-u}^q}\leq
 \frac{O(\theta_0)}{\veee{x-y}^q\veee{x-u}^{2q}},
\end{align}
and
\begin{align}\lbeq{bb2-bd}
\sum_{b:\bb=y}\tau_b\,Q''_{\Lambda;u,v}(\tb,x)&\leq\sum_z\frac{O(\theta_0)}
 {\veee{z-y}^q}\,\frac1{\veee{x-z}^q\veee{u-z}^q\veee{x-u}^q\veee{v-z}^q
 \veee{x-v}^q}\nn\\
&\quad+\sum_z\frac{O(\theta_0)}{\veee{v-y}^q}\,\frac{O(\theta_0)}{\veee{z-
 v}^q}\,\frac1{\veee{x-z}^{2q}\veee{u-z}^q\veee{x-u}^q}\nn\\
&\leq\frac{O(\theta_0)}{\veee{v-y}^q\veee{x-v}^q\veee{x-u}^{2q}}.
\end{align}
This completes bounding the building blocks.

Now we prove the bounds on $\pi_\Lambda^{\sss(j)}(x)$ for $j\ge1$.  For
the bounds on $\pi_\Lambda^{\sss(j)}(x)$ for $j\ge2$, we simply apply
\refeq{P'0-bd} and \refeq{bb1-bd}--\refeq{bb2-bd} to the diagrammatic
bound \refeq{piNbd}.  Then, we obtain
\begin{align}\lbeq{piNgeq2-prebd}
\pi_\Lambda^{\sss(j)}(x)\leq\sum_{\substack{u_1,\dots,u_j\\ v_1,\dots,
 v_j}}\frac{O(1)}{\veee{u_1}^{2q}\veee{v_1}^q\veee{u_1-v_1}^q}&\bigg(
 \prod_{i=1}^{j-1}\frac{O(\theta_0)}{\veee{v_{i+1}-u_i}^q\veee{u_{i+1}-
 v_{i+1}}^q\veee{u_{i+1}-v_i}^{2q}}\bigg)\nn\\
&\times\frac{O(\theta_0)}{\veee{x-u_j}^q\veee{x-v_j}^{2q}}\qquad(j\ge2).
\end{align}
First, we consider the sum over $u_j$ and $v_j$.  By successive
applications of Proposition~\ref{prp:conv-star}(ii) (with $x=x'$ or
$y=y'$), we obtain (see Figure~\ref{fig:star}(b))
\begin{align}\lbeq{succ-appl}
&\sum_{v_j}\sum_{u_j}\frac{O(\theta_0)}{\veee{v_j-u_{j-1}}^q\veee{u_j
 -v_j}^q\veee{u_j-v_{j-1}}^{2q}}\,\frac{O(\theta_0)}{\veee{x-u_j}^q
 \veee{x-v_j}^{2q}}\\
&\leq\sum_{v_j}\frac{O(\theta_0)^2}{\veee{v_j-u_{j-1}}^q\veee{v_{j-1}
 -v_j}^q\veee{x-v_{j-1}}^q\veee{x-v_j}^{2q}}\leq\frac{O(\theta_0)^2}
 {\veee{x-u_{j-1}}^q\veee{x-v_{j-1}}^{2q}},\nn
\end{align}
and thus
\begin{align}
\pi_\Lambda^{\sss(j)}(x)\leq\sum_{\substack{u_1,\dots,u_{j-1}\\ v_1,
 \dots,v_{j-1}}}\frac{O(1)}{\veee{u_1}^{2q}\veee{v_1}^q\veee{u_1-v_1}^q}
 &\bigg(\prod_{i=1}^{j-2}\frac{O(\theta_0)}{\veee{v_{i+1}-u_i}^q\veee{u_{i
 +1}-v_{i+1}}^q\veee{u_{i+1}-v_i}^{2q}}\bigg)\nn\\
&\times\frac{O(\theta_0)^2}{\veee{x-u_{j-1}}^q\veee{x-v_{j-1}}^{2q}}.
\end{align}
Repeating the application of Proposition~\ref{prp:conv-star}(ii) as
in \refeq{succ-appl}, we end up with
\begin{align}\lbeq{piNgeq2-bd}
\pi_\Lambda^{\sss(j)}(x)&\leq\sum_{u_1,v_1}\frac{O(1)}{\veee{u_1}^{2
 q}\veee{v_1}^q\veee{u_1-v_1}^q}\,\frac{O(\theta_0)^j}{\veee{x-u_1}^q
 \veee{x-v_1}^{2q}}\leq\frac{O(\theta_0)^j}{\veee{x}^{3q}}.
\end{align}

For the bound on $\pi_\Lambda^{\sss(1)}(x)$, we use the following bound,
instead of \refeq{P'0-bd}:
\begin{align}\lbeq{P'0-dec}
P_{\Lambda;v}^{\prime{\sss(0)}}(o,u)=\delta_{o,u}\delta_{o,v}+(1-\delta_{
 o,u}\delta_{o,v})\,P_{\Lambda;v}^{\prime{\sss(0)}}(o,u)\leq\delta_{o,u}
 \delta_{o,v}+\frac{O(\theta_0^2)}{\veee{u}^{2q}\veee{v}^q\veee{u-v}^q}.
\end{align}
In addition, instead of using \refeq{bb1-bd}, we use
\begin{align}\lbeq{bb1-dec}
\sum_{b:\bb=u}\tau_b\,Q'_{\Lambda;v}(\tb,x)&\leq\sum_z\frac{O(\theta_0)}
 {\veee{z-u}^q}\bigg(\delta_{z,v}\delta_{z,x}+(1-\delta_{z,x}\delta_{z,
 v})\,P_{\Lambda;v}^{\prime{\sss(0)}}(z,x)+\sum_{j\ge1}P_{\Lambda;v}^{
 \prime{\sss(j)}}(z,x)\bigg)\nn\\
&\leq\frac{O(\theta_0)}{\veee{x-u}^q}\,\delta_{v,x}+\sum_z\frac{O(
 \theta_0^3)}{\veee{z-u}^q\veee{x-z}^{2q}\veee{v-z}^q\veee{x-v}^q}\nn\\
&\leq\frac{O(\theta_0)}{\veee{x-u}^q}\,\delta_{v,x}+\frac{O(\theta_0^3)}
 {\veee{x-u}^q\veee{x-v}^{2q}},
\end{align}
due to \refeq{tildeG-bd}, \refeq{P'j-bd} and \refeq{P'0-dec}.  Applying
\refeq{P'0-dec}--\refeq{bb1-dec} to \refeq{piNbd} for $j=1$ and then
using Proposition~\ref{prp:conv-star}(ii), we end up with
\begin{align}
\pi_\Lambda^{\sss(1)}(x)&\leq O(\theta_0)\,\delta_{o,x}+\frac{O(\theta_0^3)}
 {\veee{x}^{3q}}+\sum_{u,v}\frac{O(\theta_0^2)}{\veee{u}^{2q}\veee{v}^q
 \veee{u-v}^q}\bigg(\frac{O(\theta_0)\,\delta_{v,x}}{\veee{x-u}^q}+\frac{
 O(\theta_0^3)}{\veee{x-u}^q\veee{x-v}^{2q}}\bigg)\nn\\
&\leq O(\theta_0)\,\delta_{o,x}+\frac{O(\theta_0^3)}{\veee{x}^{3q}}.
\end{align}

To complete the proof of Proposition~\ref{prp:GimpliesPix}, it thus
remains to show \refeq{P'j-bd}--\refeq{P''j-bd}.  The inequality
\refeq{P'j-bd} for $j=1$ immediately follows from the definition
\refeq{P'1-def} of $P_{\Lambda;u}^{\prime\sss(1)}$ (see also
Figure~\ref{fig:P-def}) and the bound \refeq{psi-bd} on
$\psi_\Lambda-\delta$.  To prove \refeq{P''j-bd} for $j=1$,
we first recall the definition \refeq{P''1-def} of
$P_{\Lambda;u,v}^{\prime\prime\sss(1)}$ (and
Figure~\ref{fig:P-def}).  Note that, by \refeq{GGpsi-bd},
$\sum_{v'}G(v'-y)\,G(z-v')\,\psi_\Lambda(v',v)$
obeys the same bound on $\sum_{v'}G(v'-y)\,G(z-v')$ (with a
different $O(1)$ term).  That is, the effect of an additional
$\psi_\Lambda$ is not significant.  Therefore, the bound on
$P_{\Lambda;u,v}^{\prime\prime\sss(1)}$ is identical, with a possible
modification of the $O(1)$ multiple, to the bound on
$P_{\Lambda;u}^{\prime\sss(1)}$ (or $P_{\Lambda;v}^{\prime\sss(1)}$)
with $v$ (resp., $u$) ``being embedded'' in one of the bubbles
consisting of $\psi_\Lambda-\delta$.  By \refeq{psi-bd},
$\psi_\Lambda(y,x)-\delta_{y,x}$ with $v$ being embedded in one of
its bubbles is bounded as
\begin{align}\lbeq{psipsi-bd}
&\sum_{k=1}^\infty\sum_{l=1}^k\sum_{y',x'}\big(\tilde G_\Lambda^2
 \big)^{*(l-1)}(y,y')\,\tilde G_\Lambda(y',x')\Big(\Exp{\varphi_{y'}
 \varphi_v}_\Lambda\tilde G_\Lambda(v,x')+\tilde G_\Lambda(y',x')\,
 \delta_{v,x'}\Big)\big(\tilde G_\Lambda^2\big)^{*(k-l)}
 (x',x)\nn\\
&=\sum_{y',x'}\psi_\Lambda(y,y')\,\tilde G_\Lambda(y',x')\Big(
 \Exp{\varphi_{y'}\varphi_v}_\Lambda\tilde G_\Lambda(v,x')+\tilde
 G_\Lambda(y',x')\,\delta_{v,x'}\Big)\psi_\Lambda(x',x)\nn\\
&\leq\sum_{y',x'}\frac{O(1)}{\veee{y'-y}^{2q}}\,\frac{O(\theta_0)}
 {\veee{x'-y'}^q}\,\frac{O(\theta_0)}{\veee{v-y'}^q\veee{x'-v}^q}\,
 \frac{O(1)}{\veee{x-x'}^{2q}}\leq\frac{O(\theta_0^2)}
 {\veee{x-y}^q\veee{v-y}^q\veee{x-v}^q}.
\end{align}
By this observation and using \refeq{IR-xbd} to bound the remaining
two two-point functions consisting of
$P_{\Lambda;u,v}^{\prime\prime\sss(1)}$ (recall \refeq{P''1-def}),
we obtain \refeq{P''j-bd} for $j=1$.

For \refeq{P'j-bd}--\refeq{P''j-bd} with $j\ge2$, we first note that,
by applying \refeq{IR-xbd} and \refeq{psi-bd} to the
definition \refeq{Pj-def} of $P_\Lambda^{\sss(j)}(y,x)$, we have
\begin{gather}
P_\Lambda^{\sss(j)}(y,x)\leq\sum_{\substack{v_2,\dots,v_j\\ v'_1,\dots,
 v'_{j-1}}}\frac{O(\theta_0^2)}{\veee{v'_1-y}^{2q}\veee{v_2-y}^q\veee{v'_1
 -v_2}^q}\prod_{i=2}^{j-1}\frac{O(\theta_0^2)}{\veee{v'_i-v_i}^{2q}\veee{
 v_{i+1}-v'_{i-1}}^q\veee{v'_i-v_{i+1}}^q}\nn\\
\times\frac{O(\theta_0^2)}{\veee{x-v_j}^{2q}\veee{x-v'_{j-1}}^q}.\lbeq{Pj-bd}
\end{gather}
By definition, the bound on $P_{\Lambda;u}^{\prime\sss(j)}(y,x)$ is
obtained by ``embedding $u$'' in one of the $2j-1$ factors of
$\veee{\cdots}^q$ (not $\veee{\cdots}^{2q}$) and then summing over all
these $2j-1$ choices.  For example, the contribution from the case in which
$\veee{v_2-y}^q$ is replaced by $\veee{u-y}^q\veee{v_2-u}^q$ is bounded,
similarly to \refeq{piNgeq2-bd}, by
\begin{align}
&\sum_{v_2,v'_1}\frac{O(\theta_0^2)}{\veee{v'_1-y}^{2q}\veee{u-y}^q\veee{v_2
 -u}^q\veee{v'_1-v_2}^q}\,\frac{O(\theta_0^2)^{j-1}}{\veee{x-v'_1}^q\veee{x
 -v_2}^{2q}}\nn\\
&\leq\sum_{v'_1}\frac{O(\theta_0^2)^j}{\veee{v'_1-y}^{2q}\veee{u-y}^q\veee{
 x-u}^q\veee{x-v'_1}^{2q}}\leq\frac{O(\theta_0^2)^j}{\veee{x-y}^{2q}\veee{u
 -y}^q\veee{x-u}^q}.
\end{align}
The other $2j-2$ contributions can be estimated in a similar way, with the
same form of the bound.  This completes the proof of \refeq{P'j-bd}.

By \refeq{psipsi-bd}, the bound on
$P_{\Lambda;u,v}^{\prime\prime\sss(j)}(y,x)$ is also obtained by
``embedding $u$ and $v$'' in one of the $2j-1$ factors of $\veee{\cdots}^q$
and one of the $j$ factors of $\veee{\cdots}^{2q}$ in \refeq{Pj-bd}, and
then summing over all these combinations.  For example, the contribution
from the case in which $\veee{v_2-y}^q$ and $\veee{v'_1-y}^{2q}$ in
\refeq{Pj-bd} are replaced, respectively, by $\veee{u-y}^q\veee{v_2-u}^q$
and $\veee{v'_1-y}^q\veee{v-y}^q\veee{v'_1-v}^q$, is bounded by
\begin{align}
&\sum_{v_2,v'_1}\frac{O(\theta_0^2)}{\veee{v'_1-y}^q\veee{v-y}^q\veee{v'_1
 -v}^q\veee{u-y}^q\veee{v_2-u}^q\veee{v'_1-v_2}^q}\,\frac{O(\theta_0^2)^{j
 -1}}{\veee{x-v'_1}^q\veee{x-v_2}^{2q}}\nn\\
&\leq\sum_{v'_1}\frac{O(\theta_0^2)^j}{\veee{v'_1-y}^q\veee{v-y}^q\veee{v'_1
 -v}^q\veee{u-y}^q\veee{x-u}^q\veee{x-v'_1}^{2q}}\nn\\
&\leq\frac{O(\theta_0^2)^j}{\veee{x-y}^q\veee{u-y}^q\veee{x-u}^q
 \veee{v-y}^q\veee{x-v}^q}.
\end{align}
The other $(2j-1)j-1$ contributions can be estimated similarly, with the
same form of the bound.  This completes the proof of \refeq{P''j-bd} and
thus Proposition~\ref{prp:GimpliesPix}.
\end{proof}

\subsection{Bounds for finite-range models}\label{ss:proof-nn}
First, we prove \refeq{pi-sumbd} and
Proposition~\ref{prp:exp-bootstrap}(iii) assuming \refeq{IR-kbd}.
Then, we prove \refeq{pi-kbd} assuming \refeq{IR-kbd} and
\refeq{IR-xbdNN} to complete the proof of
Propositions~\ref{prp:GimpliesPik}.

\begin{proof}[Proof of \refeq{pi-sumbd} assuming \refeq{IR-kbd}]
By applying \refeq{G-delta-bd} to the bound \refeq{piNbd} on $\pi_\Lambda^{\sss(0)}(x)$,
it is easy to show that, for $r=0,2$,
\begin{align}\lbeq{pi0-rthmombd}
\sum_x|x|^r\pi_\Lambda^{\sss(0)}(x)\leq\delta_{r,0}+\sum_{x\ne o}
 |x|^r\Exp{\varphi_o\varphi_x}_\Lambda^3&\leq\delta_{r,0}+\Big(
 \sup_{x\ne o}|x|^rG(x)\Big)\sum_{x\ne o}(\tau D*G)(x)\,G(x)\nn\\
&\leq\delta_{r,0}+(d\sigma^2)^{\delta_{r,2}}O(\theta_0)^2.
\end{align}

For $i\ge1$, by using the diagrammatic bound \refeq{piNbd} and
translation invariance, we have
\begin{align}\lbeq{dec-bd}
\sum_x\pi_\Lambda^{\sss(i)}(x)\leq\bigg(\sum_{v,x}P_{\Lambda;v}
 ^{\prime{\sss(0)}}(o,x)\bigg)\bigg(\sup_y\sum_{z,v,x}\tau_{y,z}
 Q''_{\Lambda;o,v}(z,x)\bigg)^{i-1}\bigg(\sup_y\sum_{z,x}\tau_{y,
 z}Q'_{\Lambda;o}(z,x)\bigg).
\end{align}
The proof of the bound on $\sum_x\pi_\Lambda^{\sss(i)}(x)$ for
$i\ge1$ is completed by showing that
\begin{align}\lbeq{block-sumbd}
\bigg(\sum_{v,x}P_{\Lambda;v}^{\prime{\sss(0)}}(o,x)-1\bigg)\vee
 \bigg(\sup_y\sum_{z,v,x}\tau_{y,z}Q''_{\Lambda;o,v}(z,x)\bigg)
 \vee\bigg(\sup_y\sum_{z,x}\tau_{y,z}Q'_{\Lambda;o}(z,x)\bigg)
 =O(\theta_0).
\end{align}
The key idea to obtain this estimate is that the bounding diagrams
for the Ising model are similar to those for self-avoiding walk
(cf., Figure~\ref{fig:piN-bd}).  The diagrams for self-avoiding walk
are known to be bounded by products of bubble diagrams (see, e.g.,
\cite{ms93}), and we can apply the same method to bound the diagrams
for the Ising model by products of bubbles.

For example, consider
\begin{align}\lbeq{tau*Q'-rewr}
\sum_{z,x}\tau_{y,z}Q'_{\Lambda;o}(z,x)=\sum_{z',x}\bigg(\sum_z
 \tau_{y,z}\big(\delta_{z,z'}+\tilde G_\Lambda(z,z')\big)\bigg)
 P'_{\Lambda;o}(z',x).
\end{align}
The factor of $\theta_0$ is due to the nonzero line segment
$\sum_z\tau_{y,z}(\delta_{z,z'}+\tilde G_\Lambda(z,z'))$, because
\begin{gather}
\sum_z\tau_{o,z}\big(\delta_{z,x}+\tilde G_\Lambda(z,x)\big)=\tau
 D(x)+\tau\sum_zD(z)\,\tilde G_\Lambda(z,x)\leq O(\theta_0)+\tau
 \sup_x\tilde G_\Lambda(o,x),\lbeq{tau*delta+G-bd}\\
\tilde G_\Lambda(o,x)\leq\tau D(x)+\tau\sum_{y\ne o}G(y)\,D(x-y)\leq
 O(\theta_0)+\tau\sup_{y\ne o}G(y)=O(\theta_0),\lbeq{tildeG-bdnn}
\end{gather}
where we have used translation invariance,  \refeq{IR-kbd} and
$\sup_xD(x)=O(\theta_0)$.  By \refeq{P'P''-def},
\begin{align}\lbeq{tau*Q'-rewrbd}
\refeq{tau*Q'-rewr}\leq O(\theta_0)\sum_{z',x}P'_{\Lambda;o}(z',x)
 =O(\theta_0)\sum_{z',x}\bigg(P_{\Lambda;o}^{\prime{\sss(0)}}(z',x)
 +\sum_{j\ge1}P_{\Lambda;o}^{\prime{\sss(j)}}(z',x)\bigg).
\end{align}
Similarly to \refeq{pi0-rthmombd} for $r=0$, the sum of
$P_{\Lambda;v}^{\prime{\sss(0)}}(z',x)$ is easily estimated as
$1+O(\theta_0)$.  We claim that the sum of
$P_{\Lambda;o}^{\prime{\sss(j)}}(z',x)$ for $j\geq1$ is
$(2j-1)\,O(\theta_0)^j$, since $P_{\Lambda;o}^{\prime{\sss(j)}}(z',x)$
is a sum of $2j-1$ terms, each of which contains $j$ chains of nonzero
bubbles; each chain is $\psi_\Lambda(v,v')-\delta_{v,v'}$ for some $v,v'$
and satisfies
\begin{align}
\sum_{v'}\big(\psi_\Lambda(v,v')-\delta_{v,v'}\big)\leq\sum_{l\ge1}
 \Big(\tau^2\big(D*(D*G^{*2})\big)(o)\Big)^l=\sum_{l\ge1}
 O(\theta_0)^l=O(\theta_0).
\end{align}
For example,
\begin{align}
\sum_{z',x}P_{\Lambda;o}^{\prime{\sss(4)}}(z',x)&=\raisebox{-1.2pc}
 {\includegraphics[scale=0.15]{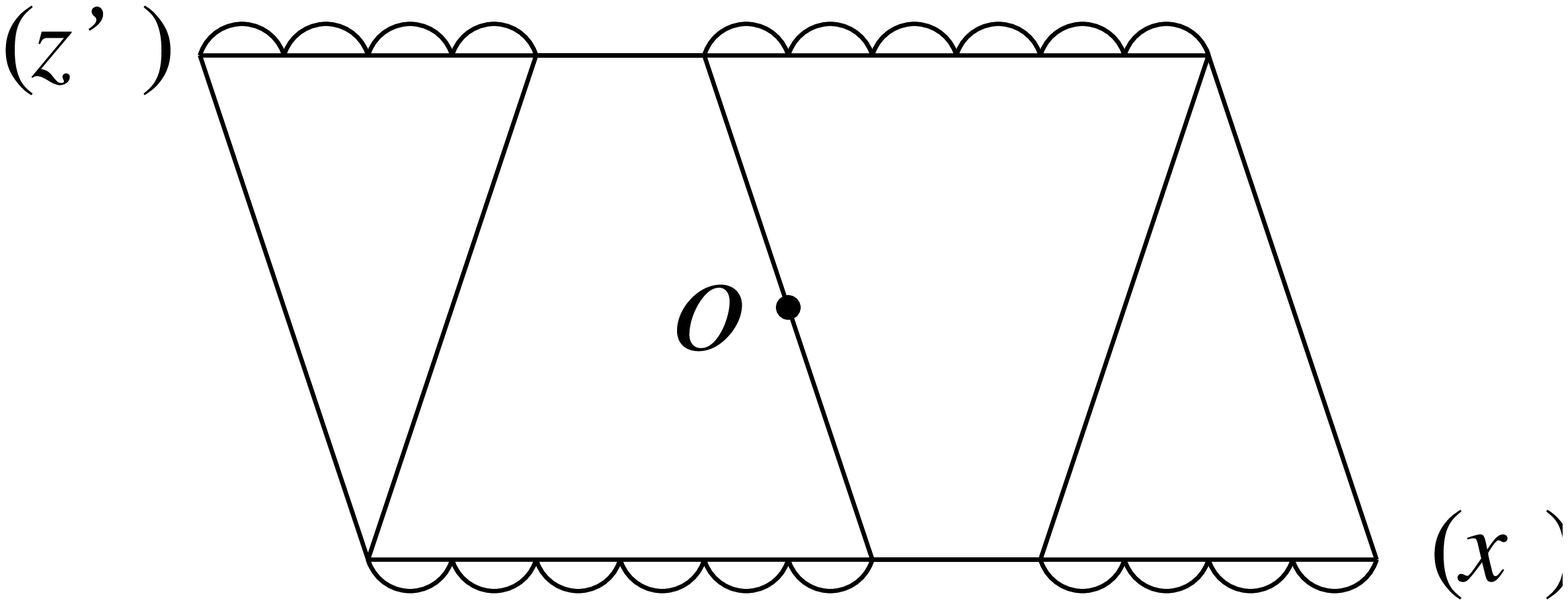}}~+6\text{ other possibilities},
\end{align}
which can be estimated, by translation invariance, as
\begin{align}
\raisebox{-1.5pc}{\includegraphics[scale=0.15]{Pprime4}}&\leq
 ~\raisebox{-1.8pc}{\includegraphics[scale=0.15]{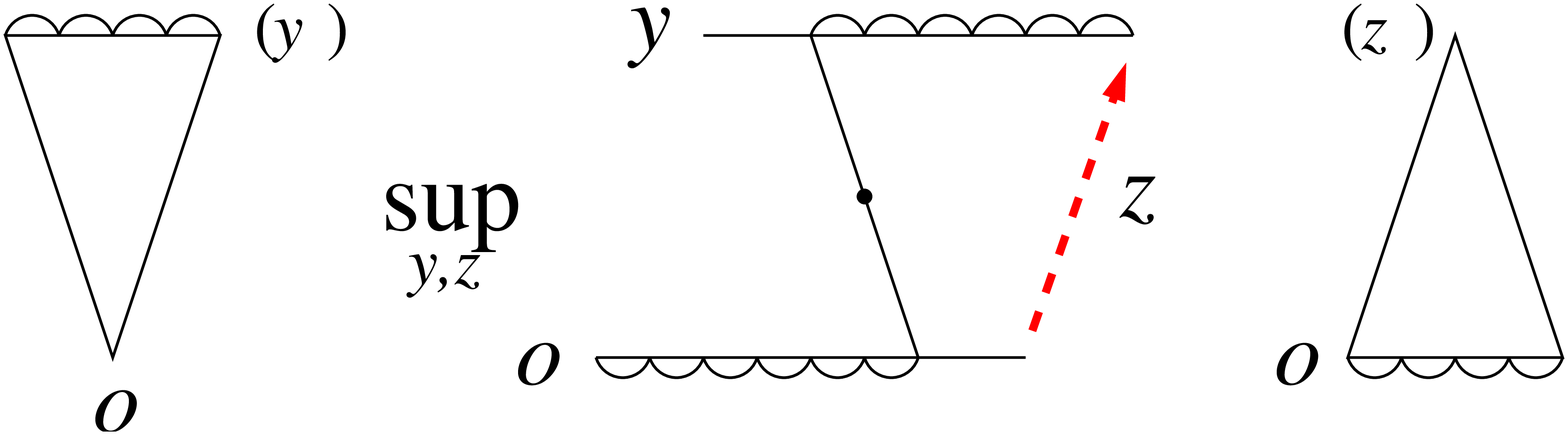}}\nn\\[5pt]
&\leq\bigg(\sum_y\big(\psi_\Lambda(o,y)-\delta_{o,y}\big)\bigg)^4\big(
 \bar W^{\sss(0)}\big)^4=O(\theta_0)^4,
\end{align}
where $\bar W^{\sss(t)}$ is given by \refeq{GbarWbar}.

The sum of $\tau_{y,z}Q''_{\Lambda;o,v}(z,x)$ in \refeq{block-sumbd}
is estimated similarly \cite{sNN}.  We complete the proof of the
bound on $\sum_x\pi_\Lambda^{\sss(j)}(x)$ for $j\ge1$.

\bigskip

To estimate $\sum_x|x|^2\pi_\Lambda^{\sss(j)}(x)$ for $j\ge1$, we
recall that, in each bounding diagram, there are at least three
distinct paths between $o$ and $x$: the uppermost path (i.e., $o\to
b_1\to v_2\to b_3\to\cdots\to x$ in \refeq{piNbd}; see also
Figure~\ref{fig:piN-bd}), the lowermost path (i.e., $o\to v_1\to
b_2\to v_3\to\cdots\to x$) and a middle zigzag path.  We use the
lowermost path to bound $|x|^2$ as
\begin{align}\lbeq{x2-bd}
|x|^2=\sum_{n=0}^j|a_n|^2+2\sum_{0\leq m<n\leq j}a_m\cdot a_n
 \leq(j+1)\sum_{n=0}^j|a_n|^2,
\end{align}
where $a_0=v_1$,
$a_1=\bb_2-v_1$ ,$a_2=v_3-\bb_2,\dots$, and $a_j=x-v_j$ or
$x-\bb_j$ depending on the parity of $j$.
\begin{figure}[t]
\begin{center}
\begin{gather*}
\includegraphics[scale=0.16]{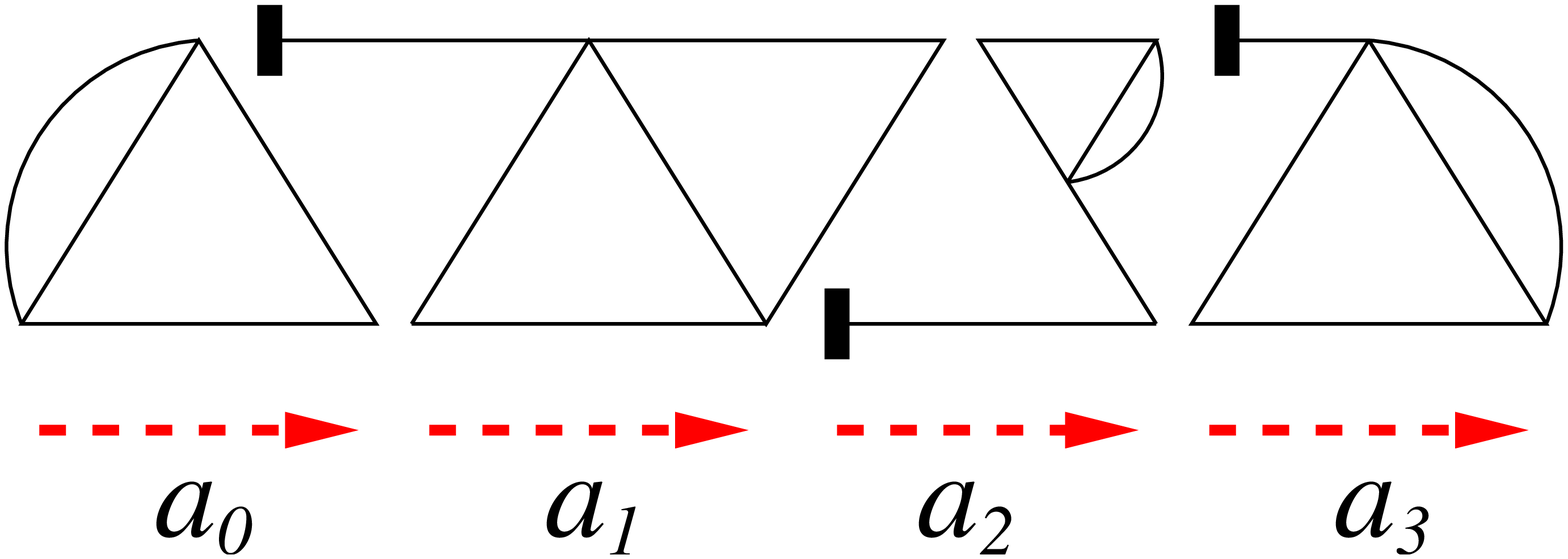}\\[1pc]
\text{(i)}\quad\raisebox{-1.2pc}{\includegraphics[scale=0.12]
 {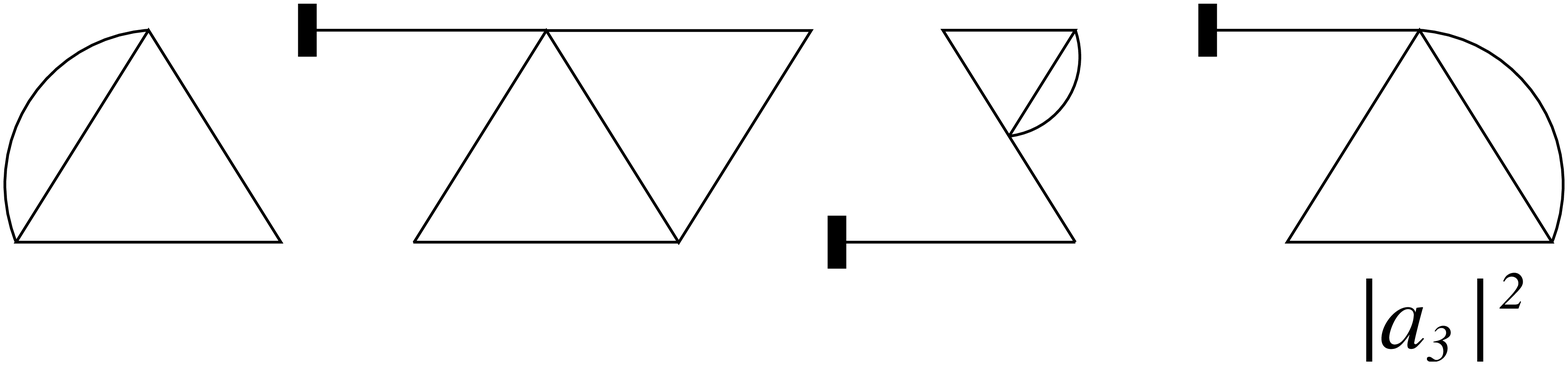}}\qquad\qquad
\text{(ii)}\quad\raisebox{-1.2pc}{\includegraphics[scale=0.12]
 {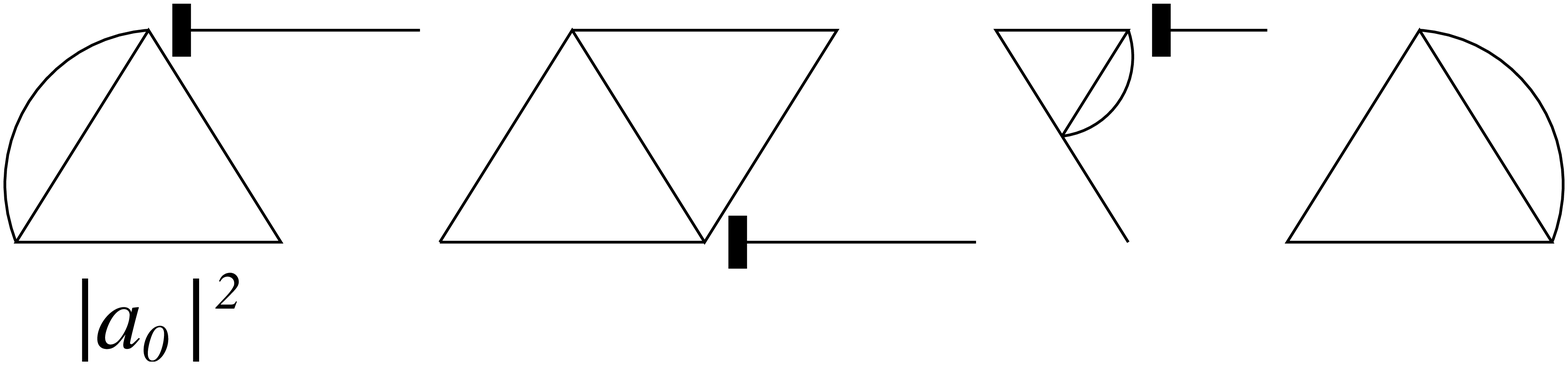}}\\[5pt]
\text{(iii)}\quad\raisebox{-1.2pc}{\includegraphics[scale=0.12]
 {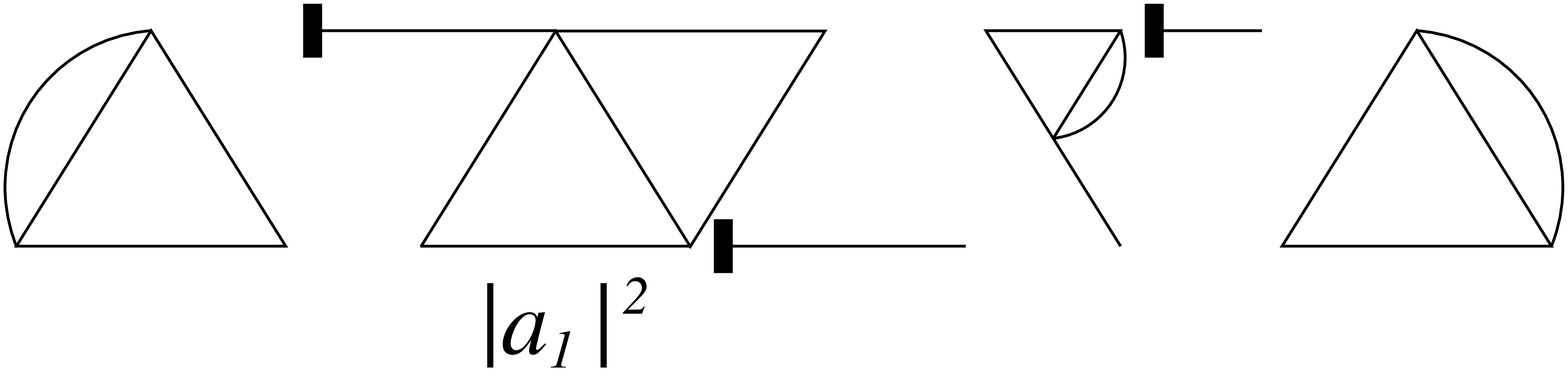}}\qquad~~~\&~~\qquad\raisebox{-1.2pc}{\includegraphics
 [scale=0.12]{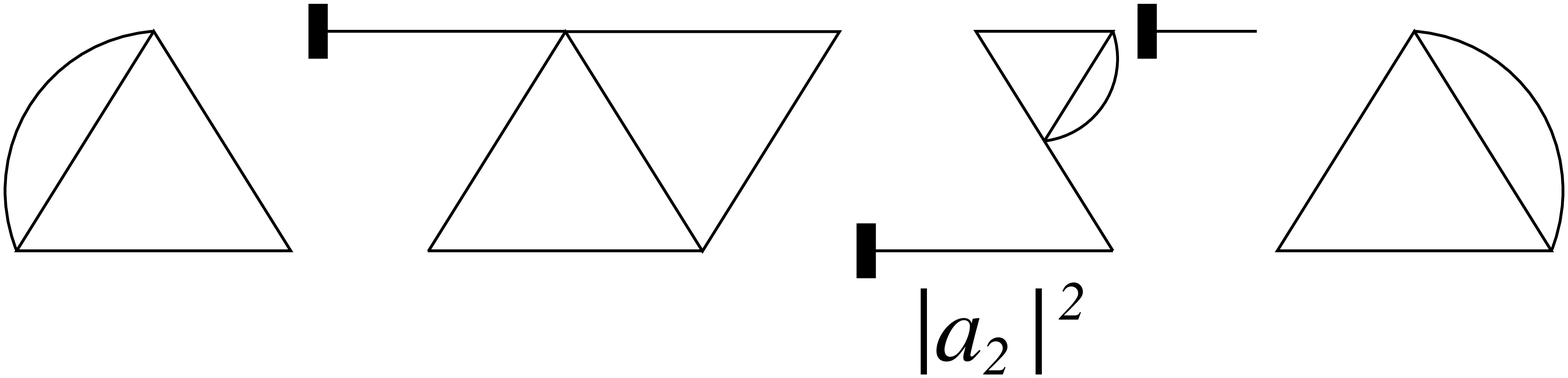}}
\end{gather*}
\caption{\label{fig:pi3-dec}One of the leading diagrams for
$\sum_x|x|^2\pi_\Lambda^{\sss(3)}(x)$ and its decompositions
depending on whether the assigned weight is (i)~$|a_3|^2$,
(ii)~$|a_0|^2$ and (iii)~$|a_n|^2$ for $n=1,2$, respectively.}
\end{center}
\end{figure}
We discuss the contributions to
$\sum_x|x|^2\pi_\Lambda^{\sss(j)}(x)$ from (i)~$|a_j|^2$,
(ii)~$|a_0|^2$ and (iii)~$|a_n|^2$ for $n\ne0,j$, separately (cf.,
Figure~\ref{fig:pi3-dec}).

\smallskip

(i) The contribution from $|a_j|^2$ is bounded by
\begin{align}\lbeq{2nddec-bd:n=j}
&\bigg(\sum_{v,y}P_{\Lambda;v}^{\prime{\sss(0)}}(o,y)\bigg)\bigg(\sup_y
 \sum_{\substack{b,v,z\\ \bb=y}}\tau_bQ''_{\Lambda;o,v}(\tb,z)\bigg)^{j
 -1}\nn\\
&\qquad\qquad\times\bigg(\sup_y\sum_{\substack{b,x\\ \bb=y}}\Big(|x|^2
 \ind{j\text{ odd}}+|x-\bb|^2\ind{j\text{ even}}\Big)\,\tau_b
 Q'_{\Lambda;o}(\tb,x)\bigg)\nn\\
&\leq O(\theta_0)^{j-1}\sup_y\sum_{z,z',x}\Big(|x|^2\ind{j\text{ odd}}
 +|x-y|^2\ind{j\text{ even}}\Big)\,\tau_{y,z}\big(\delta_{z,z'}+\tilde
 G_\Lambda(z,z')\big)P'_{\Lambda;o}(z',x).
\end{align}
By \refeq{P'0-def}, the leading contribution from
$P_{\Lambda;o}^{\prime{\sss(0)}}(z',x)$ for an odd $j$
can be estimated as
\begin{align}\lbeq{2nddec-bd:n=jbd}
&\sup_y\sum_{z,z',x}|x|^2\tau_{y,z}\big(\delta_{z,z'}+\tilde
 G_\Lambda(z,z')\big)P_{\Lambda;o}^{\prime{\sss(0)}}(z',x)\nn\\
&=\sup_y\sum_{z,z',x}\tau_{y,z}\big(\delta_{z,z'}+\tilde G_\Lambda(
 z,z')\big)\,\Exp{\varphi_{z'}\varphi_o}_\Lambda\,\Exp{\varphi_{z'}
 \varphi_x}_\Lambda^2\,|x|^2\Exp{\varphi_o\varphi_x}_\Lambda\nn\\
&\leq\sup_y\Big((\tau D*G)(y)+(\tau D*G)^{*2}(y)\Big)\,G^{*2}(o)\,
 \bar G^{\sss(2)}=d\sigma^2O(\theta_0)^2,
\end{align}
where $\bar G^{\sss(s)}$ is given by \refeq{GbarWbar}.  The other
contributions from $P_{\Lambda;o}^{\prime{\sss(i)}}(z',x)$ for
$i\ge1$ and from the even-$j$ case can be estimated similarly; if
$j$ is even, then, by using $|x-y|^2\leq2|z'-y|^2+2|x-z'|^2$ and
estimating the contributions from $|z'-y|^2$ and $|x-z'|^2$
separately, we obtain that the supremum in \refeq{2nddec-bd:n=j} is
$d\sigma^2O(\theta_0)$.  Consequently, \refeq{2nddec-bd:n=j} is
$d\sigma^2O(\theta_0)^{2\lfloor{\sss\frac{j+1}2}\rfloor}$.

\smallskip

(ii) To bound the contributions to $\sum_x|x|^2\pi_\Lambda^{\sss(j)}(x)$
from $|a_n|^2$ for $n<j$, we define (cf., Figure~\ref{fig:tildeQ''})
\begin{align}\lbeq{tildeQ''-def}
\tilde Q''_{\Lambda;u,v}(y,x)=\sum_b\bigg(P''_{\Lambda;u,v}(y,
 \bb)+\sum_{y'}\tilde G_\Lambda(y,y')\,P'_{\Lambda;u}(y',\bb)
 \,\psi_\Lambda(y,v)\bigg)\,\tau_b\big(\delta_{\tb,x}+\tilde
 G_\Lambda(\tb,x)\big).
\end{align}
By translation invariance and a similar argument to show
\refeq{block-sumbd}, we can easily prove
\begin{align}\lbeq{tildeQ''-bd}
\sup_z\sum_{y,v}\tilde Q''_{\Lambda;o,v}(y,v+z)=\sum_{y,v}\tilde
 Q''_{\Lambda;v,o}(y,z)=O(\theta_0).
\end{align}
Therefore, the contribution from $|a_0|^2$ to $\sum_x|x|^2
\pi_\Lambda^{\sss(j)}(x)$ is bounded by
\begin{align}\lbeq{2nddec-bd:n=0}
&\bigg(\sup_y\sum_{v,b}|v|^2P_{\Lambda;o}^{\prime{\sss(0)}}(v,\bb)\,
 \tau_b\big(\delta_{\tb,y}+\tilde G_\Lambda(\tb,y)\big)\bigg)\bigg(
 \sup_z\sum_{y,v}\tilde Q''_{\Lambda;v,o}(y,z)\bigg)^{j-1}\bigg(
 \sum_{z,x}P'_{\Lambda;o}(z,x)\bigg)\nn\\
&\quad\leq d\sigma^2O(\theta_0)^{j+1}.
\end{align}
\begin{figure}[tn]
\begin{center}
\includegraphics[scale=0.15]{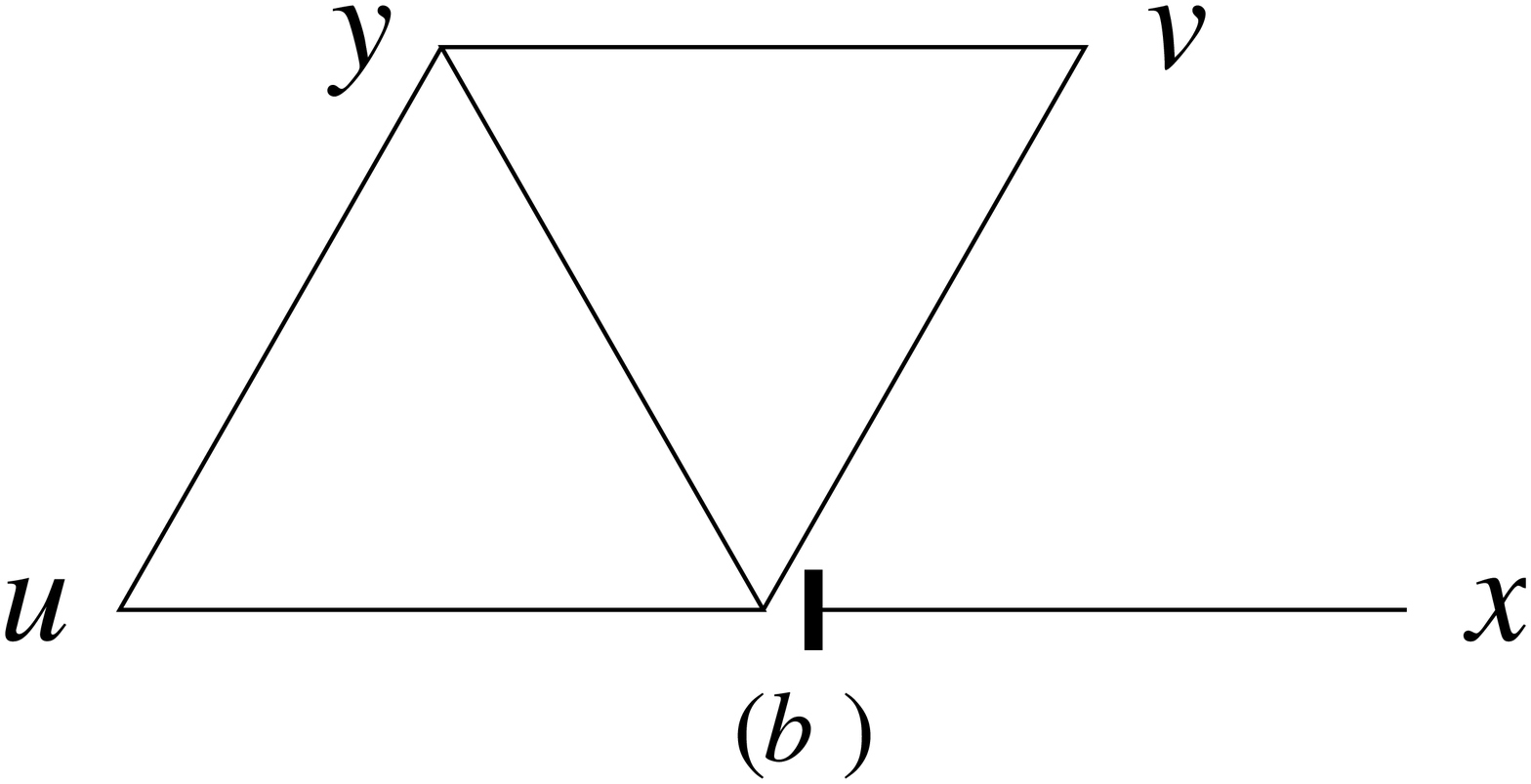}\hspace{7pc}
\includegraphics[scale=0.15]{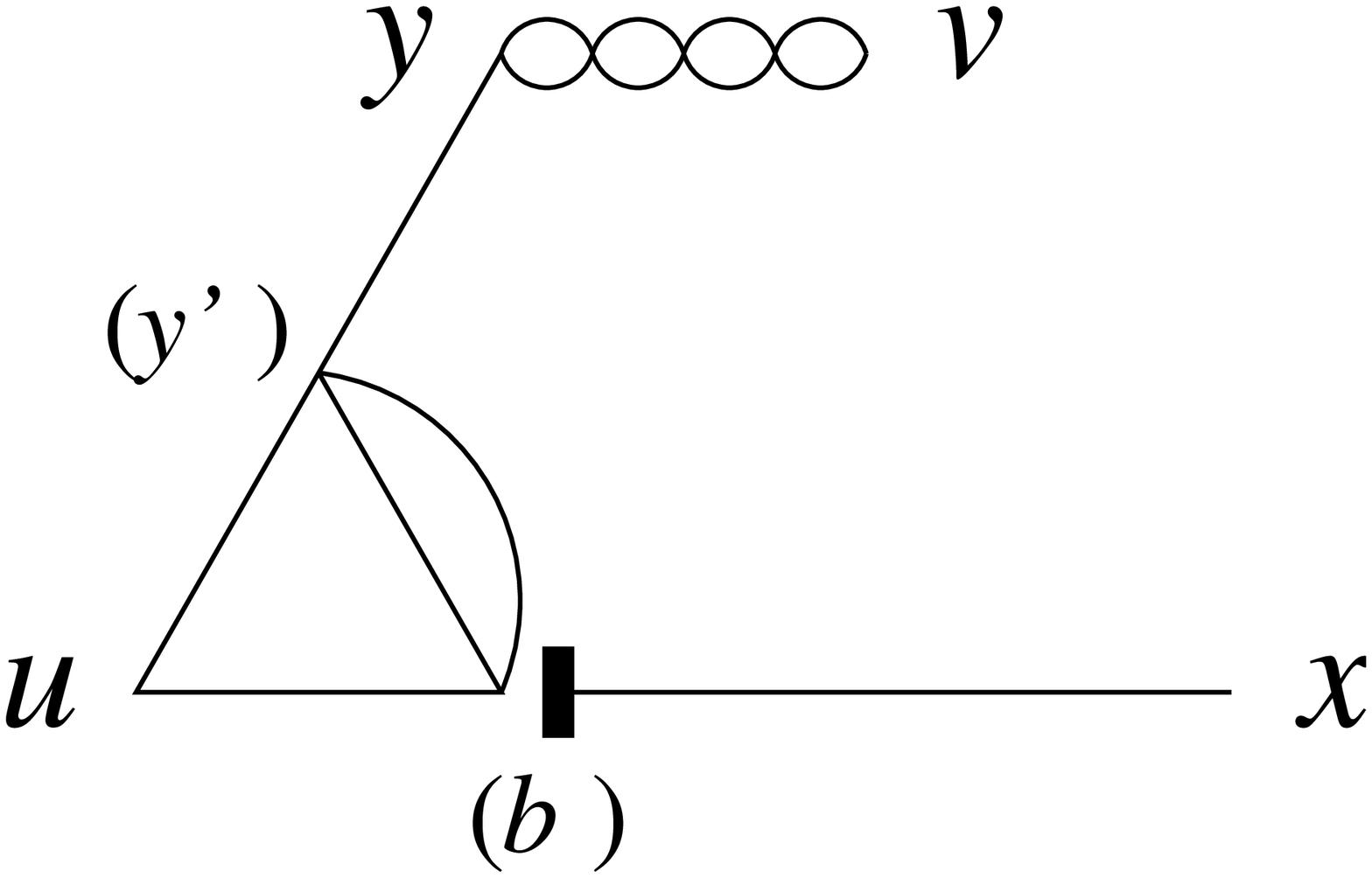}
\caption{\label{fig:tildeQ''}The leading diagrams of $\tilde
Q''_{\Lambda;u,v}(y,x)$, due to
$P_{\Lambda;u,v}^{\prime\prime{\sss(0)}}(y,\bb)$ and
$P_{\Lambda;u}^{\prime{\sss(0)}}(y',\bb)$ in \refeq{tildeQ''-def},
respectively.}
\end{center}
\end{figure}

\smallskip

(iii) By translation invariance and
\refeq{tildeQ''-def}--\refeq{tildeQ''-bd}, the contribution from
$|a_n|^2$ for an $n\ne0,j$ is bounded by
\begin{align}\lbeq{2nddec-bd:0<n<j}
&\bigg(\sum_{v,y}P_{\Lambda;v}^{\prime{\sss(0)}}(o,y)\bigg)\bigg(\sup_y
 \sum_{\substack{b,v,z\\ \bb=y}}\tau_bQ''_{\Lambda;o,v}(\tb,z)\bigg)^{n
 -1}\bigg(\sup_z\sum_{y,v}\tilde Q''_{\Lambda;v,o}(y,z)\bigg)^{j-1-n}
 \bigg(\sum_{z,x}P'_{\Lambda;o}(z,x)\bigg)\nn\\
&\times\bigg(\sup_{y,z}\sum_{\substack{b,b',v\\ \bb=y}}\Big(|\bb'
 |^2\ind{n\text{ odd}}+|v-\bb|^2\ind{n\text{ even}}\Big)\,\tau_b
 Q''_{\Lambda;o,v}(\tb,\bb')\,\tau_{b'}\big(\delta_{\tbps,v+z}
 +\tilde G_\Lambda(\tbp,v+z)\big)\bigg),
\end{align}
where the first line is $O(\theta_0)^{j-2}$.  The leading contribution
to the second line from $P_{\Lambda;o,v}^{\prime\prime{\sss(0)}}$ and
$P_{\Lambda;o}^{\prime{\sss(0)}}$ in $Q''_{\Lambda;o,v}$ for an odd $n$
is bounded, due to translation invariance, by
\begin{align}\lbeq{2nddec-bd:0<n<jbd}
&\bar G^{\sss(2)}\sup_{y,z}\Bigg(~\raisebox{-1.4pc}{\includegraphics
 [scale=0.14]{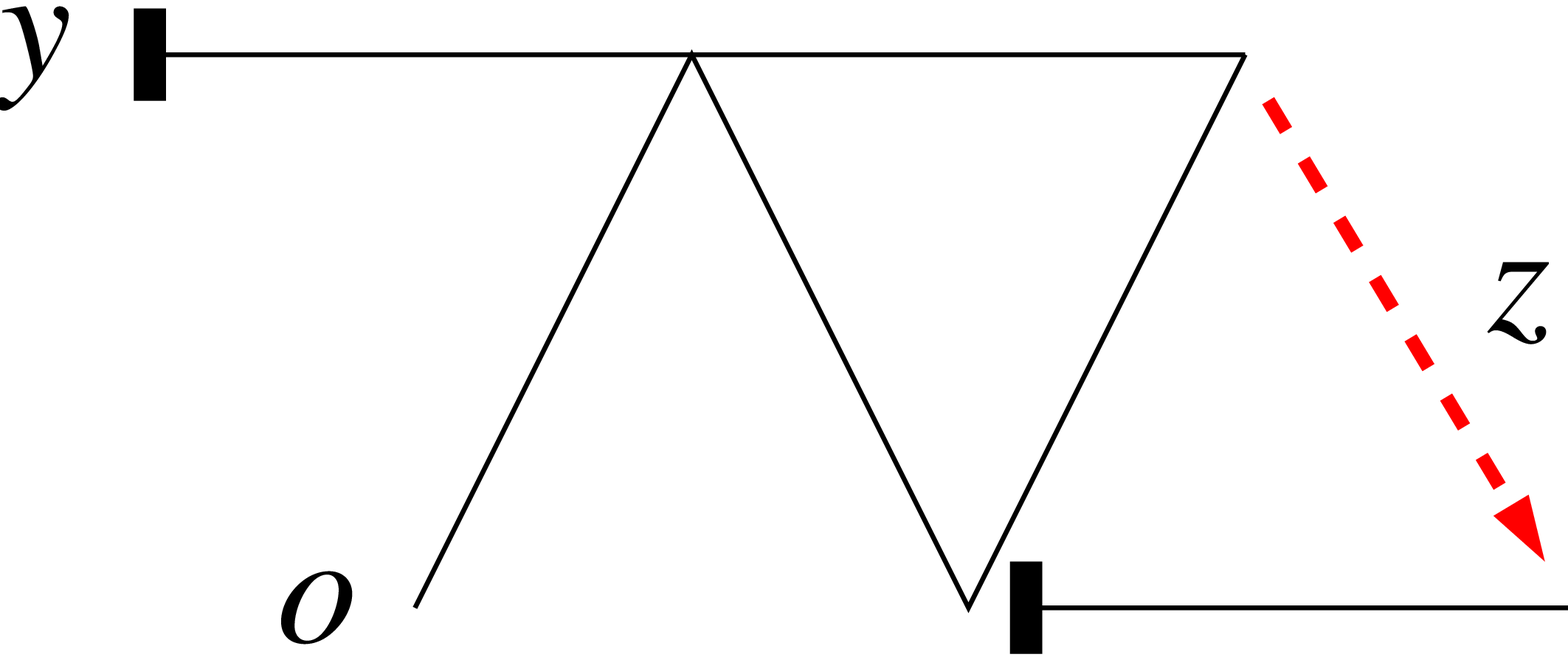}}~+\raisebox{-1.4pc}{\includegraphics
 [scale=0.14]{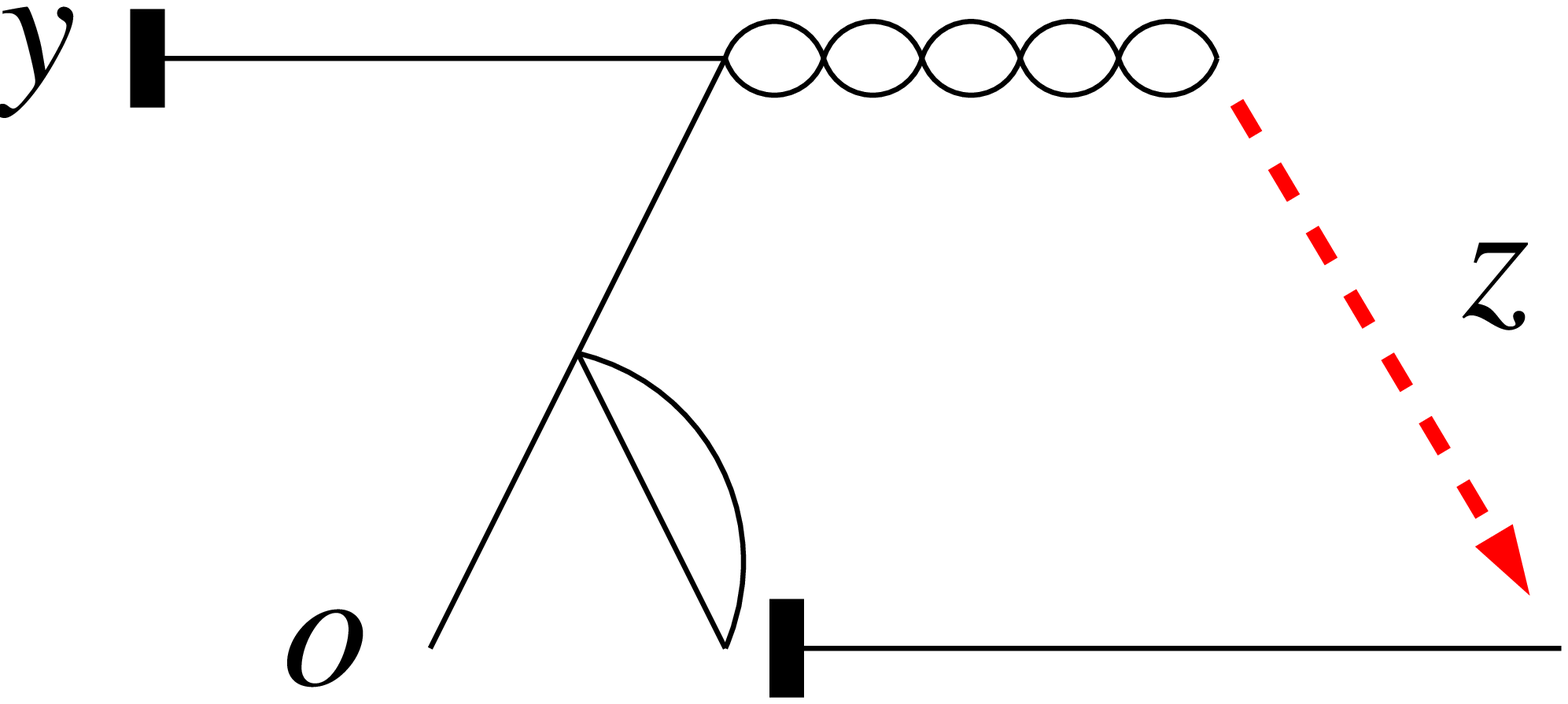}}~\Bigg)\nn\\
&\leq d\sigma^2O(\theta_0)^2\sup_z\Bigg(~\raisebox{-1.4pc}
 {\includegraphics[scale=0.14]{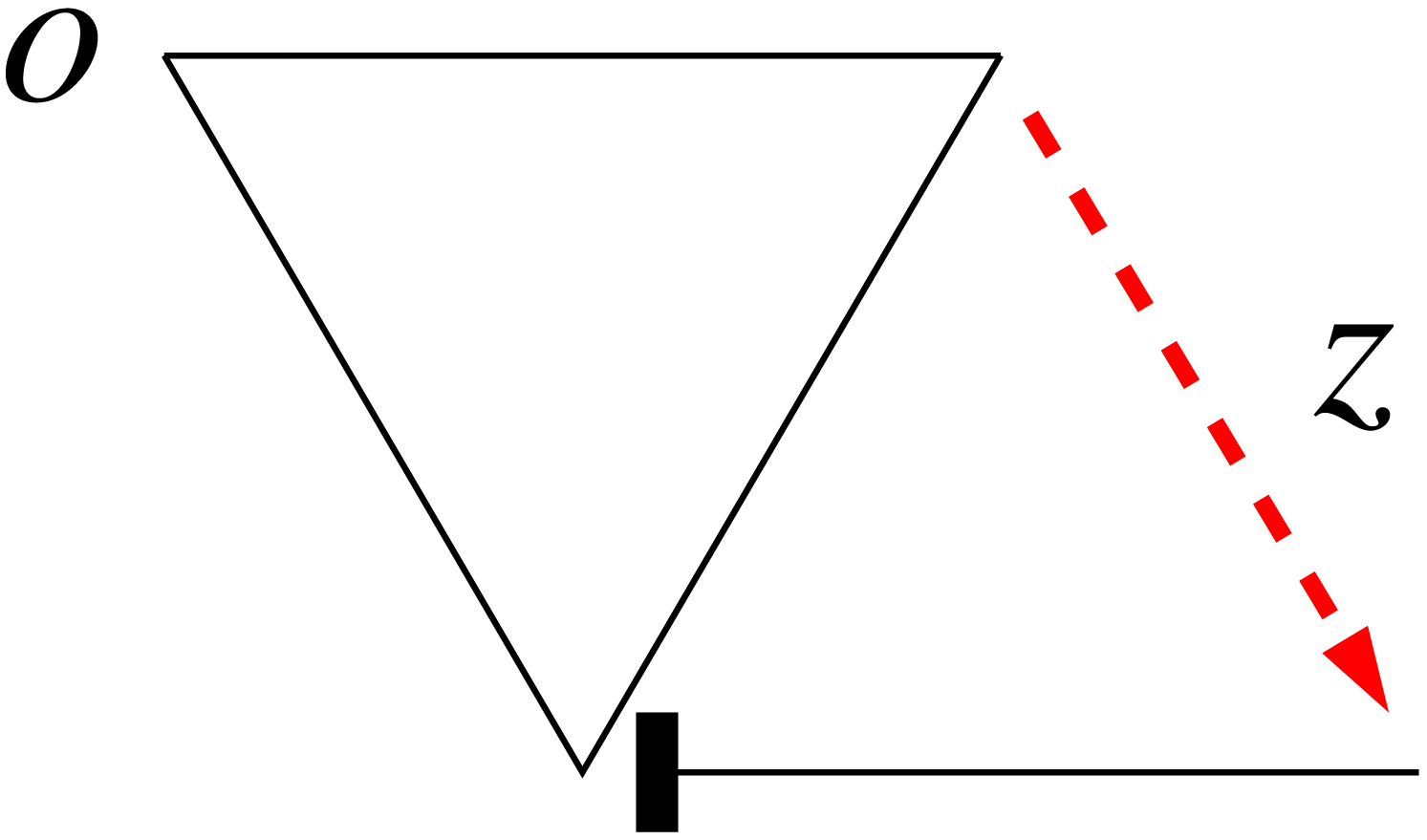}}~+~\raisebox{-1.4pc}
 {\includegraphics[scale=0.14]{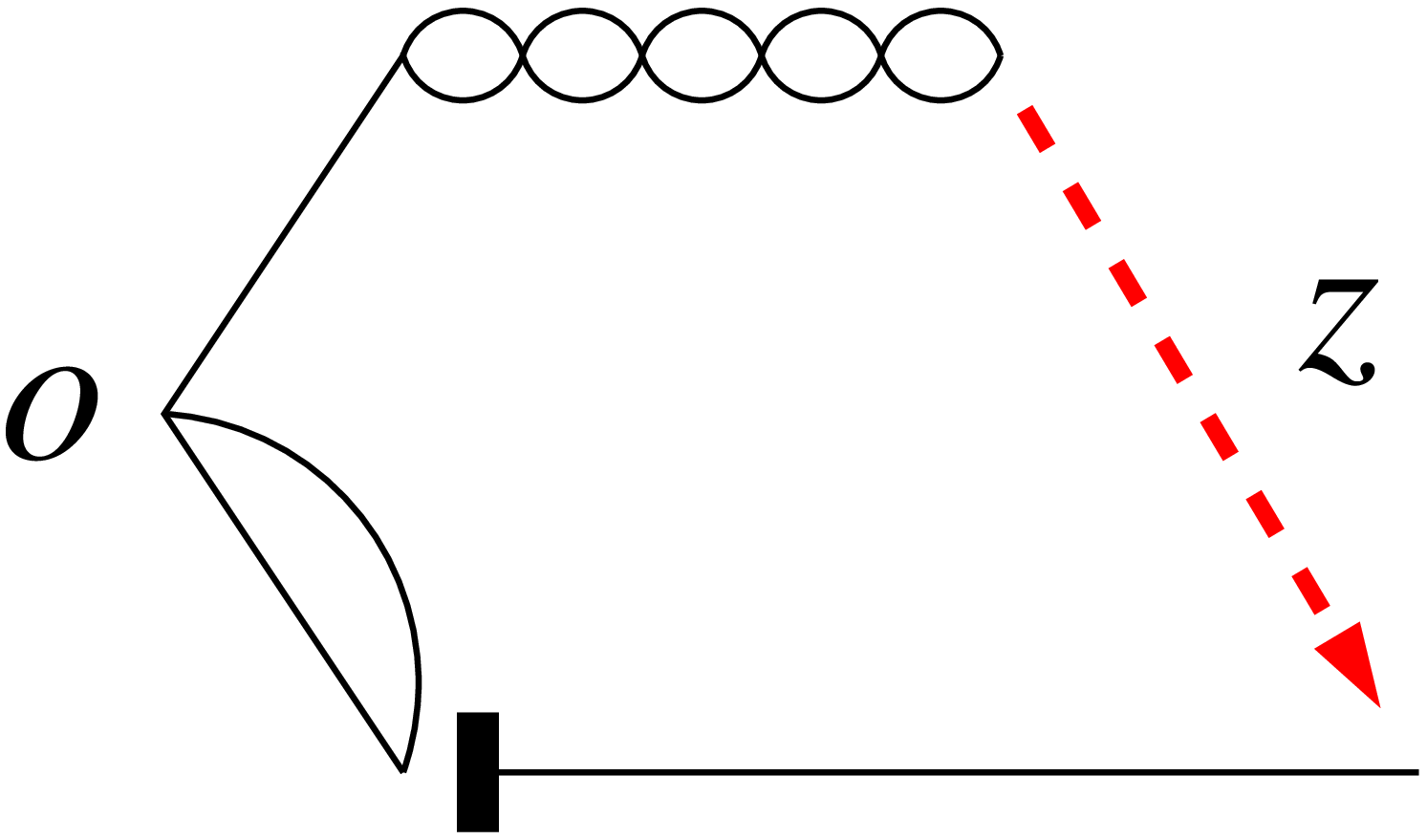}}~\Bigg)\leq d\sigma^2
 O(\theta_0)^3.
\end{align}
The other contributions from $P_{\Lambda;o,v}^{\prime\prime{\sss(i)}}$
and $P_{\Lambda;o}^{\prime{\sss(i)}}$ for $i\ge1$ and from the
even-$n$ case can be estimated similarly; if $n$ is even, then the
second supremum in \refeq{2nddec-bd:0<n<jbd} is $O(\theta_0)$.
Therefore, \refeq{2nddec-bd:0<n<j} is
$d\sigma^2O(\theta_0)^{2\lfloor{\sss\frac{j+1}2}\rfloor}$.

\smallskip

Summarizing the above (i)--(iii) and using
$2\lfloor{\sss\frac{j+1}2}\rfloor\geq j\vee2$ for $j\ge1$, we have
\begin{align}
\frac1{j+1}\sum_x|x|^2\pi_\Lambda^{\sss(j)}(x)\leq d\sigma^2\Big(
 jO(\theta_0)^{2\lfloor{\sss\frac{j+1}2}\rfloor}+O(\theta_0)^{j+1}
 \Big)\leq d\sigma^2(j+1)\,O(\theta_0)^{j\vee2}.
\end{align}
This together with \refeq{pi0-rthmombd} complete the proof of
\refeq{pi-sumbd}.
\end{proof}

\begin{proof}[Proof of Proposition~\ref{prp:exp-bootstrap}(iii)
assuming \refeq{IR-kbd}]
It is easy to see that
\begin{align}\lbeq{pi0-t+2ndmombd}
\sum_x|x|^{t+2}\pi_\Lambda^{\sss(0)}(x)\leq\sum_x|x|^{t+2}G(x)^3\leq
 \bar G^{\sss(2)}\sum_x|x|^tG(x)^2\leq d\sigma^2\theta_0\bar W^{\sss
 (t)}.
\end{align}
We show below that, for $j\ge1$,
\begin{align}\lbeq{pij-t+2ndmombd}
\sum_x|x|^{t+2}\pi_\Lambda^{\sss(j)}(x)\leq d\sigma^2\bar W^{\sss(t)}
 (j+1)^{t+3}O(\theta_0)^{j\vee2-1},
\end{align}
where the bound is independent of $\Lambda$.  Due to these uniform
bounds, we conclude that the sum of $|x|^{t+2}|\Pi(x)|$ is finite if
$\theta_0\ll1$.

Now we explain the main idea of the proof of \refeq{pij-t+2ndmombd}.
First we recall that, in the proof of the bound on
$\sum_x|x|^2\pi_\Lambda^{\sss(j)}(x)$, we distribute
$|x|^2$ along the lowermost path of each bounding diagram.  To bound
$\sum_x|x|^{t+2}\pi_\Lambda^{\sss(j)}(x)$, we again use the lowermost
path in the same way to distribute $|x|^2$, and use the uppermost path
to distribute the remaining $|x|^t$.  More precisely, we use
\begin{align}\lbeq{|x|-max}
|x|\leq(j+1)\max_{n=0,1,\dots,j}|a'_n|,
\end{align}
where $a'_0,a'_1,\dots,a'_j$ are the displacements along the uppermost
path: $a'_0=\bb_1$, $a'_1=v_2-\bb_1$, $a'_2=\bb_3-v_2,\dots$, and
$a'_j=x-v_j$ or $x-\bb_j$ depending on the parity of $j$.  Let $m$ be
such that $|a'_m|=\max_n|a'_n|$.

For the contribution to $\sum_x|x|^{t+2}\pi_\Lambda^{\sss(j)}(x)$ from
$|a_n|^2$ in \refeq{x2-bd} for $n\ne m$, we simply follow the same
strategy as explained above in the paragraphs~(i)--(iii) to prove
the bound on $\sum_x|x|^2\pi_\Lambda^{\sss(j)}(x)$.  The only
difference is that one of the bubbles $\bar W^{\sss(0)}$ contained in
the bound on the $m^\text{th}$ block is now replaced by $\bar W^{\sss(t)}$.

The contribution from $|a_m|^2$ in \refeq{x2-bd} can be estimated
in a similar way, except for a few complicated cases, due to
$P_{\Lambda;u}^{\prime{\sss(i)}}$ and
$P_{\Lambda;u,v}^{\prime\prime{\sss(i)}}$ for $i\ge1$ contained
in the $m^\text{th}$ block.  For example, let $j$ be even and
let $m=j$ (cf., the second line of \refeq{2nddec-bd:n=j}).
The following are two possibile diagrams in the contribution
from $P_{\Lambda;o}^{\prime{\sss(4)}}(f,x)$ to
$\sum_{z,x}|x-y|^2|x|^t\tau_{y,z}Q'_{\Lambda;o}(z,x)$:
\begin{align}\lbeq{IRSchwarz}
\text{(i)}\quad\raisebox{-1.5pc}{\includegraphics[scale=0.14]
 {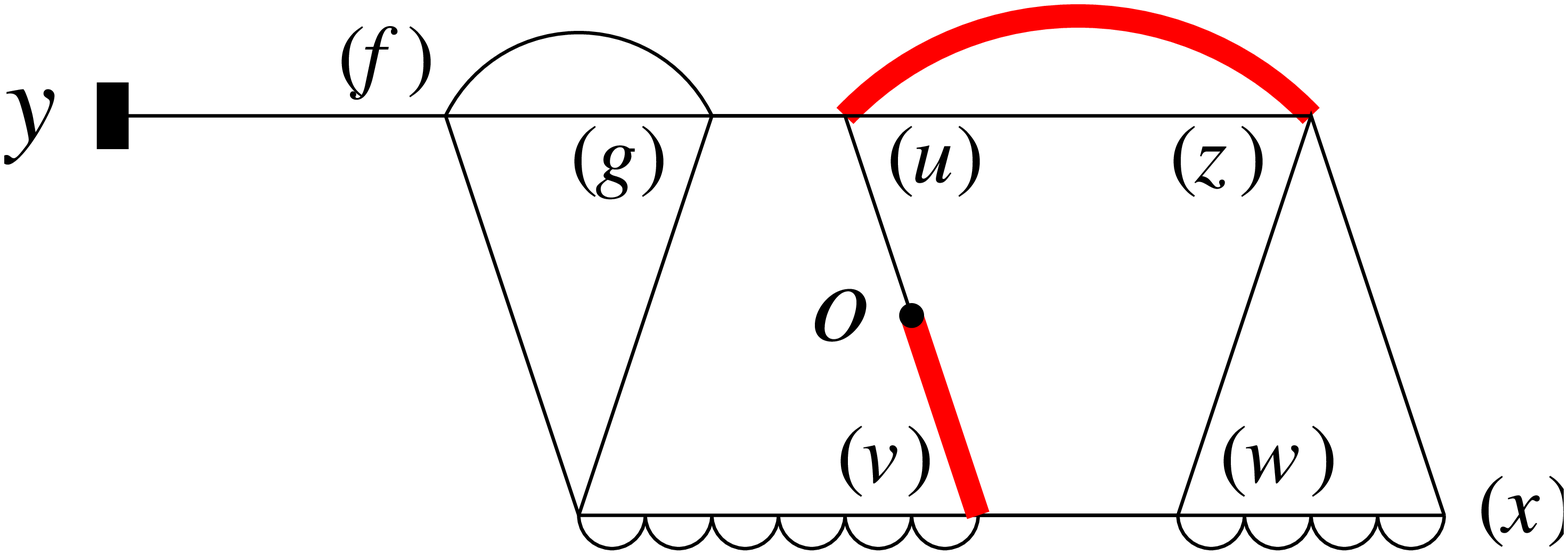}}\hspace{5pc}
\text{(ii)}\quad\raisebox{-1.5pc}{\includegraphics[scale=0.14]
 {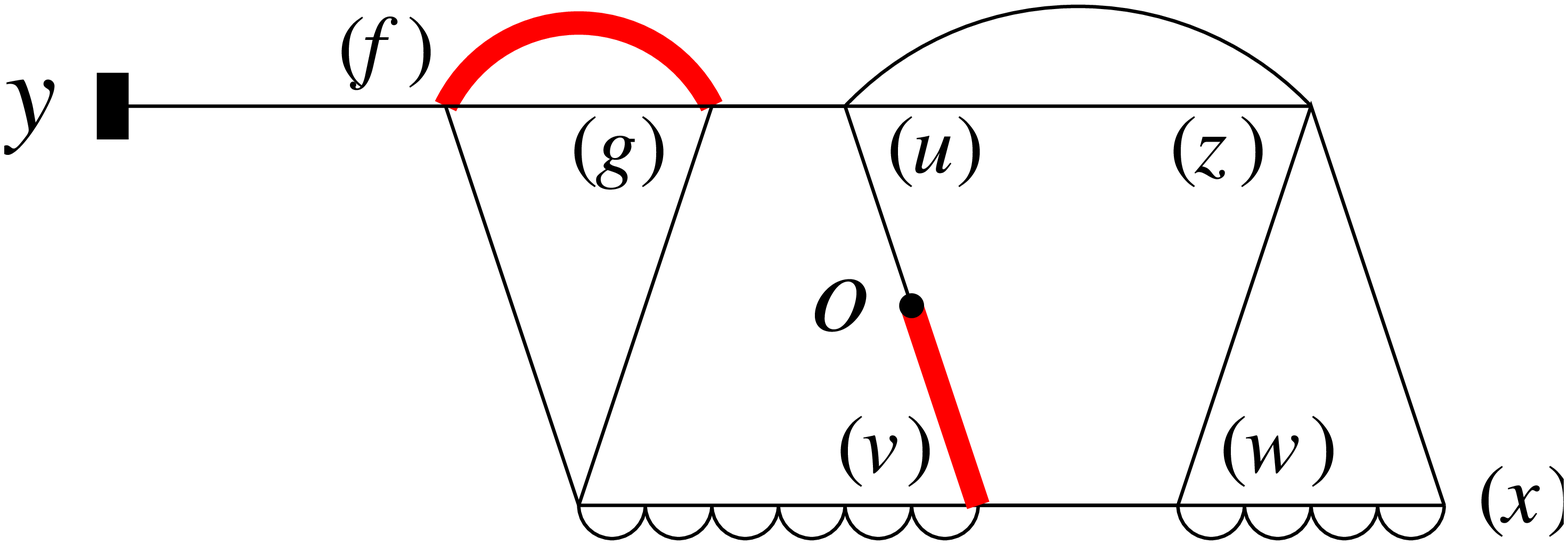}}
\end{align}
where, for simplicity, $\psi_\Lambda(f,g)-\delta_{f,g}$ and
$\psi_\Lambda(u,z)-\delta_{u,z}$ are reduced to $\tilde
G_\Lambda(f,g)^2$ and $\tilde G_\Lambda(f,g)^2$, respectively. We
suppose that $|v|$ is bigger than $|w-v|$ and $|x-w|$ along the
lowermost path from $o$ to $x$ through $v$ and $w$, so that $|x|^t$
is bounded by $3^t|v|^t$.  We also suppose that $|z-u|$ in
(\ref{eq:IRSchwarz}.i) (resp., $|g-f|$ in (\ref{eq:IRSchwarz}.ii))
is bigger than the end-to-end distance of any of the other four
segments along the uppermost path from $y$ to $x$ through $f,g,u$
and $z$. Therefore, we can bound $|x-y|^2$ by $5^2|z-u|^2$ in
(\ref{eq:IRSchwarz}.i) (resp., $5^2|g-f|^2$ in
(\ref{eq:IRSchwarz}.ii)) and bound the weighted arc between $u$ and
$z$ (resp., between $f$ and $g$) by $5^2\bar G^{\sss(2)}$.  By
translation invariance, the remaining diagram of
(\ref{eq:IRSchwarz}.i) is easily bounded as
\begin{align}\lbeq{IRSchwarz-bdi}
\sum_{f',g,u',v}\!\raisebox{-1.4pc}{\includegraphics[scale=0.14]
 {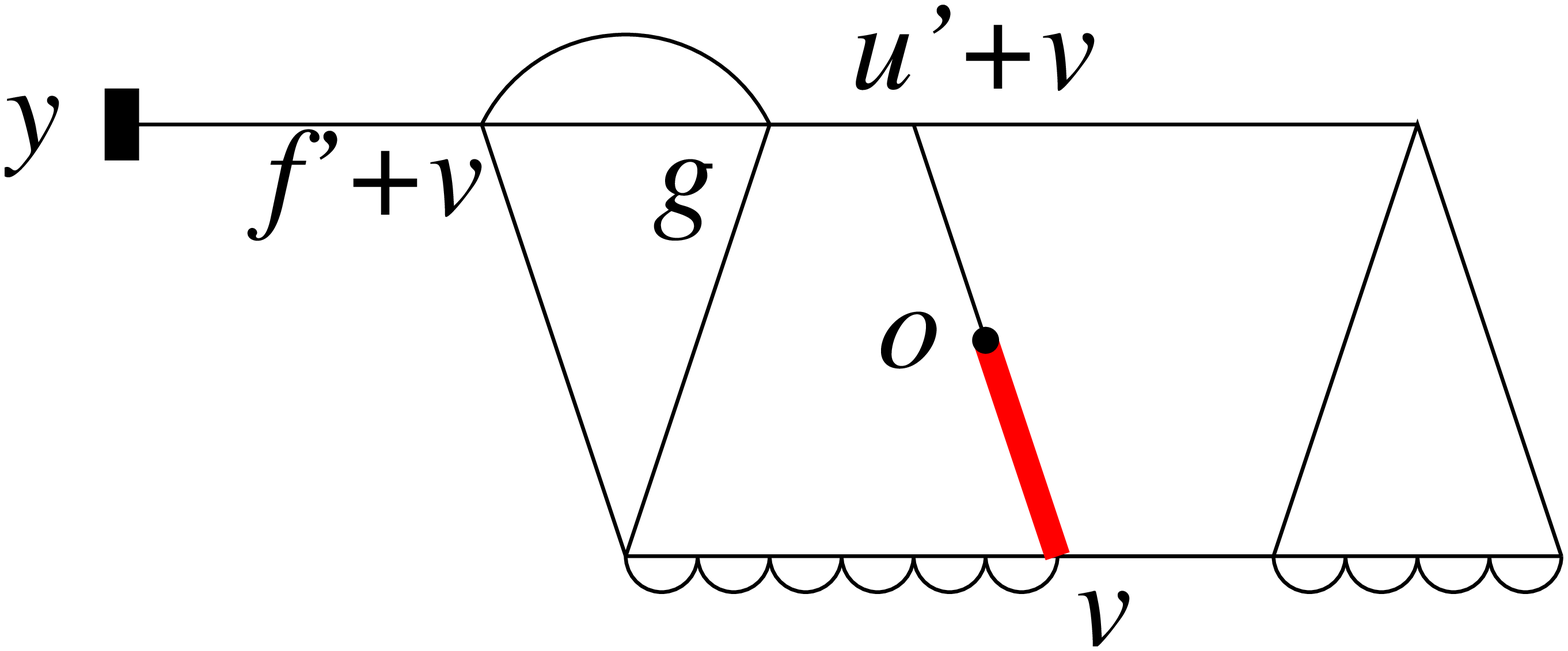}}\;=\sup_{f',g,u'}\,\raisebox{-1.4pc}
 {\includegraphics[scale=0.14]{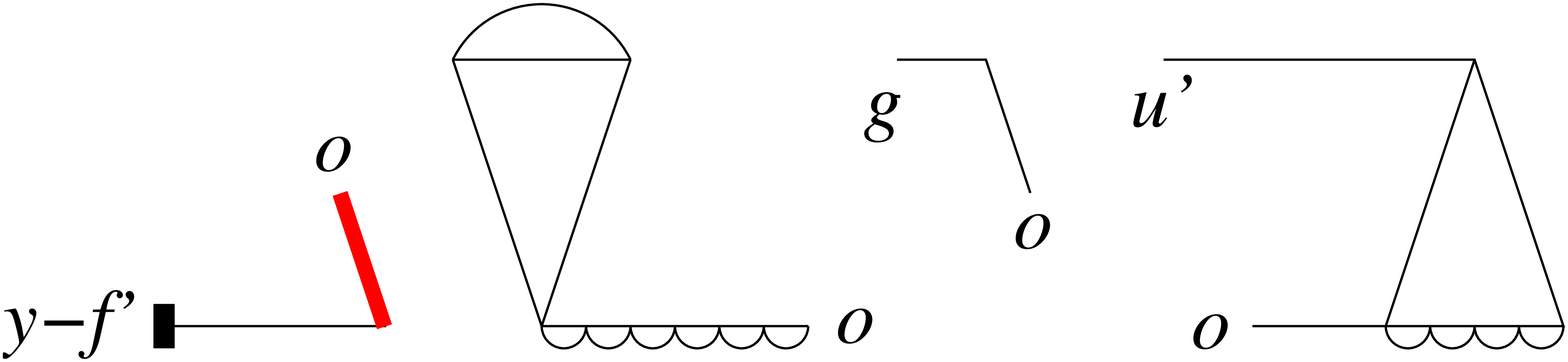}}\leq\bar
 W^{\sss(t)}\,O(\theta_0)^4,
\end{align}
where the power 4 (not 3) is due to the fact that the segment from $u'$
in the last block is nonzero.

To bound the remaining diagram of (\ref{eq:IRSchwarz}.ii) is a little
trickier.  We note that at least one of $|u|,|z-u|,|w-z|$ and $|v-w|$
along the path from $o$ to $v$ through $u,z,w$ is bigger than
$\frac14|v|$.  Suppose $|v-w|\ge\frac14|v|$, so that
$|v|^t\leq2^t|v-w|^{t/2}|v|^{t/2}$.  Then, by using the Schwarz
inequality, we obtain
\begin{align}\lbeq{IRSchwarz-bdii}
\raisebox{-1.3pc}{\includegraphics[scale=0.13]{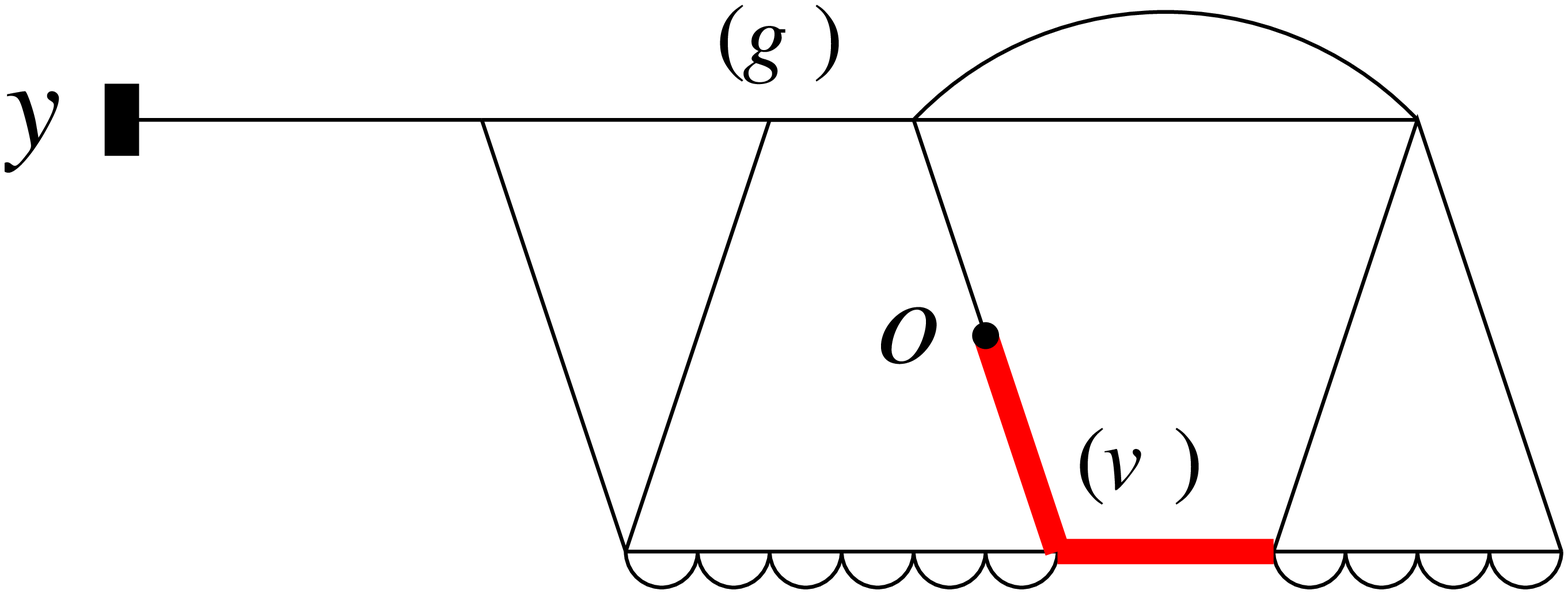}}~~\leq~~
 \Bigg(~~\raisebox{-2pc}{\includegraphics[scale=0.12]{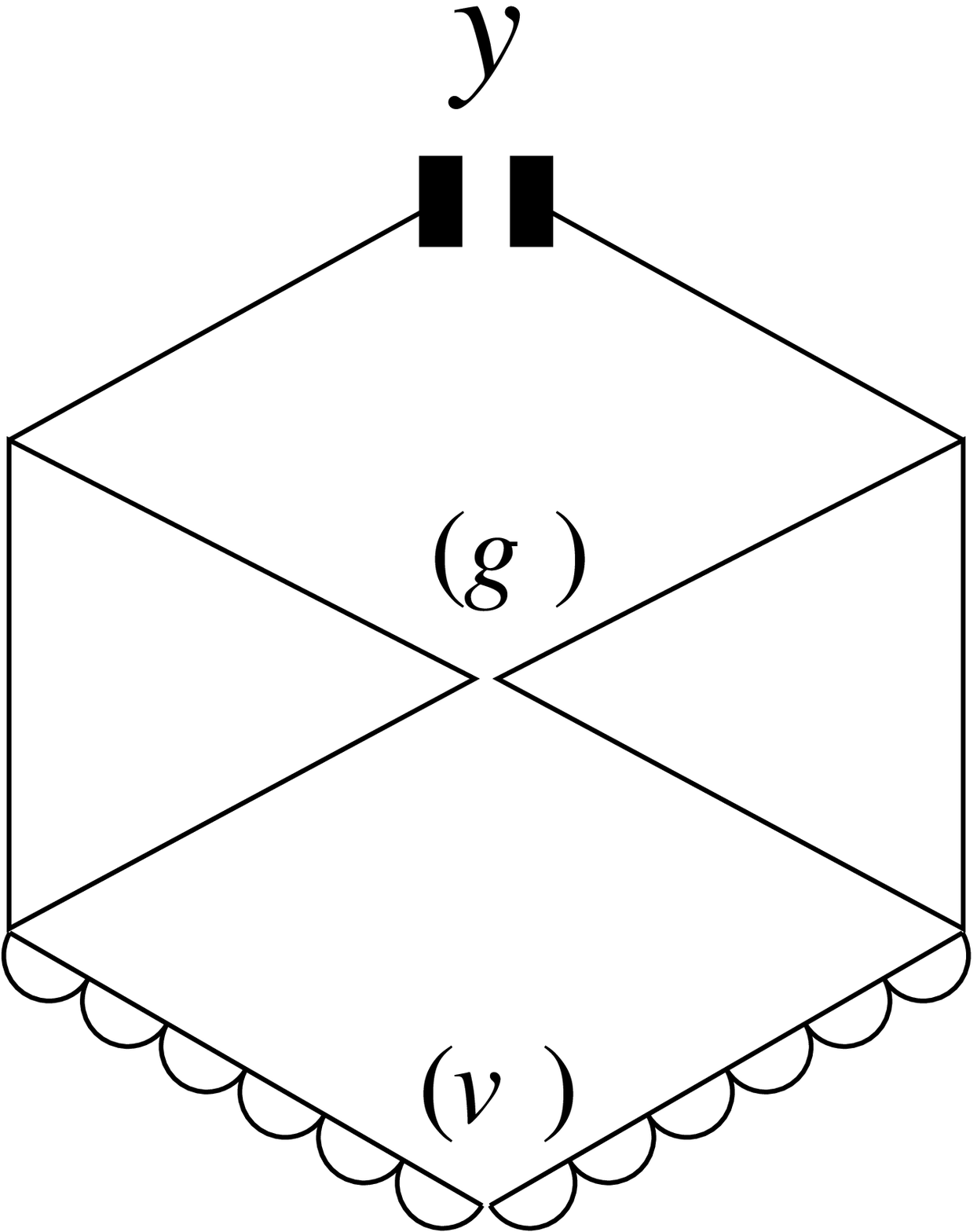}}
 ~~\Bigg)^{1/2}~\Bigg(~~\raisebox{-1.5pc}{\includegraphics[scale=0.12]
 {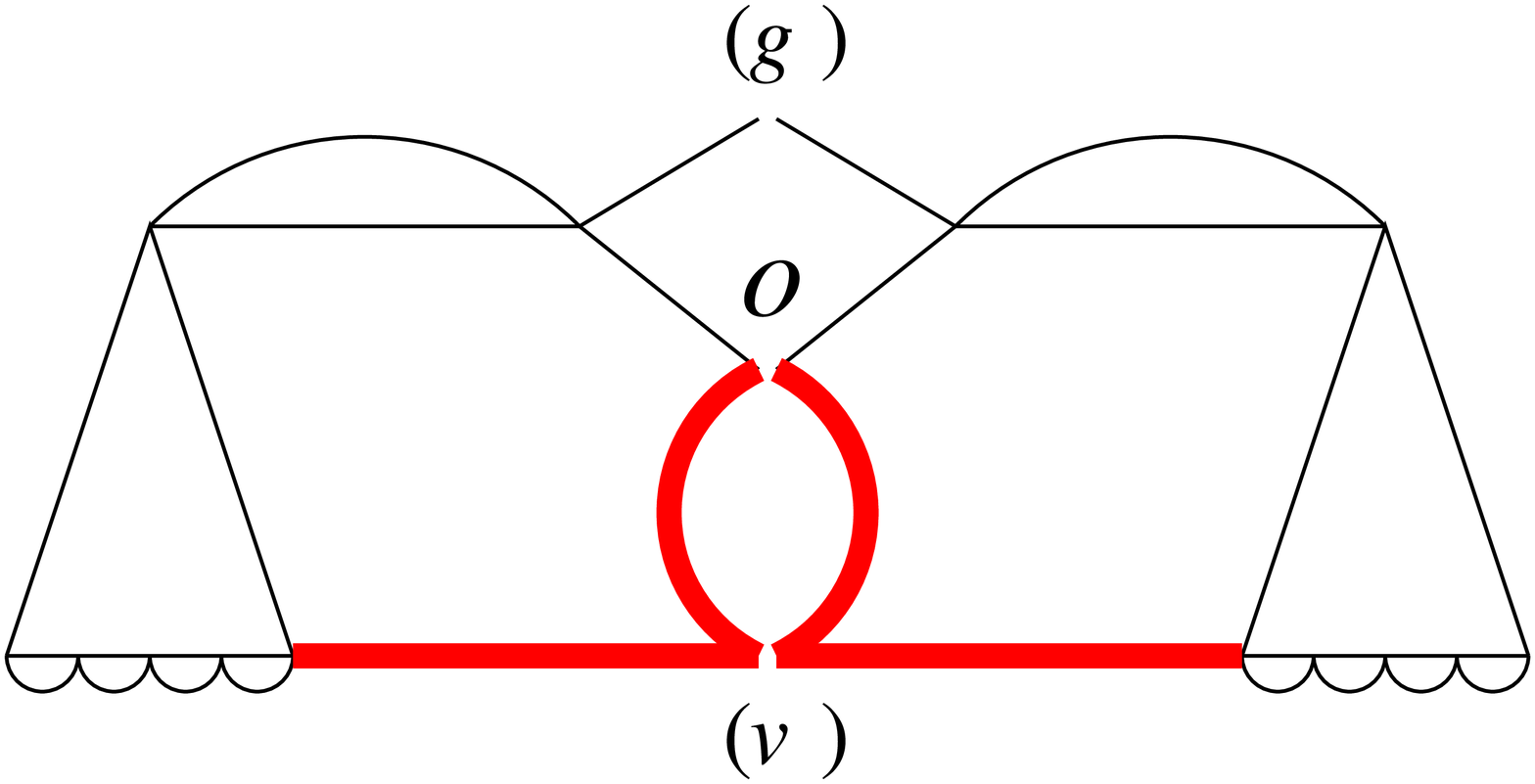}}~~\Bigg)^{1/2},
\end{align}
where the two weighted arcs between $o$ and $v$ in the second term is
$|v|^tG(v)^2\equiv(|v|^{t/2}G(v))^2$.  By translation invariance and
the fact that the north-east and north-west segments from $g$ in the
first term are nonzero, we obtain
\begin{align}
\raisebox{-2pc}{\includegraphics[scale=0.13]{IRSchwarzdec2}}~
&\leq\bigg(\sup_z\raisebox{-0.9pc}{\includegraphics[scale=0.12]
 {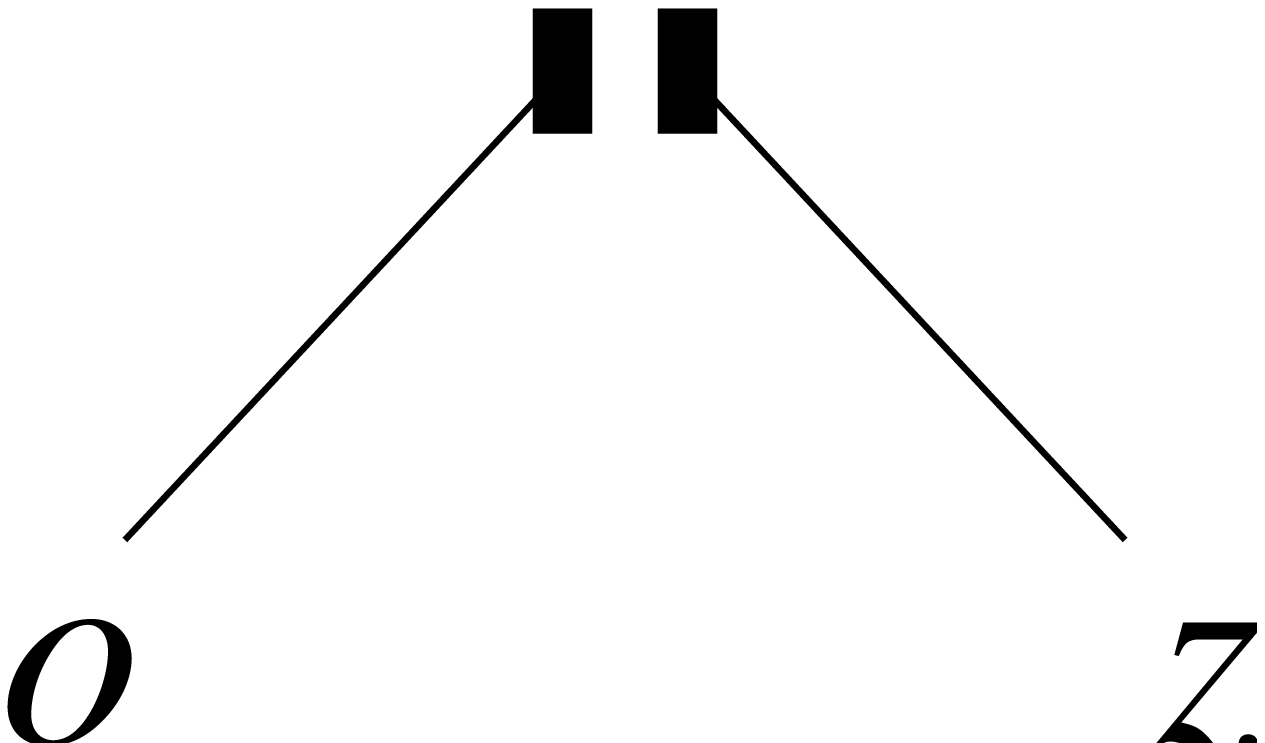}}~\bigg)\Big(\sup_{g'}\tau(D*G^{*2})(g')\Big)^2\bar
 W^{\sss(0)}\bigg(\sum_v\big(\psi_\lambda(o,v)-\delta_{o,v}\big)
 \bigg)^2\nn\\
&\leq O(\theta_0)^5.
\end{align}
With the help of
$(\bar W^{\sss(t/2)})^2\leq\bar W^{\sss(0)}\bar W^{\sss(t)}$
(due to the Schwarz inequality), we also obtain
\begin{align}
\raisebox{-1.8pc}{\includegraphics[scale=0.13]{IRSchwarzdec3}}~~\leq
 \,\bar W^{\sss(0)}\,\bar W^{\sss(t)}\Bigg(\sup_v\raisebox{-1.1pc}
 {\includegraphics[scale=0.13]{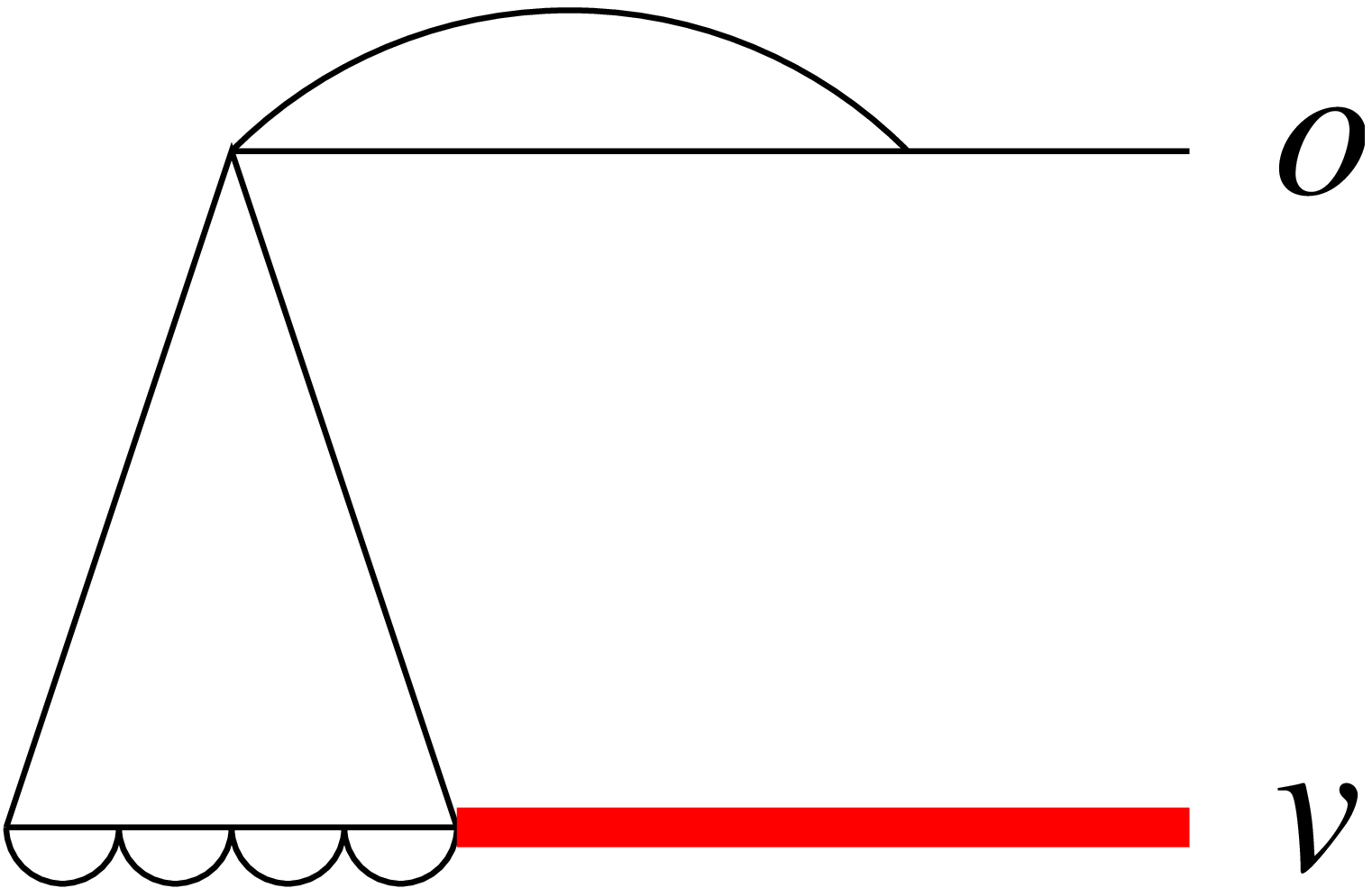}}~\Bigg)^2\leq\big(\bar
 W^{\sss(t)}\big)^2O(\theta_0)^4.
\end{align}
Therefore, \refeq{IRSchwarz-bdii} is bounded by
$\bar W^{\sss(t)}O(\theta_0)^{9/2}$.

The other cases can be estimated similarly \cite{sNN}.  As a result,
we obtain
\begin{align}
\sum_x|x|^{t+2}\pi_\Lambda^{\sss(j)}(x)\leq\sum_{m=0}^jd\sigma^2
 \bar W^{\sss(t)}(j+1)^{t+2}O(\theta_0)^{j\vee2-1},
\end{align}
which implies \refeq{pij-t+2ndmombd}.  This completes the proof of
Proposition~\ref{prp:exp-bootstrap}(iii).
\end{proof}

\begin{proof}[Proof of \refeq{pi-kbd} assuming \refeq{IR-kbd} and
\refeq{IR-xbdNN}] If $x=o$, then we simply use the bound on the sum
in \refeq{pi-sumbd} to obtain $\pi_\Lambda^{\sss(i)}(o)\leq
O(\theta_0)^i$ for any $i\ge0$. It is also easy to see that
$\pi_\Lambda^{\sss(0)}(x)$ with $x\ne o$ obeys \refeq{pi-kbd}, due
to \refeq{IR-xbdNN} and the diagrammatic bound \refeq{piNbd}.  It
thus remains to show \refeq{pi-kbd} for $\pi_\Lambda^{\sss(j)}(x)$
with $x\ne o$ and $j\ge1$.

The idea of the proof is somewhat similar to that of
Proposition~\ref{prp:exp-bootstrap}(iii) explained above.  First,
we take $|a_m|\equiv\max_n|a_n|$ from the lowermost path and
$|a'_l|\equiv\max_n|a'_n|$ from the uppermost path of a bounding
diagram.  Note that, by \refeq{|x|-max}, $|a_m|$ and $|a'_l|$ are
both bigger than $\frac1{j+1}|x|$.  That is, $|a_m|^{-q}$ and
$|a'_l|^{-q}$ are both bounded from above by $(j+1)^q|x|^{-q}$.  If
the path corresponding to $a_m$ in the $m^\text{th}$ block consists
of $N$ segments, we take the ``longest'' segment whose end-to-end
distance is therefore bigger than $\frac1{N(j+1)}|x|$.  That is,
the corresponding two-point function is bounded by $\lambda_0
N^q(j+1)^q|x|^{-q}$.  Here, $N$ depends on the parity of $m$,
as well as on $i\ge0$ for $P_{\Lambda;u,v}^{\prime\prime{\sss(i)}}$
(or $P_{\Lambda;u}^{\prime{\sss(i)}}$ if $m=0$ or $j$) and the
location of $u,v$ in each diagram, and is at most $N\leq O(i+1)$.
However, the number of nonzero chains of bubbles contained in each
diagram of $P_{\Lambda;u}^{\prime{\sss(i)}}$ and
$P_{\Lambda;u,v}^{\prime\prime{\sss(i)}}$ is $O(i)$, and
hence their contribution would be $O(\theta_0)^{O(i)}$.
This compensates the growing factor of $N^q$,
and therefore we will not have to take the effect of $N$ seriously.
The same is true for $a'_l$, and we refrain from repeating the same
argument.

Next, we take the ``longest'' segment, denoted $a''$, among those
which together with $a'_l$ (or a part of it) form a ``loop''; a
similar observation was used to obtain \refeq{IRSchwarz-bdii}. The
loop consists of segments contained in the $l^\text{th}$ block and
possibly in the $(l-1)^\text{st}$ block, and hence the number of
choices for $a''$ is at most $O(i_{l-1}+i_l+1)$, where $i_l$ is the
index of $P_{\Lambda}^{\prime{\sss(i_l)}}$ or
$P_{\Lambda}^{\prime\prime{\sss(i_l)}}$ in the $l^\text{th}$ block
($i_{-1}=0$ by convention).  By \refeq{|x|-max}, we have $|a''|\ge
O(i_{l-1}+i_l+1)^{-1}|a'_l|$, and the corresponding two-point
function is bounded by $\lambda_0O(i_{l-1}+i_l+1)^q
(j+1)^q|x|^{-q}$.  As explained above, the effect of
$O(i_{l-1}+i_l+1)^q$ would not be significant after summing over
$i_{l-1}$ and $i_l$.

We have explained how to extract three ``long'' segments
from each bounding diagram, which provide the factor
$\lambda_0^3(j+1)^{3q}|x|^{-3q}$ in \refeq{pi-kbd}; the extra
factor of $(j+1)^2$ in \refeq{pi-kbd} is due to the number of
choices of $m,l\in\{0,1,\dots,j\}$.  Therefore, the remaining
task is to control the rest of the diagram.

Suppose, for example, $0<m<l<j$ (so that $j\ge3$).  Using
$\tilde Q''_\Lambda$ defined in \refeq{tildeQ''-def},
we can reorganize the diagrammatic bound \refeq{piNbd} on
$\pi_\Lambda^{\sss(j)}(x)$ as (cf., \refeq{2nddec-bd:0<n<j})
\begin{align}\lbeq{diagbd-reorg}
\pi_\Lambda^{\sss(j)}(x)&\leq\sum_{\substack{b_m,v_m\\ y_{l+1},
 v_{l+1}}}\bigg(\sum_{\substack{b_1,\dots,b_{m-1}\\ v_1,\dots,
 v_{m-1}}}P_{\Lambda;v_1}^{\prime{\sss(0)}}(o,\bb_1)\prod_{i=
 1}^{m-1}\tau_{b_i}Q''_{\Lambda;v_i,v_{i+1}}(\tb_i,\bb_{i+1})
 \bigg)\nn\\
&\qquad\times\sum_{b_{l+1}}\bigg(\sum_{\substack{b_{m+1},\dots,
 b_l\\ v_{m+1},\dots,v_l}}\prod_{i=m}^l\tau_{b_i}Q''_{\Lambda;
 v_i,v_{i+1}}(\tb_i,\bb_{i+1})\bigg)\tau_{b_{l+1}}\big(\delta_{
 \tb_{l+1},y_{l+1}}+\tilde G_\Lambda(\tb_{l+1},y_{l+1})\big)\nn\\
&\qquad\times\Bigg(\sum_{\substack{y_{l+2},\dots,y_j\\ v_{l+2},
 \dots,v_j}}\bigg(\prod_{i=l+1}^{j-1}\tilde Q''_{\Lambda;v_i,
 v_{i+1}}(y_i,y_{i+1})\bigg)P'_{\Lambda;v_j}(y_j,x)\Bigg).
\end{align}
As explained above, we bound three ``long'' two-point functions
contained in the second line of \refeq{diagbd-reorg}; let $Y_{m,l}$
be the supremum of what remains in the second line over
$b_m,v_m,y_{l+1},v_{l+1}$.  Then we can perform the sum of the first
line over $b_m,v_m$ and the sum of the third line over
$y_{l+1},v_{l+1}$ independently; the former is $O(\theta_0)^{m-1}$
and the latter is $O(\theta_0)^{j-1-l}$, due to \refeq{block-sumbd}
and \refeq{tildeQ''-bd}, respectively. Finally, we can bound
$Y_{m,l}$ using the Schwarz inequality by $O(\theta_0)^{l-m}$, where
$l-m$ is the number of nonzero segments in the second line of
\refeq{diagbd-reorg} (i.e.,
$\sum_{b_i}\tau_{b_i}(\delta_{\tb_i,y_i}+\tilde
G_\Lambda(\tb_i,y_i))$ for some $y_m,\dots,y_{l+1}$) minus 2 (= the
maximum number of those along the uppermost and lowermost paths that
are extracted to obtain the aformentioned $|x|$-decaying term).  For
example, one of the leading contributions to $Y_{m,m+4}$ is bounded,
by using translation invariance and the Schwarz inequality, as
\begin{align}
\sup_{u,v,y}\raisebox{-1pc}{\includegraphics[scale=0.14]{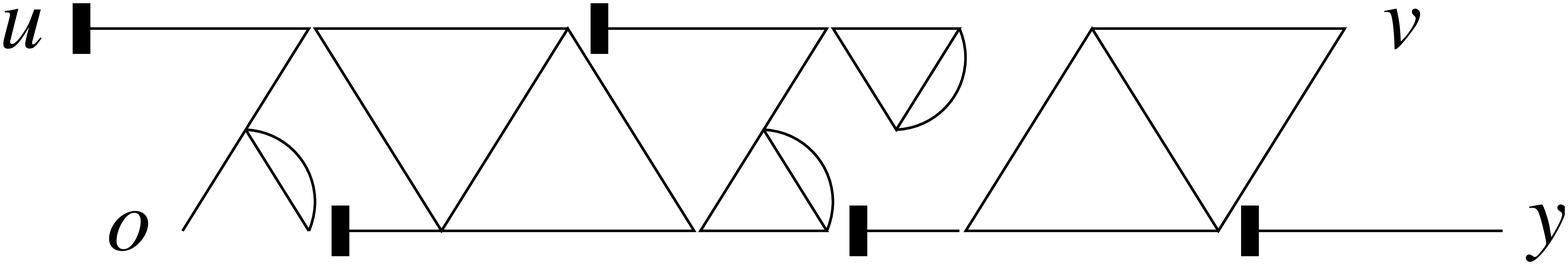}}~\leq
 O(\theta_0)~\sup_{u,z}\raisebox{-1pc}{\includegraphics[scale=0.14]
 {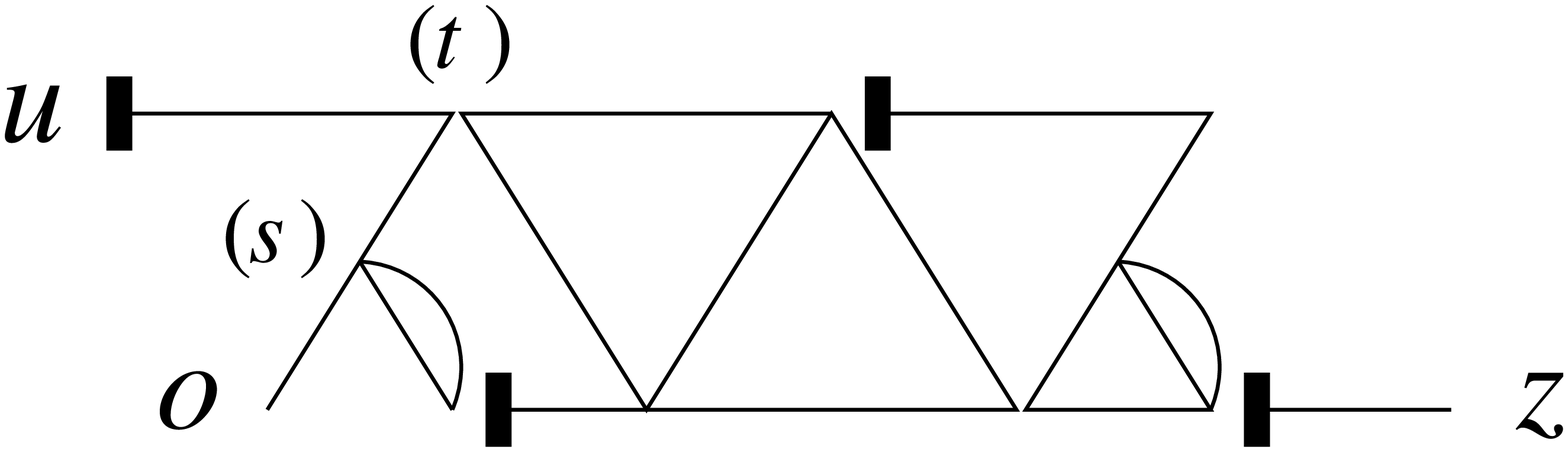}}\\
\leq O(\theta_0)^{3/2}\left(\raisebox{-1.9pc}{\includegraphics[scale
 =0.14]{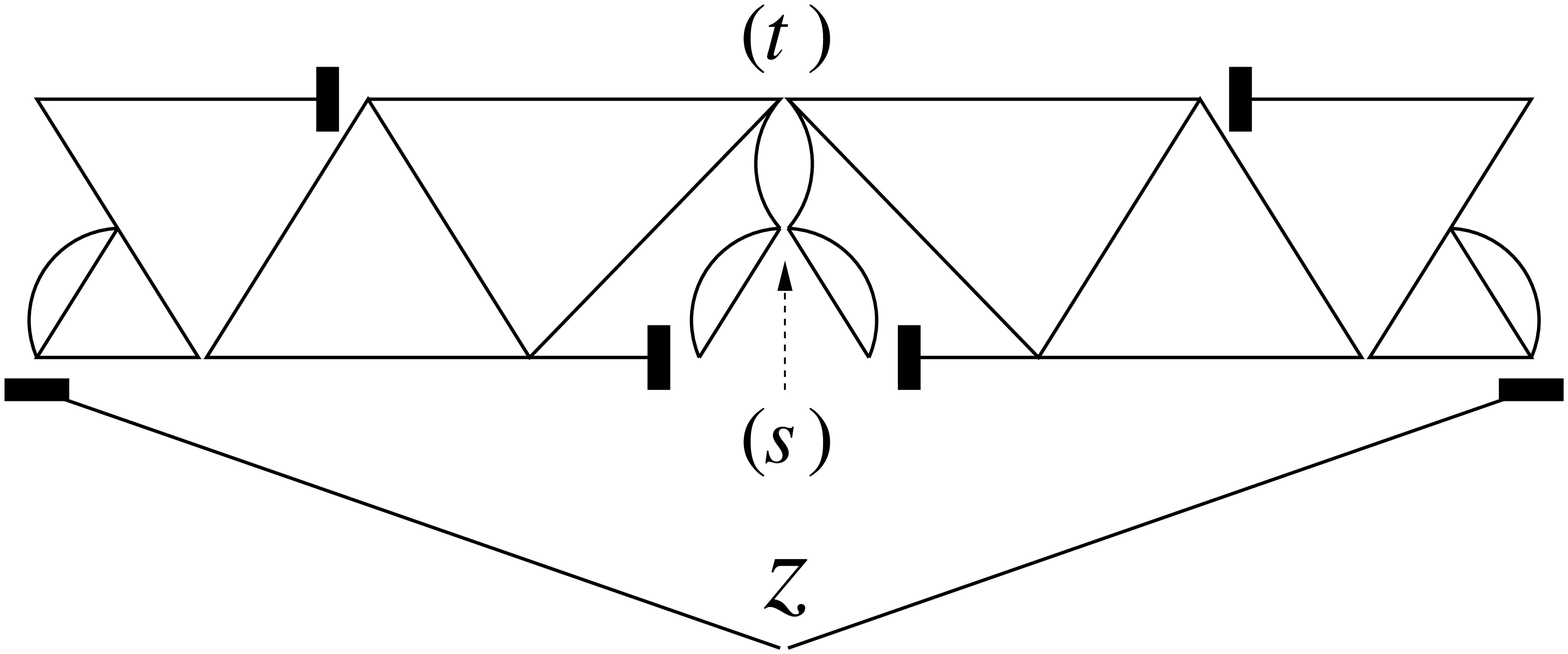}}\right)^{1/2}\leq O(\theta_0)^2~\sup_{s'}\raisebox{-1pc}
 {\includegraphics[scale=0.14]{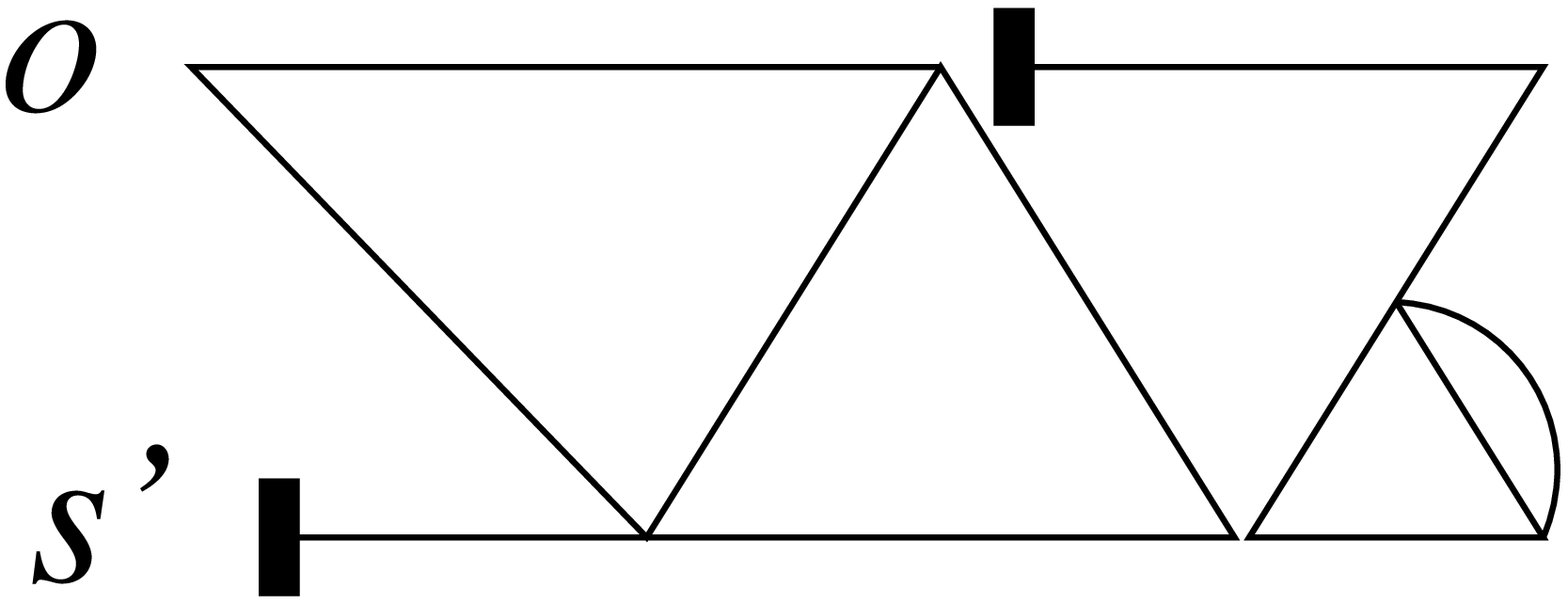}}~&\leq O(\theta_0)^4.\nn
\end{align}

The other cases can be estimated similarly \cite{sNN}.  This
completes the proof of \refeq{pi-kbd}.
\end{proof}

\section*{Acknowledgements}
First of all, I am grateful to Masao~Ohno for having drawn my
attention to the subject of this paper.  I would like to thank
Takashi~Hara for stimulating discussions and his hospitality during
my visit to Kyushu University in December~2004 and April~2005.  I
would also like to thank Aernout~van~Enter for useful discussions on
reflection positivity.  Special thanks go to Mark~Holmes and
John~Imbrie for continual encouragement and valuable comments to the
former versions of the manuscript, and Remco~van~der~Hofstad for his
constant support in various aspects.  This work was supported in
part by the Postdoctoral Fellowship of EURANDOM, and in part by the
Netherlands Organization for Scientific Research (NWO).

\end{document}